\documentclass[twoside,12pt]{article}
\usepackage{epsfig}

\pdfoutput=1
\usepackage{graphicx,color,float,amssymb,multirow}

\newcommand{\imag}{\Im {\rm m}}
\newcommand{\real}{\Re {\rm e}}
\newcommand{\wt}{\widetilde}

\newcommand{\lsim}{\raisebox{-0.13cm}{~\shortstack{$<$ \\[-0.07cm] $\sim$}}~}
\newcommand{\gsim}{\raisebox{-0.13cm}{~\shortstack{$>$ \\[-0.07cm] $\sim$}}~}

\newcommand{\Li}{\,{\rm Li}}
\usepackage{bbding} 
\newcommand{\yes}{\Checkmark}

\usepackage{ulem}

\def\ss{\mbox{\boldmath $\sigma$}}
\newcommand{\be}{\begin{equation}}
\newcommand{\ee}{\end{equation}}
\newcommand{\bea}{\begin{eqnarray}}
\newcommand{\eea}{\end{eqnarray}}

\begin{document}

\title{ \vspace{1cm} Decays of Higgs Bosons in
the Standard Model and Beyond}
\author{
Seong Youl Choi,$^{1}$\footnote{sychoi@jbnu.ac.kr}~
Jae Sik Lee,$^{2,3,4}$\footnote{jslee@jnu.ac.kr}~
Jubin Park$^{3,4}$\footnote{honolov77@gmail.com}
\\ \\
$^1$ Department of Physics and RIPC,
Jeonbuk National University, Jeonju 54896, Korea\\
$^2$ Department of Physics, Chonnam National University,
Gwangju 61186, Korea \\
$^3$ IUEP, Chonnam National University,
Gwangju 61186, Korea \\
$^4$ APCTP, Pohang, Gyeongbuk 37673, Korea
}
\maketitle
%
%

\begin{abstract}
\noindent
We make an updated review and a systematic
and comprehensive analysis of the decays of Higgs bosons in the Standard Model (SM)
and its three well-defined prototype extensions such as
the complex singlet extension of the SM (cxSM), the four types of
two Higgs-doublet models (2HDMs) without
tree-level Higgs-mediated flavor-changing neutral current (FCNC)
and the minimal supersymmetric extension of the SM (MSSM).
We summarize the theoretical predictions for the decay widths
of the SM Higgs boson and those of Higgs bosons
appearing in its extensions taking account
of all possible decay modes. We incorporate them to study and analyze
decay patterns of CP-even, CP-odd, and CP-mixed neutral Higgs bosons and charged ones.
We put special focus on the properties of a neutral Higgs boson with mass
about 125 GeV discovered at the LHC and present constraints obtained
from precision analysis of it.
%
%
%
This review is intended to be self-contained and consolidated
by coherently integrating relevant physics information for
studying decays of Higgs bosons in the SM and beyond.

\end{abstract}
\eject
\tableofcontents
\section{Introduction}
\label{sec:introduction}

Since the discovery of a resonance with a mass of approximately 125 GeV at the Large
Hadron Collider (LHC) in 2012~\cite{Aad:2012tfa, Chatrchyan:2012ufa},
the substantial subsequent studies of its properties
have been carried out with the data set collected during
the LHC Run 1 period from 2009 to  2012 and
the LHC Run 2 period from 2015 to 2018.
They have firmly confirmed the compatibility of the resonance with
the spin-zero and parity-even SM Higgs boson which appears in
the spontaneously broken gauge theory where the electroweak
interactions are governed by the SU(2)$_L \times {\rm U(1)}_Y$
gauge symmetry
group~\cite{Glashow:1961tr,Weinberg:1967tq,Salam:1968rm}.
\footnote{
Note that the gauge and fermion sectors of the SM have been already well probed
with great precision both theoretically and experimentally as can be checked
with the particle physics reference
book~Ref.$\,$\cite{PDG2020}.}
At the present time
we are on a watershed peak for exploring a new territory of particle physics
through the Higgs landscape.

\medskip

The total and differential rate
measurements of
all the possible production and decay channels of the resonance state
so far are consistent with those predicted in the SM within
experimental and theoretical uncertainties~\cite{Aad:2019mbh,Sirunyan:2018koj}.
The mass of the Higgs boson has been measured at
the per-mille precision level, mainly through the high-resolution
decay modes with four-lepton and di-photon final
states~\cite{Aad:2015zhl,Sirunyan:2020xwk}.
Furthermore, the couplings of the Higgs boson
to the gauge bosons and
the charged fermions of the third
generation~\cite{Aad:2015vsa,Sirunyan:2017khh,Sirunyan:2018kst,Aaboud:2018gay,
Aaboud:2018urx} and, recently, to the
muons~\cite{Aad:2020xfq,Sirunyan:2020two} were established
independently and unambiguously. Based on the observational
facts, we  call the discovered resonance particle
as the SM-like Higgs boson wherever appropriate in the following.

\medskip

Nevertheless, the couplings of the SM-like Higgs boson to the electrons
and lighter quarks of the first and second generations and its
cubic and quartic self-couplings
defining the profile of the Higgs potential are yet to be
established and measured independently.
Furthermore,
more complex Higgs sectors associated with additional states
have not been ruled out. Therefore, it is not yet firmly established whether
the SM-like Higgs boson is indeed the only elementary scalar state
as in the SM, whether there exist additional
elementary scalar particles, or even
whether it is a composite particle with internal structure or not.

\medskip

Conceptually, the SM with the Higgs boson could be
weakly interacting well above the weak scale of $v=246$ GeV without
violating unitarity and so with no need for new physics.
However the Higgs boson mass is influenced subtly
by the presence of heavy particles and it receives quantum corrections destabilizing
the weak scale and requiring a delicate fine-tuning of
apparently unrelated parameters. This so-called naturalness
or hierarchy problem~\cite{Gildener:1976ih,Weinberg:1975gm,Susskind:1978ms}
has been the key argument for expecting new physics to be revealed
at the TeV scale. To mention just a few, new theoretical frameworks
based on a fermion-boson symmetry called
supersymmetry~\cite{Golfand:1971iw,Volkov:1973ix,Wess:1974tw}, a collective
symmetry between the SM particles and heavier partners as in Little Higgs
theories~\cite{ArkaniHamed:2001nc,ArkaniHamed:2002pa,ArkaniHamed:2002qx}
or an effective reduction of the Planck scale to the TeV scale as in
extra-dimension models~\cite{ArkaniHamed:1998rs,Antoniadis:1998ig,Randall:1999ee,
Randall:1999vf} have been proposed and intensively investigated.

\medskip

In addition to alleviating the hierarchy problem, new scenarios involving
extensions of the Higgs sector generically have been proposed and investigated
to account for the dark matter (DM) abundance~\cite{Jungman:1995df,Bertone:2004pz},
the matter-antimatter asymmetry
of the Universe with new sources of the charge-parity (CP) symmetry
breakdown~\cite{Dine:2003ax,Morrissey:2012db,Buchmuller:2005eh},
the tiny but non-vanishing neutrino masses~\cite{GonzalezGarcia:2007ib},
inflation~\cite{Lyth:1998xn}, etc.
Such models with additional scalars can provide us with solid platforms for
exploring new Higgs boson signals concretely and comprehensively
since, in each scenario, Higgs bosons exhibit
their own distinctive features in their couplings to gauge bosons,
fermions and those among  themselves.

\medskip

After the successful completion of Run 1 and Run 2, the LHC is presently in the
second long shut down period while undergoing important upgrades for its high
luminosity phase.
Much larger data sets are to be collected during
the Run 3 period and, ultimately, during the operation
period of the high-luminosity LHC (HL-LHC) and they
will enable us to explore new physics beyond the SM (BSM)
by performing more challenging as well as more
precise measurements.
In light of such promising experimental prospects
at the LHC and at other future high-energy and high-precision experiments
\cite{Shiltsev:2019rfl,Gray:2021jij}
it is quite timely and worthwhile to perform
a systematic and comprehensive review
and analysis of the decays of Higgs bosons
including the theoretical calculations known up to now,
not only in the SM but also in various BSM scenarios
with unique features in their extended Higgs sectors.

\medskip

Certainly it is formidable to review all theoretical and
experimental aspects of Higgs sectors in all BSM models proposed so far in
a single report with limited space. Unavoidably, we restrict ourselves in this review
to the SM and the three well-defined prototype BSM models with
extended Higgs sectors possessing their own
characteristic features and broad implications.
Specifically, in addition to the SM, we consider the following
representative examples:
$(i)$
the complex singlet extension of the SM (cxSM)~\cite{Barger:2008jx,
Gonderinger:2012rd,Coimbra:2013qq,Jiang:2015cwa,Sannino:2015wka,Costa:2015llh,
Muhlleitner:2017dkd,Chiang:2017nmu,Azevedo:2018llq},
$(ii)$
the four types of two Higgs doublet models (2HDMs)
\cite{
Lee:1973iz,Lee:1974jb,Peccei:1977hh,Fayet:1974fj,Inoue:1982ej,Flores:1982pr,
Gunion:1984yn,Botella:1994cs,Branco:1999fs,Carena:2002es,Branco:2011iw}
with natural flavor conservation at the tree level
and the so-called $\rho$ parameter close to unity abiding by
the stringent experimental constraint on it,
and
$(iii)$
the minimal supersymmetric extension of the SM (MSSM)
\cite{Fayet:1976cr,Nilles:1983ge, Haber:1984rc,Barbieri:1987xf,
Martin:1997ns,Haber:1997if,
Chung:2003fi,Drees:2004jm,Baer:2006rs,Binetruy:2006ad}.
For Higgs sectors in BSM models beyond cxSM, 2HDMs and MSSM,
see, for example, Refs.~\cite{Maniatis:2009re,Ellwanger:2009dp,
Cheung:2010ba,Weinberg:1976hu,Branco:1985pf,Ivanov:2017dad}.

\medskip

Several related previous reviews on Higgs physics in the SM and
the MSSM can be found in 
Refs.$\,$\cite{Gunion:1989we,Gunion:1992hs,Carena:2002es,
Djouadi:2005gi,Djouadi:2005gj,Accomando:2006ga,
Dittmaier:2011ti,Dittmaier:2012vm,Heinemeyer:2013tqa,deFlorian:2016spz,
Dawson:2013bba,Spira:1997dg,Spira:2016ztx,Dawson:2018dcd}.
A few tailor-made sophisticated computational packages have been
developed for the mass spectra and decay widths of neutral and
charged Higgs bosons in the SM and the MSSM with
real parameters~\cite{Djouadi:1997yw,Djouadi:2018xqq,Bredenstein:2006rh} and
with explicit CP violation~\cite{Lee:2003nta,Lee:2007gn,Lee:2012wa,Frank:2006yh,
Hahn:2014mla}.
This review updates the previous works substantially by
including two  popular BSM
models  in addition to the MSSM and also by
allowing for complex parameters leading to CP-violating
phenomena~\cite{Accomando:2006ga,Ibrahim:2007fb}.
We perform a systematic
and comprehensive analysis for
the decays of neutral and charged Higgs bosons in
those three prototype BSM models as well as in the SM.
We take account of all possible decay modes
of these models including those into non-SM particles
among which some are invisible~\cite{Shrock:1982kd,Li:1985hy}
and/or exotic~\cite{Curtin:2013fra}.
%
We anticipate
more complete reviews on the Higgs sectors of many other BSM scenarios
to come out timely in step with more advanced experimental developments.

\medskip

In this review, we try to contain all the relevant information needed to
implement the
up-to-date theoretical calculations of Higgs decays.
We aim to make it be of pedagogical and
practical use especially for incorporating corrections
beyond the leading order (LO).
We elaborate on how the partial decay widths of neutral and
charged Higgs bosons are calculated at LO
and how we treat QCD and electroweak (ELW) corrections in
each decay mode.
\footnote{
The ELW corrections considered in this review
are mostly SM ones since the BSM ELW corrections,
compared to universal QCD corrections,
are still subleading, complicated, and  strongly dependent
on specific BSM models under consideration.}
We intend to make this review stand-alone, self-contained, and consolidated
by integrating relevant physics information for Higgs decays coherently.
Incidentally, we try to make it be as model-independent and analytic as possible
in order for our approach to be useful and easily applicable in studying
Higgs decays even in the BSM models not explicitly mentioned in this review.

\medskip

This review is organized as follows.
Section 2 is devoted to reviewing
the Higgs sectors of the SM and three extended scenarios
- cxSM, 2HDMs and MSSM - without imposing any constraints on
the model parameters.
We work out the analytic structure of the Higgs potential and mixing.
Also worked out are the Higgs interactions with
gauge bosons, the SM fermions, and new scalars and fermions
as well as the Higgs-boson self interactions.
We review and update the decays of
neutral Higgs bosons in Section~\ref{sec:decays_of_a_generic_neutral_higgs_boson}
and those of charged Higgs bosons in
Section~\ref{sec:decays_of_a_charged_higgs_boson}.
We provide explicit analytical
expressions of the individual partial decay
widths as precisely as possible by
including the state-of-the-art theoretical calculations.
In Section~\ref{sec:Higgcision}, we present
the constraints on the couplings of the SM-like
Higgs boson weighing about 125 GeV obtained from
global fits to the LHC precision Higgs data.
Conclusions are made in Section~\ref{sec:conclusions}.
To make this review self-contained, various supplemental
materials are provided in six appendices.
Appendix~\ref{app:smpara} includes a summary of
the SM parameters used for the numerical estimates of the Higgs decay widths
and a description of the running of the strong coupling constant
and heavy quark masses.
In Appendix~\ref{app:dsdp}, the supersymmetric contributions to the loop-induced couplings
of the Higgs boson to two gluons, two photons and $Z\gamma$ are presented
and Appendix~\ref{app:csfcpf} is devoted to the presentation
of the QCD corrections to the partial width of
the Higgs-boson decay to two photons.
In Appendix \ref{app:2hdminput}, we present expressions for
the most general 2HDM potential parameters in terms of
the masses of charged and neutral Higgs bosons and the elements
of the orthogonal matrix describing the mixing among neutral Higgs bosons
and, in Appendix \ref{app:cubicself},
we apply them for deriving cubic Higgs-boson self-couplings.
Finally, Appendix~\ref{app:bsmtools} is added as a guide
to numerical packages for calculating
precise SM and full BSM-dependent ELW corrections.
\section{Standard Model and Beyond}
\label{sec:standard_model_and_beyond}

In this section, we derive and describe the basic form of Higgs
boson masses and mixing as well as their interactions
in the SM, cxSM, 2HDMs and MSSM.
The derived analytical results are utilized comprehensively in the sequential
sections for the detailed review of the decays of neutral and charged Higgs
bosons and also for the model-independent precision
study of the SM-like neutral Higgs boson which has been
extensively probed at the LHC since its discovery in 2012.

\medskip

We note that there are codes such as
{\tt FeynRules}
\cite{Christensen:2008py,Degrande:2011ua,Alloul:2013bka},
{\tt Sarah}
\cite{Staub:2008uz,Staub:2012pb,Staub:2013tta,Staub:2015kfa},
and
{\tt LanHEP}
\cite{Semenov:1996es,Semenov:1997qm,Semenov:1998eb,
Semenov:2002jw,Semenov:2008jy,Semenov:2010qt,Semenov:2014rea}
for the automatic generation of all the Feynman rules
in the BSM models as well as in the SM.
Rather than simply using these codes, we start by presenting interaction
Lagrangians in order for the readers to
work out independently the analytic structure and parametric dependence
of the partial decay widths of neutral and charged Higgs bosons
and to more deeply understand the theoretical
and phenomenological aspects of Higgs physics in the SM and beyond.

\subsection {Standard Model}
\label{subsec:standard_model}

The self-interactions of the SM Higgs boson and its
interactions with the massive vector bosons are derived from
the Higgs Lagrangian:
\begin{equation}
{\cal L}_{\rm Higgs} =
\left(D^\mu\Phi\right)^\dagger \left(D_\mu\Phi\right) \ - \ V_{\rm SM}(\Phi)\,,
\label{eq:higgs_lagrangian}
\end{equation}
where $\Phi$ denotes a complex SU(2)$_L$ doublet Higgs field
with hypercharge $Y=1/2$ and its covariant derivative is defined as
\begin{eqnarray}
D_\mu\Phi &=&
\left(\partial_\mu-ig\frac{\tau_a}{2}W_\mu^a-ig^\prime\frac{1}{2}B_\mu
\right)\Phi\nonumber \\[2mm]
&=&
\left(\begin{array}{cc}
\partial_\mu -\frac{i}{2}(gW_\mu^3+g^\prime B_\mu) &
-\frac{ig}{2}(W_\mu^1-iW_\mu^2) \\
-\frac{ig}{2}(W_\mu^1+iW_\mu^2) &
\partial_\mu +\frac{i}{2}(gW_\mu^3-g^\prime B_\mu)
\end{array}\right)\,\Phi\,,
\end{eqnarray}
in terms of
the SU(2)$_L$ and U(1)$_Y$ gauge couplings $g$ and $g'$, respectively,
the three SU(2)$_L$ gauge bosons $W^{1,2,3}_\mu$, and
the single U(1)$_Y$ gauge boson $B_\mu$ with the usual three
$2\times 2$ Pauli matrices
\begin{equation}
\tau_1=\left(\begin{array}{cc} 0 & 1 \\ 1 & 0 \end{array}\right)\,, \ \
\tau_2=\left(\begin{array}{cc} 0 & -i \\ i & 0 \end{array}\right)\,, \ \
\tau_3=\left(\begin{array}{cc} 1 & 0 \\ 0 & -1 \end{array}\right)\,.
\end{equation}
And the renormalizable SM Higgs potential $V_{\rm SM}(\Phi)$ is given by
\begin{equation}
V_{\rm SM}(\Phi)=\mu^2 (\Phi^\dagger\Phi) \ + \ \lambda(\Phi^\dagger \Phi)^2\,,
\end{equation}
with $\mu^2 <0$ leading to the spontaneous breakdown of the electroweak
gauge symmetry.

\medskip

Taking $\Phi=(0,v+H)^T/\sqrt{2}$ with the vacuum expectation value (vev)
$v=\sqrt{-\mu^2/\lambda}$ and the real scalar field $H$ after rotating away three
Goldstone modes and using $W^\pm_\mu=(W^1_\mu\mp i W_\mu^2)/\sqrt{2}$ and
$Z_\mu=(gW^3_\mu - g^\prime B_\mu)/\sqrt{g^2+g^{\prime 2}}$,
we can render the kinetic term of the Higgs Lagrangian in
Eq.$\,$(\ref{eq:higgs_lagrangian}) into the form expanded as
\begin{eqnarray}
\left(D^\mu\Phi\right)^\dagger \left(D_\mu\Phi\right) &=&
\frac{1}{2} (\partial_\mu H) (\partial^\mu H)
+M_W^2 W_\mu^+W^{\mu -} +\frac{1}{2} M_Z^2 Z_\mu Z^\mu \\
&+&
gM_W \left(W_\mu^+W^{\mu -}+\frac{1}{2c_W^2}Z_\mu Z^\mu\right)H  +
\frac{1}{v^2}\left(M_W^2W_\mu^+W^{\mu -}+\frac{M_Z^2}{2}Z_\mu Z^\mu\right)H^2\,,
\nonumber
\end{eqnarray}
in the unitary gauge. We use the abbreviation $s_W\equiv\sin\theta_W$ for
the sine of the weak mixing angle $\theta_W$
and $c_W\equiv\cos\theta_W$, $t_W\equiv\sin\theta_W/\cos\theta_W$, etc.
The masses of the massive gauge bosons $W$ and $Z$ are given by
$M_W=gv/2$ and $M_Z=M_W/c_W$ with
$v=\left(\sqrt{2}G_F\right)^{-1/2} \approx 246\, {\rm GeV}$
fixed by the Fermi constant $G_F$, which is determined with
a precision of $0.6\, {\rm ppm}$ from muon decay measurements
\cite{Webber:2010zf,Tishchenko:2012ie}.
Incidentally, the SU(2)$_L$ and U(1)$_Y$ gauge couplings are
$g=e/s_W$ and $g^\prime=g\,t_W=e/c_W$, respectively, where
the magnitude of the electron electric charge
$e=2\sqrt{\pi\alpha}$ with $\alpha$ being the fine structure constant.
On the other hand, the SM Higgs potential takes the form of
\begin{equation}
\label{eq:SMPotential}
V_{\rm SM}(H)=-\frac{1}{8}v^2M_H^2+\frac{1}{2}M_H^2\,H^2
+\frac{1}{3!}\left(\frac{3M_H^2}{v^2}\right)v\,H^3
+\frac{1}{4!}\left(\frac{3M_H^2}{v^2}\right)\,H^4\,,
\end{equation}
which is completely fixed in terms of $v$ and the Higgs mass $M_H$
with the replacements of $\mu^2=-\lambda v^2$ and $\lambda=M_H^2/2v^2$.

\medskip

The Higgs interactions with the SM fermions are derived
by considering the following Yukawa interactions
\begin{eqnarray}
-{\cal L}_Y&=& \overline{U_R}\,{\bf h}_u\, Q^T\,(i\tau_2)\,\Phi
- \overline{D_R}\,{\bf h}_d\, Q^T\,(i\tau_2)\,\widetilde\Phi
- \overline{E_R}\,{\bf h}_e\, L^T\,(i\tau_2)\,\widetilde\Phi
\ + \ {\rm h.c.}\,,
\end{eqnarray}
where $\widetilde\Phi=i\tau_2 \Phi^*$ and
$Q^T=(U_L\,,D_L)$ and $L^T=(\nu_L\,,E_L)$
with $U$ and $D$ standing
for the three up- and down-type quarks, respectively, and
$\nu$ and $E$ for the three neutrinos and
charged leptons, respectively, in the weak eigenstate basis.
And the $3\times 3$ Yukawa matrices are denoted by ${\bf h}_{u,d,e}$.
Taking $\Phi=(0,v+H)^T/\sqrt{2}$ again, we have
\begin{equation}
-{\cal L}_{H\bar{f}f} = \sum_{f=u,d,c,s,t,b,e,\mu,\tau}\,
\frac{m_f}{v}\,H\,\overline f f\,,
\end{equation}
with the masses $m_f=h_f\,v/\sqrt{2}$ in the fermion
mass eigenstate basis diagonalizing the Higgs-fermion interactions.

\subsection{Complex Singlet Extension of the SM}
\label{subsec:complex_singlet_extension_of_the_sm}
In this subsection, as the first BSM example,
we consider a model in which the SM is extended by adding a
complex SU(2)$_L \times$ U(1)$_Y$ singlet (cxSM).

\subsubsection {\it Potential and mixing}

When a complex scalar singlet field $\mathbb{S}$ is added to the SM
Higgs sector \cite{Barger:2008jx,Gonderinger:2012rd,Coimbra:2013qq,Jiang:2015cwa,
Sannino:2015wka,Costa:2015llh,Muhlleitner:2017dkd,Chiang:2017nmu,Azevedo:2018llq},
the most general renormalizable scalar potential takes the
form~\cite{Barger:2008jx}
\begin{eqnarray}
V(\Phi\,,\mathbb{S})&=&\mu^2 (\Phi^\dagger\Phi) \ + \ \lambda(\Phi^\dagger \Phi)^2\nonumber \\
& + &
\left(\delta_1 \Phi^\dagger\Phi\,\mathbb{S}+c.c.\right) \ + \
\delta_2 \Phi^\dagger\Phi\,\left|\mathbb{S}\right|^2 \ + \
\left(\delta_3 \Phi^\dagger\Phi\, \mathbb{S}^2+ h.c.\right)\nonumber \\
&+&
\left(a_1\,\mathbb{S} + c.c.\right) \ + \
\left(b_1\,\mathbb{S}^2 + c.c.\right) \ + \
b_2\left|\mathbb{S}\right|^2 \ + \
\left(c_1\,\mathbb{S}^3 + c.c.\right) \ + \
\left(c_2\,\mathbb{S}\left|\mathbb{S}\right|^2 + c.c.\right)\nonumber \\
&+&
\left(d_1\,\mathbb{S}^4 + c.c.\right) \ + \
d_2\,\left|\mathbb{S}\right|^4 \ + \
\left(d_3\,\mathbb{S}^2 \left|\mathbb{S}\right|^2 + c.c.\right)\,.
\label{eq:cxsm_scalar_potential}
\end{eqnarray}
Imposing a global U(1) symmetry eliminates all terms
containing complex coefficients.
One may allow a soft
U(1)-breaking $b_1$ term to avoid a massless CP-odd Goldstone boson which is
not phenomenologically viable. And then, in order to avoid the
cosmological domain wall problem caused by the presence of
the $b_1$ term, one may additionally include the linear
$a_1$ term which breaks the global U(1) and
a discrete $\mathbb{Z}_2$ symmetry under $\mathbb{S}\to -\mathbb{S}$.
The resulting cxSM scalar potential takes the form
\begin{eqnarray}
V(\Phi\,,\mathbb{S})=\mu^2 (\Phi^\dagger\Phi)  +  \lambda(\Phi^\dagger \Phi)^2  +
\delta_2 \Phi^\dagger\Phi\,\left|\mathbb{S}\right|^2  +
b_2\left|\mathbb{S}\right|^2  +
d_2\left|\mathbb{S}\right|^4  +
\left(a_1\,\mathbb{S} + b_1\,\mathbb{S}^2 + c.c.\right)\,,
\end{eqnarray}
in terms of the original couplings in Eq.$\,$(\ref{eq:cxsm_scalar_potential}),
or, alternatively~\cite{LEE:2020dop},
\begin{eqnarray}
V(\Phi\,,\mathbb{S})=\mu_1^2 (\Phi^\dagger\Phi)  +  \mu_2^2\left|\mathbb{S}\right|^2  +
\lambda_1(\Phi^\dagger \Phi)^2  +
\lambda_2\left|\mathbb{S}\right|^4  +
\lambda_3 \Phi^\dagger\Phi\,\left|\mathbb{S}\right|^2  +
\left(a_1\,\mathbb{S} + b_1\,\mathbb{S}^2 + c.c.\right)\,,
\end{eqnarray}
in terms of a more systematic parameter set of 5 real parameters of
$\mu^2_{1,2}$ and $\lambda_{1,2,3}$ and
2 complex massive parameters of $a_1=|a_1|{\rm e}^{i\phi_a}$ and
$b_1=|b_1|{\rm e}^{i\phi_b}$.

\medskip

By parameterizing the SU(2)$_L$ doublet $\Phi$ and singlet $\mathbb{S}$ as
\begin{equation}
\Phi=\left(\begin{array}{c}
G^+ \\ \frac{1}{\sqrt{2}}\,(v+\phi+iG^0)
\end{array}\right)\,; \ \ \
\mathbb{S}=\frac{{\rm e}^{i\xi}}{\sqrt{2}}\,
\left[v_s+s + i( v_a +a )\right]\,,
\end{equation}
we obtain the following three tadpole conditions for minimizing the
potential:
\begin{eqnarray}
&&\left[\mu_1^2+\lambda_1\, v^2+\frac{1}{2}\lambda_3\, v_{sa}^2\right]\,v  =  0\,,
\nonumber \\[2mm]
&&
\left[\mu_2^2+\lambda_2\, v_{sa}^2+\frac{1}{2}\lambda_3\, v^2
+2\,\real(b_1{\rm e}^{2i\xi})\right]\,v_s
-2\,\imag(b_1{\rm e}^{2i\xi})\,v_a + \sqrt{2}\, \real(a_1{\rm e}^{i\xi})  =  0\,,
\nonumber \\[2mm]
&&
\left[\mu_2^2+\lambda_2\, v_{sa}^2+\frac{1}{2}\lambda_3\, v^2
-2\,\real(b_1{\rm e}^{2i\xi})\right]\,v_a
-2\,\imag(b_1{\rm e}^{2i\xi})\,v_s - \sqrt{2}\, \imag(a_1{\rm e}^{i\xi})  =  0\,,
\label{eq:tadpole_conditions_cxsm}
\end{eqnarray}
with the abbreviation $v_{sa}^2=v_s^2+v_a^2$.
The mass terms of the scalar states are given  by
\begin{equation}
\label{eq:cxsm_mass}
V_{\rm cxSM\,,mass}=\frac{1}{2} (\phi \ s^\prime  \ a^\prime)\,
{\cal M}^2_0\,
\left(\begin{array}{c}
\phi \\ s^\prime  \\ a^\prime\end{array}\right)\,,
\end{equation}
in terms of a real and symmetric $3\times 3$
mass-squared matrix ${\cal M}^2_0$ decomposed into the two parts:
\begin{eqnarray}
{\cal M}^2_0 &=&
\left(\begin{array}{ccc}
X_1+2\lambda_1 v^2 & \lambda_3 v v_{sa} &  0\\[2mm]
\lambda_3 v v_{sa}  & X_2+2\lambda_2 v_{sa}^2  & 0 \\[2mm]
0 & 0 & X_2 \end{array}\right)  \nonumber \\[3mm]
&+& 2\,
\left(\begin{array}{ccc}
0 & 0 &  0\\[2mm]
0 &
~~\real{(b_1{\rm e}^{2i\xi})}c_{2\gamma}
-\imag{(b_1{\rm e}^{2i\xi})}s_{2\gamma} &
-\real{(b_1{\rm e}^{2i\xi})}s_{2\gamma}
-\imag{(b_1{\rm e}^{2i\xi})}c_{2\gamma} \\[2mm]
0 &
-\real{(b_1{\rm e}^{2i\xi})}s_{2\gamma}
-\imag{(b_1{\rm e}^{2i\xi})}c_{2\gamma} &
-\real{(b_1{\rm e}^{2i\xi})}c_{2\gamma}
+\imag{(b_1{\rm e}^{2i\xi})}s_{2\gamma}
\end{array}\right)\,.
\end{eqnarray}
The two parameters of $X_{1,2}$ appearing in
the diagonal components of the first term are defined by
\begin{equation}
X_1 \equiv \mu_1^2+\lambda_1\, v^2+\frac{1}{2}\lambda_3\, v_{sa}^2\,, \ \ \
X_2 \equiv \mu_2^2+\lambda_2\, v_{sa}^2+\frac{1}{2}\lambda_3\, v^2\,.
\end{equation}
In Eq.~(\ref{eq:cxsm_mass}),
the primed scalar fields $s^\prime$ and $a^\prime$
are related to the original
scalar fields $s$ and $a$ through the rotation
\begin{equation}
\left(\begin{array}{c} s^\prime \\ a^\prime \end{array}\right) =
\left(\begin{array}{rr}
c_\gamma & s_\gamma \\ -s_\gamma & c_\gamma \end{array}\right)\,
\left(\begin{array}{c} s \\ a \end{array}\right) \,,
\end{equation}
with $c_\gamma=\cos\gamma=v_s/v_{sa}$ and
$s_\gamma=\sin\gamma=v_a/v_{sa}$.
Note that $X_1=0$ always to have the non-zero
vev of $v$, as can be
checked with the first tadpole condition in
Eq.$\,$(\ref{eq:tadpole_conditions_cxsm}). On the other hand,
only in the U(1)-conserving case with both $a_1=0$ and $b_1=0$,
$X_2=0$ to have the non-vanishing vevs of $v_a$ and $v_s$
giving rise to the massless Goldstone boson $a^\prime$.
In this case, the singlet vacuum takes the U(1) symmetric
vev of $v_{sa}=\sqrt{v_s^2+v_a^2}$ while each of the vevs
remains undetermined.

\medskip

In some cases, instead of the discrete $\mathbb{Z}_2$ symmetry,
a different discrete $\mathbb{Z}^\prime_2$ symmetry
under the interchange $\mathbb{S}\leftrightarrow \mathbb{S}^*$ is imposed.
In this case, the U(1)-breaking part of the scalar potential reads
\begin{equation}
\left.\left(a_1\mathbb{S}+b_1\mathbb{S}^2
+c.c\right)\right|_{\mathbb{Z}^\prime_2}
=2\left[\real{(a_1)}\,S + \real{(b_1)}\,(S^2-A^2)\right]\,,
\end{equation}
with $\mathbb{S}=S+iA=\left[(v_s+s)+i(v_a+a)\right]/\sqrt{2}$.
Note that, if the potential has the $\mathbb{Z}^\prime_2$ symmetry,
just two real parameters are sufficient for parameterizing
the U(1)-breaking part of the potential. Assuming $a_1$ and $b_1$
to be real with no loss of generality, the tadpole conditions
of the scalar potential become
\begin{eqnarray}
X_1\,v=0\,, \ \ \
(X_2+2b_1)\,v_s + \sqrt{2}\, a_1  =  0\,, \ \ \
(X_2-2b_1)\,v_a = 0\,.
\label{eq:tadpole_conditions_2nd_case}
\end{eqnarray}
Assuming $v\neq 0$, which forces $X_1=0$,
the mass-squared matrix is simplified into the form
\begin{equation}
{\cal M}^2_0 =
\left(\begin{array}{ccc}
2\lambda_1 v^2 & \lambda_3 v v_{sa} &  0\\[2mm]
\lambda_3 v v_{sa}  & X_2+2\lambda_2 v_{sa}^2 + 2b_1c_{2\gamma}
& -2b_1s_{2\gamma} \\[2mm]
0 & -2b_1s_{2\gamma}
& X_2-2b_1c_{2\gamma} \end{array}\right)\,.  \nonumber \\[3mm]
\label{eq:mass_squared_matrix_cxsm}
\end{equation}
When $v_s\neq0$ and $v_a\neq0$ guaranteeing $s_{2\gamma}\neq 0$,
the CP symmetry is spontaneously broken and all the three states
mix. In this case, one may parameterize the potential with a set of
7 parameters of $\{\lambda_1,\lambda_2,\lambda_3,v,v_{sa},\tan\gamma,b_1\}$ with
the relations
\begin{equation}
\mu_1^2= -\lambda_1\, v^2-\frac{1}{2}\lambda_3\, v_{sa}^2 \,, \ \ \
X_2=2b_1\,, \ \ \
a_1 = -2\sqrt{2}\,b_1v_{s}\,,
\end{equation}
where the second relation is solved to give
$\mu_2^2= 2b_1 -\lambda_2\, v_{sa}^2-\frac{1}{2}\lambda_3\, v^2$.

\medskip

When $v_s\neq0$ and $v_a=0$, the angle $\gamma=0$, i.e.
$s^\prime=s$ and $a^\prime=a$ and the scalar mixing occurs only
between the two states of $\phi$ and $s$
with the pseudoscalar mass-squared
\begin{eqnarray}
M_a^2=-\frac{\sqrt{2}a_1}{v_s}-4b_1\,.
\end{eqnarray}
In this 2-state mixing case, the scalar potential
can be parameterized with a set of 7 parameters of
$\{\lambda_1,\lambda_2,\lambda_3,v,v_s,a_1,b_1\}$ with the relations
\begin{equation}
\mu_1^2= -\lambda_1\, v^2-\frac{1}{2}\lambda_3\, v_{sa}^2 \,, \ \ \
X_2=-\frac{\sqrt{2}a_1}{v_s}-2b_1\,,
\end{equation}
where the second relation is solved to give
$\mu_2^2= -\sqrt{2}a_1/v_s-2b_1
-\lambda_2\, v_{sa}^2-\frac{1}{2}\lambda_3\, v^2$.
Incidentally, if $v_s=0$ in addition to $v_a=0$, the parameter
$a_1$ should vanish due to the second tadpole condition in
Eq.$\,$(\ref{eq:tadpole_conditions_2nd_case}) and $X_2=\mu_2^2+\lambda_3 v^2/2$
giving the squares of three masses as
\begin{equation}
M_\phi^2=2\lambda_1 v^2\,, \ \ \
M_s^2=\mu_2^2+\lambda_3 v^2/2 + 2 b_1\,, \ \ \
M_a^2=\mu_2^2+\lambda_3 v^2/2 - 2 b_1\,,
\end{equation}
and the 6 parameters $\{\lambda_1,\lambda_2,\lambda_3,v,\mu_2^2,b_1\}$
can be employed for describing the scalar potential.
Various types of vacua in the cxSM  are summarized 
in Table~\ref{tab:cxSMvacua}.

\begin{table}[!h]
\caption{Vacua in the cxSM
imposing $\mathbb{Z}_2^\prime$: the parameters $a_1$ and $b_1$ are
real and $v\neq0$  is taken. From Ref.~\cite{LEE:2020dop}.}
\label{tab:cxSMvacua}
\setlength{\tabcolsep}{1.5ex}%
\begin{center}
\vspace{-2mm}
\begin{tabular}{c|cll}
\hline\hline
vacua & $X_2$ & a possible set of inputs &
miscellaneous relations\\
\hline
$v_a\neq0$ \& $v_s\neq0$ & $2b_1$ &
$\{\lambda_1,\lambda_2,\lambda_3,v,v_{sa},\tan\gamma,b_1\}$ &
$a_1=-2\sqrt{2}b_1v_s$ \\
$v_a\neq0$ \& $v_s=0$ & $2b_1$ &
$\{\lambda_1,\lambda_2,\lambda_3,v,v_{a},b_1\}$ &
$a_1=0\,; v_{sa}=v_a\,,c_\gamma=0\,,s_\gamma=1$ \\
\hline
$v_a=0$ \& $v_s\neq0$ & $-{\sqrt{2}a_1}/{v_s}-2b_1$ &
$\{\lambda_1,\lambda_2,\lambda_3,v,v_{s},a_1,b_1\}$ &
$v_{sa}=v_s\,,c_\gamma=1\,,s_\gamma=0$ \\
$v_a=0$ \& $v_s=0$ & $\mu_2^2+\lambda_3v^2/2$ &
$\{\lambda_1,\lambda_2,\lambda_3,v,\mu_2^2,b_1\}$ &
$a_1=0\,; v_{sa}=0\,,c_{\gamma}\to 1\,,s_{\gamma}\to 0$ \\
\hline\hline
\end{tabular}
\end{center}
\end{table}

\medskip

Finally,
without loss of generality,
the orthogonal $3\times 3$ mixing matrix $O$ diagonalizing the
real and symmetric
mass-squared matrix in Eq.~(\ref{eq:mass_squared_matrix_cxsm})
is defined through
\begin{eqnarray}
(\phi,s^\prime,a^\prime)^T_\alpha&=&O_{\alpha i} (H_1,H_2,H_3)^T_i\,,
\end{eqnarray}
such that $O^T {\cal M}_0^2 O={\rm diag}(M_{H_1}^2,M_{H_2}^2,M_{H_3}^2)$
with the increasing ordering of $M_{H_1}\leq M_{H_2}\leq M_{H_3}$.

\subsubsection{\it Higgs-boson interactions}
\label{subsec:higgs-boson_interactions}

The interactions of the three Higgs bosons with the SM fermions and
the massive vector bosons in the cxSM are given by
\begin{eqnarray}
-{\cal L}_{H_i\bar f f} &=& \sum_{f=u,d,c,s,t,b,e,\mu,\tau}\, \frac{m_f}{v}\,
g^S_{H_i\bar f f}\, H_i\,\overline f f \,;\nonumber \\
{\cal L}_{HVV} &=&  g\,M_W \, \left(W^+_\mu W^{- \mu}\ + \
\frac{1}{2c_W^2}\,Z_\mu Z^\mu\right) \, \sum_{i=1}^3 \,g_{_{H_iVV}}\, H_i\,;
\nonumber \\
{\cal L}_{HHVV}&=&
\frac{1}{v^2}\left(M_W^2 W_\mu^+ M^{\mu -}
+\frac{M_Z^2}{2} Z_\mu Z^\mu
\right) \sum_{i,j=1}^3 g_{_{H_iH_jVV}} H_i H_j\,,
\end{eqnarray}
with the normalized dimensionless couplings simply given by
\begin{equation}
g^S_{H_i\bar f f} = g_{_{H_iVV}} = O_{\phi i}\,, \ \ \
g_{_{H_iH_jVV}}=O_{\phi i} O_{\phi j}\,.
\end{equation}
And the cubic and quartic couplings are given by the
self-interaction term of the scalar potential:
\begin{eqnarray}
-{\cal L}_{\rm self}&=&\lambda_1 v\phi^3+2\lambda_3 v_{sa}\phi^2 s^\prime
+2\lambda_3v\phi (s^{\prime2}+a^{\prime2})
+4\lambda_2 v_{sa} s^\prime(s^{\prime2}+a^{\prime2})\nonumber \\
&+&\frac{1}{4}\lambda_1\phi^4 +\lambda_3\phi^2(s^{\prime2}+a^{\prime2})
+\lambda_2(s^{\prime4}+2s^{\prime2}a^{\prime2}+a^{\prime4})\nonumber \\
&\equiv &
v\,\sum_{i\geq j\geq k=1}^3 g_{_{H_iH_jH_k}}\, H_i H_j H_k
\: +\: \sum_{i\geq j\geq k\geq l =1}^3
  g_{_{H_iH_jH_kH_l}}\, H_i H_j H_k H_l\,,
\end{eqnarray}
where the normalized cubic and quartic couplings of
the three Higgs mass eigenstates are
\footnote{Here,
the indices $\alpha$, $\beta$, $\gamma$, and $\delta$ count the Higgs weak
eigenstates of $\phi$, $s^\prime$, and $a^\prime$ and the
inequalities among them
imply that the cubic and quartic terms in the Higgs  potential are ordered in
the weak eigenstates.}
\begin{eqnarray}
\label{eq:cxsmself}
g_{_{H_iH_jH_k}} &=& \sum_{\alpha\leq \beta\leq \gamma=1}^3
\left\{O_{\alpha i} O_{\beta j} O_{\gamma  k} \right\}\,
g_{\alpha\beta\gamma} \,, \nonumber \\
g_{_{H_iH_jH_kH_l}} &=&  \sum_{\alpha\leq \beta\leq \gamma \leq \delta=1}^3
\left\{O_{\alpha i} O_{\beta j} O_{\gamma  k} O_{\delta  l} \right\}\,
g_{\alpha\beta\gamma\delta} \,,
\end{eqnarray}
with $i,j,k,l=1,2,3$, the cubic weak-eigenstate couplings
\begin{eqnarray}
g_{\phi\phi\phi}&=& \lambda_1\,,  \ \ \
g_{\phi\phi s^\prime}= 2\lambda_3\,\frac{v_{sa}}{v}\,, \nonumber \\
g_{\phi s^\prime s^\prime}&=& g_{\phi a^\prime a^\prime} =2\lambda_3\,, \ \ \
g_{s^\prime s^\prime s^\prime}= g_{s^\prime a^\prime a^\prime} =
4\lambda_2\,\frac{v_{sa}}{v}\,, \nonumber \\
\end{eqnarray}
and the quartic weak-eigenstate couplings
\begin{eqnarray}
g_{\phi\phi\phi\phi}&=& \lambda_1/4\,, \ \ \
g_{\phi\phi s^\prime s^\prime}=
g_{\phi\phi a^\prime a^\prime}= \lambda_3\,, \nonumber \\
g_{s^\prime s^\prime s^\prime s^\prime}&=&
g_{s^\prime s^\prime a^\prime a^\prime}/2=
g_{a^\prime a^\prime a^\prime a^\prime}= \lambda_2\,.
\end{eqnarray}
In Eq.~(\ref{eq:cxsmself}), the expressions
within the curly brackets $\{ \cdots \}$ need to be symmetrized with respect to
the indices $i,j,k,l$ and divided by the corresponding symmetry factors in
cases where
two or more indices are the same. For example, $\left\{ O_{\alpha i} O_{\beta j} O_{\gamma
k} \right\}$ can explicitly be evaluated as follows:
\begin{eqnarray}
\label{eq:cxsm.selfsym}
\left\{   O_{\alpha i} O_{\beta j} O_{\gamma  k} \right\} &\equiv&
      \frac{1}{N_S}\, \Big(
          O_{\alpha i} O_{\beta j} O_{\gamma  k}
         +O_{\alpha i} O_{\beta k} O_{\gamma  j}
         +O_{\alpha j} O_{\beta i} O_{\gamma  k}
         +O_{\alpha j} O_{\beta k} O_{\gamma  i}
\nonumber \\ &&
         +O_{\alpha k} O_{\beta i} O_{\gamma  j}
         +O_{\alpha k} O_{\beta j} O_{\gamma  i}\Big)\,,
\end{eqnarray}
with $N_S=3!=6$ when $i=j=k$, $N_S=1$ when $(i,j,k)=(3,2,1)$, and
$N_S=2!=2$ in all the other cases.

\subsection {Two Higgs Doublet Models}
\label{sec:two_higgs_doublet_models}

In this subsection, we give a detailed description of the models
in which the SM is extended by adding one more SU(2)$_L$ doublet
(2HDMs)
while taking the same gauge group SU(3)$_C\times{\rm SU(2)}_L\times{\rm U(1)}_Y$ as
in the SM.

\subsubsection {\it Potential and mixing}
\label{subsec:poltential_mixing_2hdm}

The general 2HDM scalar potential containing two
complex SU(2)$_L$ doublets of $\Phi_1$ and $\Phi_2$ with the same hypercharge $Y=1/2$
may be given by 
\begin{eqnarray}
\label{V2HDM}
V_{\rm 2HDM} &=&
\mu_1^2 (\Phi_1^{\dagger} \Phi_1)
+\mu_2^2 (\Phi_2^{\dagger} \Phi_2)
+m_{12}^2 (\Phi_1^{\dagger} \Phi_2)
+m_{12}^{*2}(\Phi_2^{\dagger} \Phi_1) \nonumber \\
&&+ \lambda_1 (\Phi_1^{\dagger} \Phi_1)^2 + \lambda_2
(\Phi_2^{\dagger} \Phi_2)^2 + \lambda_3 (\Phi_1^{\dagger}
\Phi_1)(\Phi_2^{\dagger} \Phi_2) + \lambda_4 (\Phi_1^{\dagger}
\Phi_2)(\Phi_2^{\dagger} \Phi_1) \nonumber \\
&&+ \lambda_5 (\Phi_1^{\dagger} \Phi_2)^2 +
\lambda_5^{*} (\Phi_2^{\dagger} \Phi_1)^2 + \lambda_6
(\Phi_1^{\dagger} \Phi_1) (\Phi_1^{\dagger} \Phi_2) + \lambda_6^{*}
(\Phi_1^{\dagger} \Phi_1)(\Phi_2^{\dagger} \Phi_1) \nonumber \\
&& + \lambda_7 (\Phi_2^{\dagger} \Phi_2) (\Phi_1^{\dagger} \Phi_2) +
\lambda_7^{*} (\Phi_2^{\dagger} \Phi_2) (\Phi_2^{\dagger} \Phi_1)\; ,
\end{eqnarray}
in terms of 2 real and 1 complex dimensionful quadratic
couplings and
4 real and 3 complex dimensionless quartic couplings.
With the parameterization of two scalar doublets $\Phi_{1,2}$ as
\begin{equation}
\Phi_1=\left(\begin{array}{c}
\phi_1^+ \\ \frac{1}{\sqrt{2}}\,(v_1+\phi_1+ia_1)
\end{array}\right)\,; \ \ \
\Phi_2={\rm e}^{i\xi}\,\left(\begin{array}{c}
\phi_2^+ \\ \frac{1}{\sqrt{2}}\,(v_2+\phi_2+ia_2)
\end{array}\right)\,,
\end{equation}
and denoting $v_1=v \cos\beta=vc_\beta$ and $v_2=v \sin\beta=vs_\beta$
with $v=\sqrt{v_1^2+v_2^2}$,
one may remove $\mu_1^2$, $\mu_2^2$,
and $\imag(m_{12}^2{\rm e}^{i\xi})$ from the 2HDM potential using three tadpole
conditions:
\begin{eqnarray}
\label{eq:2hdmtadpole}
\mu_1^2 &=& - v^2\left[\lambda_1c_\beta^2+\frac{1}{2}\lambda_3s_\beta^2+
c_\beta s_\beta\real(\lambda_6{\rm e}^{i\xi})\right]+s_\beta^2 M_{H^\pm}^2\,,
\nonumber \\
\mu_2^2 &=& - v^2\left[\lambda_2s_\beta^2+\frac{1}{2}\lambda_3c_\beta^2+
c_\beta s_\beta\real(\lambda_7{\rm e}^{i\xi})\right]+c_\beta^2 M_{H^\pm}^2\,,
\nonumber \\
\imag(m_{12}^2{\rm e}^{i\xi}) &=& -
\frac{v^2}{2}\left[
2\ c_\beta s_\beta\imag(\lambda_5{\rm e}^{2i\xi})+
c_\beta^2\imag(\lambda_6{\rm e}^{i\xi})+
s_\beta^2\imag(\lambda_7{\rm e}^{i\xi})
\right]\,,
\end{eqnarray}
with the square of the charged Higgs-boson mass
\begin{equation}
\label{eq:2hdm_mch2}
M_{H^\pm}^2=
-\frac{\real(m_{12}^2{\rm e}^{i\xi})}{c_\beta s_\beta}
-\frac{v^2}{2c_\beta s_\beta}\left[\lambda_4 c_\beta s_\beta+
2\ c_\beta s_\beta\real(\lambda_5{\rm e}^{2i\xi})+
c_\beta^2\real(\lambda_6{\rm e}^{i\xi})+
s_\beta^2\real(\lambda_7{\rm e}^{i\xi})
\right]\,.
\end{equation}
Then, including the vacuum expectation value $v$, in general
we need the following 13 parameters
plus 1 sign:
\footnote{Note that all the parameters are
neither basis independent nor physical.
There are only 11 physical degrees of freedom in the potential
as counted in, for example, Ref.~\cite{Branco:2011iw}.}
\begin{eqnarray}
\label{eq:2hdmpara}
&& v \,, t_\beta\,, |m_{12}| \,;  \nonumber \\
&& \lambda_1\,, \lambda_2\,,\lambda_3\,,\lambda_4\,,
|\lambda_5|\,,|\lambda_6|\,,|\lambda_7|\,; \nonumber \\
&&
\phi_5+2\xi\,,\phi_6+\xi\,,\phi_7+\xi\,,{\rm sign}[\cos(\phi_{12}+\xi)]\,,
\end{eqnarray}
to fully specify the general 2HDM scalar potential
in a form given by Eq.~(\ref{V2HDM}).
Here $m_{12}^2=|m_{12}^2|\,e^{i\phi_{12}}$ and
$\lambda_{5,6,7}=|\lambda_{5,6,7}|\, e^{i\phi_{5,6,7}}$ and
we note that $\sin(\phi_{12}+\xi)$ is fixed by the CP-odd tadpole condition
if the CP phases $\phi_5+2\xi\,,\phi_6+\xi$ and $\phi_7+\xi$ are
given together with $|m_{12}|$, $|\lambda_{5,6,7}|$, $v$, and $t_\beta$
and, accordingly,
$\cos(\phi_{12}+\xi)$ is determined up to a two-fold ambiguity.
One may take the convention with $\xi=0$ corresponding to
re-defining the 1 quadratic and 3 quartic complex parameters without loss of generality

\medskip

The 2HDM Higgs potential includes the mass
terms which can be cast into the form consisting of two parts
\begin{equation}
V_{\rm 2HDM\,,mass}=
M_{H^\pm}^2 H^+ H^- \
+ \ \frac{1}{2}
(\phi_1 \ \phi_2  \ a)\,{\cal M}^2_0\,
\left(\begin{array}{c}
\phi_1 \\ \phi_2  \\ a \end{array}\right)\,,
\end{equation}
in terms of the charged Higgs boson $H^-$, two
neutral scalars $\phi_{1,2}$, and one neutral pseudoscalar $a$
after absorbing the charged and neutral Goldstone bosons $G^-$ and
$G^0$ in the 2-state mixings of the two charged scalars and
two neutral pseudoscalars as
\begin{equation}
\left(\begin{array}{c} \phi_1^- \\ \phi_2^- \end{array}\right) =
\left(\begin{array}{rr}
c_\beta & -s_\beta \\ s_\beta & c_\beta \end{array}\right)\,
\left(\begin{array}{c} G^- \\ H^- \end{array}\right) \,; \ \ \
\left(\begin{array}{c} a_1 \\ a_2 \end{array}\right) =
\left(\begin{array}{rr}
c_\beta & -s_\beta \\ s_\beta & c_\beta \end{array}\right)\,
\left(\begin{array}{c} G^0 \\ a \end{array}\right) \,.
\end{equation}
And the $3\times 3$ real and symmetric
mass-squared matrix of the neutral Higgs bosons ${\cal M}_0^2$ is given by
\begin{equation}
\label{eq:2hdmm0sq}
{\cal M}^2_0 = M_A^2 \left(\begin{array}{ccc}
s_\beta^2 & -s_\beta c_\beta & 0 \\
-s_\beta c_\beta & c_\beta^2 & 0 \\
0 & 0 & 1 \end{array}\right) \ + \ {\cal M}^2_\lambda\,,
\end{equation}
with (reinstating the relative phase $\xi$ for the sake of generality)
\begin{eqnarray}
\label{eq:2hdmL4L5}
M_A^2&=&M_{H^\pm}^2+
\left[\frac{1}{2}\lambda_4 -\real(\lambda_5{\rm e}^{2i\xi})\right]v^2\,,
\end{eqnarray}
and the second part expressed in terms of the quartic couplings as
\begin{equation}
\frac{{\cal M}^2_\lambda}{v^2} = \left(\begin{array}{lll}
2\lambda_1 c_\beta^2 +2\real(\lambda_5{\rm e}^{2i\xi})s_\beta^2 &
2\lambda_{34}c_\beta s_\beta + \real(\lambda_6{\rm e}^{i\xi}) c_\beta^2 &
-\imag(\lambda_5{\rm e}^{2i\xi})s_\beta \\
+2\real(\lambda_6{\rm e}^{i\xi}) s_\beta c_\beta &
+\real(\lambda_7{\rm e}^{i\xi}) s_\beta^2 &
-\imag(\lambda_6{\rm e}^{i\xi}) c_\beta \\[5mm]
2\lambda_{34}c_\beta s_\beta + \real(\lambda_6{\rm e}^{i\xi}) c_\beta^2 &
2\lambda_2 s_\beta^2 +2\real(\lambda_5{\rm e}^{2i\xi})c_\beta^2 &
-\imag(\lambda_5{\rm e}^{2i\xi})c_\beta \\
+\real(\lambda_7{\rm e}^{i\xi}) s_\beta^2 &
+2\real(\lambda_7{\rm e}^{i\xi}) s_\beta c_\beta &
-\imag(\lambda_7{\rm e}^{i\xi}) s_\beta \\[3mm]
-\imag(\lambda_5{\rm e}^{2i\xi})s_\beta &
-\imag(\lambda_5{\rm e}^{2i\xi})c_\beta &
~~~~~~~~~0 \\
-\imag(\lambda_6{\rm e}^{i\xi}) c_\beta &
-\imag(\lambda_7{\rm e}^{i\xi}) s_\beta &
\end{array}\right)\,,
\end{equation}
where the abbreviation $\lambda_{34}=(\lambda_3+\lambda_4)/2$ and,
in passing, we note $v=2M_W/g$, $a=-s_\beta a_1+c_\beta a_2$ and
$H^+=-s_\beta \phi_1^++c_\beta \phi_2^+$.
We need to specify, therefore, the
13 parameters plus 1 sign listed in Eq.~(\ref{eq:2hdmpara})
to fix the mass-squared matrix.

\medskip

Once the real and symmetric mass-squared matrix
${\cal M}_0^2$ is given,
the orthogonal $3\times 3$ mixing matrix $O$
is defined through
\begin{eqnarray}
(\phi_1,\phi_2,a)^T_\alpha&=&O_{\alpha i} (H_1,H_2,H_3)^T_i\,,
\end{eqnarray}
such that $O^T {\cal M}_0^2 O={\rm diag}(M_{H_1}^2,M_{H_2}^2,M_{H_3}^2)$
with the increasing ordering of $M_{H_1}\leq M_{H_2}\leq M_{H_3}$.

\subsubsection{\it Interactions of Higgs bosons with massive vector bosons}

The cubic interactions of the neutral and charged Higgs bosons
with the massive gauge bosons $Z$ and $W^\pm$
are described by the three interaction Lagrangians:
\begin{eqnarray}
{\cal L}_{HVV} & = & g\,M_W \, \left(W^+_\mu W^{- \mu}\ + \
\frac{1}{2c_W^2}\,Z_\mu Z^\mu\right) \, \sum_i \,g_{_{H_iVV}}\, H_i
\,,\nonumber\\[3mm]
{\cal L}_{HHZ} &=& \frac{g}{2c_W} \sum_{i>j} g_{_{H_iH_jZ}}\, Z^{\mu}
(H_i\, \!\stackrel {\leftrightarrow} {\partial}_\mu H_j) \,, \nonumber\\[3mm]
{\cal L}_{HH^\pm W^\mp} &=& -\frac{g}{2} \, \sum_i \, g_{_{H_iH^+
W^-}}\, W^{-\mu} (H_i\, i\!\stackrel{\leftrightarrow}{\partial}_\mu
H^+)\, +\, {\rm h.c.}\,,
\end{eqnarray}
respectively, where $X\stackrel{\leftrightarrow}{\partial}_\mu Y
=X\partial_\mu Y -(\partial_\mu X)Y$, $i,j =1,2,3$ and the normalized couplings
$g_{_{H_iVV}}$, $g_{_{H_iH_jZ}}$ and $g_{_{H_iH^+
W^-}}$ are given in terms of the neutral Higgs-boson $3\times 3$
mixing matrix $O$ by (note that det$(O)=\pm1$ for any orthogonal matrix $O$):
\begin{eqnarray}
\label{eq:2hdmhvvetc}
g_{_{H_iVV}} &=& c_\beta\, O_{\phi_1 i}\: +\: s_\beta\, O_{\phi_2 i}
\, ,\nonumber \\
g_{_{H_iH_jZ}} &=& {\rm sign} [{\rm det}(O)] \, \, \varepsilon_{ijk}\,
g_{_{H_kVV}}\,,  \nonumber \\
g_{_{H_iH^+ W^-}} &=& c_\beta\, O_{\phi_2 i} - s_\beta\, O_{\phi_1 i}
- i O_{ai} \, ,
\end{eqnarray}
leading to the following sum rules:
\begin{equation}
\label{eq:2hdmsumrule}
\sum_{i=1}^3\, g_{_{H_iVV}}^2\ =\ 1\,\quad{\rm and}\quad
g_{_{H_iVV}}^2+|g_{_{H_iH^+ W^-}}|^2\ =\ 1\,\quad {\rm for~ each}~
i=1,2,3\,.
\end{equation}
On the other hand, the quartic interactions of
the neutral and charged Higgs bosons
with the massive gauge bosons $Z$ and $W^\pm$ and
massless photons are given by
\begin{equation}
{\cal L}_{HHVV}=\frac{1}{v^2}\left(M_W^2 W_\mu^+ M^{\mu -} +\frac{M_Z^2}{2} Z_\mu Z^\mu
\right) \sum_{i,j=1}^3 g_{_{H_iH_jVV}} H_i H_j\,,
\end{equation}
with $g_{_{H_i H_j VV}}=\delta_{ij}$ and
\begin{eqnarray}
{\cal L}_{H^+ H^- V V} & = &
\left(
\frac{g^2}{2} W_\mu^+W^{\mu -}
+ \frac{g_Z^2 c_{2W}^2}{4} Z^\mu Z_\mu  + e^2 A^\mu A_\mu
+ {e\, g_Z\, c_{2W}^2} A^\mu Z_\mu \right)
H^+ H^-\,, \nonumber \\
{\cal L}_{H^\pm H ZW^\mp} & = & \frac{g_Z\,g\, s_W^2}{2}  \left(
Z_\mu W^{-\,\mu} \sum_{i=1}^3  g_{_{Z W^- H^+ H_i}}  H^+ H_i
+{\rm h.c.} \right)\,, \nonumber \\
{\cal L}_{H^\pm H AW^\mp} & = & -\frac{e\, g}{2} \left(
 A_\mu W^{-\,\mu} \sum_{i=1}^3 g_{_{A W^- H^+ H_i}}  H^+ H_i + {\rm h.c.}
\right)\,,
\end{eqnarray}
with $g_{_{ZW^-H^+H_i}} = g_{_{A W^- H^+ H_i}}
= s_\beta\, O_{\phi_1 i}\: -\: c_\beta\, O_{\phi_2 i}\, - i O_{a i}$,
$c_{2W}=\cos2\theta_W$, and $g_Z=g/c_W=e/(s_Wc_W)$.

\subsubsection{\it Interactions of Higgs bosons with the SM fermions}
Without loss of generality, the Yukawa couplings in 2HDMs
could be cast into the form~\cite{Cheung:2013rva}:
\begin{eqnarray}
-{\cal L}_Y&=& h_u\, \overline{u_R}\, Q^T\,(i\tau_2)\,\Phi_2
-h_d\, \overline{d_R}\, Q^T\,(i\tau_2)\,
\left(\eta_1^d\,\widetilde\Phi_1 +\eta_2^d\,\widetilde\Phi_2\right)
\nonumber \\[2mm] &-&
h_l\, \overline{l_R}\, L^T\,(i\tau_2)\,
\left(\eta_1^l\,\widetilde\Phi_1 +\eta_2^l\,\widetilde\Phi_2\right)
\ + \ {\rm h.c.}\,,
\end{eqnarray}
where $\widetilde\Phi_i=i\tau_2 \Phi_i^*$ and
$Q^T=(u_L\,,d_L)$ and $L^T=(\nu_L\,,l_L)$
with $u$ and $d$ standing for three up- and down-type quarks,
respectively, and $l$ for three charged leptons.
We note that there is a freedom to
redefine the two linear combinations of $\Phi_2$ and $\Phi_1$ to eliminate
the coupling of the up-type quarks to $\Phi_1$~\cite{Davidson:2005cw}. The 2HDMs
are classified according to the values of $\eta_{1,2}^l$ and $\eta_{1,2}^d$
as in Table~\ref{tab:2hdtype}.
\begin{table}[!hbt]
\caption{\label{tab:2hdtype}
Classification of 2HDMs satisfying the Glashow-Weinberg condition
\cite{Glashow:1976nt}
which guarantees the absence of
tree-level Higgs-mediated
flavor-changing neutral current (FCNC).
}
\begin{center}
\begin{tabular}{|l|cccc|}
\hline
\hline
         & \hspace{0.5cm} 2HDM I\hspace{0.5cm} & 2HDM II\hspace{0.5cm}
& 2HDM III\hspace{0.5cm} & 2HDM IV\hspace{0.5cm}  \\
\hline
$\eta_1^d$  & $0$ & $1$ & $0$ & $1$  \\
$\eta_2^d$  & $1$ & $0$ & $1$ & $0$  \\
\hline
\hline
$\eta_1^l$  & $0$ & $1$ & $1$ & $0$  \\
$\eta_2^l$  & $1$ & $0$ & $0$ & $1$  \\
\hline
\hline
\end{tabular}
\end{center}
\end{table}

\medskip

By identifying the couplings in terms of the vev $v$ and the mixing
angle $\beta$ as
\footnote{
Here we take the convention with $\xi=0$ and the couplings $h_{u,d,l}$ are
supposed to be real without loss of generality.}
\begin{equation}
h_u = \frac{\sqrt{2}m_u}{v}\,\frac{1}{s_\beta}\,; \ \
h_d = \frac{\sqrt{2}m_d}{v}\,\frac{1}{\eta_1^dc_\beta+\eta_2^d s_\beta}\,; \ \
h_l = \frac{\sqrt{2}m_l}{v}\,\frac{1}{\eta_1^lc_\beta+\eta_2^l s_\beta}\,,
\end{equation}
we obtain the following Lagrangians
\begin{eqnarray}
\label{eq:nhff.2hdm}
-{\cal L}_{H_i\bar{f}f} &=&
\frac{m_u}{v}\left[
\bar{u}\, \left(
\frac{O_{\phi_2 i}}{s_\beta} -i\,\frac{c_\beta}{s_\beta}O_{ai}\,\gamma_5\,
\right)\,u \right]\,H_i
\nonumber \\[2mm] &+&
\frac{m_d}{v}\left[
\bar{d}\, \left(
\frac{\eta_1^dO_{\phi_1 i}+\eta_2^dO_{\phi_2
i}}{\eta_1^dc_\beta+\eta_2^ds_\beta}
-i\,\frac{\eta_1^ds_\beta-\eta_2^dc_\beta}{\eta_1^dc_\beta+\eta_2^ds_\beta}
O_{ai}\,\gamma_5\,
\right)\,d \right]\,H_i
\nonumber \\[2mm] &+&
\frac{m_l}{v}\left[
\bar{l}\, \left(
\frac{\eta_1^lO_{\phi_1 i}+\eta_2^lO_{\phi_2
i}}{\eta_1^lc_\beta+\eta_2^ls_\beta}
-i\,\frac{\eta_1^ls_\beta-\eta_2^lc_\beta}{\eta_1^lc_\beta+\eta_2^ls_\beta}
O_{ai}\,\gamma_5\,
\right)\,l \right]\,H_i\,,
\end{eqnarray}
for the interactions of neutral Higgs bosons with fermion pairs.
For the interactions of the charged Higgs boson with fermions,
\begin{eqnarray}
\label{eq:chff.2hdm}
-{\cal L}_{H^\pm\bar{u}d} &=&
-\frac{\sqrt{2}m_u}{v}
\left(\frac{c_\beta}{s_\beta}\right) \,\bar{u}\,P_L\,d\,H^+
- \frac{\sqrt{2}m_d}{v}\left(\,
\frac{\eta_1^ds_\beta-\eta_2^dc_\beta}{\eta_1^dc_\beta+\eta_2^ds_\beta}
\right)\,\bar{u}\,P_R\,d\,H^+
\nonumber \\ &&
- \frac{\sqrt{2}m_l}{v}\left(\,
\frac{\eta_1^ls_\beta-\eta_2^lc_\beta}{\eta_1^lc_\beta+\eta_2^ls_\beta}
\right)\,\bar{\nu}\,P_R\,l\,H^+
\ + \ {\rm h.c.}\,,
\end{eqnarray}
where $P_R=(1+\gamma_5)/2$ and $P_{L}=(1-\gamma_5)/2$.

\subsubsection{\it Higgs-boson self-interactions}
Given the orthogonal mixing matrix $O$ diagonalizing
the mass-squared matrix of the neutral Higgs bosons,
the cubic and quartic Higgs-boson self-couplings
are given in terms of the Higgs mass eigenstates
by~\cite{Choi:1999uk,Choi:2001pg,Choi:2002zp,Carena:2002bb}:
%
\begin{eqnarray} \label{hselfcoup}
-{\cal L}_{3H} & = &v\,\sum_{i\geq j\geq k=1}^3 g_{_{H_iH_jH_k}}\, H_i
  H_j H_k\: +\: v\,\sum_{i=1}^3 g_{_{H_iH^+H^-}}\,H_iH^+H^-\,,\\[3mm]
-{\cal L}_{4H} & =& \sum_{i\geq j\geq k\geq l =1}^3
  g_{_{H_iH_jH_kH_l}}\, H_i H_j H_k H_l\
+\ \sum_{i\geq j =1}^3 g_{_{H_iH_j H^+H^-}}\, H_i H_j H^+
  H^-\nonumber\\
&&+\  g_{_{H^+H^-H^+H^-}}\, (H^+ H^-)^2\,,\
\end{eqnarray}
where the normalized cubic and quartic weak-eigenstate
couplings are
\footnote{Here, the indices
$\alpha$, $\beta$, and $\gamma$ count the Higgs weak eigenstates of
$\phi_1$, $\phi_2$, and $a$ and the inequalities among them imply that
the cubic and quartic terms in the Higgs  potential are ordered in
the weak eigenstate basis.}
\begin{eqnarray}
\label{eq:2hdmSelf}
g_{_{H_iH_jH_k}} &=& \sum_{\alpha\leq \beta\leq \gamma=1}^3
\left\{O_{\alpha i} O_{\beta j} O_{\gamma  k} \right\}\,
g_{\alpha\beta\gamma} \,, \quad
g_{_{H_iH^+H^-}}\ =\ \sum_{\alpha =1}^3 O_{\alpha i}\,
g_{_{\alpha H^+H^-}}\,,\quad
\label{eq:2hdm22}\\[3mm]
g_{_{H_iH_jH_kH_l}} &=&  \sum_{\alpha\leq \beta\leq \gamma \leq \delta=1}^3
\left\{O_{\alpha i} O_{\beta j} O_{\gamma  k} O_{\delta  l} \right\}\,
g_{\alpha\beta\gamma\delta} \,, \nonumber\\[3mm]
g_{_{H_iH_j H^+H^-}} & =& \sum_{\alpha\leq \beta =1}^3
\left\{O_{\alpha i} O_{\beta j} \right\}\,
g_{\alpha\beta {\scriptscriptstyle H^+H^-}}\,.
\label{eq:2hdm23}
\end{eqnarray}
We note again that, in Eqs.~(\ref{eq:2hdm22}) and (\ref{eq:2hdm23}),
the expressions within
the curly brackets $\{ \cdots \}$ need to be symmetrized fully
with respect to
the indices $i,j,k,l$ and divided by the corresponding symmetry factors in
cases where two or more indices are the same as in, for example,
Eq.~(\ref{eq:cxsm.selfsym}).

\medskip

For the sake of completeness, we present all the effective cubic
and quartic Higgs--boson self--couplings of the Higgs weak eigenstates.
The cubic self-couplings of the neutral Higgs
bosons are given by
\footnote{
The relative phase $\xi$ could be reinstated by replacing
$\lambda_5$ with $\lambda_5{\rm e}^{2i\xi}$ and
$\lambda_{6,7}$ with $\lambda_{6,7}{\rm e}^{i\xi}$, if necessary.}
\begin{eqnarray}
\label{eq:2hdmTriSelf}
g_{_{\phi_1\phi_1\phi_1}} \!&=&\!\! c_\beta\lambda_1\: +\:
\frac{1}{2}\,s_\beta\,\real\lambda_6\,, \nonumber\\
g_{_{\phi_1\phi_1\phi_2}} \!&=&\!\! s_\beta\, \lambda_{34}\:
+\: s_\beta\, \real\lambda_5\: +\: \frac{3}{2}\,
c_\beta\, \real\lambda_6\,,\nonumber\\
g_{_{\phi_1\phi_2\phi_2}} \!&=&\!\! c_\beta\, \lambda_{34}\:
+\: c_\beta\, \real\lambda_5\: +\: \frac{3}{2}\,
s_\beta\, \real\lambda_7\,,\nonumber\\
g_{_{\phi_2\phi_2\phi_2}} \!&=&\!\! s_\beta \lambda_2\: +\:
\frac{1}{2}\, c_\beta\,\real\lambda_7\,,\nonumber\\
g_{_{\phi_1\phi_1 a}} \!&=&\!\! -s_\beta c_\beta\, \imag\lambda_5\: -\:
\frac{1}{2}\, (1 + 2c^2_\beta)\,\imag\lambda_6\,,\nonumber\\
g_{_{\phi_1\phi_2 a}} \!&=&\!\! -2\imag\lambda_5\: -\: s_\beta c_\beta\,
\imag\,(\lambda_6 + \lambda_7)\,,\nonumber\\
g_{_{\phi_2\phi_2 a}} \!&=&\!\! -s_\beta c_\beta\,\imag\lambda_5\:
-\: \frac{1}{2}\, (1 + 2s^2_\beta)\, \imag\lambda_7\,,\nonumber\\
g_{_{\phi_1 aa}} \!&=&\!\! s^2_\beta c_\beta\lambda_1\: +\:
c^3_\beta\,\lambda_{34}\: -\: c_\beta (1 +
s^2_\beta)\, \real\lambda_5\: +\: \frac{1}{2}\, s_\beta (s^2_\beta -
2c^2_\beta)\,\real\lambda_6\: +\:
\frac{1}{2} s_\beta c^2_\beta\, \real\lambda_7\,,\nonumber\\
g_{_{\phi_2 aa}} \!&=&\!\! s_\beta c^2_\beta\lambda_2\: +\:
s^3_\beta\, \lambda_{34}\: -\: s_\beta (1 +
c^2_\beta)\, \real\lambda_5\: +\: \frac{1}{2} s^2_\beta
c_\beta\, \real\lambda_6\: +\: \frac{1}{2}\, c_\beta (c^2_\beta -
2s^2_\beta)\,\real\lambda_7\,,\nonumber\\
g_{_{aaa}} \!&=&\!\! s_\beta c_\beta\, \imag\lambda_5\: -\:
\frac{1}{2}\,s^2_\beta\, \imag\lambda_6\: -\: \frac{1}{2}\,
c^2_\beta\, \imag\lambda_7\, ,
\end{eqnarray}
with the abbreviation $\lambda_{34} = \frac{1}{2}\, (\lambda_3
+ \lambda_4)$. The  effective cubic couplings
$g_{_{\alpha H^+H^-}}$ read:
\begin{eqnarray}
\label{eq:2hdmTriSelfCH}
g_{_{\phi_1H^+H^-}}  \!\!&=& 2s^2_\beta c_\beta\lambda_1\: +\:
c^3_\beta\lambda_3\: -\: s^2_\beta c_\beta \lambda_4\: -\:
2s^2_\beta c_\beta\, \real \lambda_5\:
+\: s_\beta (s^2_\beta - 2c^2_\beta)\, \real\lambda_6\nonumber\\
&&+\: s_\beta c^2_\beta \real \lambda_7\, ,\nonumber\\
g_{_{\phi_2 H^+H^-}} \!\!&=& 2s_\beta c^2_\beta \lambda_2\: +\:
s^3_\beta\lambda_3\: -\: s_\beta c^2_\beta \lambda_4\: -\:
2s_\beta c^2_\beta \, \real \lambda_5\:
+\: s^2_\beta c_\beta \, \real\lambda_6\nonumber\\
&&+\: c_\beta (c^2_\beta - 2s^2_\beta)\, \real \lambda_7\, ,\nonumber\\
g_{_{aH^+H^-}} \!\!&=& 2s_\beta c_\beta\, \imag\lambda_5\: -\:
s^2_\beta\, \imag\lambda_6\: -\: c^2_\beta\, \imag\lambda_7\, .
\end{eqnarray}
On the other hand, the quartic couplings for the neutral Higgs bosons are
\begin{eqnarray}
  \label{4H}
g_{_{\phi_1\phi_1\phi_1\phi_1}} \!&=&\! \frac{1}{4}\, \lambda_1\,,\qquad
g_{_{\phi_1\phi_1\phi_1\phi_2}} \ =\  \frac{1}{2}\, \real\lambda_6\,,\qquad
g_{_{\phi_1\phi_1\phi_2\phi_2}} \ =\
  \frac{1}{2}\,\lambda_{34}\: +\: \frac{1}{2}\,\real\lambda_5\,,\nonumber\\
g_{_{\phi_1\phi_2\phi_2\phi_2}} \!&=&\! \frac{1}{2}\, \real\lambda_7\,,\qquad
g_{_{\phi_2\phi_2\phi_2\phi_2}} \ =\ \frac{1}{4}\,\lambda_2\,,\nonumber\\
g_{_{\phi_1\phi_1\phi_1a}} \!&=&\!
      -\frac{1}{2}\,c_\beta\, \imag\lambda_6\,,\qquad
g_{_{\phi_1\phi_1\phi_2 a}} \ =\
   -c_\beta\,\imag\lambda_5\: -\: \frac{1}{2}\,s_\beta\,
\imag\lambda_6\,,\nonumber\\
g_{_{\phi_1\phi_2\phi_2 a}} \!&=&\!
     -s_\beta\,\imag\lambda_5\: -\: \frac{1}{2}\,c_\beta\,
\imag\lambda_7\,,\qquad
g_{_{\phi_2\phi_2\phi_2 a}}\ =\
-\frac{1}{2}\,s_\beta\,\imag\lambda_7\,,\nonumber\\
g_{_{\phi_1\phi_1 aa}} \!&=&\! \frac{1}{2}\,s^2_\beta\, \lambda_1\: +\:
\frac{1}{2}\, c^2_\beta\, \lambda_{34}\: -\: \frac{1}{2}\, c^2_\beta\,
\real\lambda_5\:
-\: \frac{1}{2}\, s_\beta c_\beta\, \real\lambda_6\,,\nonumber\\
g_{_{\phi_1\phi_2 aa}} \!&=&\! -2s_\beta c_\beta\, \real\lambda_5\: +\:
\frac{1}{2}\, s^2_\beta\, \real\lambda_6\: +\: \frac{1}{2}\,
c^2_\beta\, \real\lambda_7\,,\nonumber\\
g_{_{\phi_2\phi_2 aa}} \!&=&\! \frac{1}{2}\,c^2_\beta\, \lambda_2\: +\:
\frac{1}{2}\, s^2_\beta\, \lambda_{34}\: -\: \frac{1}{2}\, s^2_\beta\,
\real\lambda_5\:
-\: \frac{1}{2}\, s_\beta c_\beta\, \real\lambda_7\,,\nonumber\\
g_{_{\phi_1 aaa}} \!&=&\! s_\beta c^2_\beta\, \imag\lambda_5\: -\:
\frac{1}{2}\, s^2_\beta c_\beta\, \imag\lambda_6\: -\:
\frac{1}{2}\, c^3_\beta\, \imag\lambda_7\,,\nonumber\\
g_{_{\phi_2 aaa}} \!&=&\! s^2_\beta c_\beta\, \imag\lambda_5\: -\:
\frac{1}{2}\, s^3_\beta\, \imag\lambda_6\: -\:
\frac{1}{2}\, s_\beta c^2_\beta\, \imag\lambda_7\,,\nonumber\\
g_{_{aaaa}} \!&=&\! \frac{1}{4}\, g_{_{H^+H^-H^+H^-}}\; ,
\end{eqnarray}
together with the quartic coupling of the charged Higgs bosons given by
\begin{eqnarray}
  \label{4Hplus}
g_{_{H^+H^-H^+H^-}} &=&
s^4_\beta \lambda_1\: +\: c^4_\beta \lambda_2\:
+\: s^2_\beta c^2_\beta (\lambda_3 + \lambda_4)\: +\:
2s^2_\beta c^2_\beta \real\lambda_5\:
-\: 2s^3_\beta c_\beta \real\lambda_6
-\: 2s_\beta c^3_\beta \real\lambda_7\,.
\end{eqnarray}
Finally, the remaining quartic couplings involving the charged Higgs
boson pairs, $g_{\alpha\beta {\scriptscriptstyle H^+H^-}}$, are given
by
\begin{eqnarray}
  \label{2Hplus}
g_{_{\phi_1\phi_1 H^+H^-}} \!&=&\! s^2_\beta\, \lambda_1\: +\:
\frac{1}{2}\, c^2_\beta \, \lambda_3\: -\: s_\beta c_\beta\,
\real\lambda_6\,,\nonumber\\
g_{_{\phi_1\phi_2 H^+H^-}} \!&=&\! -\,s_\beta c_\beta\,\lambda_4\: -\:
2s_\beta c_\beta\, \real\lambda_5\: +\: s^2_\beta\,
\real\lambda_6\: +\: c^2_\beta\, \real\lambda_7\,,\nonumber\\
g_{_{\phi_2\phi_2 H^+H^-}} \!&=&\! c^2_\beta\, \lambda_2\: +\:
\frac{1}{2}\, s^2_\beta \, \lambda_3\: -\: s_\beta c_\beta\,
\real\lambda_7\,,\nonumber\\
g_{_{\phi_1 a H^+H^-}} \!&=&\!
2 s_\beta c^2_\beta\, \imag\lambda_5\: -\: s^2_\beta c_\beta\,
\imag\lambda_6\: -\: c^3_\beta \imag\lambda_7\,,\nonumber\\
g_{_{\phi_2 a H^+H^-}} \!&=&\!
2 s^2_\beta c_\beta\, \imag\lambda_5\: -\: s^3_\beta\,
\imag\lambda_6\: -\: s_\beta c^2_\beta \imag\lambda_7\,,\nonumber\\
g_{_{aa H^+H^-}} \!&=&\! g_{_{H^+H^-H^+H^-}}\; .
\end{eqnarray}

\subsection{Minimal Supersymmetric Extension of the SM }

In this subsection, we give a brief review of the minimal supersymmetric extension
of the SM (MSSM) with our particular focus on the MSSM Higgs
sector. For general reviews on the MSSM, see
Refs.$\,$\cite{Fayet:1976cr,Nilles:1983ge,Haber:1984rc,Barbieri:1987xf,
Martin:1997ns,Haber:1997if,Chung:2003fi,Drees:2004jm,Baer:2006rs,Binetruy:2006ad}.

\subsubsection{\it Potential and mixing}
The superpotential of the MSSM is written as
\footnote{Here and in the following, we
introduce a product between SU(2)$_L$ doublets as defined by
$A\cdot B=A^T i\tau_2 B=\epsilon_{ab}A^a B^b$ with $\epsilon_{12}=-\epsilon_{21}=+1$.}
\begin{equation}
W_{\rm MSSM}=
\widehat{U}^C {\bf h}_u \widehat{Q} \cdot \widehat{H}_u\:
+\:   \widehat{D}^C {\bf h}_d \widehat{H}_d  \cdot \widehat{Q}  \: +\:
\widehat{E}^C {\bf h}_e \widehat{H}_d \cdot \widehat{L} \: +\: \mu
\widehat{H}_u  \cdot \widehat{H}_d\ ,
\end{equation}
where $\widehat{H}_{u,d}$  are the  two Higgs chiral  superfields, and
$\widehat{Q}$,  $\widehat{L}$,  $\widehat{U}^C$,  $\widehat{D}^C$  and
$\widehat{E}^C$ are the left-handed doublet  and right-handed singlet
superfields related to up- and  down-type quarks and  charged leptons.
The  Yukawa couplings ${\bf  h}_{u,d,e}$ are in general $3\times  3$ complex
matrices describing  the charged-lepton   and   quark    masses   and
their  mixings. The superpotential contains one supersymmetry
preserving mass parameter, the  $\mu$ parameter that mixes the two
Higgs supermultiplets,
which has to be of the electroweak order for a natural realization of
the electroweak symmetry breaking mechanism without any significant fine
tuning.
See Table~\ref{tab:mssm_particles} for the full particle contents
appearing in the MSSM.
\begin{table}[!h]
\caption{\label{tab:mssm_particles}
Particle contents in the
minimal supersymmetric extension of the SM (MSSM).
Note the relations $Q=T_3+Y$ and
$\widetilde W^\pm=(\widetilde W^1\mp i \widetilde W^2)/\sqrt{2}$  and
$\widetilde W^0=\widetilde W^3$.
}
\begin{center}
\begin{tabular}{|c|c|c|c|c|c|c|c|c|}
\hline  &&&&&&&&\\[-3mm]
spin-0 & spin-$\frac{1}{2}$ & spin-1 &
$~$color$~$ & $T$ & $T_3$ & $Y$ & $B$ & $L$ \\[2mm]
\hline  &&&&&&&&\\ [-3mm]
$\widetilde Q_L=\left(\begin{array}{c} \widetilde U_L \\ \widetilde D_L\end{array}\right)$ &
$Q_L=\left(\begin{array}{c} U_L \\ D_L\end{array}\right)$ & & $3$ &
$~~\frac{1}{2}~~$ & $\left(\begin{array}{c} +1/2 \\ -1/2 \end{array}\right)$ &
$~+\frac{1}{6}~$ &  $~+\frac{1}{3}~$ & $~0~$ \\&&&&&&&&\\[-3mm]
$\widetilde U_R^*$ & $\left(U^C\right)_L$ & & $3^*$ & $0$ & $0$ & $-\frac{2}{3}$ &
$-\frac{1}{3}$ & $0$ \\&&&&&&&&\\[-3mm]
$\widetilde D_R^*$ & $\left(D^C\right)_L$ & & $3^*$ & $0$ & $0$ & $+\frac{1}{3}$ &
$-\frac{1}{3}$ & $0$ \\&&&&&&&&\\[-3mm]
$\widetilde L_L=\left(\begin{array}{c} \widetilde \nu_L \\ \widetilde E_L\end{array}\right)$ &
$L_L=\left(\begin{array}{c} \nu_L \\ E_L\end{array}\right)$ & & $1$ &
$\frac{1}{2}$ & $\left(\begin{array}{c} +1/2 \\ -1/2 \end{array}\right)$ &
$-\frac{1}{2}$ &  $0$ & $1$ \\&&&&&&&&\\[-3mm]
$\widetilde E_R^*$ & $\left(E^C\right)_L$ & & $1$ & $0$ & $0$ & $+1$ & $0$ &
$-1$ \\[2mm]
\hline  &&&&&&&&\\[-3mm]
 & $\widetilde g$ & $g$ & $8$ & $0$ & $0$ & $0$ & $0$ & $0$ \\&&&&&&&&\\[-3mm]
& $\left(\begin{array}{c} \widetilde W^+ \\ \widetilde W^0 \\ \widetilde W^- \end{array}\right)$ &
$\left(\begin{array}{c} W^+ \\ W^0 \\  W^- \end{array}\right)$ & $1$ & $1$ &
$\left(\begin{array}{c} +1 \\ ~~0 \\ -1 \end{array}\right)$ & $0$ & $0$ & $0$ \\&&&&&&&&\\[-3mm]
 & $\widetilde B$ & $B$ & $1$ & $0$ & $0$ & $0$ & $0$ & $0$ \\ &&&&&&&&\\ [-3mm]
\hline  &&&&&&&&\\ [-3mm]
$H_{1(d)} =\left(\begin{array}{c} H_1^0 \\ H_1^- \end{array}\right)$ &
$\widetilde H_{1L}=\left(\begin{array}{c} \widetilde H_1^0 \\ \widetilde H_1^-\end{array}\right)_L$ &
& $1$ & $\frac{1}{2}$ & $\left(\begin{array}{c} +1/2 \\ -1/2 \end{array}\right)$ &
$-\frac{1}{2}$ &  $0$ & $0$ \\ &&&&&&&&\\[-3mm]
$H_{2(u)} =\left(\begin{array}{c} H_2^+ \\ H_2^0 \end{array}\right)$ &
$\widetilde H_{2L}=\left(\begin{array}{c} \widetilde H_2^+ \\ \widetilde H_2^0\end{array}\right)_L$ &
& $1$ & $\frac{1}{2}$ & $\left(\begin{array}{c} +1/2 \\ -1/2 \end{array}\right)$ &
$+\frac{1}{2}$ &  $0$ & $0$ \\[5mm]
\hline
\end{tabular}
\end{center}
\end{table}

\medskip

In an  unconstrained version of the  MSSM, there are a  large number of
different  mass  parameters   present  in  the  soft  SUSY-breaking
Lagrangian
\begin{eqnarray}
  \label{Lsoft}
-{\cal L}_{\rm soft} &=& \frac{1}{2}\, \Big(\, M_1\,
 \widetilde{B}\widetilde{B}\: +\: M_2\, \widetilde{W}^i\widetilde{W}^i\:
+\: M_3\, \tilde{g}^a\tilde{g}^a\,  \ +\ {\rm h.c.}\Big)\: +\:
 \widetilde{Q}^\dagger
\widetilde{\bf M}^2_Q \widetilde{Q}\: +\:  \widetilde{L}^\dagger
\widetilde{\bf M}^2_L \widetilde{L}\: +\: \widetilde{U}^\dagger
\widetilde{\bf M}^2_U \widetilde{U}\nonumber\\
&&+\: \widetilde{D}^\dagger \widetilde{\bf M}^2_D \widetilde{D}\:
+\: \widetilde{E}^\dagger \widetilde{\bf M}^2_E \widetilde{E}\:
+\: M^2_{H_u} H^\dagger_u H_u\: +\: M^2_{H_d} H^\dagger_d H_d\:
+\: \Big(B\mu\, H_u \cdot H_d\ +\ {\rm h.c.}\Big)\nonumber\\
&&+\: \Big(  \widetilde{U}^\dagger {\bf a}_u \widetilde{Q} \cdot H_u  \:
+\: \widetilde{D}^\dagger {\bf a}_d H_d \cdot \widetilde{Q}  \: +\:
\widetilde{E}^\dagger {\bf a}_e H_d  \cdot \widetilde{L} \ +\ {\rm h.c.}\Big)\;.
\end{eqnarray}
Here $M_{1,2,3}$ are the soft SUSY-breaking masses associated with
one U(1)$_Y$ gaugino $\widetilde{B}$, three SU(2)$_L$ gauginos
$\widetilde{W}^i$, and eight SU(3)$_C$ gauginos $\widetilde{g}^a$, respectively.
In addition, $M^2_{H_{u,d}}$ and  $B\mu$ are the  soft masses related to  the Higgs
doublets $H_{u,d}$ and their bilinear mixing. Finally, $\widetilde{\bf
M}^2_{Q,L,D,U,E}$ are  the $3\times  3$ soft mass-squared  matrices of
squarks  and sleptons,  and  ${\bf a}_{u,d,e}$  are the  corresponding
$3\times  3$ soft  Yukawa  mass matrices.~\footnote{Alternatively,  the
soft  Yukawa mass  matrices ${\bf  a}_{u,d,e}$ may  be defined  by the
relation:  $({\bf a}_{u,d,e})_{ij}  =  ({\bf h}_{u,d,e})_{ij}\,  ({\bf
A}_{u,d,e})_{ij}$, where  the parameters $({\bf  A}_{u,d,e})_{ij}$ are
generically of order $M_{\rm  SUSY}$ in gravity-mediated SUSY breaking
models.} Hence, in addition
to the  $\mu$ term, the  unconstrained CP-violating MSSM  contains
109 real mass parameters including 46 CP phases:
\begin{eqnarray}
M_{1,2,3} \ &:& \ 3 \times ~\,2\,(1) = ~\,6\,(3) \,,
\nonumber \\
\widetilde{M}^2_{Q,L,D,U,E} \ &:& \ 5 \times ~\,9\,(3) = 45\,(15)\,,
\nonumber \\
{\bf a}_{u,d,e} \ &:& \ 3 \times 18\,(9) = 54\,(27)\,,
\nonumber \\
M_{H_{u,d}}^2 &:& \ 2 \times ~\,1\,(0) = ~\,2\,(0)\,,
\nonumber \\
B \ &:& \ \hphantom{1 \times}~\, ~\,2\,(1) = ~\,2\,(1)\,,
\end{eqnarray}
where, in each line,
the number of CP phases is separately counted in parentheses
for the corresponding soft-SUSY breaking parameters.
A caution is made in order to avoid misinterpreting
that the 109 parameters are all physical.
The number of new physical parameters is actually
105. Including them, the MSSM contains in total
124 physical parameters, as counted systematically
in, for example, Ref. \cite{Haber:1997if}.

\medskip

By identifying $H_u=\Phi_2$ and
$H_d=\widetilde\Phi_1=i\tau_2\Phi_1^*=(\phi_1^{0*}\,,-\phi_1^-)^T$,
\footnote{We recall $\Phi_1\ =\ \left(
\phi^+_1 \,, \ \ \frac{1}{\sqrt{2}} ( v_1\, +\, \phi_1\, +\, ia_1)
\right)^T$ and $\Phi_2\ =\ e^{i\xi}\, \left(
\phi^+_2 \,, \  \ \frac{1}{\sqrt{2}}\, ( v_2 \, +\, \phi_2\, +\, ia_2 )
\right)^T$ together with $a=-s_\beta a_1+c_\beta a_2$.}
one may obtain the same form of the Higgs potential as in the 2HDMs:
\begin{eqnarray}
V_{\rm MSSM} &=& \mu^2_1 (\Phi_1^\dagger\Phi_1)\, +\,
\mu^2_2 (\Phi_2^\dagger\Phi_2)\, +\, m^2_{12} (\Phi_1^\dagger \Phi_2)\,
+\, m^{*2}_{12} (\Phi_2^\dagger \Phi_1)\, +\,
\lambda_1 (\Phi_1^\dagger \Phi_1)^2\, +\,
\lambda_2 (\Phi_2^\dagger \Phi_2)^2\,\nonumber\\
&& +\, \lambda_3 (\Phi_1^\dagger \Phi_1)(\Phi_2^\dagger \Phi_2)\, +\,
\lambda_4 (\Phi_1^\dagger \Phi_2)(\Phi_2^\dagger \Phi_1)\, +\,
\lambda_5 (\Phi_1^\dagger \Phi_2)^2\, +\,
\lambda^*_5 (\Phi_2^\dagger \Phi_1)^2\, \\
&&+\, \lambda_6 (\Phi_1^\dagger \Phi_1)(\Phi_1^\dagger \Phi_2)\, +
\lambda^*_6 (\Phi_1^\dagger \Phi_1)(\Phi_2^\dagger \Phi_1)\,
+\, \lambda_7 (\Phi_2^\dagger \Phi_2)(\Phi_1^\dagger \Phi_2)\, +
\lambda^*_7 (\Phi_2^\dagger \Phi_2)(\Phi_2^\dagger \Phi_1)\,,
\nonumber
\end{eqnarray}
with the potential parameters given in terms of the $\mu$ and
the soft SUSY-breaking parameters as well as the SU(2)$_L$ and U(1)$_Y$ gauge
couplings by
\footnote{Note that $H_u\cdot H_d= H_u^T i\tau_2 H_d =
-\Phi_2^T\Phi_1^* =-\Phi_1^\dagger \Phi_2$.}
\begin{eqnarray}
  \label{LVpar}
\mu^2_1 &=& M_{H_d}^2 + |\mu|^2\, ,\qquad \mu^2_2\ =\ M_{H_u}^2 + |\mu|^2\, , \qquad
m_{12}^2\ = \ -B\mu\, , \nonumber \\
\lambda_1\ &=&\ \lambda_2\ =\ \frac{1}{8}\, (g^2 + g'^2)\, , \qquad
\lambda_3 = \frac{1}{4}\, (g^2 -g'^2)\, ,\qquad \lambda_4\ =\ -\frac{1}{2}\, g^2\, ,
\nonumber \\ \lambda_5\ &=&\ \lambda_6\ =\ \lambda_7\ =\ 0\,.
\end{eqnarray}
Note that the quartic couplings $\lambda_{1,2,3,4}$ are solely
determined by the gauge couplings and $\lambda_{5,6,7}$ are vanishing
at the tree level.
However, the quartic couplings $\lambda_{5,6,7}$ receive significant
radiative corrections from scalar-top and scalar-bottom loops and,
especially in the presence of CP-violating phases in the
soft SUSY-breaking terms, the CP-violating mixing among the
three neutral Higgs states are induced
\cite{Pilaftsis:1998pe,Pilaftsis:1998dd,Pilaftsis:1999qt,Demir:1999hj,
Choi:2000wz,Carena:2000yi,Carena:2001fw}. In this case,
as in the 2HDMs, the orthogonal $3\times 3$ mixing matrix $O$
has to be introduced for diagonalizing the $3\times 3$
real and symmetric
mass-squared matrix of three neutral Higgs states through
\begin{eqnarray}
(\phi_1,\phi_2,a)^T_\alpha&=&O_{\alpha i} (H_1,H_2,H_3)^T_i
\end{eqnarray}
with the increasing ordering of $M_{H_1}\leq M_{H_2}\leq M_{H_3}$.

\subsubsection{\it Interactions of Higgs bosons with
the SM particles and self-interactions}
The interactions of Higgs bosons with massive vector bosons and
those among themselves in the MSSM are formally
the same as in the 2HDMs.

\medskip

The tree-level
Higgs-boson interactions with the SM fermions are the same as those
in the type-II 2HDM
without including the finite loop-induced threshold corrections
mediated by the exchange of gluinos and charginos
\cite{Hempfling:1993kv,Hall:1993gn,Carena:1994bv,Pierce:1996zz,Borzumati:1999sp,
Babu:1998er}.
Including the threshold corrections, the couplings of neutral and
charged Higgs bosons to down-type fermions could be significantly
modified for large values of $t_\beta$.
More explicitly,
by resumming potentially large $\tan\beta$-enhanced
effects due to the threshold corrections,
the  down-type quark Yukawa couplings may take the form
\begin{equation}
h_{q_\downarrow}\ =\ \frac{\sqrt{2}\,m_{q_\downarrow}}{v\,c_\beta}\ \frac{1}{1+\Delta_{q_\downarrow}\,t_\beta}
\end{equation}
where  ${q_\downarrow}=d,s,b$ and the $t_\beta$-enhanced
threshold corrections enter through the one-loop quantities
\footnote{We note that the one-loop quantities are uncertain
at the level of about 10\% depending on scale choices for $\alpha_s$.}
\begin{eqnarray}
\Delta_d \!&=&\! \frac{2\alpha_s}{3\pi}\,\mu^*M_3^*\,
I(M_{\widetilde{Q}_1}^2,M_{\widetilde{D}_1}^2,|M_3|^2)\,, \nonumber \\
\Delta_s \!&=&\! \frac{2\alpha_s}{3\pi}\,\mu^*M_3^*\,
I(M_{\widetilde{Q}_2}^2,M_{\widetilde{D}_2}^2,|M_3|^2)\,, \nonumber \\
\Delta_b \!&=&\! \frac{2\alpha_s}{3\pi}\,\mu^*M_3^*\,
I(M_{\widetilde{Q}_3}^2,M_{\widetilde{D}_3}^2,|M_3|^2)
\ + \ \frac{|h_t|^2}{16\pi^2}\ \mu^* A_t^*
I(M_{\widetilde{Q}_3}^2,M_{\widetilde{U}_3}^2,|\mu|^2)\,,
\end{eqnarray}
with the loop function $I(a,b,c)$ defined as
\begin{equation}
I(a,b,c)\ =\
\frac{ab\ln(a/b) + bc\ln(b/c) + ac\ln(c/a)}
{(a-b)(b-c)(a-c)}\ .
\end{equation}
In the presence of CP-violating mixing in the neutral
Higgs sector,
the effective Lagrangian describing the Higgs interactions with up- and
down-type quarks is given by
\begin{eqnarray}
{\cal L}_{H\bar qq} &=&
-\sum_{i=1}^3 \sum_{q=q_\uparrow,q_\downarrow}\frac{m_q}{v}\left[ \bar{q}
\left(g^S_{_{H_i\bar{q}q}}+ig^P_{_{H_i\bar{q}q}}\gamma_5\right)q\right] H_i
\nonumber \\[2mm] &&
+\sum_{(q_\uparrow,q_\downarrow)}
\left\{\frac{\sqrt{2}\,m_{q_\uparrow}}{v} \left[ \bar{q}_\uparrow
\left(g^S_{_{H^+\bar{q}_\uparrow q_\downarrow}}+
ig^P_{_{H^+\bar{q}_\uparrow q_\downarrow}}\gamma_5\right) q_\downarrow\right] H^+
\ \ +{\rm h.c.}\right\}\,,
\end{eqnarray}
where $(q_\uparrow,q_\downarrow)=(u,d),(c,s),(t,b)$.
At the tree level, as in the type-II 2HDM,
\begin{equation}
\label{eq:mssm_hqq_tree}
g^S_{_{H_i\bar{q}q}}=O_{\phi_1\,i}/c_\beta\,, \ \
g^P_{_{H_i\bar{q}q}}=-t_\beta\,O_{a\,i}\,;  \ \ \
g^S_{_{H^+\bar{q}_\uparrow q_\downarrow}}\ =\
\frac{1}{2}\left[\frac{1}{t_\beta}+\frac{m_{q_\downarrow}}{m_{q_\uparrow}}\,
  t_\beta\right]\,, \ \ \
g^P_{_{H^+\bar{q}_\uparrow q_\downarrow}}\ =\
\frac{i}{2}\left[\frac{1}{t_\beta}-\frac{m_{q_\downarrow}}{m_{q_\uparrow}}\,
  t_\beta\right]\,,
\end{equation}
for the  neutral and charged Higgs bosons, respectively,
see Eqs.~(\ref{eq:nhff.2hdm}) and (\ref{eq:chff.2hdm}).
While, in the presence of $t_\beta$-enhanced threshold corrections, the couplings
$g^S_{_{H_i\bar{q}_\downarrow q_\downarrow }}$,
$g^P_{_{H_i\bar{q}_\downarrow q_\downarrow }}$,
$g^S_{_{H^+\bar{q}_\uparrow q_\downarrow}}$ and
$g^P_{_{H^+\bar{q}_\uparrow q_\downarrow}}$
are given by~\cite{Pilaftsis:2002fe,Borzumati:2004rd,Lee:2009up}:
\begin{eqnarray}
g^S_{H_i\bar{q}_\downarrow q_\downarrow } & =& \real\, \bigg(\,
\frac{1}{1\, +\, \Delta_{q_\downarrow}\,t_\beta}\,\bigg)\,
\frac{O_{\phi_1 i}}{c_\beta}
\ +\ \real\, \bigg(\, \frac{\Delta_{q_\downarrow}}{1\, +\,
\Delta_{q_\downarrow}\, t_\beta}\,\bigg)\
\frac{O_{\phi_2 i}}{c_\beta}
+\: \imag\, \bigg[\,
\frac{ \Delta_{q_\downarrow}\, (t^2_\beta\, +\, 1)}{1\, +\,
\Delta_{q_\downarrow}\, t_\beta}\,\bigg]\
O_{ai}\, , \nonumber\\[0.35cm]
g^P_{H_i\bar{q}_\downarrow q_\downarrow } & =& -\, \real\, \bigg(\,
\frac{ t_\beta\, -\, \Delta_{q_\downarrow}}{1\, +\, \Delta_{q_\downarrow} t_\beta}\,\bigg)\, O_{ai}
\ +\ \imag\, \bigg(\, \frac{\Delta_{q_\downarrow}\,t_\beta}{1\, +\,
\Delta_{q_\downarrow}\, t_\beta}\,\bigg)\
\frac{O_{\phi_1 i}}{c_\beta}
-\: \imag\, \bigg(\,
\frac{\Delta_{q_\downarrow}}{1\, +\, \Delta_{q_\downarrow}\, t_\beta}\,\bigg)\
\frac{O_{\phi_2 i}}{c_\beta}\ ;
\nonumber \\[0.35cm]
g^S_{_{H^+\bar{q}_\uparrow q_\downarrow}} &=&
\frac{1}{2}\left[\frac{1}{t_\beta}+\frac{m_{q_\downarrow}}{m_{q_\uparrow}}
\left(
\frac{t_\beta\,-\,\Delta_{q_\downarrow}^*}{1\,+\,\Delta_{q_\downarrow}^*\,t_\beta}
\right)\right]\,, \ \ \
g^P_{_{H^+\bar{q}_\uparrow q_\downarrow}} =
\frac{i}{2}\left[\frac{1}{t_\beta}-\frac{m_{q_\downarrow}}{m_{q_\uparrow}}
\left(
\frac{t_\beta\,-\,\Delta_{q_\downarrow}^*}{1\,+\,\Delta_{q_\downarrow}^*\,t_\beta}
\right)\right]\,.
\end{eqnarray}
Note that the above couplings approach to the tree-level ones
in Eq.~(\ref{eq:mssm_hqq_tree}) in the limit of $\Delta_{q_\downarrow}=0$.
The size of the resummed threshold corrections could be significant
enough to make the down-type quark Yukawa coupling $h_{q_\downarrow}$
as comparably large as the top-quark Yukawa coupling
when $\real(\Delta_{q_\downarrow})<0$  and $\tan\beta$ is large.
Some extreme cases with $|h_b| \sim 1$ and $|h_s| \sim 1$ are discussed,
for example, in Ref.~\cite{Borzumati:2004rd} and
Ref.~\cite{Lee:2009up}, respectively, taking account of
the CP-violating  mixing in neutral Higgs sector.

\subsubsection{\it Interactions of Higgs bosons with
the SUSY particles}
\label{sec:SUSYinteractions}

For the sake of completeness and reference, although no serious analyses on them
are presented in the present review,  we fix the convention for the interactions
of Higgs bosons with the supersymmetric (SUSY) particles
such as charginos, neutralinos and sfermions.

\medskip

The interactions of neutral Higgs bosons with charginos, which
are mixtures of charged gauginos and higgsinos, are described by the following
Lagrangian:
\begin{eqnarray}
{\cal L}_{H^0\wt{\chi}^+\wt{\chi}^-}
&=&-\frac{g}{\sqrt{2}}\sum_{i,j=1}^2\sum_{k=1}^3 H_k
\overline{\wt{\chi}_i^-}
\left(g_{H_k\tilde{\chi}^+_i\tilde{\chi}^-_j}^{S}+i\gamma_5
g_{H_k\tilde{\chi}^+_i\tilde{\chi}^-_j}^{P}\right)
\wt{\chi}_j^-\,,
\end{eqnarray}
with the normalized Higgs-chargino-chargino couplings
\begin{eqnarray}
g_{H_k\tilde{\chi}^+_i\tilde{\chi}^-_j}^{S}&=&\frac{1}{2}\left\{
[(C_R)_{i1}(C_L)^*_{j2}G^{\phi_1}_k+(C_R)_{i2}(C_L)^*_{j1}G^{\phi_2}_k]
+[i\leftrightarrow j]^* \right\}\,,
\nonumber \\
g_{H_k\tilde{\chi}^+_i\tilde{\chi}^-_j}^{P} &=&\frac{i}{2}\left\{
[(C_R)_{i1}(C_L)^*_{j2}G^{\phi_1}_k+(C_R)_{i2}(C_L)^*_{j1}G^{\phi_2}_k]
-[i\leftrightarrow j]^* \right\}\,,
\end{eqnarray}
where $G^{\phi_1}_k=(O_{\phi_1 k}-is_\beta O_{ak})$ and
$G^{\phi_2}_k=(O_{\phi_2 k}-ic_\beta O_{ak})$.
The two different unitary $2\times 2$ chargino mixing matrices
$(C_L)_{i\alpha}$ and $(C_R)_{i\alpha}$ are required to diagonalize
the chargino mass matrix
\begin{eqnarray}
{\cal M}_C = \left(\begin{array}{cc}
     M_2              & \sqrt{2} M_W\, c_{\beta} \\[2mm]
\sqrt{2} M_W\, s_{\beta} & \mu
             \end{array}\right)\, ,
\end{eqnarray}
in the $(\tilde{W}^-,\tilde{H}^-)_{L}$
and $(\tilde{W}^+,\tilde{H}^+)_{L}$ bases with the convention
$\tilde{H}^-_L=\tilde{H}^-_1$ and $\tilde{H}^+_{L}=\tilde{H}^+_2$
in such a way that
\begin{equation}
C_R{\cal M}_C C_L^\dagger ={\sf diag}\{m_{\tilde{\chi}^\pm_1},\,
m_{\tilde{\chi}^\pm_2}\}\,,
\end{equation}
with the increasing ordering of
$m_{\tilde{\chi}^\pm_1} \leq m_{\tilde{\chi}^\pm_2}$.
Explicitly, the mixing matrices
relate the electroweak eigenstates to the mass eigenstates, via
%
\begin{eqnarray}
    \tilde{W}^-_{L} &=& \sum_{i=1,2} (C_L)^*_{i1 } \tilde{\chi}_{iL}^-\,,\qquad
    \tilde{H}^-_{L}  = \sum_{i=1,2} (C_L)^*_{i2 } \tilde{\chi}_{iL}^-
                        \,,\nonumber\\
    \tilde{W}^-_{R} &=& \sum_{i=1,2} (C_R)^*_{i1 } \tilde{\chi}_{iR}^-\,,\qquad
    \tilde{H}^-_{R}  =  \sum_{i=1,2} (C_R)^*_{i2 } \tilde{\chi}_{iR}^-\,.
\end{eqnarray}
Note that the convention $\tilde{H}^{-}_{L(R)} = \tilde{H}^{-}_{1(2)}$
is adopted with the subscripts 1 and 2 being associated with the Higgs
supermultiplets leading to the tree-level mass generation of the down- and
up-type quarks, respectively,
see Table~\ref{tab:mssm_particles}.
We recall that we take the following abbreviations throughout this paper:
$s_\beta\equiv\sin\beta$, $c_\beta\equiv\cos\beta$,
$t_\beta\equiv\tan\beta$, $s_{2\beta}\equiv\sin\,2\beta$,
$c_{2\beta}\equiv\cos\,2\beta$, $s_W\equiv\sin\theta_W$,
$c_W\equiv\cos\theta_W$, etc.

\medskip

The interactions of three neutral Higgs bosons with
neutralinos, which are mixtures of 2 neutral gauginos and 2 neutral higgsinos,
are described by the following Lagrangian:
\begin{eqnarray}
  {\cal L}_{H^0\wt{\chi}^0\wt{\chi}^0}
= -\frac{g}{2}\sum_{i,j=1}^4\sum_{k=1}^3 H_k
   \overline{\wt{\chi}_i^0}
   \left(g_{H_k\tilde{\chi}^0_i\tilde{\chi}^0_j}^{S}
  +i\gamma_5
   g_{H_k\tilde{\chi}^0_i\tilde{\chi}^0_j}^{P}\right)
   \wt{\chi}_j^0 \,,
\end{eqnarray}
with the normalized Higgs-neutralino-neutralino couplings
\begin{eqnarray}
    g_{H_k\tilde{\chi}^0_i\tilde{\chi}^0_j}^{S}
&=& \frac{1}{2}\real{[(N_{j2}^*-t_W N_{j1}^*)
    (N_{i3}^*G^{\phi_1}_k-N_{i4}^*G^{\phi_2}_k)+(i\leftrightarrow j)]}\,,
    \nonumber \\
    g_{H_k\tilde{\chi}^0_i\tilde{\chi}^0_j}^{P}
&=&-\frac{1}{2}\imag{[(N_{j2}^*-t_W N_{j1}^*)
    (N_{i3}^*G^{\phi_1}_k-N_{i4}^*G^{\phi_2}_k)+(i\leftrightarrow j)]}\,,
\end{eqnarray}
where $i,j=1$-$4$ for the four neutralino states and $k=1$-$3$ for
the three neutral Higgs bosons.
One unitary $4\times 4$ neutralino mixing matrix is
required to render the $4\times 4$ symmetric neutralino mass
matrix expressed as
\begin{eqnarray}
{\cal M}_N=\left(\begin{array}{cccc}
  M_1       &      0          &  -M_Z c_\beta s_W  & M_Z s_\beta s_W \\[2mm]
   0        &     M_2         &   M_Z c_\beta c_W  & -M_Z s_\beta c_W\\[2mm]
-M_Z c_\beta s_W & M_Z c_\beta c_W &       0       &     -\mu        \\[2mm]
 M_Z s_\beta s_W &-M_Z s_\beta c_W &     -\mu      &       0
                  \end{array}\right)\,,
\end{eqnarray}
in the $(\widetilde{B},\widetilde{W}^3,\widetilde{H}^0_1,\widetilde{H}^0_2)_L$
basis into a diagonal matrix as
\begin{equation}
N^* {\cal M}_N N^\dagger = {\sf diag}\,
(m_{\widetilde{\chi}_1^0},m_{\widetilde{\chi}_2^0},m_{\widetilde{\chi}_3^0},m_{\widetilde{\chi}_4^0})
\,,
\end{equation}
with the increasing mass ordering of
$m_{\widetilde{\chi}_1^0} \leq m_{\widetilde{\chi}_2^0} \leq m_{\widetilde{\chi}_3^0}
\leq m_{\widetilde{\chi}_4^0}$.
The single neutralino mixing matrix $N_{i\alpha}$
relates the left-handed and right-handed electroweak eigenstates to
the left-handed and right-handed mass eigenstates via
\begin{eqnarray}
    (\widetilde{B},\widetilde{W}^3,\widetilde{H}^0_1,\widetilde{H}^0_2)^T_{\alpha L}
&=& N_{i\alpha}^*
    ({\widetilde{\chi}_1^0},{\widetilde{\chi}_2^0},{\widetilde{\chi}_3^0},
    {\widetilde{\chi}_4^0})^T_{iL} \ \  \ {\rm and} \nonumber\\
   (\widetilde{B},\widetilde{W}^3,\widetilde{H}^0_1,\widetilde{H}^0_2)^T_{\alpha R}
&=& N_{i\alpha}
    ({\widetilde{\chi}_1^0},{\widetilde{\chi}_2^0},{\widetilde{\chi}_3^0},
    {\widetilde{\chi}_4^0})^T_{iR}\,,
\end{eqnarray}
respectively.

\medskip

The interactions of the charged Higgs bosons
$H^\pm$ with charginos and neutralinos
are described by the following Lagrangian:
\begin{eqnarray}
  {\cal L}_{H^\pm\wt{\chi}_i^0\wt{\chi}_j^\mp}
= -\frac{g}{\sqrt{2}}\sum_{i=1}^4\sum_{j=1}^2 H^+
\,\overline{\wt{\chi}_i^0}
\left(g_{H^+\widetilde{\chi}^0_i\widetilde{\chi}^-_j}^{S}
+i\gamma_5
g_{H^+\widetilde{\chi}^0_i\widetilde{\chi}^-_j}^{P}\right)
\wt{\chi}_j^-
+{\rm h.c.}\,,
\end{eqnarray}
with the normalized couplings of the charged Higgs boson $H^+$
with a chargino and a neutralino
\begin{eqnarray}
    g_{H^+\widetilde{\chi}^0_i\widetilde{\chi}^-_j}^{S}
&=& \frac{1}{2}\left\{s_\beta\left[\sqrt{2}N_{i3}^*(C_L)_{j1}^*-(N_{i2}^*
    +t_W N_{i1}^*)(C_L)_{j2}^*\right] \right. \nonumber \\
&&\left. ~  +c_\beta\left[\sqrt{2}N_{i4}(C_R)_{j1}^*+(N_{i2}
   +t_W N_{i1})(C_R)_{j2}^*\right]\right\} \,, \nonumber \\
    g_{H^+\widetilde{\chi}^0_i\widetilde{\chi}^-_j}^{P}
&=&\frac{i}{2}\left\{s_\beta\left[\sqrt{2}N_{i3}^*(C_L)_{j1}^*-(N_{i2}^*
   +t_W N_{i1}^*)(C_L)_{j2}^*\right] \right. \nonumber \\
&&\left. ~  -c_\beta\left[\sqrt{2}N_{i4}(C_R)_{j1}^*+(N_{i2}
   +t_W N_{i1})(C_R)_{j2}^*\right]\right\} \,,
\end{eqnarray}
expressed in terms of the chargino and neutralino mixing matrices.

\medskip

The neutral Higgs--sfermion--sfermion interactions can be written
in terms of the sfermion mass eigenstates as
\begin{equation}
{\cal L}_{H\widetilde{f}\widetilde{f}}=v\sum_{f=u,d,l}\,\sum_{i=1}^3\sum_{j,k=1,2}\,
g_{H_i\widetilde{f}^*_j\widetilde{f}_k}
(H_i\,\widetilde{f}^*_j\,\widetilde{f}_k)\,,
\end{equation}
where the couplings of the Higgs bosons with sfermions in the mass eigenstate
basis
\begin{equation}
 v\,g_{H_i\widetilde{f}^*_j\widetilde{f}_k}
=\sum_{\alpha=\phi_1,\phi_2,a}~\sum_{\beta,\gamma=L,R}
\left(\Gamma^{\alpha\widetilde{f}^*\widetilde{f}}\right)_{\beta\gamma}
O_{\alpha i}U^{\widetilde{f}*}_{\beta j} U^{\widetilde{f}}_{\gamma k}\,,
\end{equation}
expressed in terms of the $2\times 2$ scalar--sfermion--sfermion
coupling matrix $\Gamma^{\alpha\widetilde{f}^*\widetilde{f}}$ in the weak-eigenstate
basis, of which the explicit form is given later
in Eq.~(\ref{eq:nhsfsf}),
and the $3\times 3$ Higgs and
$2\times 2$ sfermion mixing matrices,
$O$ and $U^{\widetilde{f}}$, with the convention of $\alpha=(\phi_1,\phi_2,a)=(1,2,3)$,
$\beta,\gamma = L, R$, $i=(H_1,H_2,H_3)=(1,2,3)$ and $j,k=1,2$
chosen properly for the sake of notational convenience.
Likewise, the charged Higgs-boson interactions with up-- and down--type
sfermions are given by
\begin{equation}
{\cal L}_{H^\pm\widetilde{f}\widetilde{f'}}=v\,
\sum_{(f,f')=(u,d),(\nu,l)}~\sum_{j,k=1,2}
g_{H^+\widetilde{f}^*_j\widetilde{f'}_k}
(H^+\,\widetilde{f}^*_j\,\widetilde{f'}_k) + {\rm h.c.},
\end{equation}
where the couplings of the charged Higgs boson $H^+$
with sfermions in the mass eigenstate basis
\begin{equation}
v\, g_{H^+\widetilde{f}^*_j\widetilde{f'}_k}\ =\ \sum_{\beta,\gamma=L,R}
\left(\Gamma^{H^+\widetilde{f}^*\widetilde{f'}}\right)_{\beta\gamma}\,
U^{\widetilde{f}*}_{\beta j}\, U^{\widetilde{f'}}_{\gamma k}\,,
\end{equation}
expressed in terms of the $2\times 2$ charged Higgs--sfermion--sfermion
coupling matrix $\Gamma^{H^+\widetilde{f}^*\widetilde{f}}$ in the weak-eigenstate
basis, of which the explicit form is given later
in Eqs.~(\ref{eq:chsqsq}) and (\ref{eq:chslsl}),
and the $2\times 2$ sfermion mixing
matrix $U^{\widetilde{f}}$ with the convention of $\beta,\gamma = L, R$ and
$j,k=1,2$.
The unitary $2\times 2$ sfermion mixing matrix $U^{\widetilde{f}}$
is obtained by diagonalizing
the $2\times 2$ sfermion mass matrix $\widetilde{\cal M}^2_f$
for $f=t, b$ and $\tau$ in such a way that
\begin{equation}
U^{\widetilde{f}\dagger} \, \widetilde{\cal M}^2_f \,
U^{\widetilde{f}} ={\sf diag}(m_{\widetilde{f}_1}^2,m_{\widetilde{f}_2}^2)\,,
\end{equation}
with the increasing mass ordering of
$m_{\widetilde{f}_1}^2 \leq m_{\widetilde{f}_2}^2$.  The mixing matrix
$U^{\widetilde{f}}$ relates the sfermion electroweak eigenstates
$\widetilde{f}_{L,R}$ to the sfermion mass eigenstates $\widetilde{f}_{1,2}$
via
\begin{equation}
(\widetilde{f}_L,\widetilde{f}_R)^T_\alpha\,=\,
U^{\widetilde{f}}_{\alpha i} \,
(\widetilde{f}_1,\widetilde{f}_2)^T_i\,.
\end{equation}
Explicitly, the stop and sbottom mass-squared matrices are written in the
$\left(\widetilde{q}_L, \widetilde{q}_R\right)$ electroweak basis as
\begin{equation}
  \label{Mscalar}
\hspace{-0.5 cm}
\widetilde{\cal M}^2_{q=t,b}  = \left( \begin{array}{cc}
M^2_{\widetilde{Q}_3}\, +\, m^2_q\, +\, c_{2\beta} M^2_Z\, ( T^q_z\, -\,
Q_q s_W^2 ) & h_q^* v_q (A^*_q - \mu R_q )/\sqrt{2}\\
h_q v_q (A_q - \mu^* R_q)/\sqrt{2} & \hspace{-0.2cm}
M^2_{\widetilde{R}_3}\, +\, m^2_q\, +\, c_{2\beta} M^2_Z\, Q_q s^2_W
\end{array}\right)\, ,
\end{equation}
with the third-generation left- and right-sfermion
soft SUSY-breaking mass-squared $M^2_{\widetilde{Q}_3}$
and $M^2_{\widetilde{R}_3=\widetilde{U}_3\,,\widetilde{D}_3}$,
a cubic soft-breaking term $A_q$,
$T^t_z = - T^b_z = 1/2$,
$Q_t = 2/3$, $Q_b = -1/3$,
$v_b=v_1$, $v_t=v_2$,
$R_b = \tan\beta = v_2/v_1$, $R_t = \cot\beta$, and
the Yukawa coupling $h_q$ of the quark $q$.
Similarly, the stau mass-squared matrix is written in the $\left(\widetilde{\tau}_L,
\widetilde{\tau}_R\right)$ electroweak basis as
\begin{eqnarray}
  \label{Mstau}
\hspace{-0.5 cm}
\widetilde{\cal M}^2_\tau  = \left( \begin{array}{cc}
M^2_{\widetilde{L}_3}\, +\, m^2_\tau\, +\, c_{2\beta} M^2_Z\, (s_W^2-1/2 )
   & h_\tau^* v_1 (A^*_\tau - \mu \tan\beta )/\sqrt{2}\\
h_\tau v_1 (A_\tau - \mu^* \tan\beta)/\sqrt{2} & \hspace{-0.2cm}
M^2_{\widetilde{E}_3}\, +\, m^2_\tau\, -\, c_{2\beta} M^2_Z\, s^2_W
\end{array}\right)\,,
\end{eqnarray}
derived directly from the sbottom mass-squared mass matrix by replacing
$b$ by $\tau$ and $\widetilde{Q}_3, \widetilde{R}_3$
by $\widetilde{L}_3, \widetilde{E}_3$, and taking $Q_\tau=-1$.
Incidentally,  the mass of the tau sneutrino
$\widetilde{\nu_\tau}$ is simply given by $m_{\widetilde\nu_\tau}
= \sqrt{ M^2_{\widetilde{L}_3} + \frac{1}{2}
c_{2\beta} M_Z^2 }$, as it has no right--handed counterpart in the
MSSM unlike the squark and charged slepton cases.

\medskip

For the sake of completeness and explicit analytic and numerical
calculations, we present the explicit form of
the Higgs--sfermion--sfermion couplings in the electroweak-interaction basis
for the third-generation sfermions.
The $2\times 2$ coupling matrices $\Gamma^{\alpha\widetilde{f}^*\widetilde{f}}$
are given in the $(\widetilde{f}_L, \widetilde{f}_R)$ basis
with $f=t, b, \tau, \nu_\tau$ and $\alpha=a, \phi_1, \phi_2$ by
\begin{eqnarray}
%
%
\label{eq:nhsfsf}
\Gamma^{a\widetilde{b}^*\widetilde{b}} &=& \frac{1}{\sqrt{2}}\left(
\begin{array}{cc}
0 & i\,h_b^*(s_\beta A_b^*+c_\beta \mu) \\
-i\,h_b(s_\beta A_b+c_\beta \mu^*) & 0
\end{array} \right)\,,
\nonumber \\
\Gamma^{\phi_1\widetilde{b}^*\widetilde{b}} &=& \left(
\begin{array}{cc}
-|h_b|^2vc_\beta+ \frac{1}{4}\left(g^2+\frac{1}{3}g^{\prime 2}\right)vc_\beta&
-\frac{1}{\sqrt{2}}h_b^*A_b^* \\
-\frac{1}{\sqrt{2}}h_bA_b &
-|h_b|^2vc_\beta+ \frac{1}{6}g^{\prime2} vc_\beta
\end{array} \right)\,,
\nonumber \\
\Gamma^{\phi_2\widetilde{b}^*\widetilde{b}} &=& \left(
\begin{array}{cc}
- \frac{1}{4}\left(g^2+\frac{1}{3}g^{\prime 2}\right)vs_\beta&
\frac{1}{\sqrt{2}}h_b^*\mu \\
\frac{1}{\sqrt{2}}h_b\mu^* &
-\frac{1}{6}g^{\prime2} vs_\beta
\end{array} \right)\,,
\nonumber \\
%
%
\Gamma^{a\widetilde{t}^*\widetilde{t}} &=& \frac{1}{\sqrt{2}}\left(
\begin{array}{cc}
0 & i\,h_t^*(c_\beta A_t^*+s_\beta \mu) \\
-i\,h_t(c_\beta A_t+s_\beta \mu^*) & 0
\end{array} \right)\,,
\nonumber \\
\Gamma^{\phi_1\widetilde{t}^*\widetilde{t}} &=& \left(
\begin{array}{cc}
- \frac{1}{4}\left(g^2-\frac{1}{3}g^{\prime 2}\right)vc_\beta&
\frac{1}{\sqrt{2}}h_t^*\mu \\
\frac{1}{\sqrt{2}}h_t\mu^* &
-\frac{1}{3}g^{\prime2} vc_\beta
\end{array} \right)\,,
\nonumber \\
\Gamma^{\phi_2\widetilde{t}^*\widetilde{t}} &=& \left(
\begin{array}{cc}
-|h_t|^2vs_\beta+ \frac{1}{4}\left(g^2-\frac{1}{3}g^{\prime 2}\right)vs_\beta&
-\frac{1}{\sqrt{2}}h_t^*A_t^* \\
-\frac{1}{\sqrt{2}}h_tA_t &
-|h_t|^2vs_\beta+ \frac{1}{3}g^{\prime2} vs_\beta
\end{array} \right)\,,
\nonumber \\
%
%
\Gamma^{a\widetilde{\tau}^*\widetilde{\tau}} &=& \frac{1}{\sqrt{2}}\left(
\begin{array}{cc}
0 & i\,h_\tau^*(s_\beta A_\tau^*+c_\beta \mu) \\
-i\,h_\tau(s_\beta A_\tau+c_\beta \mu^*) & 0
\end{array} \right)\,,
\nonumber \\
\Gamma^{\phi_1\widetilde{\tau}^*\widetilde{\tau}} &=& \left(
\begin{array}{cc}
-|h_\tau|^2vc_\beta+ \frac{1}{4}\left(g^2-g^{\prime 2}\right) vc_\beta&
-\frac{1}{\sqrt{2}}h_\tau^*A_\tau^* \\
-\frac{1}{\sqrt{2}}h_\tau A_\tau &
-|h_\tau|^2vc_\beta+ \frac{1}{2}g^{\prime2} vc_\beta
\end{array} \right)\,,
\nonumber \\
\Gamma^{\phi_2\widetilde{\tau}^*\widetilde{\tau}} &=& \left(
\begin{array}{cc}
  -\frac{1}{4}\left(g^2- g^{\prime 2}\right) vs_\beta&
\frac{1}{\sqrt{2}}h_\tau^*\mu \\
\frac{1}{\sqrt{2}}h_\tau\mu^* &
-\frac{1}{2}g^{\prime2} vs_\beta
\end{array} \right)\,,
\nonumber\\
\Gamma^{a\widetilde{\nu}^*_\tau\widetilde{\nu}_\tau} &=& 0\,,
\qquad
\Gamma^{\phi_1\widetilde{\nu}^*_\tau\widetilde{\nu}_\tau} = -\frac{1}{4}
\left( g^2 + g^{\prime 2} \right) v c_\beta,
\qquad
\Gamma^{\phi_2\widetilde{\nu}^*_\tau\widetilde{\nu}_\tau} = \frac{1}{4} \left(
g^2 + g^{\prime 2} \right) v s_\beta \, .
\end{eqnarray}
The $2\times 2$ coupling matrix
$\Gamma^{H^+\widetilde{u}^*\widetilde{d}}$ is given in the $(\widetilde{u}_L, \widetilde{d}_R)$
basis by
\begin{eqnarray}
\label{eq:chsqsq}
  \Gamma^{H^+\widetilde{u}^*\widetilde{d}}\
= \ \left(
\begin{array}{cc} \frac{1}{\sqrt{2}}\,(|h_u|^2 + |h_d|^2
- g^2)\,v s_\beta c_\beta & h_d^*\, ( s_\beta A^*_d + c_\beta \mu )\\
h_u\,( c_\beta A_u + s_\beta \mu^*) &
\frac{1}{\sqrt{2}}\, h_u h^*_d\, v \end{array}\right)\,,
\end{eqnarray}
and the couplings of the charged Higgs boson with
a tau sneutrino and a stau given by
\begin{eqnarray}
\label{eq:chslsl}
\Gamma^{H^+\widetilde{\nu}_\tau^*\widetilde{\tau}_L}\ &=& \
 \frac{1}{\sqrt{2}}\,(|h_\tau|^2 - g^2)\,v s_\beta\, c_\beta\,,\qquad
\Gamma^{H^+\widetilde{\nu}_\tau^*\widetilde{\tau}_R}\ = \
  h^*_\tau \left(s_\beta A^*_\tau+c_\beta \mu\right)\,.
\end{eqnarray}

\section{Decays of a Generic Neutral Higgs Boson}
\label{sec:decays_of_a_generic_neutral_higgs_boson}

Without loss of generality, the Lagrangian describing
the interactions of a generic neutral Higgs boson $H$
with two fermions, which is applicable for all the models
described in the previous section, can be written as
\begin{equation}
\label{eq:hff}
{\cal L}_{H\bar ff}=- \frac{m_f}{v}\, H \bar{f}
\left(g^S_{H\bar ff}+ig^P_{H\bar ff}\gamma_5\right)f\,,
\end{equation}
in terms of the normalized scalar and pseudoscalar couplings of
$g^S_{H\bar ff}$ and $g^P_{H\bar ff}$ with $m_f$ denoting the fermion mass and
$v\approx 246\,{\rm GeV}$. The Lagrangian describing the interactions
of the neutral Higgs boson $H$ with massive gauge bosons
$Z$ and $W^\pm$ can be written as
\begin{equation}
{\cal L}_{HVV}  =  g\,M_W \, \left(
g_{_{HWW}}\,W^+_\mu W^{- \mu}\ + \
g_{_{HZZ}}\,\frac{1}{2c_W^2}\,Z_\mu Z^\mu\right) \,  H\,,
\end{equation}
in terms of the normalized couplings of $g_{_{HWW}}$ and $g_{_{HZZ}}$
with  $g=e/s_W$ the SU(2)$_L$ gauge coupling,
$s_W\equiv\sin\theta_W$, $c_W\equiv\cos\theta_W$,
$t_W\equiv\sin\theta_W/\cos\theta_W$, etc.
And, if not mentioned otherwise, we set
$g_{_{HWW}}=g_{_{HZZ}}=g_{_{HVV}}$ in the following.

\medskip

In the presence of Higgs/scalar bosons $\varphi$'s lighter
than $H$, the neutral Higgs boson $H$ can decay into a lighter
Higgs boson and a massive vector boson and also into two lighter Higgs/scalar
bosons. The interaction Lagrangian describing these types of
decays could be cast into the expressions:
\begin{eqnarray}
{\cal L}_{H\varphi Z} &=& \frac{g}{2c_W} g_{_{H\varphi Z}}\, Z^{\mu}
(H\, \!\stackrel {\leftrightarrow} {\partial}_\mu \varphi) \,,
\nonumber \\
{\cal L}_{H\varphi^\pm W^\mp} & =& -\frac{g}{2} \,  g_{_{H\varphi^+ W^-}}\,
W^{-\mu} (H\, i\!\stackrel{\leftrightarrow}{\partial}_\mu
\varphi^+)\, +\, {\rm h.c.}\,,
\nonumber \\
{\cal L}_{\rm self}
\supset -v\,\sum_{i\geq j} g_{_{H\varphi_i\varphi_j}}\, H \varphi_i \varphi_j
&=&-v\left[
g_{_{H\varphi_1\varphi_1}}\, H \varphi_1^2  +
\left(g_{_{H\varphi_2\varphi_1}}\, H \varphi_2 \varphi_1 +
g_{_{H\varphi_2\varphi_2}}\, H \varphi_2^2\right) + \cdots
\right]\,.
\end{eqnarray}
Note that the scalar states $\varphi_i$ and $\varphi_j$ are
ordered in the last expression so as to avoid the couplings
such as $g_{H\varphi_1\varphi_2}$ with a wrong ordering of the scalar states.

\medskip

It is noteworthy that we are assuming the neutral Higgs boson $H$
to be a general CP-mixed state, i.e. a scalar-pseudoscalar mixture.

%
\subsection {Decays into two fermions: $H\to f\bar f$}
Including the radiative corrections known up to now, the Higgs decay width into
fermions can be organized as~\cite{Spira:2016ztx}:
\begin{eqnarray}
\label{eq:ghff}
\Gamma(H\rightarrow f \bar{f}) &= &
N_C^f\frac{m_f^2}{v^2}\frac{\beta_f M_H}{8\pi}
\left[
\beta_f^2|g^S_{H\bar{f}f}|^2
\left(1+\delta_{\rm QCD}+\delta^{f:S}_t
+\delta^f_{\rm mixed}\right)
\left(1+\delta^f_{\rm elw}\right)
\right.
\nonumber \\[2mm]
&&\hspace{2.40cm} \ + \
\left.
|g^P_{H\bar{f}f}|^2
\left(1+\delta_{\rm QCD}+\delta^{f:P}_t\right) \right]\,,
\end{eqnarray}
where $\beta_f \equiv \sqrt{1-4\kappa_f}$ with
$\kappa_f=M_f^2/M_H^2$ and the color factor $N_C^f=3$ for quarks and 1 for
leptons.~\footnote{In this review,
we denote the pole mass of the fermion $f$ by $M_f$ and its running mass
by $m_f(\mu)$.}
The lepton pole mass is taken for $m_f$ while, for the Higgs decays into
quarks, the $\overline{\rm MS}$ quark mass $\overline{m}_q(M_H)$
is used.
In passing, we recall that $1/v^2=\sqrt{2}G_F$.
%
%
Note that, in the scalar part, the QCD and electroweak (ELW) corrections are factorized.
This factorization is supported by the reduction of the mixed corrections
by a factor of 3~\cite{Mihaila:2015lwa}.
In the pseudoscalar part, we neglect the ELW corrections.
A pseudoscalar component appears in BSM models and the corresponding
ELW corrections, compared to the QCD corrections,
are much more complicated and
depend on specific BSM models under consideration.
In this review, we do not consider the ELW corrections for the decay processes
involving multi-Higgses, CP-odd or charged Higgses
sacrificing the precision
in the calculation of the partial widths for those decays.
For the calculation of full BSM-dependent ELW corrections and also
of more precise SM ones,
we provide Appendix~\ref{app:bsmtools} in which
we make a brief introduction to relevant numerical packages.

\medskip
The pure QCD corrections for the Higgs decays into a quark pair $q\bar{q}$
consist of a universal part of $\delta_{\rm QCD}$ and
two types of flavor- and parity-dependent
parts of $\delta_t^{q:S}$ and $\delta_t^{q:P}$ which are
given by
\cite{Braaten:1980yq,Sakai:1980fa,Inami:1980qp,Drees:1990dq,Drees:1989du,
Gorishnii:1983cu,Gorishnii:1990zu,Gorishnii:1991zr,Kataev:1993be,Surguladze:1994gc,
Chetyrkin:1996sr,Melnikov:1995yp,Chetyrkin:1995pd,Larin:1995sq}
\begin{eqnarray}
\label{eq:ghff_qcd}
\delta_{\rm QCD}&=&
5.67 \frac{\alpha_s (M_H)}{\pi} + (35.94 - 1.36
N_F) \left( \frac{\alpha_s (M_H)}{\pi} \right)^2
\nonumber \\ &&
+ (164.14 - 25.77 N_F + 0.259 N_F^2) \left( \frac{\alpha_s(M_H)}{\pi} \right)^3
\nonumber \\ &&
+(39.34-220.9 N_F+9.685 N_F^2-0.0205 N_F^3) \left(
\frac{\alpha_s(M_H)}{\pi} \right)^4\,, \nonumber \\[2mm]
\delta^{q:S}_t &=&\frac{g^S_{H\bar t t}}{{g^S_{H\bar q q}}}\,
\left(\frac{\alpha_s (M_H)}{\pi}\right)^2 \left[ 1.57 -
\frac{2}{3} \log \frac{M_H^2}{M_t^2} + \frac{1}{9} \log^2
\frac{\overline{m}_q^2 (M_H)}{M_H^2} \right]\,, \nonumber \\[2mm]
\delta^{q:P}_t &=&\frac{g^P_{H\bar t t}}{{g^P_{H\bar q q}}}\,
\left(\frac{\alpha_s (M_H)}{\pi}\right)^2 \left[ 3.83 -
\log \frac{M_H^2}{M_t^2} + \frac{1}{6} \log^2
\frac{\overline{m}_q^2 (M_H)}{M_H^2} \right]\,,
\end{eqnarray}
where $N_F$ counts the flavor number of quarks
lighter than $H$. The QCD coupling strength $\alpha_s$
and the running $\overline{\rm MS}$ quark mass
$\overline{m}_q(M_H)$ are defined at the scale of the Higgs mass
to absorb large mass logarithms.

\medskip

For the electroweak corrections
\cite{Fleischer:1980ub,Bardin:1990zj,Dabelstein:1991ky,Kniehl:1991ze},
we adopt the approximation~\cite{Djouadi:1991uf,Spira:2016ztx}
\begin{equation}
\label{eq:ghff_elw}
\delta^f_{\rm elw}=
\frac{3}{2} \frac{\alpha}{\pi}Q_f^2 \left(\frac{3}{2} -
\log \frac{M_H^2}{M_f^2} \right) + \frac{G_F}{8 \sqrt{2} \pi^2}
\left\{ k_f M_t^2 + M_W^2 \left[ - 5 + \frac{3}{s_W^2} \log c_W^2 \right]
- 8\,M_Z^2 (6 v_{Z\bar f f}^2 - a_{Z\bar f f}^2) \right\}\,,
\end{equation}
where $v_{Z\bar{f}f}=I_3^f/2-Q_f s_W^2$ and $a_{Z\bar{f}f}=I_3^f/2$
with $I_3^f$ denoting the third component of the electroweak isospin
and $Q_f$ the electric charge of the fermion $f$.
We refer to Eq.~(\ref{eq:zff}) for the $Z$ couplings with fermions.
The large logarithm $\log M_H^2/M_f^2$
can be absorbed in the running fermion mass as in the QCD corrections.
For decays into leptons and light quarks,
the coefficient $k_f=7$ while it is 1 for $b$ and $t$ quarks.
The electroweak corrections are below the percent level for $f=b,c$ while
they are of ${\cal O}$(1-5)\% for $f=\tau,\mu$.
For the more precise evaluations
of SM and full BSM-dependent
ELW corrections, see Appendix~\ref{app:bsmtools}.

\medskip

The mixed corrections evaluated by means of low-energy theorems
could be cast into the expressions
\cite{Kniehl:1994ju,Kwiatkowski:1994cu,Chetyrkin:1996wr}:
\begin{eqnarray}
\delta^q_{\rm mixed}&=&-\frac{G_FM_t^2}{8\sqrt{2}\pi^2}
\left(\frac{3}{2}+\zeta_2\right)
\frac{\alpha_s(M_t)}{\pi}
\hspace{1cm}\mbox{for light quarks}\,,\nonumber \\[2mm]
\delta^{b,t}_{\rm mixed}&=&-\frac{G_FM_t^2}{8\sqrt{2}\pi^2}\
4\left(1+\zeta_2\right)\frac{\alpha_s(M_t)}{\pi}
\hspace{1cm}\mbox{for $b$ and $t$}\,,
\end{eqnarray}
at next-to-next-to-leading-order (NNLO) with $\zeta_2 = \pi^2/6$.
%
Also available are the full mixed
QCD-electroweak corrections $\delta^b_{\rm mixed}$ for $H\to b\bar b$
which amount to about $-0.08$\% for $M_H=125.09$ GeV
\cite{Kataev:1997cq,Mihaila:2015lwa}.

\medskip

\begin{figure}[t!]
\begin{center}
\includegraphics[width=7.7cm]{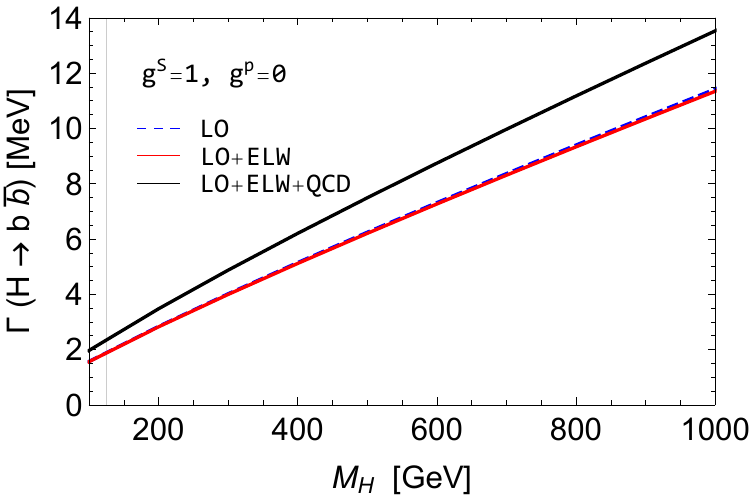}
\includegraphics[width=7.7cm]{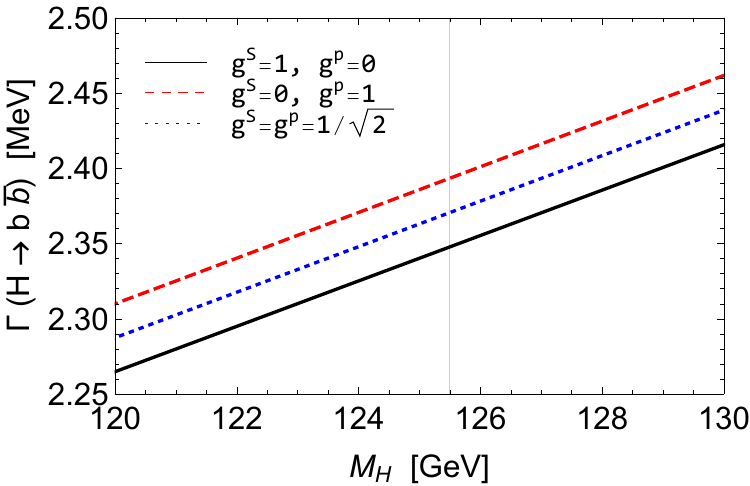}
\includegraphics[width=7.7cm]{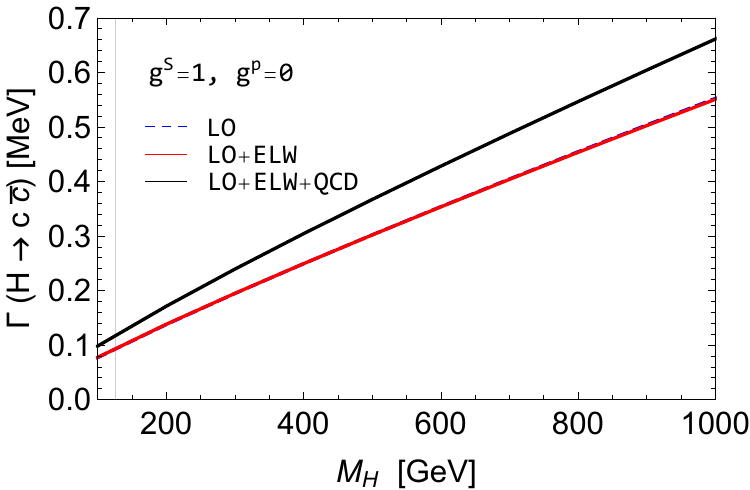}
\includegraphics[width=7.7cm]{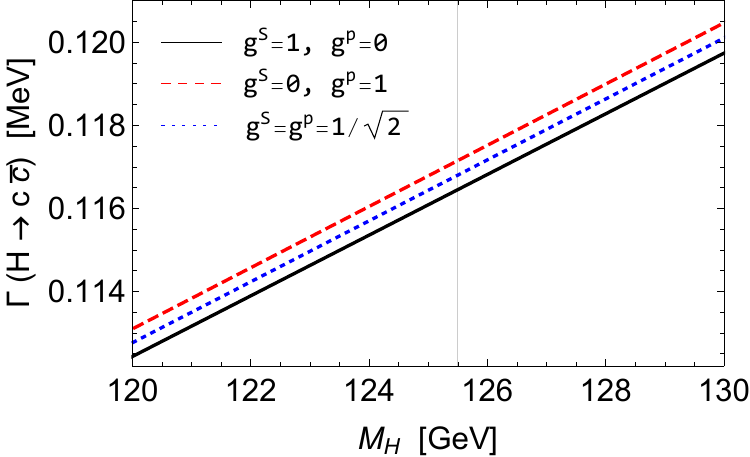}
\includegraphics[width=7.7cm]{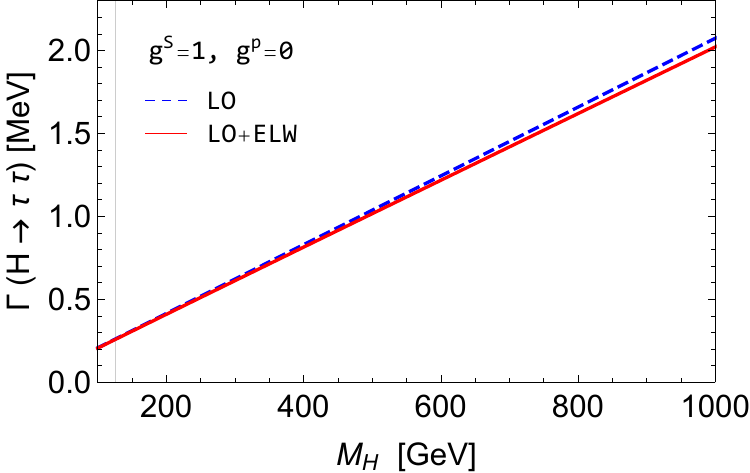}
\includegraphics[width=7.7cm]{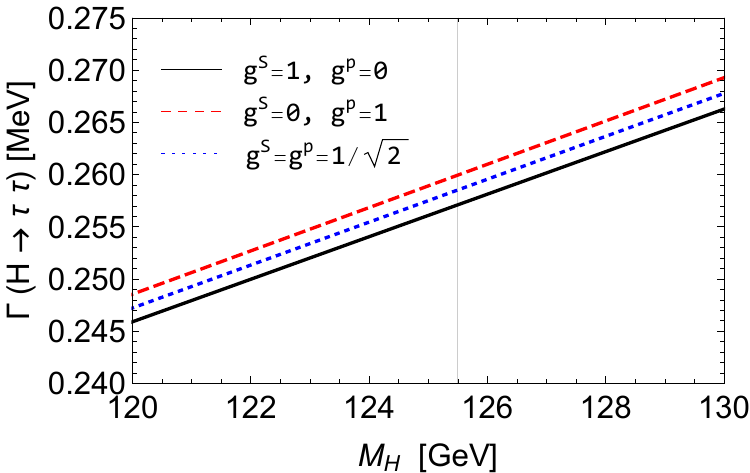}
\includegraphics[width=7.7cm]{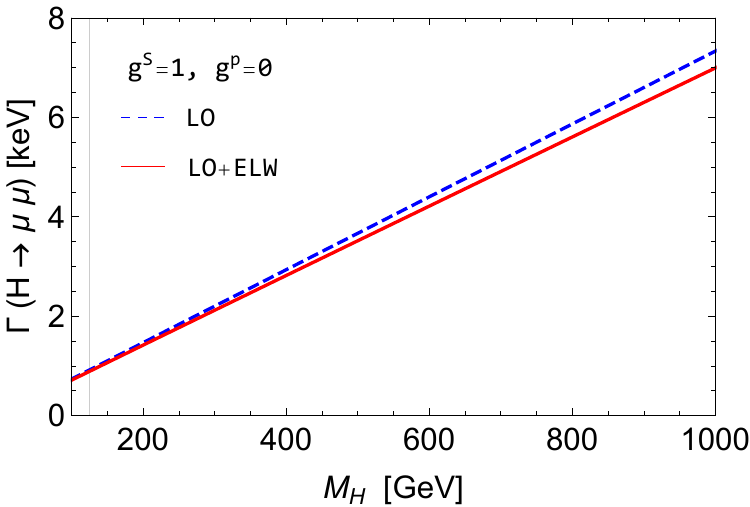}
\includegraphics[width=7.7cm]{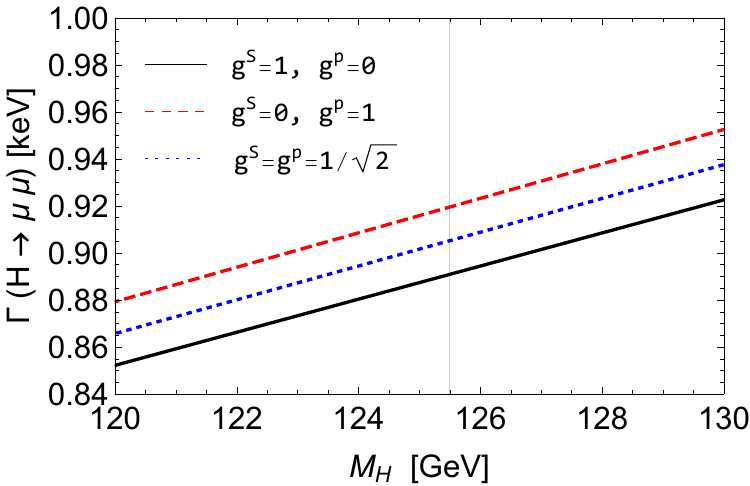}
\end{center}
\caption{
(Left) Decay widths of a
neutral Higgs boson with mass $M_H$
into $b\bar b$, $c\bar c$, $\tau^+\tau^-$, and $\mu^+\mu^-$
from top to bottom taking
$g^S_{H\bar f f} =1$ and $g^P_{H\bar f f} =0$.
See the main text for details.
(Right) In each row, the corresponding full decay widths
are shown for the three choices of
$(g^S_{H\bar f f}\,,g^P_{H\bar f f})=$
$(1,0)$ (solid), $(0,1)$ (dashed), and
$(1/\sqrt{2},1/\sqrt{2})$ (dotted) in the low mass region
around $M_H=125$ GeV.
The vertical lines locate the positions of $M_H=125.5$ GeV.
Note that the muonic decay widths are in units of keV.
}
\label{fig:h2ff}
\end{figure}
\medskip
In the left column of Fig.~\ref{fig:h2ff},
taking $g^S_{H\bar f f} =1$ and $g^P_{H\bar f f} =0$,
we show the decay widths of a Higgs boson
$H$ into a pair of $b$ quarks, $c$ quarks, tau leptons, and
muons for varying $M_H$.
For the decays $H\to b \bar b$ and $H\to c \bar c$,
the lower dashed lines are for the decay widths
at leading order (LO) while
the upper (black) solid
lines are for those taking full account of
the QCD and electroweak (ELW) corrections. The decay widths including
only the electroweak corrections are denoted
by the lower (red) solid lines.
For the decays $H\to \tau^+\tau^-$ and $H\to \mu^+\mu^-$,
the dashed lines are for the decay widths at LO and
the sold lines are for the decay widths including the electroweak corrections.
The behavior does not alter much for other choices of
$(g^S_{H\bar f f}\,,g^P_{H\bar f f})$ as far as
$|g^S_{H\bar f f}|^2+|g^P_{H\bar f f}|^2=1$
since the QCD correction $\delta_{\rm QCD}$, which is common in
the scalar and pseudoscalar contributions to
the Higgs decay width into quarks, dominates.

\medskip

In each frame of the right column of Fig.~\ref{fig:h2ff},
the corresponding full decay widths are shown
in the low mass region of $120\, {\rm GeV} <M_H < 130\, {\rm GeV}$
for the three choices of $(g^S_{H\bar f f}\,,g^P_{H\bar f f})=$
$(1,0)$, $(0,1)$, and $(1/\sqrt{2},1/\sqrt{2})$
denoted by the solid, dashed and dotted lines, respectively.
The pure scalar case with $(g^S_{H\bar f f}\,,g^P_{H\bar f f})=(1,0)$
has a slightly smaller width compared to the pure
pseudoscalar case with $(g^S_{H\bar f f}\,,g^P_{H\bar f f})=(0,1)$ due to the
kinematical suppression factor of $\beta_f^2$. 
%

\medskip
At leading order (LO),
taking the consideration of double off-shell effects,
the decay width of a Higgs boson into a top-quark pair
$t\bar{t}$, each of which  subsequently decays into $bW^+$
and $\bar{b}W^-$, is given by
\cite{Grau:1990uu,Moretti:1994ds}
\begin{equation}
\label{eq:GHtt0}
\Gamma^{\rm LO}(H\to t(p_t)\,\bar t(\bar p_t)\to b \bar b W^+W^-)
=N_C^t\frac{\overline{m}_t^2(M_H)}{v^2}\frac{g^4 M_H}{2^{13}\pi^5}\, {\cal F}\,,
\end{equation}
where the dimensionless quantity ${\cal F}$ is given by
an integrated function of the pole masses of
$b$ and $t$ quarks, the $W$-boson mass, and
the top-quark total width $\Gamma_t$ as~\cite{Chang:2018ult}
\begin{eqnarray}
\label{eq:calF_Htt}
{\cal F}&=&
\int \lambda_H^{1/2}\lambda_t^{1/2}\lambda_{\bar t}^{1/2}
\left[|\lambda_S|^2\left|g^S_{H\bar t t}\right|^2
+|\lambda_P|^2\left|g^P_{H\bar t t}\right|^2\right]
\left(1+\frac{p_t^2+\bar p_t^2}{2M_W^2}+\frac{p_t^2\bar p_t^2}{4M_W^4}\right)
\nonumber \\
&&\hspace{2.5cm} \times
\frac{p_t^2+M_b^2-M_W^2} {(p_t^2-M_t^2)^2+M_t^2\Gamma_t^2}\
\frac{\bar p_t^2+M_b^2-M_W^2} {(\bar p_t^2-M_t^2)^2+M_t^2\Gamma_t^2}\
d{p_t^2}\,d{\bar p_t^2}\,,
\end{eqnarray}
with the 5 dimensionless triangle functions
\begin{eqnarray}
\lambda_H&=&1+\frac{(p_{t}^2)^2}{M_H^4}+\frac{(\bar p_t^2)^2}{M_H^4}
-2\frac{p_{t}^2}{M_H^2} -2\frac{\bar p_t^2}{M_H^2}
-2\frac{p_{t}^2\bar p_t^2}{M_H^4}\,, \nonumber \\[2mm]
\lambda_t&=&\left(1-\frac{M_b^2}{p_t^2}-\frac{M_W^2}{p_t^2}\right)^2
-4\frac{M_b^2}{p_t^2}\,\frac{M_W^2}{p_t^2}\,, \nonumber \\[2mm]
\lambda_{\bar t}&=&\left(1-\frac{M_b^2}{\bar p_t^2}-\frac{M_W^2}{\bar p_t^2}\right)^2
-4\frac{M_b^2}{\bar p_t^2}\,\frac{M_W^2}{\bar p_t^2}\,, \nonumber \\[2mm]
\lambda_S&=&
\left\{\left[(1+\lambda_H^{1/2})^2-(p_t^2-\bar p_t^2)^2/M_H^4\right]^{1/2}
-\left[(1-\lambda_H^{1/2})^2-(p_t^2-\bar p_t^2)^2/M_H^4\right]^{1/2}\right\}/2\,,
\nonumber \\[2mm]
\lambda_P&=&
\left\{\left[(1+\lambda_H^{1/2})^2-(p_t^2-\bar p_t^2)^2/M_H^4\right]^{1/2}
+\left[(1-\lambda_H^{1/2})^2-(p_t^2-\bar p_t^2)^2/M_H^4\right]^{1/2}\right\}/2\,.
\end{eqnarray}
After integrating over $p_t^2$ and $\bar p_t^2$, we can
recast the LO decay width into
\begin{equation}
\Gamma^{\rm LO}(H\to t\,\bar t\to b \bar b W^+W^-) \equiv
\left|g^S_{H\bar t t}\right|^2\,{\cal T}^{\rm LO}_S +
\left|g^P_{H\bar t t}\right|^2\,{\cal T}^{\rm LO}_P\,,
\end{equation}
and, taking account of the radiative corrections as well as
double off-shell effects,
the Higgs decay width into a top quark pair
$\Gamma(H\to t^*\bar t^*)$ has been estimated as follows
\footnote{For $\delta^t_{\rm elw}$, we apply
the SM approximation Eq.~(\ref{eq:ghff_elw}) both below and above
the top-quark-pair threshold.
Note that the QCD corrections are not valid
in the threshold region due to the top-quark mass effects. For them,
we refer to~\cite{Spira:2016ztx} and references there in.}
\begin{eqnarray}
\label{eq:ghtt}
\Gamma(H\to t^*\bar t^*) =
\left|g^S_{H\bar t t}\right|^2\,{\cal T}^{\rm LO}_S
\left(1+\delta_{\rm QCD}+\delta^{t:S}_t
+\delta^t_{\rm mixed}\right)
\left(1+\delta^t_{\rm elw}\right)
+\left|g^P_{H\bar t t}\right|^2\,{\cal T}^{\rm LO}_P
\left(1+\delta_{\rm QCD}+\delta^{t:P}_t\right)\,.
\end{eqnarray}
When $M_H>2M_t$, taking $p^2_t=\bar{p}^2_t=M^2_t$ leads to
$\lambda_S=\lambda_H^{1/2}=\beta_t=(1-4M_t^2/M_H^2)^{1/2}$ and
$\lambda_P=1$ and, neglecting the kinematical $b$-quark mass in the $t\to b W$ process,
we reach the following factorized form for the LO decay width
\begin{equation}
\left.
\Gamma^{\rm LO}(H\to t\,\bar t\to b \bar b W^+W^-)
\right|_{M_H>2M_t}
=\Gamma^{\rm LO}(H\to t\bar t)\,
\left(\frac{\Gamma^{\rm LO}(t\to b W)}{\Gamma_t}\right)^2\,,
\end{equation}
using the narrow-width approximation (NWA) denoted by
\begin{equation}
\label{eq:delta}
\delta(p^2-m^2)=\lim_{\Gamma\to 0}
\frac{m\Gamma}{\pi}\frac{1}{(p^2-m^2)^2+m^2\Gamma^2}\,,
\end{equation}
and the LO decay widths for $H\to t\bar{t}$ and $t\to bW$ given by
\begin{eqnarray}
\Gamma^{\rm LO}(H\to t\bar t)&=&
N_C^t\frac{\overline{m}_t^2(M_H)}{v^2}\frac{\beta_t M_H}{8\pi}
\left[\beta_t^2\left|g^S_{H\bar t t}\right|^2+\left|g^P_{H\bar t t}\right|^2\right]\,,
\\[2mm]
\label{eq:gamtb}
\Gamma^{\rm LO}(t\to b W)&=&
\frac{g^2 M_t}{2^6\pi}\left(1-\frac{M_W^2}{M_t^2}\right)^2
\left(2+\frac{M_t^2}{M_W^2}\right)\,,
\end{eqnarray}
respectively.
%

\medskip

\begin{figure}[t!]
\begin{center}
\includegraphics[width=8.4cm]{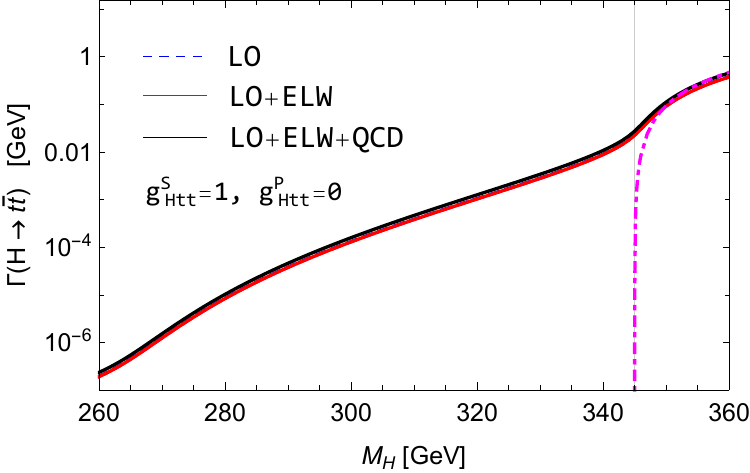}
\includegraphics[width=8.4cm]{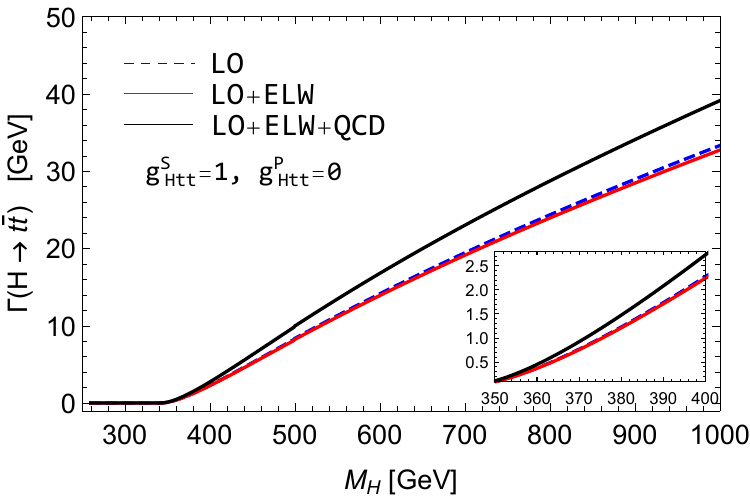}
\includegraphics[width=8.4cm]{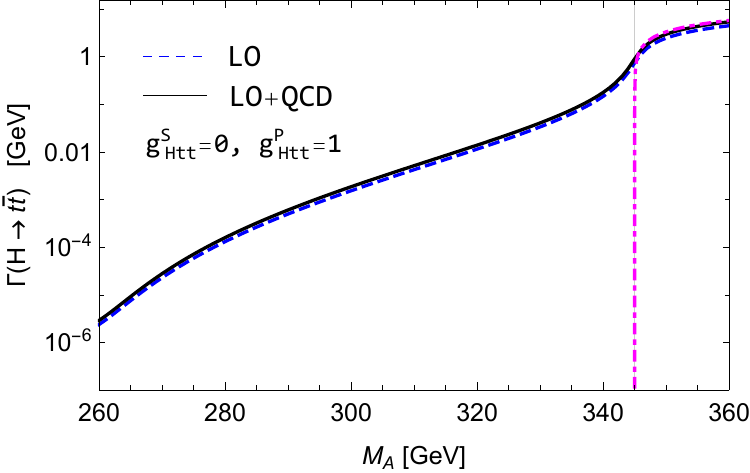}
\includegraphics[width=8.4cm]{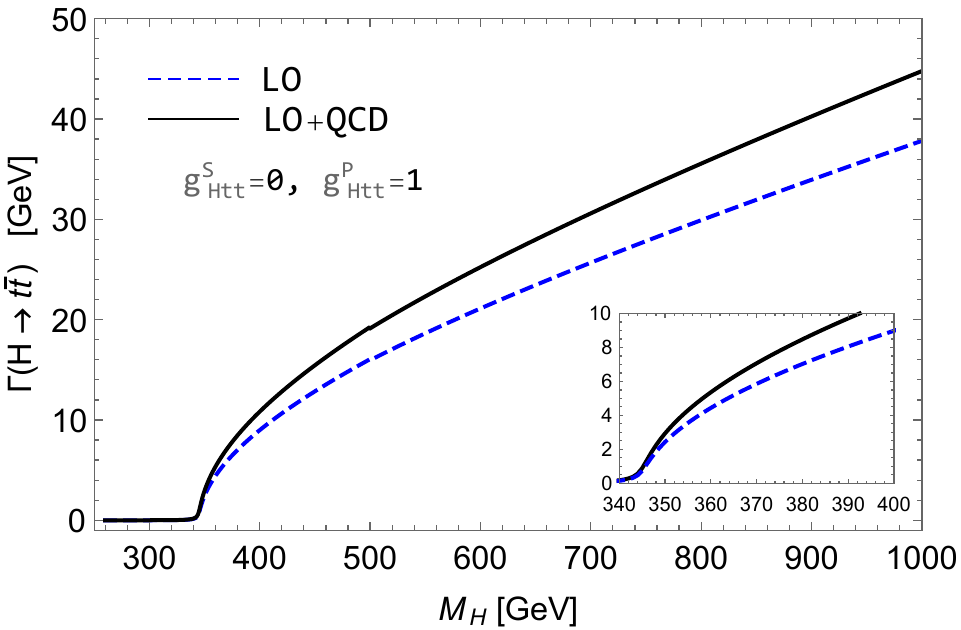}
\end{center}
\caption{
Decay widths of a neutral Higgs boson with mass
$M_H$ into $t^*\bar t^*$
or $\Gamma(H\to t^*\bar t^*)$ taking
$(g^S_{H\bar t t},g^P_{H\bar t t})=(1,0)$ (upper) and
$(g^S_{H\bar t t},g^P_{H\bar t t})=(0,1)$ (lower).
In the left panels, the vertical lines locate the top-quark-pair thresholds and
the (magenta) dash-dotted lines are for the corresponding
full 2-body decay widths of $\Gamma(H\to t\bar t)$.
Note that, in the right panels, we switch from $\Gamma(H\to t^*\bar t^*)$
to $\Gamma(H\to t\bar t)$ from $M_H=500$ GeV and above.}
\label{fig:h2tt}
\end{figure}
In Fig.~\ref{fig:h2tt},
we show the decay width $\Gamma(H\to t^*\bar t^*)$ as a function
of $M_H$, taking $(g^S_{H\bar t t},g^P_{H\bar t t})=(1,0)$ (upper) and
$(g^S_{H\bar t t},g^P_{H\bar t t})=(0,1)$ (lower), respectively.
We take $\Gamma_t=\Gamma^{\rm LO}(t\to bW)$ assuming that
a top quark decays 100\% into a $b$ quark and a $W$ boson and
$\Gamma^{\rm LO}(H\to t\bar t\to b\bar b W^+W^-)$ converges to
$\Gamma^{\rm LO}(H\to t\bar t)$ in the high $M_H$ limit.
Practically, far above the top-quark-pair threshold with $M_H>500$ GeV,
we return to Eq.~(\ref{eq:ghff}) to suppress
the contributions to $\Gamma(H\to t^* \bar t^*)$
from the kinetic edge region of $\sqrt{p_t^2}+\sqrt{\bar p_t^2} \sim M_H$,
assuming that the intermediate top quarks are reconstructed
by requiring on-shell conditions of ${p_t^2} \simeq M_t^2$ and
${\bar p_t^2} \simeq M_t^2$.
\footnote{For example, when $M_H=1000$ GeV,
we find that $\Gamma(H\to t^*\bar t^*)/\Gamma(H\to t\bar t)$ takes the
values of $1.07$ and $1.10$ for
$(g^S_{H\bar t t},g^P_{H\bar t t})=(1,0)$ and
$(g^S_{H\bar t t},g^P_{H\bar t t})=(0,1)$, respectively.}

\medskip

\begin{figure}[t!]
\vspace{ 1.0cm}
\begin{center}
\includegraphics[width=12.0cm]{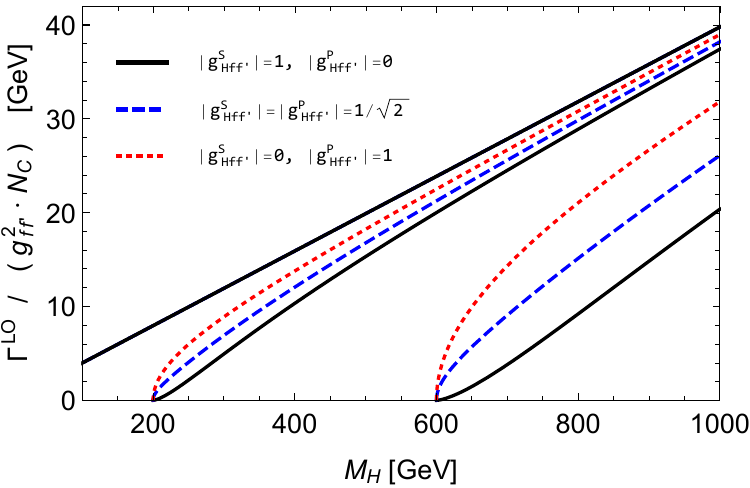}
\end{center}
\vspace{-0.5cm}
\caption{
$\Gamma^{\rm LO}(H\to f\bar f^\prime)/(g_{\bar ff^\prime}^2\,N_C^{ff^\prime})$
as functions of $M_H$
for the three choices of
$(|g^S_{_{H\bar{f}f^\prime}}|,|g^P_{_{H\bar{f}f^\prime}}|)=(1,0)$ (black solid),
$(|g^S_{_{H\bar{f}f^\prime}}|,|g^P_{_{H\bar{f}f^\prime}}|)
=(1/\sqrt{2},1/\sqrt{2})$ (blue dashed), and
$(|g^S_{_{H\bar{f}f^\prime}}|,|g^P_{_{H\bar{f}f^\prime}}|)=(0,1)$ (red dotted).
We have taken $m_f=m_{f^\prime}=0\,,100\,,300$ GeV from left to right.
Note that there is no dependence on the choice of the couplings when
$m_f=m_{f^\prime}=0$
as far as $|g^S_{_{H\bar{f}f^\prime}}|^2+|g^P_{_{H\bar{f}f^\prime}}|^2$
remains the same.
}
\label{fig:h2ffp}
\end{figure}

Finally, for decays into two different fermions such as charginos or neutralinos or
for flavor changing decays such as $H\to b\bar s$,
one may write the effective interaction as
\begin{equation}
\label{eq:hinos}
{\cal L}_{H\bar ff^\prime}=- g_{\bar ff^\prime} H \bar{f}
(g^S_{_{H\bar{f}f^\prime}}+ig^P_{_{H\bar{f}f^\prime}}\gamma_5)f^\prime
\  + \  {\rm h.c.}\,,
\end{equation}
without loss of generality.  Then, at LO, the decay width may take a form of
\begin{eqnarray}
  \Gamma^{\rm LO}(H\to f\bar f^\prime)
= N_C^{ff^\prime} \frac{g_{\bar ff^\prime}^2 M_H \lambda^{1/2}_{ff^\prime}}{8\pi}
\left[(1\!-\!\kappa\!-\!\kappa^\prime)
(|g^S_{_{H\bar{f}f^\prime}}|^2+|g^P_{_{H\bar{f}f^\prime}}|^2)
-2\sqrt{\kappa\kappa^\prime}
(|g^S_{_{H\bar{f}f^\prime}}|^2-|g^P_{_{H\bar{f}f^\prime}}|^2)\right]\,,
\end{eqnarray}
where $\kappa \equiv m_{f}^2/M_H^2$, $\kappa^\prime \equiv m_{f^\prime}^2/M_H^2$
and $\lambda_{ff^\prime}=(1-\kappa-\kappa^\prime)^2
-4\kappa\kappa^\prime $.  The color factor
$N_C^{ff^\prime}=3$ for quarks and 1 for leptons, charginos, and neutralinos.
For the decays into neutralinos $\wt{\chi}^0_j$ and $\wt{\chi}^0_k$,
we need to multiply the factor of $4/(1+\delta_{jk})$ with
$\delta_{jk}=1$ for an identical Majorana neutralino pair.
See Fig.~\ref{fig:h2ffp} for the $M_H$ dependence of the
normalized LO decay width
$\Gamma^{\rm LO}(H\to f\bar f^\prime)/(g_{\bar ff^\prime}^2\,N_C^{ff^\prime})$
for various choices of the couplings
$|g^S_{_{H\bar{f}f^\prime}}|$ and $|g^P_{_{H\bar{f}f^\prime}}|$
and the fermion masses with $m_f=m_{f^\prime}$.
When $m_f\neq m_{f^\prime}$,
we observe that the LO decay width
locates between the pseudoscalar (red dotted) and scalar (black solid)
cases if the position of the mass threshold
$m_f+m_{f^\prime}$ is the same.
In this review,
we consider the Higgs decays into SUSY particles only at LO.
For the QCD and ELW corrections to them, we refer to
Refs.~\cite{Dawson:1996xz,Eberl:1999he,Weber:2007id,
Accomando:2011jy,Heinemeyer:2014yya}.

\subsection {Decays into two massive vector bosons: $H\to VV$ with $V=Z,W$}
\begin{figure}[t!]
\begin{center}
\includegraphics[width=8.4cm,height=7.0cm]{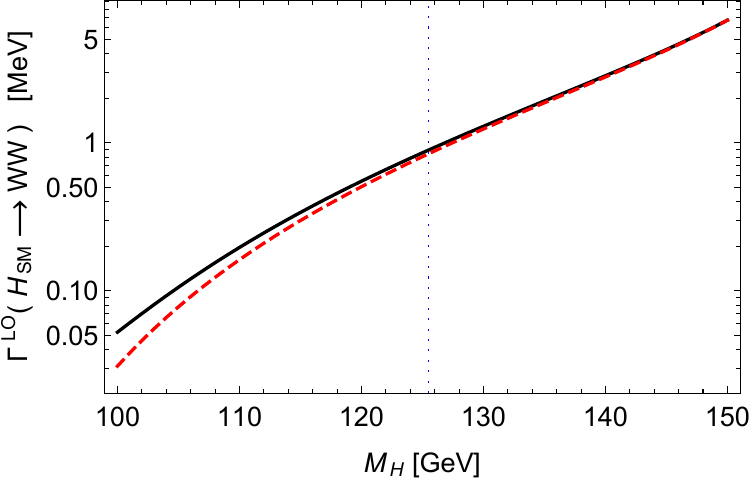}
\includegraphics[width=8.4cm,height=7.0cm]{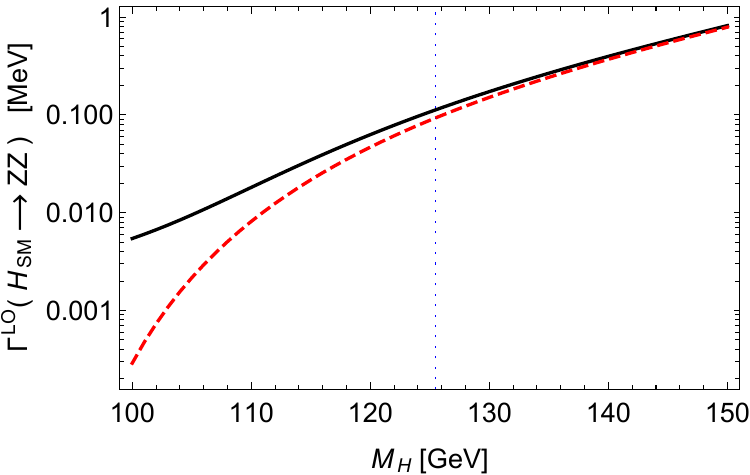}
\end{center}
\vspace{-0.5cm}
\caption{
Comparisons of $\Gamma^{\rm LO}(H\to V^*V^*)$ (solid)
and $\Gamma^{\rm LO}(H\to VV^*)$ (dashed).
The vertical lines correspond to $M_H = 125.5$ GeV and
$g_{_{HVV}}=1$ is taken.
}
\label{fig:hvv34body}
\end{figure}

Taking the full consideration of double off-shell effects, the Higgs decay
width into two massive vector bosons is given by
\cite{Cahn:1988ru,Grau:1990uu,Moretti:1994ds,Djouadi:1995gv}
\begin{eqnarray}
\label{eq:ghvv_lo}
\Gamma^{\rm LO}(H\to V^*V^*)&=&\frac{1}{\pi^2}
\int_0^{\omega_V}\frac{\epsilon_V {\rm d}y}{(y-1)^2+\epsilon_V^2}
\int_0^{(\sqrt\omega_V-\sqrt y)^2}\frac{\epsilon_V {\rm d}x}{(x-1)^2+\epsilon_V^2}
\nonumber \\[2mm]
&\times & \frac{\delta_V g_{_{HVV}}^2 G_F M_H^3}{16\sqrt{2}\pi}\,
\lambda^{1/2}\left(1,\frac{x}{\omega_V},\frac{y}{\omega_V}\right)
\left[\lambda\left(1,\frac{x}{\omega_V},\frac{y}{\omega_V}\right)+12\frac{xy}{\omega_V^2}
\right]\,,
\end{eqnarray}
where $\delta_W=2$, $\delta_Z=1$, $\omega_V=1/\kappa_V=M_H^2/M_V^2$,
$\epsilon_V=\Gamma_V/M_V$, and
$\lambda(a,b,c)=(a-b-c)^2-4bc$.
When $M_V<M_H<2M_V$ as in the case of the 125 GeV Higgs boson,
the off-shell effects of one of the two vector bosons are negligible and
the decay width reads~\cite{Keung:1984hn}
\begin{eqnarray}
\Gamma^{\rm LO}(H\to VV^*) \, &=&\, \frac{\delta_{VV^*}}{\pi}
\int_0^{(\sqrt\omega_V-1)^2}\frac{\epsilon_V {\rm d}x}{(x-1)^2+\epsilon_V^2}
\nonumber \\[2mm]
&\times&
\frac{\delta_V g_{_{HVV}}^2 G_F M_H^3}{16\sqrt{2}\pi}\,\frac{1}{\omega_V^3}\,
\lambda^{1/2}(\omega_V,x,1) \left[\lambda(\omega_V,x,1)+12x\right]\,,
\end{eqnarray}
with $\delta_{VV^*}=2$.
See Fig.~\ref{fig:hvv34body} for comparisons of
$\Gamma^{\rm LO}(H\to V^*V^*)$ and $\Gamma^{\rm LO}(H\to VV^*)$.
Incidentally, one may neglect all the off-shell effects
for a heavy Higgs boson with $M_H>2M_V$ and the decay width
takes a simple form:
\begin{equation}
\Gamma^{\rm LO}(H\to VV)=
\frac{\delta_V g_{_{HVV}}^2 G_F M_H^3}{16\sqrt{2}\pi}\,
\beta_V\left[1-4\kappa_V+12\kappa_V^2\right]\,,
\end{equation}
where $\beta_V=\sqrt{1-4\kappa_V}$ with $\kappa_V=M_V^2/M_H^2$.

\medskip

\begin{figure}[t!]
\begin{center}
\includegraphics[width=16.5cm]{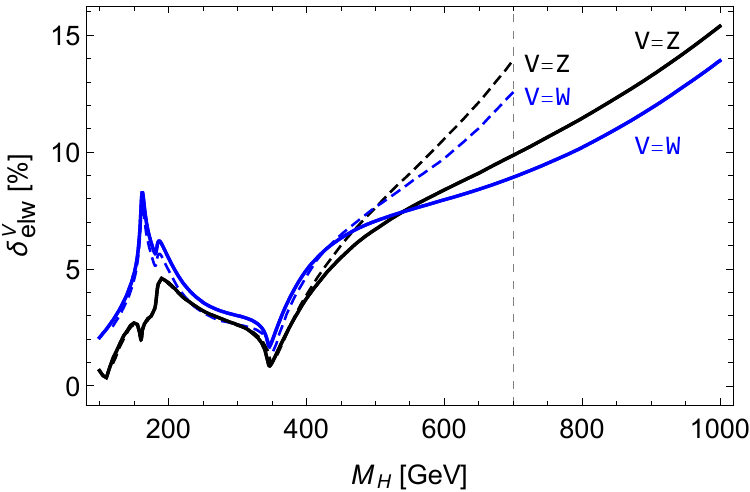}
\end{center}
\vspace{-0.5cm}
\caption{
The electroweak correction factors $\delta^{V=Z,W}_{\rm elw}$
obtained from the complete electroweak corrections
of ${\cal O}(\alpha)$ to the decay processes of
$H\to ZZ \to e^- e^+\mu^- \mu^+$ ($V=Z$) and
$H\to WW \to \nu_e e^+\mu^- \bar\nu_\mu$ ($V=W$).
The dashed lines are obtained by combining the results presented
in Figs.~7 and 8 of Ref.~\cite{Bredenstein:2006rh} while the solid lines
by using the most recent version of {\tt PROPHECY4Fv3.0}
taking the SM (or ``model$=$0") in the ``inputfile".
For our numerical analyses, we adopt the solid lines by {\tt PROPHECY4Fv3.0}.
}
\label{fig:delwV}
\end{figure}
Beyond the leading order, including radiative corrections
and the double off-shell effects above as well as below
the gauge-boson-pair thresholds,
we estimate the radiatively--corrected decay width into
two vector bosons by introducing a correction factor as
\begin{equation}
\label{eq:ghvv}
\Gamma(H\to VV\to 4f)  =  \Gamma^{\rm LO}(H\to V^*V^*)\,
\left(1+\delta^V_{\rm elw}\right)\,.
\end{equation}
For the SM electroweak correction factors $\delta^{V=Z,W}_{\rm elw}$,
we use the most recent version of {\tt PROPHECY4Fv3.0}
\cite{Bredenstein:2006rh,Bredenstein:2006nk,
Bredenstein:2006ha,Altenkamp:2017ldc,Altenkamp:2017kxk,Altenkamp:2018bcs,
Denner:2018opp} to calculate
the complete ${\cal O}(\alpha)$ electroweak corrections
to the Higgs decays into four fermions through intermediate
$W$ and $Z$ bosons, supplemented by the corrections
originating from heavy-Higgs effects and final-state
radiation. We note that these
electroweak corrections are applicable even near and below
the gauge-boson-pair thresholds where the narrow-width approximation (NWA)
is not applicable.
The corrections amount to about 3\% or less for $M_H=125$ GeV,
as can be checked in Fig.~\ref{fig:delwV}.
For the consistent implementation of the ELW corrections
using {\tt PROPHECY4F}, we note that
one may need to adopt the complex pole masses of the $W$ and $Z$ bosons for
the LO decay widths which are given by
\begin{equation}
M_V^{\rm pole}-i\Gamma^{\rm pole}_V=\frac{M_V-i\Gamma_V}{\sqrt{1+\Gamma_V^2/M_V^2}}\,.
\end{equation}
Using the complex pole masses, we find that the LO decay width 
into $W^*W^*\,(Z^*Z^*)$ increases by the amount of
about 0.4\%\,(0.5\%) taking $M_H=125.5$ GeV, 
compared to those obtained using the on-shell
masses and widths.

\medskip

It has been estimated that
missing corrections beyond ${\cal O}(\alpha)$ make the theoretical
calculations for the inclusive decay rates into four fermions
uncertain by the amount of 0.5\%~\cite{Denner:2011mq,Heinemeyer:2013tqa}.

\medskip

\begin{figure}[t!]
\begin{center}
\includegraphics[width=8.4cm]{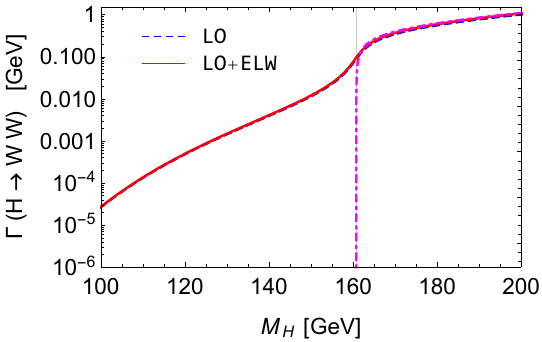}
\includegraphics[width=8.4cm]{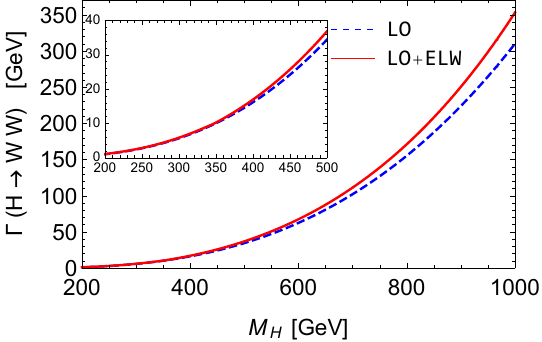}
\includegraphics[width=8.4cm]{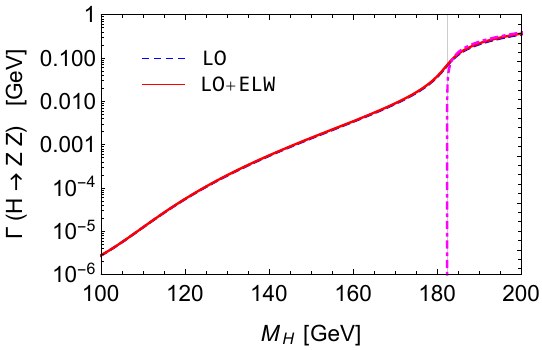}
\includegraphics[width=8.4cm]{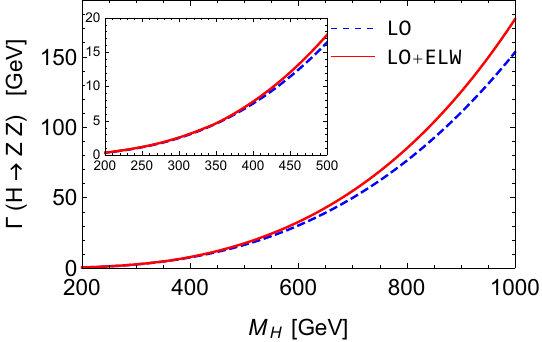}
\end{center}
\caption{
Decay widths of a neutral Higgs boson with mass $M_H$
into $W^*W^*$ (upper) and into $Z^*Z^*$ (lower) or $\Gamma(H\to VV \to 4f)$
taking $g_{_{HVV}}=1$.
In the left panels, the vertical lines locate the vector-boson-pair thresholds and
the (magenta) dash-dotted lines are for the corresponding
2-body decay widths of
$\Gamma(H\to VV)=\Gamma^{\rm LO}(H\to VV)(1+\delta_{\rm elw}^V)$.}
\label{fig:h2vv}
\end{figure}

In Fig.~\ref{fig:h2vv},  we show the decay widths of a neutral Higgs boson with
mass $M_H$ into $W^*W^*$ (upper) and into $Z^*Z^*$ (lower).
Note that we include the electroweak corrections
shown by the solid lines in Fig.~\ref{fig:delwV}.
%
We observe that the electroweak corrections grow as $M_H$ increases
and they make the partial decay widths into $WW$ and $ZZ$ as large as
350 GeV and 180 GeV, respectively, around $M_H=1$ TeV.

\subsection{Decays into a lighter scalar boson and a vector boson and
into two lighter scalar bosons:
$H\to \varphi V\,, \varphi\varphi$}
In the presence of multiple Higgs bosons,
the decay of a heavier Higgs boson $H$ into a lighter neutral
Higgs boson $\varphi$ and a massive gauge boson $Z$ may occur
and, considering the case of a virtual $Z^*$,
its decay width is given by an integral form as
\cite{Cahn:1988ru,Moretti:1994ds,Djouadi:1995gv}
\begin{equation}
\label{eq:ghpZ}
\Gamma^{\rm LO}(H\rightarrow \varphi Z^*) =
\frac{G_FM_{H}^3g_{_{H\varphi Z}}^2}{8\sqrt{2}\pi}
\int_0^{(\sqrt{\omega}-\sqrt{\omega_\varphi})^2}{\rm d}x
\frac{\epsilon_Z\lambda^{3/2}(\omega,\omega_\varphi,x)}
{\omega^3\pi[(x-1)^2+\epsilon_Z^2]}\,,
\end{equation}
with $\omega=M_{H}^2/M_Z^2$ and $\omega_\varphi=M_{\varphi}^2/M_Z^2$.
When $M_H$ is larger than $M_{\varphi}+ M_Z$,
using the $Z$-boson narrow-width approximation,
it reduces to
\begin{equation}
\Gamma^{\rm LO}(H\to \varphi Z) \ = \
\frac{G_F M_H^3}{8\sqrt{2}\pi} g_{_{H\varphi Z}}^2
\lambda^{3/2}(1,\kappa_\varphi,\kappa_Z)\,,
\end{equation}
where $\kappa_\varphi=M_{\varphi}^2/M_{H}^2$
and $\kappa_Z=M_Z^2/M_{H}^2$.
And, a heavier Higgs boson $H$ might also decay into a lighter charged
Higgs boson $\varphi^\pm$ and a massive gauge boson $W^\mp$ with its
decay width given by an integral form as
\cite{Cahn:1988ru,Moretti:1994ds,Djouadi:1995gv}
\begin{equation}
\label{eq:ghpW}
\Gamma^{\rm LO}(H\rightarrow \varphi^\pm W^{\mp *}) =
\frac{G_FM_{H}^3|g_{_{H\varphi^+W^-}}|^2}{8\sqrt{2}\pi}
\int_0^{(\sqrt{\omega}-\sqrt{\omega_\pm})^2}{\rm d}x
\frac{\epsilon_W\lambda^{3/2}(\omega,\omega_\pm,x)}
{\omega^3\pi[(x-1)^2+\epsilon_W^2]}\,,
\end{equation}
where $\Gamma^{\rm LO}(H\rightarrow \varphi^\pm W^{\mp *})
=\Gamma^{\rm LO}(H\rightarrow \varphi^+ W^{-*}) =
\Gamma^{\rm LO}(H\rightarrow \varphi^- W^{+*})$
with $\omega=M_{H}^2/M_W^2$ and $\omega_\pm=M_{\varphi^\pm}^2/M_W^2$.
When $M_{H}$ is larger than $M_{\varphi^\pm} + M_W$,
in the $W$-boson narrow-width approximation,
it reduces to
\begin{equation}
\Gamma^{\rm LO}(H\to \varphi^\pm W^{\mp}) \ = \
\frac{G_F M_{H}^3}{8\sqrt{2}\pi} |g_{_{H\varphi^+W^-}}|^2
\lambda^{3/2}(1,\kappa_\varphi,\kappa_W)\,,
\end{equation}
where $\kappa_\varphi=M_{\varphi^\pm}^2/M_{H}^2$ and $\kappa_W=M_W^2/M_{H}^2$.
%

\medskip

Finally, when a heavier Higgs boson decays into a pair of lighter neutral
Higgs bosons of $\varphi_i$ and $\varphi_j$
or into a pair of sfermions, at LO, we have for the decay width
\begin{equation}
\label{eq:ghpp}
\Gamma^{\rm LO}\ \left(H\to \varphi_i\varphi_j\,,\wt{f}^*_i\wt{f}_j \right)
=\ N_\varphi\frac{v^2 |{\cal G}|^2}{16\pi
M_{H}}\lambda^{1/2}(1,\kappa_i,\kappa_j)\,,
\end{equation}
where $(N_\varphi,{\cal G}) =
(1+\delta_{ij},g_{_{H\varphi_i\varphi_j}})$ or
$(N_C^f, g_{H\widetilde{f}^*_i\widetilde{f}_j})$  and
$\kappa_i=M_{\varphi_i,\widetilde{f_i}}^2/M_{H}^2$.
The decay width into a pair of lighter charged Higgs bosons is given
by taking $(N_\varphi,{\cal G})=(1,g_{_{HH^+H^-}})$.
Again, the Higgs decays into SUSY particles  are
considered only at LO.
In $H\to\varphi_i\varphi_j$ decay, we neglect
the off-shell effects assuming $\Gamma_\varphi \sim \Gamma_{H_{\rm SM}}$
with a caution that
the off-shell effects can not be neglected when
the decay involves a transition between two heavy states with wide widths.

\medskip

\begin{figure}[t!]
\begin{center}
\includegraphics[width=8.4cm]{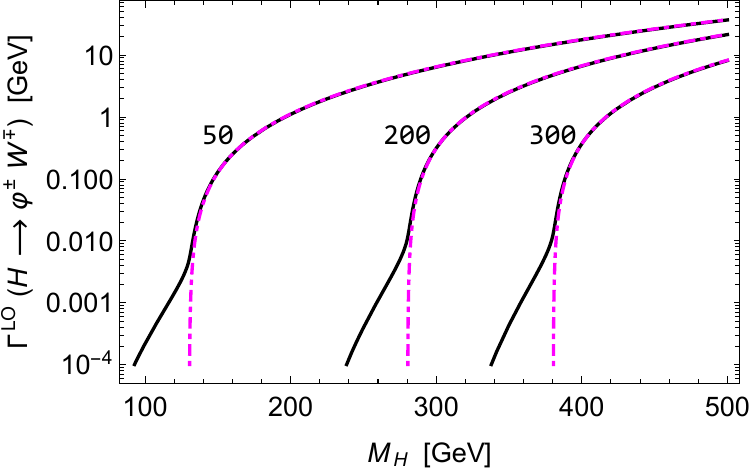}
\includegraphics[width=8.4cm]{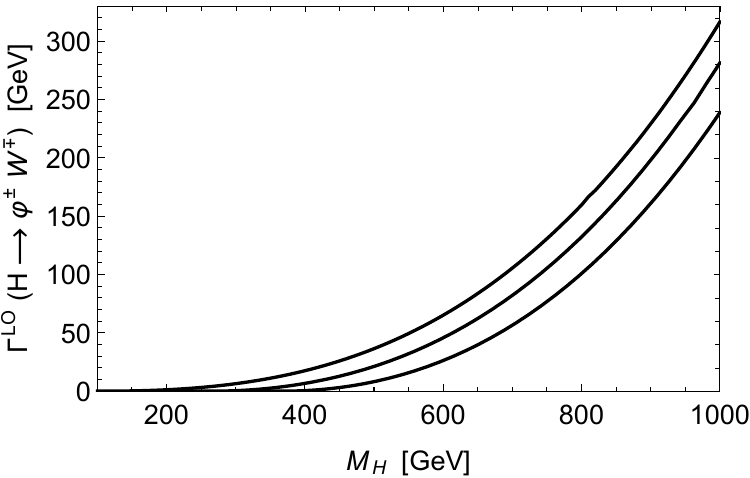}
\includegraphics[width=8.4cm]{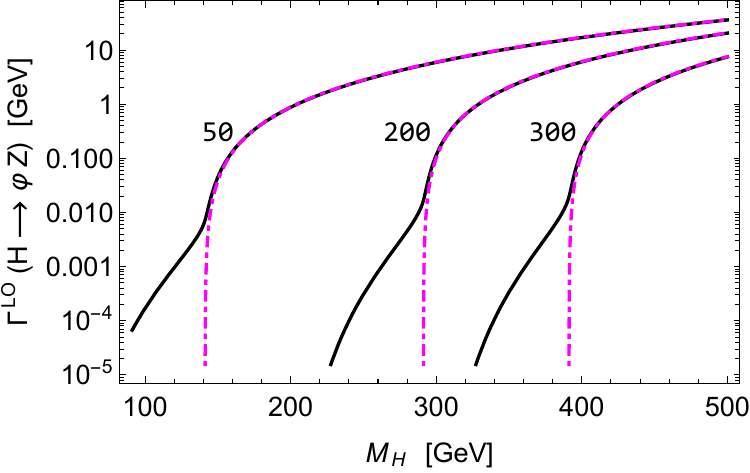}
\includegraphics[width=8.4cm]{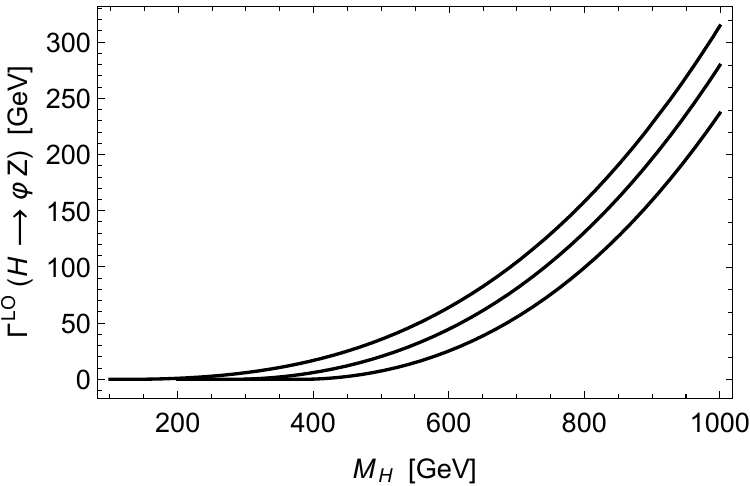}
\includegraphics[width=8.4cm]{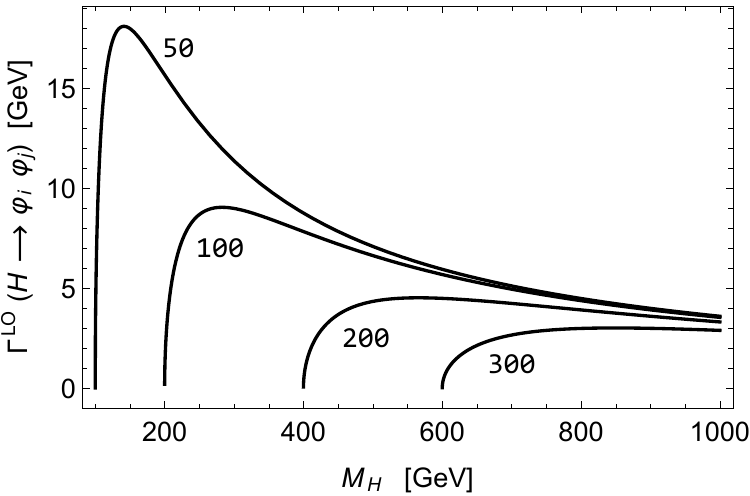}
\end{center}
\caption{
The LO decay widths of a neutral Higgs boson with mass $M_H$ into
$\varphi^- W^{+ *}$ (upper),  $\varphi Z^*$ (middle), and
$\varphi_i\varphi_j$ (lower) taking
$|g_{_{H\varphi^+ W^-}}|=1$, $g_{_{H\varphi Z}}=1$, and
$N_\varphi|{\cal G}|^2=1$, respectively.
In the upper and middle panels,
the (magenta) dash-dotted lines are for the corresponding
2-body decay widths and we are taking
$M_\varphi/{\rm GeV}=50\,, 200$\,, and $300$ from left to right.
While in the lower panel,
$M_\varphi/{\rm GeV}=50\,, 100\,, 200$\,, and $300$ are taken.
}
\label{fig:h2lighter}
\end{figure}
In Fig.~\ref{fig:h2lighter}, we show the LO
decay width of
$\Gamma^{\rm LO}(H\to \varphi^- W^{+*})$ (upper) which is the same as
$\Gamma^{\rm LO}(H\to \varphi^+ W^{-*})$,
$\Gamma^{\rm LO}(H\to \varphi Z^*)$ (middle), and
$\Gamma^{\rm LO}(H\to \varphi_i\varphi_j)$ (lower).
For the mass of a lighter scalar boson $\varphi$, we take three values of
$50$ GeV, $200$ GeV, and $300$ GeV for $H\to \varphi V^*$ for
illustration. For the decays $H\to\varphi_i \varphi_j$, we take
$M_{\varphi_i}=M_{\varphi_j}=$ $50$ GeV, $100$ GeV, $200$ GeV, and
$300$ GeV.
All the relevant couplings are taken to be $1$ in the numerical
analyses.

\subsection {Decays into two gluons: $H\to gg$}
\begin{figure}[t!]
\vspace{-1.0cm}
\begin{center}
\includegraphics[width=12.0cm]{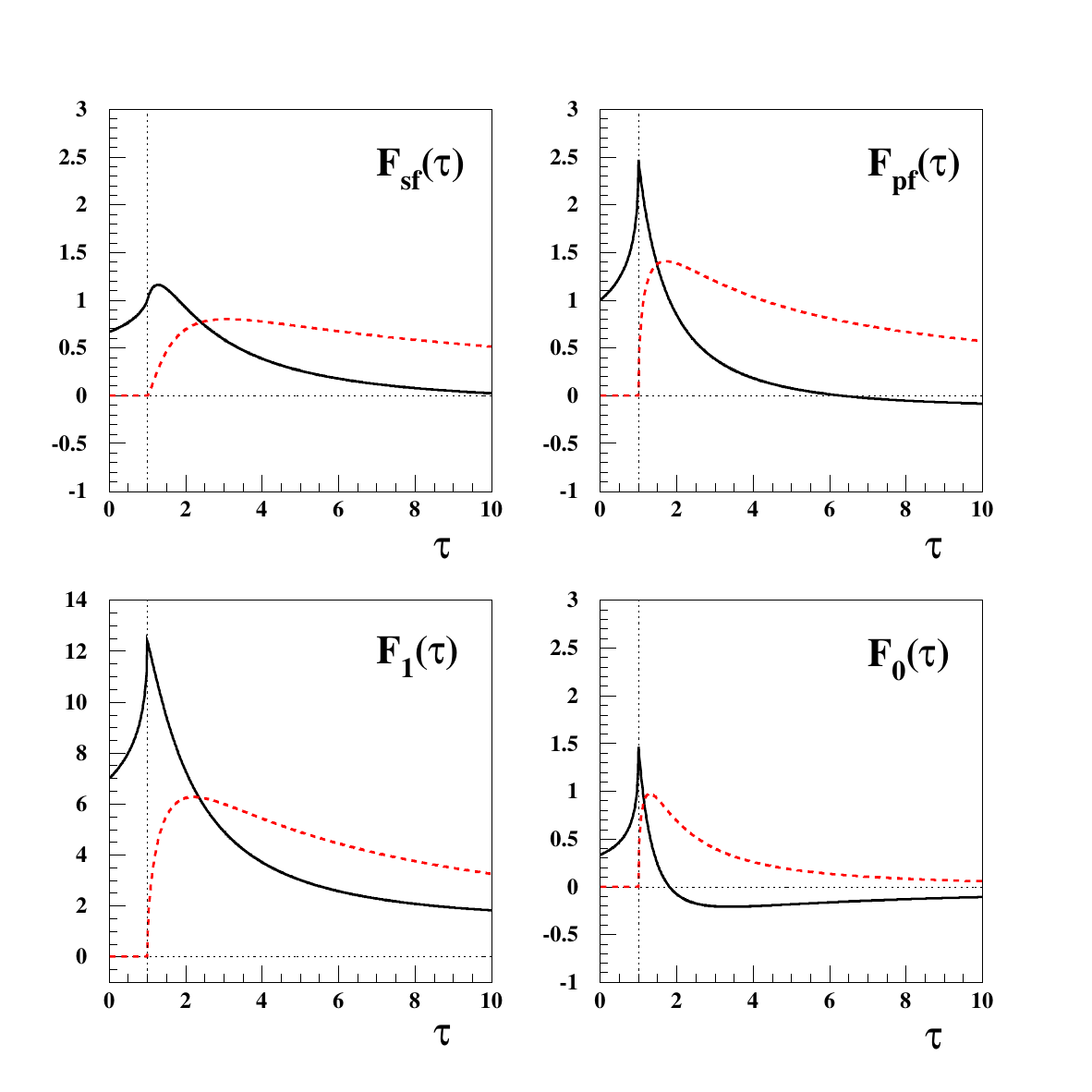}
\end{center}
\vspace{-0.5cm}
\caption{Behavior of
the real (solid) and imaginary (dashed) parts of the
four form factors of $F_{sf}$, $F_{pf}$, $F_1$ and $F_0$
versus $\tau=s/4m^2$ with $\sqrt{s}=M_H$ and $m$ being the mass
of the particle running in the triangle loops.
The form factors are defined explicitly in the main text
where they first appear,
see Eqs.~(\ref{eq:fsf_fpf}), (\ref{eq:f1}), and (\ref{eq:f0}).
The vertical lines denote the mass threshold above which
$M_H>2m$.
}
\label{fig:fplot}
\end{figure}

By introducing two form factors,
without loss of generality,
the amplitude for the decay process $H \rightarrow gg$ can be written as
\begin{eqnarray} \label{higg}
{\cal M}_{gg H}^{ab}=-\frac{\alpha_s(M_H)\,M_{H}^2\,\delta^{ab}}{4\pi\,v}
\bigg\{S^g(M_{H})
\left(\epsilon^*_{1\perp}\cdot\epsilon^*_{2\perp}\right)
 -P^g(M_{H})\frac{2}{M_{H}^2}
\langle\epsilon^*_1\epsilon^*_2 k_1k_2\rangle
\bigg\}\,,
\end{eqnarray}
where $a$ and $b$ ($a,b=1$ to 8) are indices of the eight generators in
the SU(3) adjoint representation, $k_{1,2}$ the four momenta of the
two gluons and $\epsilon_{1,2}$ the wave vectors of the corresponding gluons,
$\epsilon^\mu_{1\perp} = \epsilon^\mu_1 - 2k^\mu_1 (k_2 \cdot
\epsilon_1) / M^2_{H}$, $\epsilon^\mu_{2\perp} = \epsilon^\mu_2 -
2k^\mu_2 (k_1 \cdot \epsilon_2) / M^2_{H}$ and $\langle \epsilon_1
\epsilon_2 k_1 k_2 \rangle \equiv \epsilon_{\mu\nu\rho\sigma}\,
\epsilon_1^\mu \epsilon_2^\nu k_1^\rho k_2^\sigma$.
Retaining only the dominant contributions from third--generation quarks
and introducing $\Delta S^g$ and $\Delta P^g$ to
parameterize contributions from the triangle loops in which
non-SM colored particles are running,
the scalar and pseudoscalar form factors are given by
\footnote{See Appendix \ref{app:dsdp} for
$\Delta S^g$ and $\Delta P^g$ in the MSSM.}
\begin{eqnarray}
   S^g(M_{H})
 = \sum_{f=b,t,c}
   g^{S}_{H\bar{f}f}\,F_{sf}(\tau_{f})
   + \Delta S^g\,; \ \ \
   P^g(M_{H})
 = \sum_{f=b,t,c}
   g^{P}_{H\bar{f}f}\,F_{pf}(\tau_{f})
   + \Delta P^g\,,
\end{eqnarray}
where $\tau_f\equiv M_H^2/4M_f^2$ is defined
by using the pole masses of the bottom and top quarks.
The form factors $F_{sf}$ and $F_{pf}$ can be expressed by
\begin{eqnarray}
F_{sf}(\tau)&=&\tau^{-1}\,[1+(1-\tau^{-1}) f(\tau)]\,,~~
F_{pf}(\tau)=\tau^{-1}\,f(\tau)\,,
\label{eq:fsf_fpf}
\end{eqnarray}
in terms of a so-called scaling function $f(\tau)$ which
stands for the integral function
\begin{eqnarray}
f(\tau)=-\frac{1}{2}\int_0^1\frac{{\rm d}y}{y}\ln\left[1-4\tau y(1-y)\right]
       =\left\{\begin{array}{cl}
           {\rm arcsin}^2(\sqrt{\tau}) \,:   & \qquad \tau\leq 1\,, \\
   -\frac{1}{4}\left[\ln \left(\frac{\sqrt{\tau}+\sqrt{\tau-1}}{
                                     \sqrt{\tau}-\sqrt{\tau-1}}\right)
                    -i\pi\right]^2\,: & \qquad \tau\geq 1\,.
\end{array}\right.
\label{eq:f_function}
\end{eqnarray}
It is clear that the imaginary parts of the form factors
appear for Higgs-boson masses greater than twice the mass of
the colored particle running in the loop, i.e., $\tau\geq 1$. In the limit
$\tau\rightarrow 0$, $F_{sf}(0)=2/3$ and $F_{pf}(0)=1$, see Fig.~\ref{fig:fplot}.

\medskip

The decay width of the process $H\rightarrow gg$
may be cast into the form
\begin{eqnarray}
\label{eq:ghgg}
\Gamma(H\rightarrow gg)\ =\ \frac{M_{H}^3\alpha^2_S}{32\pi^3\,v^2}
\left[
\left|S^g(M_{H})\right|^2\,\left(1+\delta^{g:S}_{\rm QCD}\right)
\left(1+\delta^{g:S}_{\rm elw}\right)
+
\left|P^g(M_{H})\right|^2\,\left(1+\delta^{g:P}_{\rm QCD}\right)
\left(1+\delta^{g:P}_{\rm elw}\right)
\right]\,,
\end{eqnarray}
including the QCD and electroweak corrections.
The two corrections are factorized as dictated by
the universal infrared and collinear behavior of QCD corrections
and the universality of the dominant part of ELW corrections.
The QCD correction of $\delta^{g:S}_{\rm QCD}$
is known up to NLO including the full quark mass dependence \cite{Spira:1995rr}
and up to the N$^3$LO in the limit of heavy top
quarks~\cite{Djouadi:1991tka,Chetyrkin:1997iv,Baikov:2006ch}:
\begin{eqnarray}
\label{eq:ghgg_qcd_s}
\delta^{g:S}_{\rm QCD} & = &
\left(\frac{95}{4}-\frac{7}{6} N_F^L + \Delta^{g:S}_m\right)
\frac{\alpha_s^{(N_F^L)}(M_H)}{\pi}  \nonumber \\
& + & \left[370.20- 47.19 N_F^L + 0.902 (N_F^L)^2
+ (2.375 + 0.667 N_F^L)
\log\frac{M_H^2}{M_t^2}\right] \left( \frac{\alpha_s^{(N_F^L)}(M_H)}{\pi}
\right)^2 \nonumber \\
& + & \left[ 4533.46 - 1062.82 N_F^L + 52.62 (N_F^L)^2 - 0.5378 (N_F^L)^2
\phantom{\log\frac{M_H^2}{M_t^2}} \right.
\nonumber \\
&&\, + (66.66 + 14.60 N_F^L - 0.6887 (N_F^L)^2) \log\frac{M_H^2}{M_t^2}
\nonumber  \\
&&\left. + (6.53 + 1.44 N_F^L - 0.111 (N_F^L)^2) \log^2\frac{M_H^2}{M_t^2} \right]
\left( \frac{\alpha^{(N_F^L)}_{s}(M_H)}{\pi} \right)^3\,,
\end{eqnarray}
with $N_F^L=5$ counting the flavor number of light quarks
and $\Delta^{g:S}_m\approx 0.7$
\footnote{We take the value from Fig.~7(b) in Ref.~\cite{Spira:1995rr}.}
for the NLO quark-mass effects from the top, bottom and charm quarks
\cite{Spira:1995rr}.
Taking $M_H=125.5$ GeV, we find
$\delta^{g:S}_{\rm QCD} = 0.67+0.20+0.02$ for the
NLO, NNLO, and  N$^3$LO corrections.
On the other hand, the QCD corrections to the pseudoscalar part develop
a Coulomb singularity at the top-quark-pair threshold which
can be regularized by including the finite top decay width
\cite{Melnikov:1994jb,Fadin:1987wz,Fadin:1988fn,Strassler:1990nw}.
The correction $\delta^{g:P}_{\rm QCD}$ is known up to NNLO in the limit of
heavy top quarks~\cite{Chetyrkin:1998mw} while the
NLO corrections are known exactly~\cite{Spira:1995rr}:
\begin{eqnarray}
\label{eq:ghgg_qcd_p}
\delta^{g:P}_{\rm QCD}
& = & \left(\frac{97}{4}-\frac{7}{6} N_F^L + \Delta_m^{g:P}\right)
\frac{\alpha^{(N_F^L)}_{s}(M_H)}{\pi}  \nonumber \\
& + & \left(392.22- 48.58 N_F^L + 0.888 (N_F^L)^2
+ N_F^L \log\frac{M_H^2}{M_t^2}\right)
\left( \frac{\alpha^{(N_F^L)}_{s}(M_H)}{\pi} \right)^2\,,
\end{eqnarray}
with $\Delta^{g:P}_m\approx 0.04$ for $\tan\beta=1$, for example, in the MSSM.
\footnote{In the MSSM, we note that
$\Delta^{g:P}_m$ significantly depends on $\tan\beta$, see Fig.~20(b) in Ref.~\cite{Spira:1995rr}.}
Taking $M_H=125.5$ GeV, we find
$\delta^{g:P}_{\rm QCD} \simeq 0.66+0.22$ for the NLO and NNLO terms.
We note that $\delta^{g:P}_{\rm QCD}$ is derived in the heavy top-quark limit
and it does not contain the Coulomb singularity.

\medskip

\begin{figure}[t!]
\begin{center}
\includegraphics[width=14.5cm]{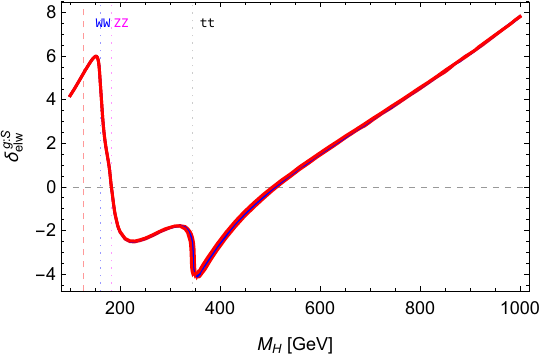}
\end{center}
\caption{
Behavior of the two-loop SM electroweak corrections of $\delta^{g:S}_{\rm elw}$
for $\Gamma(H\to gg)$ versus $M_H$.
The vertical lines locate $M_H=125.5$ GeV and
the $WW$, $ZZ$, and $t\bar t$ thresholds.
The line width represents uncertainties taking
account of $M_t=172.64 \pm 1.58$ GeV.
From the authors of Refs.~\cite{Actis:2008ts,Actis:2008ug,Actis:2008uh,Passarino:2007fp}.
See also Refs.~\cite{Degrassi:2005mc,Degrassi:2004mx,Aglietti:2004nj}.
}
\label{fig:gSelw}
\end{figure}
The electroweak corrections $\delta^{g:S}_{\rm elw}$ of
${\cal O}(G_FM_t^2)$
are given by~\cite{Chetyrkin:1996wr,Djouadi:1994ge,Chetyrkin:1996ke}
\begin{equation}
\label{eq:gSelw}
\delta^{g:S}_{\rm elw}=\frac{G_FM_t^2}{8\sqrt{2}\pi^2}\,.
\end{equation}
Note that  these ${\cal O}(G_F M_t^2)$ electroweak
corrections increase the gluonic decay width only by the amount of about $0.3$\%.
It turns out that the full electroweak corrections
\cite{Aglietti:2004nj,Aglietti:2006yd,Degrassi:2004mx,
Actis:2008ug,Actis:2008ts}
lead to much more significant enhancement of
$\delta^{g:S}_{\rm elw} \sim 5$\% below the $WW$ threshold,
see Fig.~\ref{fig:gSelw}.
On the other hand, the other electroweak corrections
$\delta^{g:P}_{\rm elw}$ may be given by~\cite{Brod:2008ct,Brod:2008ub}
\begin{equation}
\label{eq:ghgg_elw_p}
\delta^{g:P}_{\rm elw}\simeq -\left(7+\eta^{g:P}_{\rm elw}\right)
\frac{G_FM_t^2}{8\sqrt{2}\pi^2}\,,
\end{equation}
where the first factor of $7$ counts the contribution of the SU(2)$_L$ doublet
which, in the decoupling limit, plays the role of the SM SU(2)$_L$ doublet
including the 125 GeV Higgs boson.
And the second factor of $\eta^{g:P}_{\rm elw}$ denotes
the model-dependent BSM contribution.
In the type-II 2HDM, for example, it is given by
$\eta^{g:P}_{\rm elw}=10/\tan^2\beta$
in the infinite top quark mass approximation~\cite{Brod:2008ub}.
These electroweak corrections
reduce the gluonic decay width at the percent level.
Note that Eq.~(\ref{eq:ghgg_elw_p})
is a crude approximation obtained by
focusing on the top-Yukawa contributions
which are subleading in the SM.
The same arguments apply for Eq.~(\ref{eq:gSelw}) which we
are not using for our numerical study though.

\medskip

We have addressed all the known QCD and electroweak corrections
to the decay width of a neutral Higgs boson into two gluons.
The theoretical uncertainties due to the unknown higher-order
QCD and NLO electroweak corrections
are estimated as 3\% and 1\%,
respectively~\cite{Denner:2011mq,deFlorian:2016spz}.

\medskip

\begin{figure}[t!]
\begin{center}
\includegraphics[width=8.4cm]{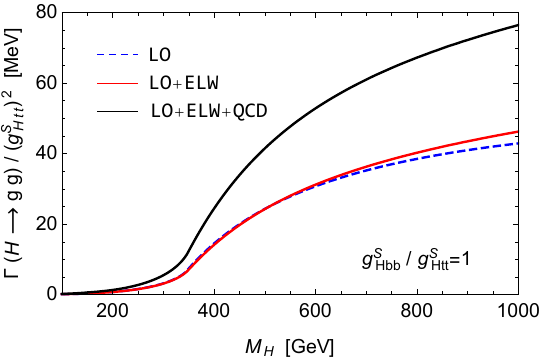}
\includegraphics[width=8.4cm]{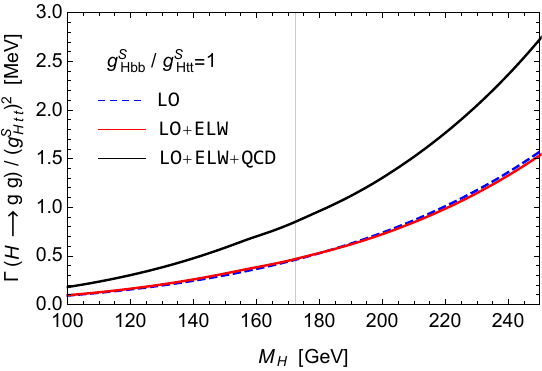}
\includegraphics[width=8.4cm]{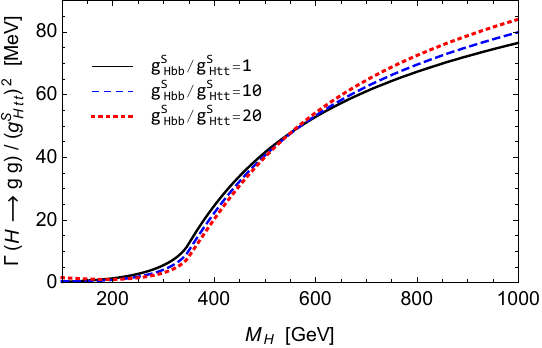}
\includegraphics[width=8.4cm]{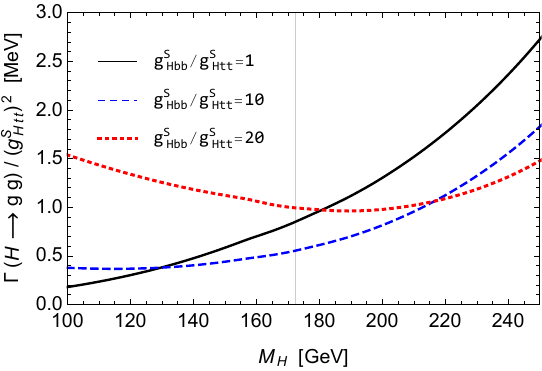}
\end{center}
\caption{
(Upper)
Normalized decay widths of a neutral Higgs boson with mass $M_H$ into $gg$
taking $g^S_{H\bar b b}/g^S_{H\bar t t}=1$, $\Delta S^g=0$, and
$g^P_{H\bar q q}=\Delta P^g=0$.
In the right panel, the low $M_H$ region is magnified.
The LO (blue dashed), LO$+$ELW (lower red solid), and
LO$+$ELW$+$QCD (upper black solid) results are separately shown.
(Lower)
The same as in the upper panels but for three values of
$g^S_{H\bar b b}/g^S_{H\bar t t}=1\,, 10$\,, and $20$ taking full
account of the electroweak and QCD corrections.
The vertical lines in the right panels locate the position $M_H=M_t$.
}
\label{fig:h2gg1}
\end{figure}
\begin{figure}[t!]
\begin{center}
\includegraphics[width=8.4cm]{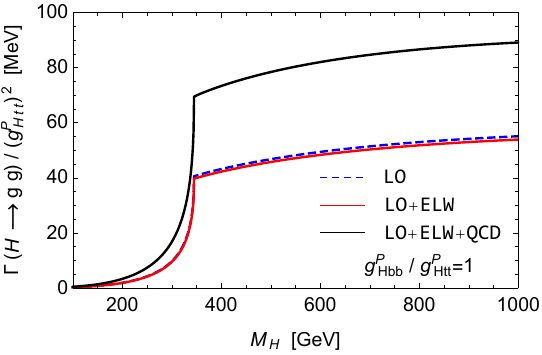}
\includegraphics[width=8.4cm]{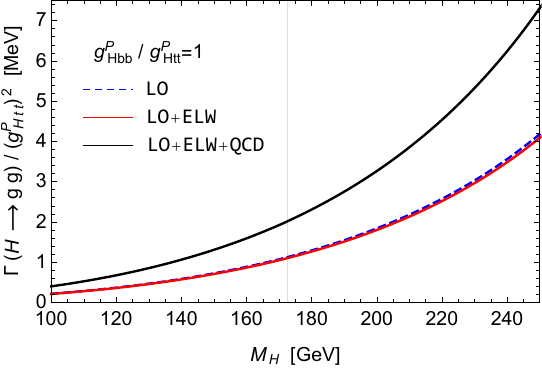}
\includegraphics[width=8.4cm]{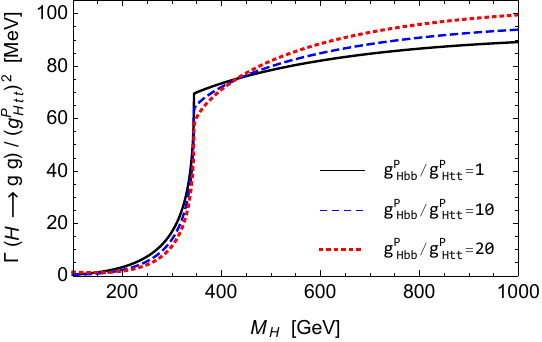}
\includegraphics[width=8.4cm]{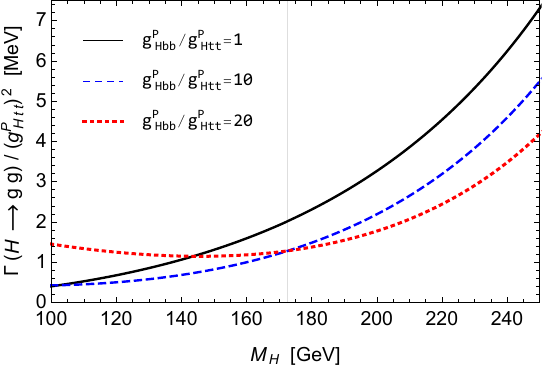}
\end{center}
\caption{
The same as in Fig.~\ref{fig:h2gg1} while
taking $g^S_{H\bar q q}=0$ and $g^P_{H\bar q q}\neq 0$
with $\Delta S^g=\Delta P^g=0$.
}
\label{fig:h2gg2}
\end{figure}
In the upper panels of Fig.~\ref{fig:h2gg1},
we show the normalized decay widths
$\Gamma(H\to gg)/(g^S_{H\bar t t})^2$ at LO (blue dashed),
including only the electroweak corrections (red solid), and
including both the electroweak and QCD corrections (black solid).
We assume that all the pseudo-scalar couplings
of $g^P_{H\bar q q}$  are vanishing and
$\Delta S^g=0$.
The electroweak corrections $\delta^{g:S}_{\rm elw}$
are directly read off from Fig.~\ref{fig:gSelw}.
%
In the right panel, we magnify the low $M_H$ region
and locate the position $M_H=M_t$ with a thin vertical line.
%
In the lower panels, we show the normalized decay widths
including the available electroweak and QCD corrections
and taking three values of $g^S_{H\bar b b}/g^S_{H\bar t t}=1$ (black solid),
$10$ (blue dashed), and $20$ (red dotted), considering the situation
in which the bottom Yukawa coupling is enhanced as in the 2HDM II for
large $\tan\beta$.
Note that the black solid lines in the lower panels are
the same as those in the upper ones.

\medskip

In Fig.~\ref{fig:h2gg2}, the alternative choice is made
to show $\Gamma(H\to gg)/(g^P_{H\bar t t})^2$
assuming all the scalar couplings of $g^S_{H\bar q q}$  are vanishing and
$\Delta P^g=0$. For the electroweak corrections $\delta^{g:P}_{\rm elw}$,
we take $\eta^{g:P}_{\rm elw}=0$.
Compared to the scalar case in which the form factor $f_{sf}(\tau)$
is involved, 
the rise near
the top-quark-pair threshold is sharper and bigger due to the behavior of
the real and imaginary parts of the form factor $f_{pf}(\tau)$ around $\tau=1$,
see Fig.~\ref{fig:fplot}.
In the low $M_H$ region, we find that the decay widths are larger
by about the factor of $[f_{pf}(0)/f_{sf}(0)]^2 \sim 2$ when
the $b$-quark contributions are neglected.
As the coupling ratios $g^{S,P}_{H\bar b b}/g^{S,P}_{H\bar t t}$ increase,
the contributions from $b$-quark loops become comparable to and larger
than those from $t$-quark loops and, in this case,
the decay widths
are nearly the same especially around $M_H = 100$ GeV
as can be checked with the dashed and dotted lines in  the lower-right panels of
Figs.~\ref{fig:h2gg1} and \ref{fig:h2gg2}.
%

\subsection {Decays into two photons: $H\to \gamma\gamma$}
\label{sec:h2aa}
The amplitude for the radiative decay process $H \rightarrow \gamma\gamma$,
playing a crucial role in the discovery of the Higgs boson at the LHC,
can be written as
\begin{eqnarray} \label{hipp}
{\cal M}_{\gamma\gamma H}=-\frac{\alpha(0) M_{H}^2}{4\pi\,v}
\bigg\{S^\gamma(M_{H})\, \left(\epsilon^*_{1\perp}\cdot\epsilon^*_{2\perp}\right)
 -P^\gamma(M_{H})\frac{2}{M_{H}^2}
\langle\epsilon^*_1\epsilon^*_2 k_1k_2\rangle
\bigg\}\,,
\end{eqnarray}
in terms of the two form factors of $S^\gamma$ and $P^\gamma$.
Here
$k_{1,2}$ and $\epsilon_{1,2}$ are the four--momenta and wave vectors
of the two photons, respectively, as in the decay $H\rightarrow gg$.
Note that the electromagnetic fine structure constant in the coupling
should be taken at the scale $q^2=0$ since the final--state photons are real.
Retaining only the dominant contributions from third--generation
fermions and the charged gauge bosons $W^\pm$
and introducing two residual form factors
$\Delta S^\gamma$ and $\Delta P^\gamma$ to parameterize contributions from the
triangle loops in which non-SM charged particles are running, the scalar and
pseudoscalar form factors are given by
\footnote{See Appendix \ref{app:dsdp} for
$\Delta S^\gamma$ and $\Delta P^\gamma$ in the MSSM.}
\begin{eqnarray}
   S^\gamma(M_{H})
&=& 2\sum_{f=b,t,\tau} N_C^f\, Q_f^2\,
    g^{S}_{H\bar{f}f}\,F_{sf}(\tau_{f}) - g_{_{HWW}} F_1(\tau_{W})
    +\Delta S^\gamma \,; \nonumber \\
   P^\gamma(M_{H})
&=& 2\sum_{f=b,t,\tau} N_C^f\,Q_f^2\,
    g^{P}_{H\bar{f}f} \,F_{pf}(\tau_{f})+\Delta P^\gamma \,,
\end{eqnarray}
where $N_C^f=3$ for quarks and $N_C^f=1$ for charged leptons,
respectively. For the $\tau$ lepton and the $W$ boson,
$\tau_\tau=M_{H}^2/4M_\tau^2$ and $\tau_{W}=M_{H}^2/4M_W^2$,
respectively. On the other hand, for quarks,
$\tau_{q}$ is defined in terms of the running quark mass
at the scale of $M_H/2$, i.e.
$\tau_{q}=M_{H}^2/4[m_q(M_H/2)]^2$ where $m_q$ is normalized as $m_q(M_q)=M_q$.
\footnote{For $m_q(\mu)$, see Appendix \ref{app:smpara}.}
Note that the choice of the scale $\mu_q=M_H/2$ correctly gives
$m_q(M_H/2)=M_q$ at the threshold where $M_H=2M_q$ and
makes the full two-loop QCD corrections remain small
in the entire range of the variable $\tau_q$
by effectively absorbing all relevant large logarithms
into the running mass.
The form factor $F_1$ is given by
\begin{eqnarray}
F_1(\tau)=2+3\tau^{-1}+3\tau^{-1} (2-\tau^{-1} )f(\tau)\,,
\label{eq:f1}
\end{eqnarray}
which takes the value of 7 in the limit $\tau\to 0$, see Fig.~\ref{fig:fplot}
for the $\tau$ dependence of the form factor.
At LO, the decay width of the radiative process is given by
\begin{eqnarray}
\label{eq:ghaa_lo}
\Gamma^{\rm LO}(H\rightarrow \gamma\gamma)=\frac{M_{H}^3\alpha^2}{256\pi^3\,v^2}
         \left[\,\left|S^\gamma(M_{H})\right|^2
              +\left|P^\gamma(M_{H})\right|^2\right]\,,
\end{eqnarray}
with the fine structure constant $\alpha=\alpha(0)\simeq 1/137$.

\medskip

\begin{figure}[t!]
\begin{center}
\includegraphics[width=8.5cm]{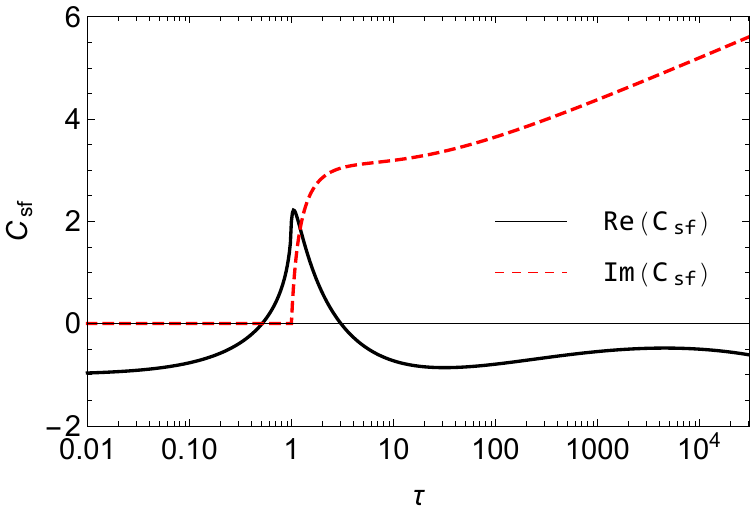}
\includegraphics[width=8.5cm]{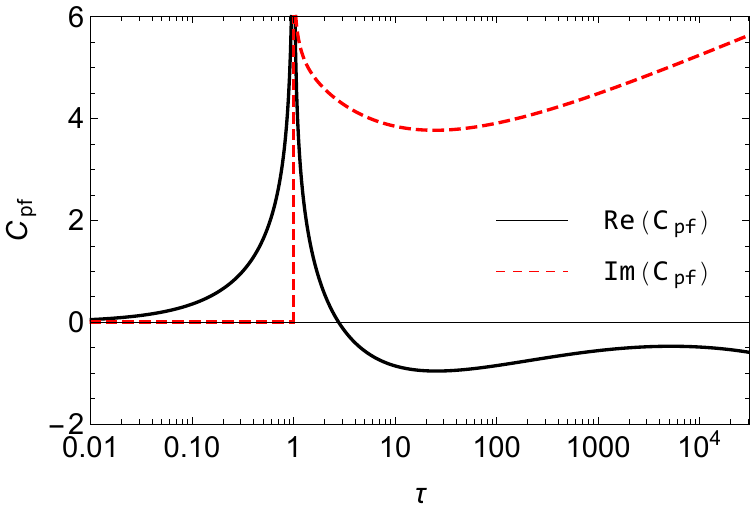}
\end{center}
\caption{
Behavior of
the real (solid) and imaginary (dashed) parts of
the scaling factors $C_{sf}(\tau)$ (left)
and $C_{pf}(\tau)$ (right) for the QCD corrections to
the decay of a Higgs boson into two photons.
}
\label{fig:csfcpf}
\end{figure}

As a gluon cannot be radiated from the colored quark
loop contributing to the $H\gamma\gamma$ vertex owing to charge conjugation
invariance and color conservation, the full two-loop QCD corrections
\cite{Spira:1995rr,Muhlleitner:2006wx,Bonciani:2007ex,
Zheng:1990qa,Djouadi:1990aj,Dawson:1992cy,Melnikov:1993tj,Djouadi:1993ji,Inoue:1994jq,
Fleischer:2004vb,Harlander:2005rq,Anastasiou:2006hc,Aglietti:2006tp,Djouadi:1996pb}
can be taken into account by simply
introducing the scaling factors in the form factors
for the $b$- and $t$-quark contributions as:
\footnote{For a detailed description of the scaling factors of
$C_{sf}(\tau)$ and $C_{pf}(\tau)$, see Appendix \ref{app:csfcpf}.}
\begin{eqnarray}
\label{eq:ghaa_qcd}
F_{sf}(\tau_{q}) &\longrightarrow& F_{sf}(\tau_{q})\,
\left[1+C_{sf}(\tau_q)\frac{\alpha_s(M_H)}{\pi}\right] \,; \nonumber \\
F_{pf}(\tau_{q}) &\longrightarrow& F_{pf}(\tau_{q})\,
\left[1+C_{pf}(\tau_q)\frac{\alpha_s(M_H)}{\pi}\right] \,.
\end{eqnarray}
The scaling factors $C_{sf}$ and $C_{pf}$ approach $-1$ and $0$, respectively,
in the limit $\tau\rightarrow 0$, see Fig.~\ref{fig:csfcpf}.

\medskip

The asymptotic values of the scalar and pseudoscalar scaling
factors in the heavy quark limit, $C_{sf}(0)=-1$ and $C_{pf}(0)=0$,
can also be deduced by means of a general low-energy theorem for amplitudes
involving soft Higgs particles.
According to the low-energy theorem, the NLO QCD corrections to the scalar
coupling to two photons in the heavy quark limit can be obtained from the
effective Lagrangian~\cite{Spira:1997dg,Spira:1995rr,Ellis:1975ap,
Shifman:1979eb,Kniehl:1995tn}
\begin{eqnarray}
\label{eq:low_e_eff_hrr}
  {\cal L}^S_{\rm eff}
=  g^S_{H\bar q q}\,\frac{Q^2_q}{4} \frac{\beta^q_{\rm QED}/\alpha}{1+\gamma_m(\alpha_s)}
  \, F^{\mu\nu}F_{\mu\nu}\,\frac{H}{v}\,,
\end{eqnarray}
where the QED $\beta$ function
$\beta^q_{\rm QED}/\alpha = 2(\alpha/\pi)[1+\alpha_s/\pi+\cdots]$ with the heavy
quark $q$ contribution and the anomalous mass dimension
$\gamma_m=2\alpha_s/\pi+\cdots$~\cite{Tarasov:2019rwk}. Therefore, the NLO expansion
of the effective Lagrangian of the scalar coupling to two photons reads
\begin{eqnarray}
\label{eq:eff_nlo_hrr_expanded}
   {\cal L}^S_{\rm eff}
= g^S_{H\bar q q}\,Q^2_q \frac{\alpha}{2\pi} F^{\mu\nu} F_{\mu\nu}
  \left[1-\frac{\alpha_s}{\pi}+{\cal O}(\alpha^2_s)\right]\,\frac{H}{v}\,,
\end{eqnarray}
which agrees with the asymptotic value of $C_{sf}(0)=-1$ in the
heavy quark limit.
On the other hand, the pseudoscalar scaling factor is zero in the
heavy quark limit, i.e., $C_{pf}(0)=0$, because the QCD corrections to
the pseudoscalar decay mode vanish in this limit due to the Adler-Bardeen
theorem~\cite{Adler:1969er}. The effective Lagrangian for the
pseudoscalar coupling to two photons can be derived from the
Adler-Bell-Jackiw (ABJ) anomaly in the divergence of the axial vector
current~\cite{Adler:1969gk,Bell:1969ts} and is given to all orders of
the perturbation theory by \cite{Spira:1997dg,Spira:1995rr}
\begin{eqnarray}
\label{eq:low_e_arr}
  {\cal L}^P_{\rm eff}
 = g^P_{H\bar q q}\, Q^2_q\, \frac{3\alpha}{4\pi}\,
   F^{\mu\nu}\widetilde{F}_{\mu\nu}\, \frac{H}{v}\,,
\end{eqnarray}
for the pseudoscalar boson, as the lowest order axial-anomaly term is not
renormalized at all orders.

\medskip

\begin{figure}[ht!]
\begin{center}
\includegraphics[width=14.5cm]{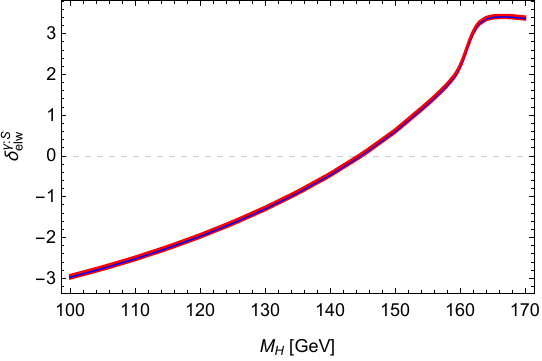}
\end{center}
\vspace{-0.6cm}
\caption{
The two-loop electroweak corrections $\delta_{\rm elw}^{\gamma:S}$
to the part of the decay width of a Higgs boson into two photons
via the scalar form factor
in the full complex-mass scheme
which has been designed to cure the unphysical infinities of two-loop amplitudes
\cite{Actis:2008ts}.
The line width represents uncertainties taking
account of $M_t=172.64 \pm 1.58$ GeV.
From the authors of
Refs.~\cite{Actis:2008ts,Actis:2008ug,Actis:2008uh,Passarino:2007fp}.
}
\label{fig:dhaaelw}
\end{figure}
The two-loop electroweak corrections of $\delta^{\gamma:S}_{\rm elw}$
to the scalar part
of the decay width via the scalar form factor
have been calculated
in Refs.~\cite{Actis:2008ts,Djouadi:1997rj,Degrassi:2005mc,Passarino:2007fp},
see Fig.~\ref{fig:dhaaelw}.
The electroweak corrections to the pseudoscalar part
are also available~\cite{Brod:2008ct,Brod:2008ub}:
\begin{equation}
\label{eq:dhaaelwP}
\delta^{\gamma:P}_{\rm elw}=
-\frac{G_FM_t^2}{8\sqrt{2}\pi^2}
\left( 4 + \eta^{\gamma :P}_{\rm elw} \right)\,,
\end{equation}
where, similarly as in the decay $H\to gg$,
the factor $4$ counts the contribution of the SM SU(2)$_L$ doublet
in the decoupling limit  and
$\eta^{\gamma :P}_{\rm elw}$ the model-dependent BSM contribution.
In the type-II 2HDM, for example,
$\eta^{\gamma :P}_{\rm elw}=7/\tan^2\beta$
in the infinite top quark mass approximation~\cite{Brod:2008ct,Brod:2008ub},
suppressed for large $\tan\beta$.
Note again that Eq.~(\ref{eq:dhaaelwP})
is a crude approximation obtained by
focusing on the top-Yukawa contributions which are subleading in the SM.

\medskip

\begin{figure}[t!]
\begin{center}
\includegraphics[width=12.5cm]{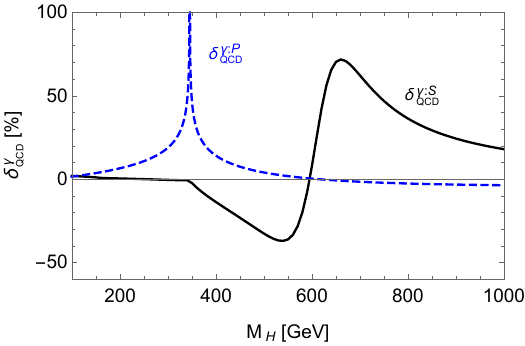}
\end{center}
\caption{
Behavior of the two-loop QCD corrections to
the decay width of a Higgs boson into two photons:
$\delta^{\gamma : S}_{\rm QCD}$ (solid) and
$\delta^{\gamma : P}_{\rm QCD}$ (dashed).
We take  $g_{_{HWW}}=g^S_{H\bar t t}=g^S_{H\bar b b}=g^S_{H\tau\tau}=1$
for $\delta^{\gamma : S}_{\rm QCD}$ while
$g^P_{H\bar b b}=g^P_{H\tau\tau}=0$ for $\delta^{\gamma : P}_{\rm QCD}$.
Both of the residual form factors
$\Delta S^\gamma$ and $\Delta P^\gamma$
are taken  to be vanishing.
}
\label{fig:dhaaqcd}
\end{figure}
Incorporating all the QCD and electroweak corrections, one may write
\begin{eqnarray}
\label{eq:ghaa}
\Gamma(H\rightarrow \gamma\gamma)=\frac{M_{H}^3\alpha^2}{256\pi^3\,v^2}
         \left[\,\left|S^\gamma(M_{H})\right|^2
\left(1+\delta^{\gamma:S}_{\rm QCD}\right)
\left(1+\delta^{\gamma:S}_{\rm elw}\right)
        +\left|P^\gamma(M_{H})\right|^2
\left(1+\delta^{\gamma:P}_{\rm QCD}\right)
\left(1+\delta^{\gamma:P}_{\rm elw}\right)\right]\,.
\end{eqnarray}
Note that the electroweak corrections are directly from
Fig.~\ref{fig:dhaaelw} and Eq.~(\ref{eq:dhaaelwP}) while the QCD corrections
enter through the scaling factors $C_{sf}(\tau_q)$ and $C_{pf}(\tau_q)$.
For $M_H=125.5$ GeV, the QCD corrections
$\delta^{\gamma:S}_{\rm QCD}$ increase the decay width into two photons
by about 2\%.
On the other hand, the electroweak corrections $\delta^{\gamma:S}_{\rm elw}$
decrease the decay width by about 2\% , almost canceling the
NLO QCD corrections to the corresponding part.
%
In Fig.~\ref{fig:dhaaqcd}, we show the QCD corrections
$\delta^{\gamma : S}_{\rm QCD}$ (solid) and
$\delta^{\gamma : P}_{\rm QCD}$ (dashed) with varying $M_H$.
For the scalar QCD correction $\delta^{\gamma : S}_{\rm QCD}$,
we take the SM values of $g_{_{HWW}}=g^S_{H\bar t t}=g^S_{H\bar b b}
=g^S_{H\tau\tau}=1$ with $\Delta S^\gamma=0$. While, for the pseudoscalar QCD correction
$\delta^{\gamma : P}_{\rm QCD}$, we assume a scenario in which
the pseudoscalar form factor $P^\gamma$ is dominated by the top-quark contribution
taking $g^P_{H\bar b b}=g^P_{H\tau\tau}=\Delta P^\gamma=0$.
\footnote{In the general case with arbitrary
$g_{_{HWW}}$ and $g^{S,P}_{H\bar f f}$ couplings,
the scaling factors $C_{sf}(\tau_q)$ and $C_{pf}(\tau_q)$ should be
taken into account at the amplitude level
to incorporate the corresponding QCD corrections properly.}
At $M_H = 2M_t$, the pseudoscalar QCD correction
$\delta^{\gamma : P}_{\rm QCD}$ diverges
due to the singular property of $C_{pf}$ at $\tau=1$.
Around $M_H=600$ GeV where the large cancellation occurs between
the $W$-boson and top-quark contributions, the scalar QCD correction
$\delta^{\gamma:S}_{\rm QCD}$ is relatively large and it
could vary between about $-0.4$ and $0.8$.

\medskip

\begin{figure}[t!]
\begin{center}
\includegraphics[width=8.4cm]{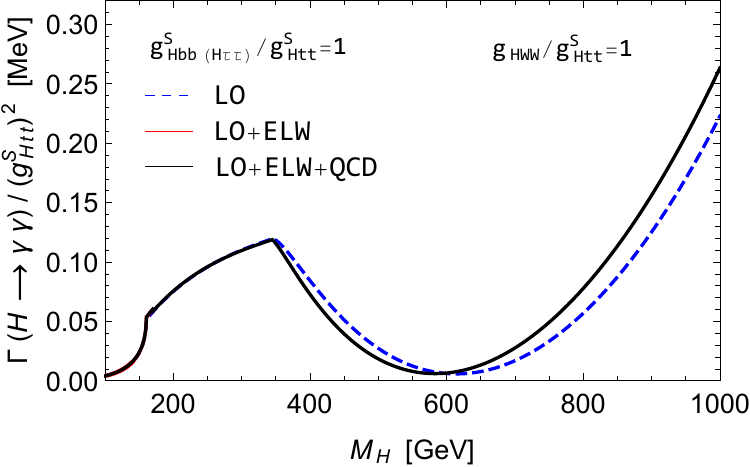}
\includegraphics[width=8.4cm]{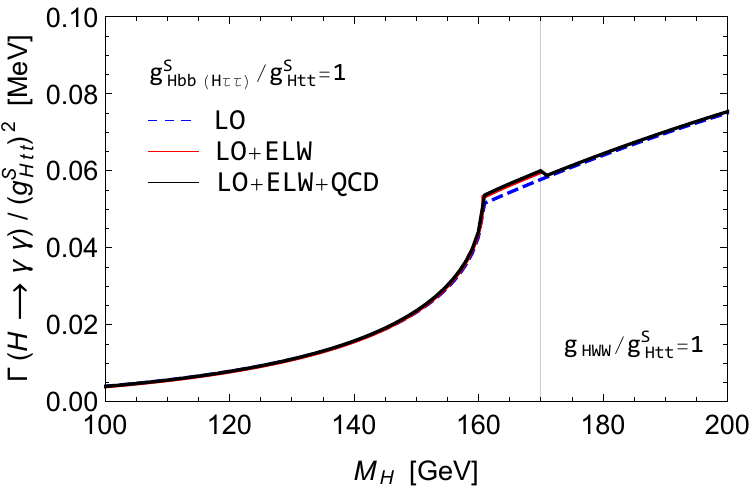}
\includegraphics[width=8.4cm]{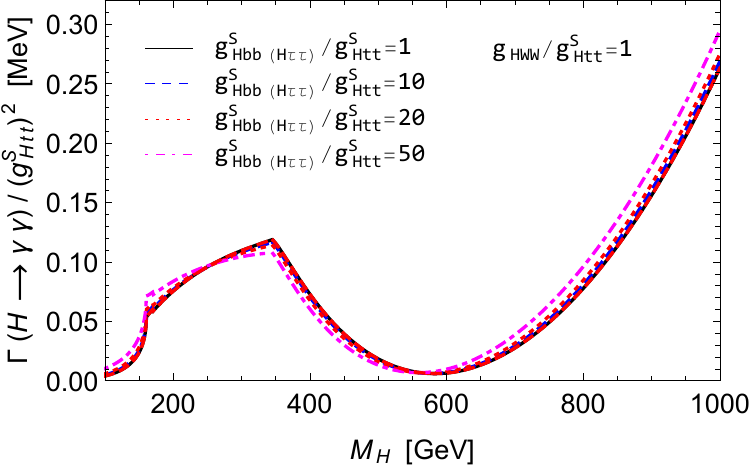}
\includegraphics[width=8.4cm]{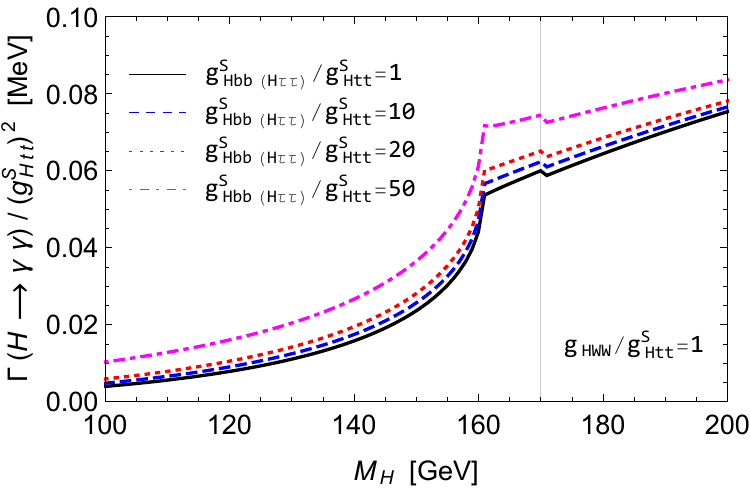}
\includegraphics[width=8.4cm]{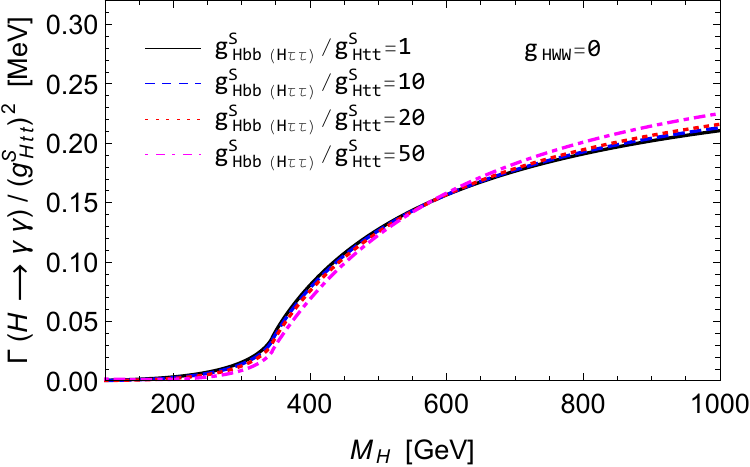}
\includegraphics[width=8.4cm]{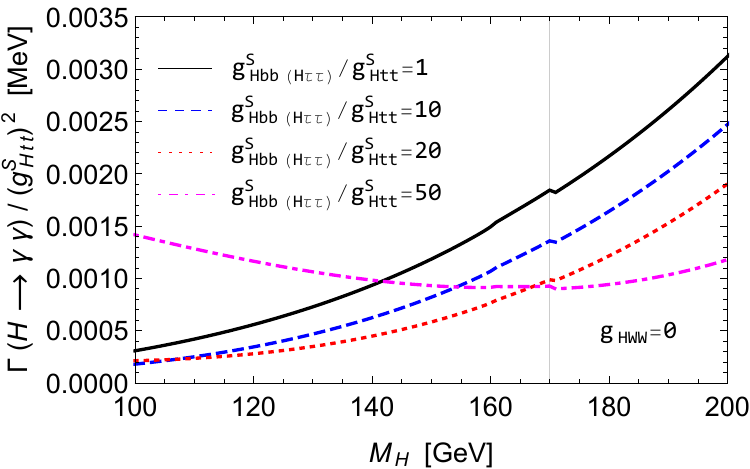}
\end{center}
\caption{
(Upper)
Normalized decay widths of a neutral Higgs boson with a mass $M_H$ into
$\gamma\gamma$
taking $g_{_{HWW}}/g^S_{H\bar t t}=g^S_{H\bar b b\,,H\tau\tau}/g^S_{H\bar t t}=1$
and $g^P_{H\bar f f}=0$.
In the right panel, the low $M_H$ region is magnified.
The LO (blue dashed), LO$+$ELW (red solid), and
LO$+$ELW$+$QCD (black solid) results are separately shown.
(Middle)
The same as in the upper panels but for the four values of
$g^S_{H\bar b b}/g^S_{H\bar t t}=1\,,10\,,20$\,, and $50$
taking full account of the electroweak and QCD corrections.
$g_{_{HWW}}/g^S_{H\bar t t}=1$ is taken.
(Lower)
The same as in the middle panels but with $g_{_{HWW}}=0$.
The vertical lines in the right panels locate the position $M_H=170$ GeV
beyond which the electroweak corrections are ignored.
In all panels, we take $\Delta S^\gamma=\Delta P^\gamma=0$.
}
\label{fig:h2aa1}
\end{figure}
\begin{figure}[t!]
\begin{center}
\includegraphics[width=8.4cm]{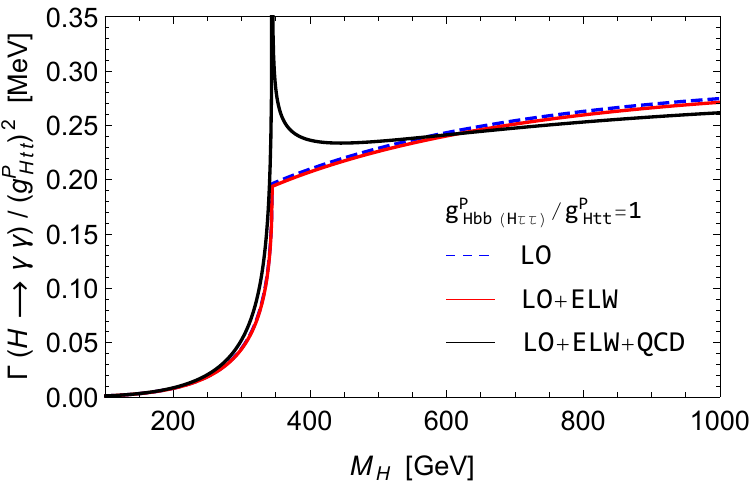}
\includegraphics[width=8.4cm]{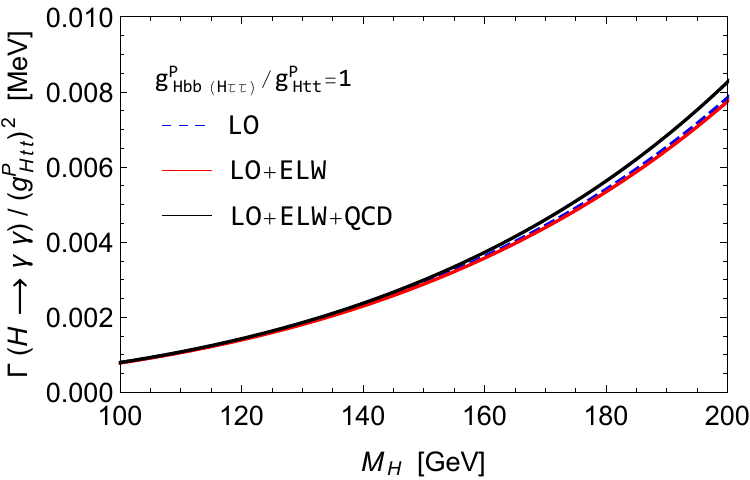}
\includegraphics[width=8.4cm]{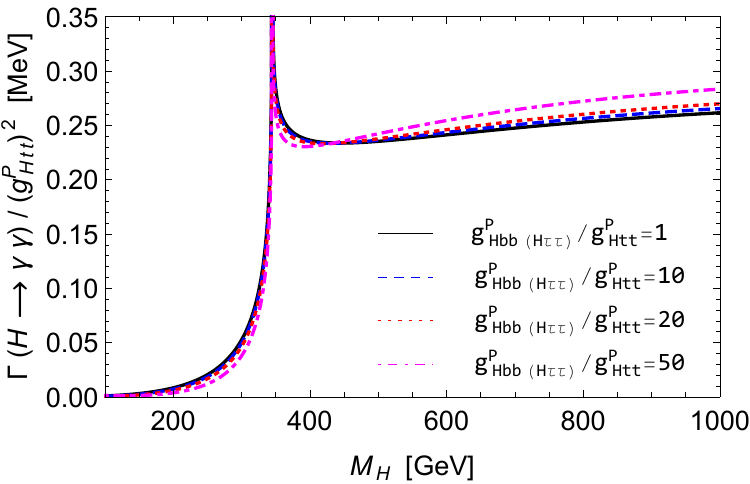}
\includegraphics[width=8.4cm]{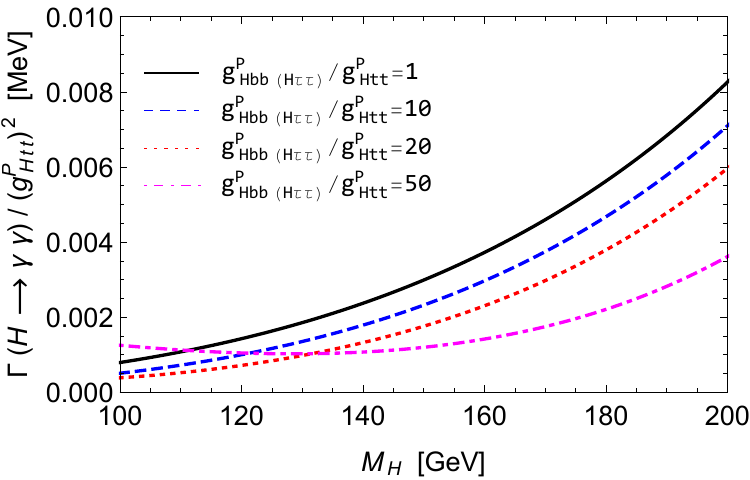}
\end{center}
\caption{
Normalized decay widths of a neutral Higgs boson with a mass $M_H$ into
$\gamma\gamma$ via the
pseudoscalar form factor $P^\gamma$ with $\Delta P^\gamma=0$. In this case,
only the fermion couplings of $g^P_{H\bar f f}$ with $f=t,b,\tau$ are relevant.
}
\label{fig:h2aa2}
\end{figure}

In the upper panels of Fig.~\ref{fig:h2aa1},
we show the decay width of a neutral Higgs boson
into two photons normalized to
the $g_{H\bar t t}$ coupling squared taking
$g_{_{HWW}}/g^S_{H\bar t t}=g^S_{H\bar b b\,,H\tau\tau}/g^S_{H\bar t t}=1$ and
$g^P_{H\bar f f}=\Delta S^\gamma=\Delta P^\gamma=0$.
This reduces to the SM decay width of a Higgs particle with mass
$M_H$ when $g_{H\bar t t}=1$.
For $M_H\leq 170$ GeV, we apply the electroweak corrections $\delta_{\rm elw}^{\gamma:S}$
directly read off from Fig.~\ref{fig:dhaaelw} and
we simply neglect them above $M_H=170$ GeV,
expecting the electroweak corrections to be evaluated with high precision for heavy
Higgs bosons if necessary.
Below the $W$-boson-pair threshold,
the $W$-loop contributions are dominant, leading to
the sharp rise as $M_H$ approaches $2M_W$. Passing $M_H=2M_W$ from below,
the real part of the $W$-loop contributions decreases but
its imaginary part starts to be developed. As a result, the Higgs
decay width continues to increase with increasing $M_H$ until the Higgs
mass meets the top-quark-pair threshold, $M_H=2 M_t$.
Passing the top-quark-pair threshold,
the real and newly-developed
imaginary parts of the $t$-quark loop contributions
cancel those of the $W$-loop contributions
\cite{Melnikov:1993tj,Djouadi:1993ji}.
We observe that, beyond $M_H=2M_t$,
the cancellation between the real and imaginary parts of the
$W$-loop and $t$-loop contributions
broadly occurs and it leads to a dip around $M_H=600$ GeV.
Specifically we find
\begin{equation}
S^\gamma (M_H=600\,{\rm GeV}) \simeq (-1.593-2.599\,i)\,g_{_{HWW}} +
(1.391 +2.131\,i) \,g_{H\bar{t}t}^S\,,
\end{equation}
with negligible contributions from the $b$-quark and $\tau$-lepton loops.
In the middle panels of Fig.~\ref{fig:h2aa1},
we show the variation depending on
$g^S_{H\bar b b}/g^S_{H\bar t t}=g^S_{H\tau\tau}/g^S_{H\bar t t}$ still taking
$g_{_{HWW}}/g^S_{H\bar t t}=1$. The (black) solid lines
represent the same case as in the upper panels taking full account of
the QCD and electroweak corrections.
And, in the lower panels, we show the results taking $g_{_{HWW}}=0$.
The last case may apply to the heavy neutral Higgs bosons appearing in
the 2HDMs and/or MSSM when their couplings to the massive vector bosons are
naturally suppressed and almost
vanishing~\cite{Gunion:2002zf,Haber:2002ey}.

\medskip

In Fig.~\ref{fig:h2aa2}, the alternative choice is made to show the
normalized decay width $\Gamma(H\to \gamma\gamma)/(g^P_{H\bar t t})^2$
assuming that all the scalar couplings of $g^S_{H\bar f f}$
are vanishing and, again,
taking $\Delta S^\gamma=\Delta P^\gamma=0$.
Note that, in this case, only the fermion loops are contributing.
For the electroweak corrections $\delta^{\gamma:P}_{\rm elw}$,
we take $\eta^{\gamma:P}_{\rm elw}=0$.
At $M_H = 2M_t$, the decay width diverges
because of the singular property of the QCD corrections.
In the lower panels, we show the dependence on
$g^P_{H\bar b b}/g^P_{H\bar t t}=g^P_{H\tau\tau}/g^P_{H\bar t t}$ for
the four values of $1\,,10\,,20$, and $50$ taking full account
of the electroweak and QCD corrections.
In the right panels, as the same as in Fig.~\ref{fig:h2aa1}, we magnify the low
$M_H$ regions.
%

\subsection {Decays into a vector boson $Z$ and a photon: $H\to Z\gamma$}
\begin{figure}[t!]
\begin{center}
\includegraphics[width=8.4cm]{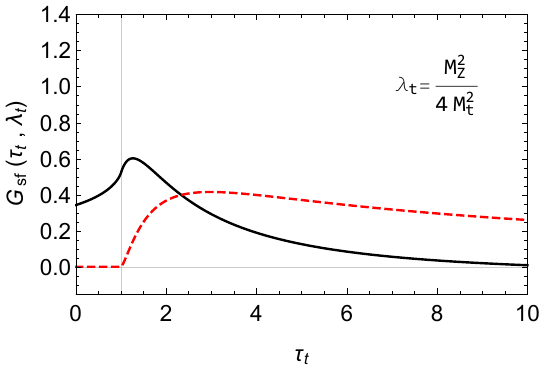}
\includegraphics[width=8.4cm]{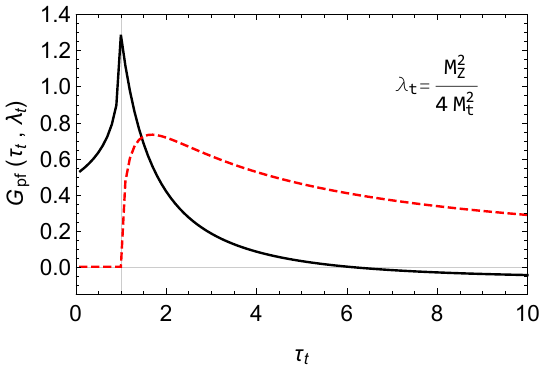}\\[3mm]
\includegraphics[width=8.4cm]{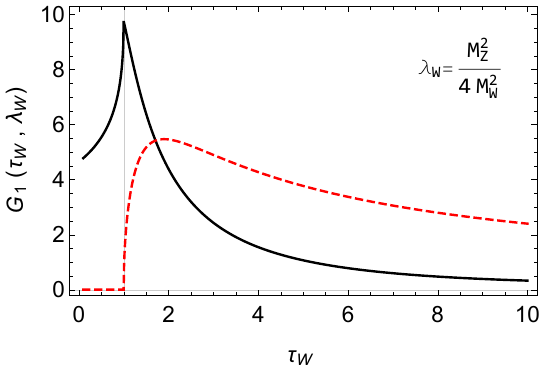}
\end{center}
\vspace{-0.5cm}
\caption{
Behavior of
the real (solid) and imaginary (dashed) parts of
the form factors $G_{sf}(\tau_t\,,\lambda_t)$ and
$G_{pf}(\tau_t\,,\lambda_t)$ (upper) and
that of $G_1(\tau_W\,,\lambda_W)$ (lower).
We recall the relations
$\tau_x=M_H^2/4M_x^2$ and $\lambda_x=M_Z^2/4M_x^2$.
The vertical lines at $\tau_x=1$ denote the mass threshold above which
$M_H>2M_x$.
%
%
%
%
}
\label{fig:Loop_gG}
\end{figure}
The amplitude for the decay process $H \to Z\gamma$ can be written
as~\cite{Djouadi:1996yq}
\begin{eqnarray}
{\cal M}_{Z\gamma H} = -\,\frac{\alpha}{2\pi v}\Bigg\{
S^{Z\gamma}(M_{H},M_Z)\,
\left[ k_1\cdot k_2\,\epsilon_1^*\cdot\epsilon_2^*
-k_1\cdot\epsilon_2^*\,k_2\cdot\epsilon_1^* \right]
- P^{Z\gamma}(M_{H},M_Z)\,
\langle \epsilon_1^*\epsilon_2^* k_1 k_2\rangle
\Bigg\}\,,
\end{eqnarray}
with the two form factors of $S^{Z\gamma}$ and $P^{Z\gamma}$.
\footnote{We follow the conventions and notations  of
Ref.~\cite{Lee:2012wa} for the form factors.}
Here $k_{1,2}$ are the momenta of the $Z$ boson and the photon
(we note that $2k_1\cdot k_2 = M_{H}^2-M_Z^2$),
$\epsilon_{1,2}$ are their polarization vectors, and
$\langle \epsilon_1\epsilon_2 k_1 k_2\rangle
\equiv \epsilon_{\mu\nu\alpha\beta}\epsilon_1^\mu\epsilon_2^\nu k_1^\alpha k_2^\beta$.
Retaining only the dominant contributions from third--generation
fermions and $W^\pm$
and introducing two residual form factors $\Delta S^{Z\gamma}$ and
$\Delta P^{Z\gamma}$ to parameterize contributions from the triangle loops in which
non-SM particles are running,
the scalar and pseudoscalar form factors are given by
\footnote{See Appendix \ref{app:dsdp} for the explicit forms of
$\Delta S^{Z\gamma}$ and $\Delta P^{Z\gamma}$ in the MSSM.}
\begin{eqnarray}
S^{Z\gamma}(M_{H},M_Z)\,
&=& 2 \sum_{f=t,b,\tau} N_C^f Q_f \frac{I_3^f-2s_W^2 Q_f}{s_Wc_W}\
g^{S}_{H\bar{f}f} \ G_{sf}(\tau_f,\lambda_f)
-\frac{1}{s_W}\,g_{_{HWW}} G_1(\tau_W,\lambda_W)
+\Delta S^{Z\gamma}\,,
\nonumber  \\[2mm]
P^{Z\gamma}(M_{H},M_Z)\,
&=& 2 \sum_{f=t,b,\tau} N_C^f Q_f \frac{I_3^f-2s_W^2 Q_f}{s_Wc_W}\
g^{P}_{H\bar{f}f} \ G_{pf}(\tau_f,\lambda_f)
+\Delta P^{Z\gamma} \,,
\end{eqnarray}
where $I_3^{\,u, \nu}=+1/2$, $I_3^{\,d, e}=-1/2$,
$\tau_x=M_H^2/4m_x^2$ and $\lambda_x=M_Z^2/4m_x^2$, respectively.
\footnote{We take the pole masses  of top and bottom quarks for
$\tau_{t,b}$ and $\lambda_{t,b}$.}
The loop functions are given by:
\begin{eqnarray}
G_{sf}(\tau_f,\lambda_f)&=&
I_2(\tau_f,\lambda_f)-I_1(\tau_f,\lambda_f)\,; \ \ \
G_{pf}(\tau_f,\lambda_f)=
I_2(\tau_f,\lambda_f) \,,  \\[2mm]
G_1(\tau_W,\lambda_W)
&=& c_W \
\Bigg\{ \left[2\tau_W(t_W^2-1)+(t_W^2-5)\right] I_1(\tau_W,\lambda_W)
+4(3-t_W^2)I_2(\tau_W,\lambda_W) \Bigg\}\,, \nonumber
\end{eqnarray}
where $I_{1,2}$ are functions of the two variables of $\tau$ and
$\lambda$ and they are expressed as
\begin{eqnarray}
I_1(\tau,\lambda)&=&
\frac{1}{2(\lambda-\tau)}+\frac{1}{2(\lambda-\tau)^2} \left[f(\tau)-f(\lambda)\right]
+\frac{\lambda}{(\lambda-\tau)^2} \left[g(\tau)-g(\lambda) \right]\,,
\nonumber \\[2mm]
I_2(\tau,\lambda)&=&-\frac{1}{2(\lambda-\tau)}\left[f(\tau)-f(\lambda)\right]\,,
\end{eqnarray}
in terms of the $f(\tau)$ function, defined in
Eq.$\,$(\ref{eq:f_function}), and the function $g(\tau)$ which is defined as
\begin{eqnarray}
g(\tau) =\left\{\begin{array}{cl}
\sqrt{\frac{1}{\tau}-1}\,{\rm arcsin}(\sqrt{\tau}) \,:   & \qquad \tau\leq 1\,, \\[2mm]
\frac{1}{2}\sqrt{\frac{\tau-1}{\tau}}
\left[\ln \left(\frac{\sqrt{\tau}+\sqrt{\tau-1}}{
                                     \sqrt{\tau}-\sqrt{\tau-1}}\right)
                    -i\pi\right]\,: & \qquad \tau\geq 1\,.
\end{array}\right.
\end{eqnarray}
The explicit $\tau_x$ dependence of the form factors $G_{sf}$ and $G_{pf}$
for $x=t$ and $G_1$ for $x=W$ is shown in Fig.~\ref{fig:Loop_gG}, clearly
exhibiting the development of their imaginary parts beyond $M_H> 2 M_x$
with $\lambda_{t,W}<1$.

\medskip

\begin{figure}[t!]
\begin{center}
\includegraphics[width=8.4cm]{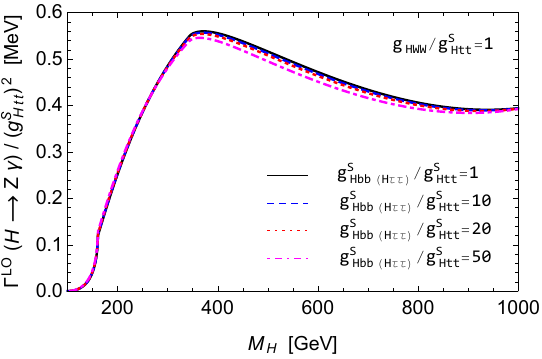}
\includegraphics[width=8.4cm]{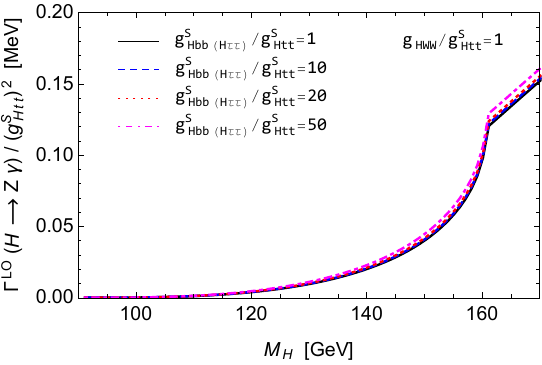}
\includegraphics[width=8.4cm]{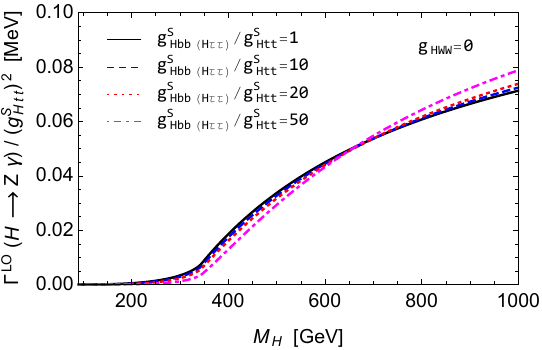}
\includegraphics[width=8.4cm]{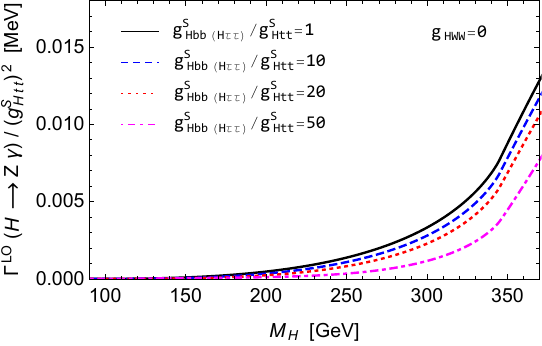}
\end{center}
\caption{
(Upper)
Normalized decay widths of a neutral Higgs boson with a mass $M_H$ into
$Z\gamma$ at LO taking $g_{_{HWW}}/g^S_{H\bar t t}=1$
and $g^P_{H\bar f f}=\Delta S^{Z\gamma}=\Delta P^{Z\gamma}=0$ for
four values of $g^S_{H\bar b b\,,H\tau\tau}/g^S_{H\bar t t}=1\,,10\,,20$,
and $50$.
In the right panel, the low $M_H$ region is magnified.
(Lower)
The same as in the upper panels but taking $g_{_{HWW}}=0$.
}
\label{fig:h2za1}
\end{figure}
\begin{figure}[t!]
\begin{center}
\includegraphics[width=8.4cm]{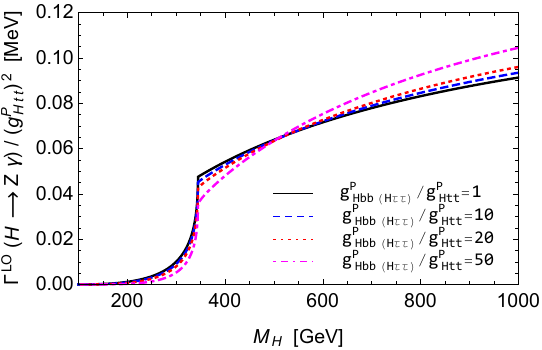}
\includegraphics[width=8.4cm]{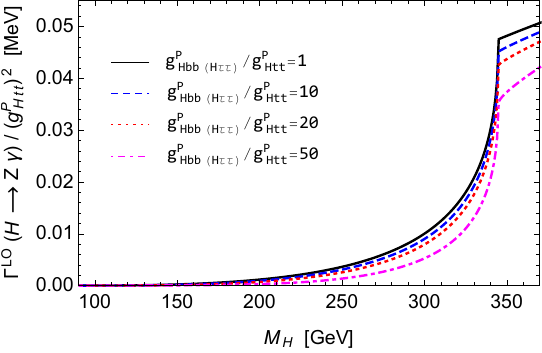}
\end{center}
\caption{
Normalized decay widths of a neutral Higgs boson with a mass $M_H$ into
$Z\gamma$ via the
pseudoscalar form factor $P^{Z\gamma}$ at LO taking $\Delta
P^{Z\gamma}=0$.
In this case,
only the fermion couplings of $g^P_{H\bar f f}$ with $f=t,b,\tau$ are relevant.
}
\label{fig:h2za2}
\end{figure}

At LO, the radiative decay width is given by
\cite{Cahn:1978nz,Bergstrom:1985hp,Gamberini:1987sv}
\begin{eqnarray}
\label{eq:ghZa}
\Gamma^{\rm LO}(H \to Z\gamma) \, =\,
\frac{\alpha(0)\, G_F^2 M_W^2 s_W^2}{64\pi^4} M_{H}^3
\left(1-\frac{M_Z^2}{M_{H}^2}\right)^3 \,\left(
\left|S^{Z\gamma}(M_{H},M_Z)\right|^2+
\left|P^{Z\gamma}(M_{H},M_Z)\right|^2
\right)\,.
\end{eqnarray}
A detailed description of the scalar and pseudoscalar form factors
in the framework of the MSSM taken as a specific BSM model is given in
Appendix \ref{app:dsdp}.

\medskip

The QCD corrections turn out to be less than $0.3$\%
\cite{Spira:1991tj,Gehrmann:2015dua,Bonciani:2015eua}.
On the other hand, the theoretical uncertainties
of the electroweak corrections have been estimated
as $\sim 5$\%~\cite{deFlorian:2016spz}
which constitutes the largest theoretical uncertainty involved in
the experimentally clean and/or dominant SM Higgs decay modes
into $b\bar b\,, WW\,, gg\,, \tau\tau\,, c\bar c\,, ZZ\,, \gamma\gamma\,, Z\gamma$\,,
and $\mu\mu$.

\medskip
Experimentally,
the $H\to Z\gamma$ mode has to be extracted from the Dalitz decays
of $H\to \gamma f\bar f$
\cite{Abbasabadi:1996ze,Abbasabadi:2004wq,Abbasabadi:2006dd,
Chen:2012ju,Dicus:2013ycd,Sun:2013rqa,Passarino:2013nka}
\footnote{Very recently, the ATLAS collaboration has reported on
an evidence for the $H\to\ell\ell\gamma$ process
for a Higgs boson with a mass of $125.09$ GeV and
a dilepton invariant mass $m_{\ell\ell}<30$ GeV
with $\ell=e$ and $\mu$~\cite{Aad:2021jqf}.
The observed significance is $3.2\sigma$ over the background-only hypothesis,
compared to the expected significance of $2.1\sigma$ for the SM prediction.
We note that, for the low values of $m_{\ell\ell}$,
the Dalitz decay is dominated by the
loop-induced $H\to \gamma^* \gamma\to \ell\ell\gamma$ process
with subleading contributions from the loop-induced
$H\to Z^*\gamma\to ee\gamma$
and tree-level $H\to\mu\mu\gamma$ processes
for $\ell=e$ and $\mu$, respectively.}
which consist of:
$(i)$ the $H\to Z^{(*)}\gamma$ decay followed by the decay $Z^{(*)}\to f\bar f$,
$(ii)$ the tree-level process $H\to f\bar f$ with a photon radiated from
the final-state fermions,
$(iii)$ the loop-induced process $H\to \gamma \gamma^*$ via triangle diagrams
followed by the decay $\gamma^*\to f\bar f$,
$(iv)$
the loop-induced process $H\to \gamma f\bar f$ via box diagrams and
$(v)$ the loop-induced process $H\to f\bar f$ via triangle diagrams
with a photon radiated from the final-state fermions.
For a clean separation of the $H\to Z\gamma$ decay from other processes,
appropriate experimental cuts have to be imposed.

\medskip

In the upper panels of Fig.~\ref{fig:h2za1},
we show the LO decay width into a vector boson $Z$ and a photon normalized to
the $g_{H\bar t t}$ coupling squared taking $g_{_{HWW}}/g^S_{H\bar t t}=1$
and $g^P_{H\bar f f}=\Delta S^{Z\gamma}=\Delta P^{Z\gamma}=0$  for
the four values of
$g^S_{H\bar b b\,,H\tau\tau}/g^S_{H\bar t t}=1,\, 10,\, 20$, and $50$.
This reduces to the SM decay width of a Higgs particle weighing $M_H$
when $g_{_{HWW}}/g^S_{H\bar t t}=g^S_{H\bar b b\,,H\tau\tau}/g^S_{H\bar t t}=1$
together with $g_{H\bar t t}=1$.
Below the $W$-boson-pair threshold, the $W$-loop contributions are
dominant, leading to the sharp rise as $M_H$ approaches $2M_W$.
Passing the $W$-pair
threshold $M_H=2M_W$ from below,
the real part of the $W$-loop contributions decreases while
the imaginary part starts to develop
leading to another mild rise,
see the lower panel of Fig.~\ref{fig:Loop_gG}.
Passing $M_H=2M_t$, the $t$-quark loop contributions
start to cancel that of the $W$-loop contributions
as shown in the upper-left panel of Fig.~\ref{fig:h2za1}.
In the lower panels, we show the results taking $g_{_{HWW}}=0$
for the four values of
$g^S_{H\bar b b\,,H\tau\tau}/g^S_{H\bar t t}=1\,,10\,,20\,,50$.
This case may apply to the heavy neutral Higgs bosons appearing in
the 2HDMs and/or MSSM when their couplings to the massive vector bosons are
suppressed and they are almost vanishing~\cite{Gunion:2002zf,Haber:2002ey}.

\medskip

In Fig.~\ref{fig:h2za2}, the alternative pseudoscalar choice
is made to show $\Gamma(H\to Z\gamma)/(g^P_{H\bar t t})^2$ taking the four values of
$g^P_{H\bar b b}/g^P_{H\bar t t}=g^P_{H\tau\tau}/g^P_{H\bar t t}=1\,,10\,,20\,,50$
with $\Delta S^{Z\gamma}=\Delta P^{Z\gamma}=0$.
Here we assume all the scalar couplings of $g^S_{H\bar f f}$  are vanishing.
Note that, in this case, only the fermion loops are contributing
as the pseudoscalar state does not couple to gauge bosons at the tree level.
In the right panel, as the same as in Fig.~\ref{fig:h2za1}, we magnify the low
$M_H$ regions. Compared to the scalar case shown in
Fig.~\ref{fig:h2za1},
it is clearly shown in
Fig.~\ref{fig:h2za2} that the scalar and pseudoscalar
decays exhibit quite distinct patterns, in particular, around the $t$-pair
threshold.
The distinct patterns between the scalar case with $g_{_{HWW}}=0$
(two lower frames of Fig.~\ref{fig:h2za1}) and the
pseudoscalar one (Fig.~\ref{fig:h2za2}) basically
come from the differences in sizes and behaviors
of the form factors $G_{sf}$ and $G_{pf}$ shown in
Fig.~\ref{fig:Loop_gG}.

\medskip

\begin{table}[t!]
\caption{
The leading $M_H$ dependence and
the ballpark values of normalized decay widths
of the neutral Higgs boson $H$ far above the mass threshold of its decay
products.  Specifically, we have taken $M_H=1$ TeV.
For the radiative loop-induced decays, the scalar (upper)
and pseudoscalar (lower) contributions are shown separately.
For the scalar contributions to $H\to \gamma\gamma$ and $H\to Z\gamma$
which are dominated by $t$-quark and $W$-boson loops, we assume
$g^S_{H\bar t t}=g_{_{HWW}}=1$.
See Table \ref{tab:runasmq}
for the $\overline{\rm MS}$ quark masses $\overline{m}_q(M_H)$
while we refer to Eq.~(\ref{eq:mqforhaa}) for $m_q(\mu=M_H/2)$.
}
\label{tab:NHdecay}
\renewcommand{\arraystretch}{1.7}
\vspace{-0.5cm}
\begin{center}
\begin{tabular}{|c|c|c|c|}
\hline
 Decay Mode  &
Leading $M_H$ dependence &
$\Gamma_{M_H=1\rm{TeV}}^{\rm \, normalized}$ [GeV] &
Reference Figure \\
\hline \hline
 $H \rightarrow  f \bar f$         & $M_H$     &
$40\left[\frac{\overline{m}_q(M_H)}{\overline{m}_t(M_H)}\right]^2$&
Figs.~\ref{fig:h2ff} and \ref{fig:h2tt}  \\ [1mm]
& & $\frac{40}{3}\left[\frac{M_{\tau\,,\mu}}{\overline{m}_t(M_H)} \right]^2$
&  \\ \hline
 $H \rightarrow WW\,(ZZ)$          & $M_H^3$  & $350\,(180)$  &
Fig.~\ref{fig:h2vv}  \\ \hline
 $H \rightarrow \varphi V$              & $M_H^3$  & $300$  &
Fig.~\ref{fig:h2lighter} \\
 $H \rightarrow \varphi \varphi$     & $1/M_H$  & $4$     &     \\ \hline\hline
\multirow{2}{*}{$H \rightarrow gg$}
& $M_H^3 \left|F_{sf}\left(\frac{M_H^2}{4M_t^2}\right)\right|^2$ & $7 \times 10^{-2}$  &
Fig.~\ref{fig:h2gg1}  \\
\cline{2-4}
& $M_H^3 \left|F_{pf}\left(\frac{M_H^2}{4M_t^2}\right)\right|^2$ & $9 \times 10^{-2}$  &
Fig.~\ref{fig:h2gg2}  \\
\hline
\multirow{2}{*}{$H \rightarrow \gamma\gamma$}
&  $M_H^3 \left|\frac{8}{3} F_{sf}\left(\frac{M_H^2}{[4m_t(M_H/2)]^2}\right)
- F_1\left(\frac{M_H^2}{4M_W^2}\right)\right|^2$
& $3 \times 10^{-4}$  & Fig.~\ref{fig:h2aa1}  \\ \cline{2-4}
&  $M_H^3 \left|F_{pf}\left(\frac{M_H^2}{[4m_t(M_H/2)]^2}\right)\right|^2$
&  $3 \times 10^{-4}$ & Fig.~\ref{fig:h2aa2} \\ \hline
\multirow{2}{*}{$H \rightarrow Z\gamma$} &
$M_H^3 \left|0.9\, G_{sf}\left(\frac{M_H^2}{4M_t^2},\frac{M_Z^2}{4M_t^2}\right)
- G_1\left(\frac{M_H^2}{4M_W^2},\frac{M_Z^2}{4M_W^2}\right)\right|^2$
& $6 \times 10^{-4}$  & Fig.~\ref{fig:h2za1} \\ \cline{2-4}
& $M_H^3 \left|G_{pf}\left(\frac{M_H^2}{4M_t^2},\frac{M_Z^2}{4M_t^2}\right)\right|^2$
& $8 \times 10^{-5}$  & Fig.~\ref{fig:h2za2}  \\ \hline
\end{tabular}
\end{center}
\end{table}
%
\begin{table}[t!]
\caption{\
The QCD and ELW corrections to the decay widths
of neutral and charged  Higgs bosons which are denoted by $H$ and $H^+$, respectively.
The abbreviations used are:
NC $=$ Not Considered,
[S] $=$ Scalar part,
[P] $=$ Pseudoscalar part.
}
\label{tab:NHcorrections}
\renewcommand{\arraystretch}{1.7}
\begin{center}
\begin{tabular}{|c|c|c|c|}
\hline
 Decay Mode  &
Expressions for a partial width &
QCD corrections &
ELW corrections\\ \hline\hline
$H\to b\bar b\,, c\bar c$ & Eq.~(\ref{eq:ghff})  &
Eq.~(\ref{eq:ghff_qcd}) &   \\[-5mm]
%
$H\to \tau^+\tau^-\,, \mu^+\mu^-$ & Eq.~(\ref{eq:ghff})  &
$-\!\!\!-\!\!\!-$ &
$\begin{array}{c} {\rm [S]}\,:\,{\rm Eq}.~(\ref{eq:ghff_elw}) \\[2mm]
\hspace{-10.4mm}{\rm [P]}\,:\,{\rm NC} \end{array}$ \\[-5mm] 
$H\to t\bar t$ &
$\begin{array}{c} {\rm Eq}.~(\ref{eq:ghtt})\,{\rm for}\,M_H\leq 500~{\rm GeV} \\[-3mm]
{\rm Eq}.~(\ref{eq:ghff})\,{\rm for}\,M_H > 500~{\rm GeV} \end{array}$  &
Eq.~(\ref{eq:ghff_qcd}) &   \\ \hline
%
$H\to WW\,, ZZ$ & Eqs.~(\ref{eq:ghvv_lo},\ref{eq:ghvv}) & $-\!\!\!-\!\!\!-$ &
Fig.~\ref{fig:delwV} ({\tt PROPHECY4Fv3.0})
\\ \hline
$H\to \varphi Z$ & Eq.~(\ref{eq:ghpZ}) & & \\ 
$H\to \varphi^\pm W^\mp$ & Eq.~(\ref{eq:ghpW}) & $-\!\!\!-\!\!\!-$ & NC \\
$H\to \varphi\varphi$ & Eq.~(\ref{eq:ghpp}) & & \\ \hline\hline 
$H\to gg$ & Eq.~(\ref{eq:ghgg}) &
$\begin{array}{c} {\rm [S]}\,:\,{\rm Eq}.~(\ref{eq:ghgg_qcd_s}) \\[-3mm]
{\rm [P]}\,:\,{\rm Eq}.~(\ref{eq:ghgg_qcd_p}) \end{array}$ &
$\begin{array}{c} {\rm [S]}\,:\,{\rm Fig}.~\ref{fig:gSelw}~({\rm Grids})\, \\[-3mm]
{\rm [P]}\,:\,{\rm Eq}.~(\ref{eq:ghgg_elw_p})~~~~~~ \end{array}$ \\ \hline
$H\to \gamma\gamma$ & Eqs.~(\ref{eq:ghaa_lo},\ref{eq:ghaa}) &
Eq.~(\ref{eq:ghaa_lo}) $\oplus$ Eq.~(\ref{eq:ghaa_qcd}) &
$\begin{array}{c} {\rm [S]}\,:\,{\rm Fig}.~\ref{fig:dhaaelw}~({\rm Grids})\, \\[-3mm]
{\rm [P]}\,:\,{\rm Eq}.~(\ref{eq:dhaaelwP})\,~~~~~~~ \end{array}$ \\ \hline
$H\to Z\gamma$  & Eq.~(\ref{eq:ghZa}) & $-\!\!\!-\!\!\!-$ & NC \\ \hline \hline
$H^+\to c\bar s$  & Eq.~(\ref{eq:gchff}) & $\delta_{\rm QCD}$ in Eq.~(\ref{eq:ghff_qcd}) &  \\
$H^+\to \nu\tau^+\,,\nu\mu^+ $  & Eq.~(\ref{eq:gchff}) & $-\!\!\!-\!\!\!-$ & NC \\
$H^+\to t \bar b$  &
$\begin{array}{c} {\rm Eq}.~(\ref{eq:gchtb})\,{\rm for}\,M_{H^\pm}\leq 500~{\rm GeV} \\[-3mm]
{\rm Eq}.~(\ref{eq:gchff})\,{\rm for}\,M_{H^\pm} > 500~{\rm GeV} \end{array}$  &
$\delta_{\rm QCD}$ in Eq.~(\ref{eq:ghff_qcd}) &   \\ \hline
$H^+\to H W^+$  & Eq.~(\ref{eq:gchhw}) & $-\!\!\!-\!\!\!-$ & NC \\ \hline
\end{tabular}
\end{center}
\end{table}
Before moving to the last subsection for numerical results obtained
by analyzing the decays of several neutral Higgs bosons,
we provide Table~\ref{tab:NHdecay}
in which the leading $M_H$ dependence and
the ballpark values of normalized decay widths at $M_H=1$ TeV
are shown for all the
decay modes elaborated on up to this subsection.
Table~\ref{tab:NHcorrections} is further provided for
a summary of the QCD and electroweak corrections considered in
Section~\ref{sec:decays_of_a_generic_neutral_higgs_boson}
and Section~\ref{sec:decays_of_a_charged_higgs_boson} which are for
the decays of $H$ and $H^+$, respectively.

\subsection {Numerical results}
%
Closing this section dedicated to a detailed study of
neutral Higgs boson decays, we present the results of two
numerical analyses of $(i)$ the decays of a
neutral Higgs boson with its mass fixed to $125.5$ GeV dictated
by the LHC discovery and $(ii)$ the decays of heavy neutral Higgs bosons
which are mixtures of CP-even and CP-odd states.

\medskip
In the first numerical analysis, we extend the SM
in a somewhat model-independent way by allowing for
the pseudoscalar as well as scalar couplings of the $125.5$ GeV Higgs boson and
perform a comprehensive analysis of its decays by estimating
all the widths and branching ratios as precise as possible.
For the second numerical analysis, we have specifically chosen the type-I 2HDM
in which the Yukawa couplings of the lightest Higgs boson
as well as its couplings to a pair of massive vector bosons
quickly approach the corresponding SM values as the masses of
the heavy neutral Higgs bosons increase
and their decouplings are not delayed
\cite{Haber:2000kq,Ferreira:2014naa,Choi:2020uum}.
In this model, there is no much need of decoupling the heavy Higgs bosons
to avoid  conflicts with the current LHC Higgs precision data.
Moreover, though assuming that the lightest neutral Higgs boson is
a purely CP-even state, the heavy neutral states could still
undergo a significant CP-violating mixing in the presence of CP-violating
phases in the Higgs potential~\cite{Ellis:2004fs,Choi:2004kq,Carena:2015uoe}.

\medskip

We emphasize that the numerical analyses performed in this subsection are solely
based on the detailed analytical and numerical results presented in this
Section~\ref{sec:decays_of_a_generic_neutral_higgs_boson} and
supplemental materials provided in Appendices.

\subsubsection{\it
Anatomy of Higgs boson decays with $M_H=125.5$ GeV
}
In this subsubsection, we present all the details of
calculating  the decay widths of
a neutral Higgs boson
by taking $M_H=125.5$ GeV~\cite{Sirunyan:2020xwk},
while allowing for nontrivial pseudoscalar as well as scalar
couplings of the Higgs boson to fermion pairs.
The SM parameters taken in this present analysis are
summarized in Appendix \ref{app:smpara}.
Note that one may apply the results to the
genuine CP-even SM Higgs boson by taking
$g^S_{H\bar f f}=g_{_{HVV}}=1$ and
$g^P_{H\bar f f}=\Delta S^{g\,,\gamma\,,Z\gamma}
=\Delta P^{g\,,\gamma\,,Z\gamma}=0$.

\medskip

For the decays of a neutral Higgs boson with
$M_H=125.5$ GeV into $b$ and $c$ quarks and
muons and tau leptons, the relevant radiative corrections
are given numerically by
\begin{eqnarray}
\delta_{\rm QCD} &=& +0.203 + 0.037 + 0.002 - 0.001 = +0.241\,; \nonumber \\
\delta^{b:S}_t &=& +0.011\,\xi^{b:S}_t\,, \ \
\delta^{c:S}_t \ \ \ =  +0.019\,\xi^{c:S}_t; \nonumber \\
\delta^b_{\rm elw} &=& -0.005 \,, \ \
\delta^c_{\rm elw} \ \ \ =  +0.004\,; \ \
\delta^\tau_{\rm elw}=  -0.0097\,, \ \
\delta^\mu_{\rm elw}=  -0.0307\,; \nonumber \\
\delta^b_{\rm mixed} &=& -0.001\,, \ \
\delta^c_{\rm mixed} =  -0.000\,; \nonumber \\
\delta^b_{\rm tot} &=& +0.244 \,, \ \
\delta^c_{\rm tot} \ \ \ =  +0.265\,; \ \
\delta^\tau_{\rm tot}\, =  -0.0097\,, \ \
\delta^\mu_{\rm tot}\, =  -0.0307\,,
\end{eqnarray}
with the rescaling factors
defined by $\xi^{q:S(P)}_t\equiv g_{H\bar t t}^{S(P)}/ g_{H\bar q q}^{S(P)}$.
Incidentally, we have $\delta^{b:P}_t = 0.018\,\xi^{b:P}_t$ and
$\delta^{c:P}_t = 0.030\,\xi^{c:P}_t$ contributing to the pseudoscalar part.

\medskip

For the decays $H\to WW,ZZ$, from Fig.~\ref{fig:delwV},
we have two electroweak corrections as
\begin{equation}
\delta^W_{\rm elw} = 0.0308\,, \ \ \
\delta^Z_{\rm elw} = 0.0152\,.
\end{equation}
For the decay $H\to gg$, the scalar and pseudoscalar form factors
and the relevant radiative corrections are given numerically apart
from the scalar and pseudoscalar Higgs--fermion--fermion couplings by
\begin{eqnarray}
\label{eq:hgg}
S^g&=&  0.688\,g_{H\bar{t}t}^S + (-0.043+0.062\,i)\,g_{H\bar{b}b}^S
+  (-0.009+0.008\,i)\,g_{H\bar{c}c}^S 
+ \Delta S^{g}  \,; \nonumber \\[2mm]
P^g&=&  1.048\,g_{H\bar{t}t}^P + (-0.049+0.063\,i)\,g_{H\bar{b}b}^P
 + (-0.010+0.008\,i)\,g_{H\bar{c}c}^P 
+ \Delta P^{g} \,; \nonumber \\[2mm]
\delta^{g:S}_{\rm QCD}&=& 0.8850\,, \ \
\delta^{g:S}_{\rm elw}=  0.0516\,;  \ \
\delta^{g:P}_{\rm QCD}=  0.8775\,, \ \
\delta^{g:P}_{\rm elw}= -0.0218 \,,
\end{eqnarray}
where $\delta^{g:S}_{\rm elw}$ is from Fig.~\ref{fig:gSelw} and
we set $\eta^{g:P}_{\rm elw}= 0$ for $\delta^{g:P}_{\rm elw}$.
In the SM, we have $S^g=0.636 + 0.070\,i$ and $P^g=0$.
Similarly for the decay $H\to \gamma\gamma$, in terms of the $HWW$ and
Higgs--fermion--fermion couplings, we obtain
the following scalar and pseudoscalar form factors
\begin{eqnarray}
\label{eq:haa}
S^\gamma&=& -8.341\,g_{_{HWW}} + 1.826\,g_{H\bar{t}t}^S
+(-0.020 + 0.024\,i)\,g_{H\bar{b}b}^S
   +(-0.023 + 0.021\,i)\,g_{H\tau\tau}^S + \Delta S^{\gamma}\,;
\nonumber \\[2mm]
P^\gamma&=& 2.772\,g_{H\bar{t}t}^P
+(-0.022 + 0.024\,i)\,g_{H\bar{b}b}^P
   +(-0.025 + 0.021\,i)\,g_{H\tau\tau}^P + \Delta P^{\gamma}\,; \nonumber \\[2mm]
\delta^{\gamma:S}_{\rm QCD}&=& 0.016\,, \ \
\delta^{\gamma:S}_{\rm elw}=  -0.016 \,; \ \ 
\delta^{\gamma:P}_{\rm QCD}= 0.028 \,, \ \
\delta^{\gamma:P}_{\rm elw}= -0.012  \,,
\end{eqnarray}
where the QCD corrections are obtained by implementing the scaling
factors $C_{sf,pf}(\tau_{t,b})$ into the form factors $F_{sf,pf}(\tau_{t,b})$
and the electroweak correction $\delta^{\gamma:S}_{\rm elw}$
is from Fig.~\ref{fig:dhaaelw}. For $\delta^{\gamma:P}_{\rm elw}$, we set
$\eta^{\gamma:P}_{\rm elw}= 0$.
For the numerical estimate of $\delta^{\gamma : S}_{\rm QCD}$, more precisely,
we take the SM values of $g_{_{HWW}}=g^S_{H\bar t t}=g^S_{H\bar b b}
=g^S_{H\tau\tau}=1$. While, for the numerical estimate of
$\delta^{\gamma : P}_{\rm QCD}$, we assume a scenario in which
the pseudoscalar form factor $P^\gamma$ is dominated by
the top-quark contribution taking $g^P_{H\bar b b}=g^P_{H\tau\tau}=0$,
i.e. neglecting the $b$-- and $\tau$--loop contributions.
In the SM, we have $S^\gamma=-6.558 + 0.047\,i$ and $P^\gamma=0$.

\medskip

Finally, for the decay $H\to Z\gamma$,
the scalar and pseudoscalar form factors are given by
\begin{eqnarray}
S^{Z\gamma}&=& -12.372\,g_{_{HWW}} +0.689\,g_{H\bar{t}t}^S
+ (-0.019+0.011\,i)\,g_{H\bar{b}b}^S + (-0.0005+0.0002\,i)\,g_{H\tau\tau}^S
+ \Delta S^{Z\gamma}\,; \nonumber \\ [2mm]
P^{Z\gamma}&=& 1.046\,g_{H\bar{t}t}^P
+ (-0.022+0.011\,i)\,g_{H\bar{b}b}^P + (-0.0006+0.0002\,i)\,g_{H\tau\tau}^P
+ \Delta P^{Z\gamma}\,,
\end{eqnarray}
in terms of the $HWW$ and Higgs--fermion--fermion couplings.
In the SM, we have $S^{Z\gamma}=-11.702 + 0.011\,i$ and
$P^{Z\gamma}=0$. The $H\to Z\gamma$ decay width is estimated at LO.

%
\begin{table}[t!]
\caption{
Partial and total decay widths of the SM Higgs in MeV taking $M_H=125.5$ GeV.
We compare our numerical estimates
with those presented in Ref.~\cite{deFlorian:2016spz}
by introducing a quantity $\delta_\Gamma$  defined by
$\delta_\Gamma \equiv \left(
\Gamma_{\rm This~Review}-\Gamma_{\rm Ref.\small{\cite{deFlorian:2016spz}}}\right)
/\Gamma_{\rm Ref.\small{\cite{deFlorian:2016spz}}}$.
Also presented are theoretical uncertainties (THUs)
of the partial and total decay widths from missing higher orders estimated around
$M_H=125$ GeV, see Tables 178 and 182 in Ref.~\cite{deFlorian:2016spz}.
}
\label{tab:SMGAM}
\begin{center}
\begin{tabular}{l|ccccc}
\hline\hline
&
$\Gamma(H\rightarrow b\bar b)$ & $\Gamma(H\to WW)$ & $\Gamma(H\to gg)$ &
$\Gamma(H\to \tau\tau)$ & $\Gamma(H\to c\bar c)$ \\
\hline
This Review &
$2.367$  & $9.185 \times 10^{-1}$  & $3.382 \times 10^{-1}$  &
$2.572 \times 10^{-1}$  & $1.173 \times 10^{-1}$  \\
\hline
Ref.~\cite{deFlorian:2016spz} &
$2.387$ & $9.222 \times 10^{-1}$ & $3.386 \times 10^{-1}$ &
$2.573 \times 10^{-1}$ & $1.185 \times 10^{-1}$ \\
\hline
$\delta_\Gamma$ [\%] & $-0.8$  & $-0.4$  & $-0.1$  & $-0.05$  & $-1.0$ \\
\hline
THU [\%] & $\pm 0.5$  & $\pm 0.5$  & $\pm 3.2$  & $\pm 0.5$  & $\pm 0.5$ \\
\hline\hline
 &
$\Gamma(H\to ZZ)$ & $\Gamma(H\to \gamma\gamma)$ & $\Gamma(H\to Z\gamma)$ &
$\Gamma(H\to \mu\mu)$ & $\Gamma_{\rm tot}$ \\
\hline
This Review & $1.139 \times 10^{-1}$  & $9.405 \times 10^{-3}$  &
$6.531 \times 10^{-3}$  & $8.913 \times 10^{-4}$  & $4.129$ \\
\hline
Ref.~\cite{deFlorian:2016spz} & $1.139 \times 10^{-1}$ & $9.438 \times 10^{-3}$ &
$6.550 \times 10^{-3}$ & $8.927 \times 10^{-4}$ & $4.156$ \\

\hline
$\delta_\Gamma$ [\%] & $-0.03$  & $-0.4$  & $-0.3$  & $-0.2$  & $-0.65$ \\
\hline
THU [\%] & $\pm 0.5$  & $\pm 1.0$  & $\pm 5.0$  & $\pm 0.5$  & $\pm 0.73$ \\
\hline\hline
\end{tabular}
\end{center}
\end{table}
%
%

\medskip

In Table ~\ref{tab:SMGAM}, we show the partial and total decay widths of
the SM Higgs boson with $M_H=125.5$ GeV
taking $g^S_{H\bar f f}=g_{_{HVV}}=1$ and
$g^P_{H\bar f f}=\Delta S^{g\,,\gamma\,,Z\gamma}
=\Delta P^{g\,,\gamma\,,Z\gamma}=0$.
For a quantitative comparison with those presented in
Ref.~\cite{deFlorian:2016spz}, we introduce
$\delta_\Gamma$'s, which are defined by
$\delta_\Gamma \equiv \left(
\Gamma_{\rm This~Review}-\Gamma_{\rm Ref.\small{\cite{deFlorian:2016spz}}}\right)
/\Gamma_{\rm Ref.\small{\cite{deFlorian:2016spz}}}$ for each decay mode,
for being contrasted with
theoretical uncertainties (THUs) given in Ref.~\cite{deFlorian:2016spz}.
\footnote{Note that we use the same values for all the
input parameters as in Ref.~\cite{deFlorian:2016spz}.}
In terms of $\delta_\Gamma/{\rm THU}$,
we find excellent agreement between our
analysis and that in Ref.~\cite{deFlorian:2016spz}
except for $H\to b\bar b$ and $H\to c\bar c$ for which 
we find $\delta_\Gamma/|{\rm THU}|\simeq -2$.
The largest contribution to the discrepancy of $\Gamma_{\rm tot}$
comes from $H\to b \bar b$ with the second (third)
largest one from $H\to W W\,(c\bar c)$.
We note that our estimations of the decay widths into quarks
are smaller than those in Ref.~\cite{deFlorian:2016spz}.
%
This might come from our  incomplete and rough implementation of
the ELW corrections of Eq.~(\ref{eq:ghff_elw}).

\begin{table}[t!]
\caption{
Branching ratios (BRs)  of the SM Higgs taking $M_H=125.5$ GeV.
We compare our numerical estimates
with those presented in Ref.~\cite{deFlorian:2016spz}.
Estimation of the total uncertainty THU$+$PU has been done
by adding linearly the THU and the total parametric uncertainty (PU)
where the latter is obtained by adding the individual PUs in quadrature, see
Tables 174, 175, 176, 177, and 178 in Ref.~\cite{deFlorian:2016spz}.
}
\label{tab:SMBR}
\begin{center}
\begin{tabular}{l|ccccc}
\hline\hline
&
${B}(H\rightarrow b\bar b)$ & ${B}(H\to WW)$ & ${B}(H\to gg)$ &
${B}(H\to \tau\tau)$ & ${B}(H\to c\bar c)$ \\
\hline
This Review &
$5.733\times 10^{-1}$  & $2.225 \times 10^{-1}$  & $8.190 \times 10^{-2}$  &
$6.230 \times 10^{-2}$  & $2.841 \times 10^{-2}$  \\
\hline
Ref.~\cite{deFlorian:2016spz} &
$5.744\times 10^{-1}$  & $2.219 \times 10^{-1}$  & $8.147 \times 10^{-2}$  &
$6.192 \times 10^{-2}$  & $2.852 \times 10^{-2}$  \\
\hline
THU$+$PU [\%] & $1.7$  & $2.1$  & $7.2$  & $2.3$  & $6.6$ \\
\hline\hline
 &
${B}(H\to ZZ)$ & ${B}(H\to \gamma\gamma)$ & ${B}(H\to Z\gamma)$ &
${B}(H\to \mu\mu)$ & $\Gamma_{\rm tot}$[MeV] \\
\hline
This Review & $2.758 \times 10^{-2}$  & $2.278 \times 10^{-3}$  &
$1.582 \times 10^{-3}$  & $2.159 \times 10^{-4}$  & $4.129$ \\
\hline
Ref.~\cite{deFlorian:2016spz} & $2.741 \times 10^{-2}$  & $2.271 \times 10^{-3}$  &
$1.576 \times 10^{-3}$  & $2.148 \times 10^{-4}$  & $4.156$ \\
\hline
THU$+$PU [\%] & $2.1$  & $2.9$  & $6.9$  & $2.4$  & $1.9$ \\
\hline\hline
\end{tabular}
\end{center}
\end{table}

\medskip

\begin{figure}[t!]
\begin{center}
\includegraphics[width=12.5cm]{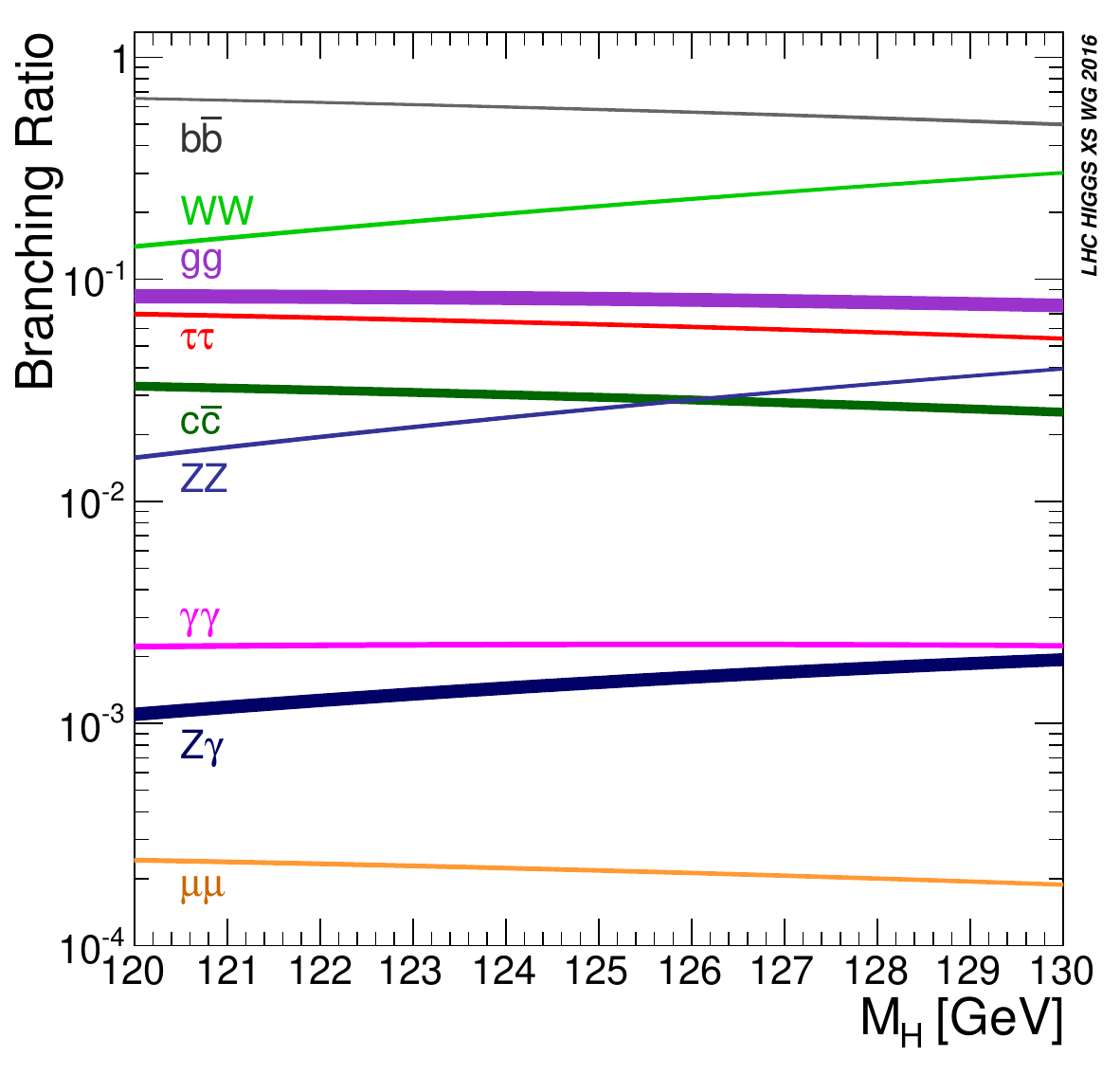}
\end{center}
\vspace{-0.5cm}
\caption{
Higgs boson branching ratios and their uncertainties for the
 mass range around $125$ GeV.
The plot is taken from Ref.~\cite{deFlorian:2016spz}.
}
\label{fig:1610.07922.SMBRs}
\end{figure}

In Table~\ref{tab:SMBR}, we show the branching ratios and
the total decay width of the SM Higgs boson taking $M_H=125.5$ GeV.
Again comparisons are made with those in Ref.~\cite{deFlorian:2016spz}
together with the total uncertainties. We pick up THUs and PUs
from
Tables 174, 175, 176, 177, and 178 in Ref.~\cite{deFlorian:2016spz}.
We note that the total uncertainty is about 2 - 3 \% for
$H\to b\bar b, WW, \tau\tau, ZZ, \gamma\gamma$ and $\mu\mu$. While
it is about 7 \% for $H\to gg, c\bar c$, and $Z\gamma$. The total decay width
is determined with about 2\% error.
In Fig.~\ref{fig:1610.07922.SMBRs}, the branching ratios (BRs) are
shown in the Higgs-boson mass range between 120 GeV and 130 GeV.
For each BR line, the band width represents the corresponding total
uncertainty.

\subsubsection{\it Decays of heavy Higgs bosons in CP-violating 2HDMs}
\label{sec:heavyHdecaysInCPV2HDMs}
In this subsubsection, we study the decays of heavy
neutral Higgs bosons appearing in BSM models. To be specific, we choose
the type-I 2HDM identifying the lightest neutral Higgs boson
as the SM-like $125.5$ GeV one discovered at the LHC.
We assume that the lightest Higgs boson is purely CP even while
the two heavier Higgs bosons exhibit nontrivial CP-violating (CPV)
mixing in the presence of the complex
quartic couplings of $\lambda_{5,6,7}$ in the Higgs potential.
In this scenario,
with no much need of decoupling the heavier Higgs bosons,
all the branching ratios and the total decay width of the
lightest Higgs boson remain
consistent with those of the SM Higgs
within the ranges allowed by the current LHC Higgs
precision data~\cite{Choi:2020uum}.
The decay widths and branching ratios are calculated as summarized in
Table~\ref{tab:NHcorrections} but the ELW corrections are neglected
for consistency. For the full model- and
parameter-dependent ELW corrections in 2HDMs,
see Appendix~\ref{app:bsmtools}.

\medskip

To fix all the relevant couplings of three neutral Higgs bosons,
one may start from
the orthogonal $3\times 3$ matrix $O$ describing the mixing among them.
For CPV scenarios in 2HDMs,  the three neutral Higgs bosons
do not carry definite CP parities and they become mixtures of CP-even and CP-odd
states.  In this case, without loss of generality,
the mixing matrix can be parameterized as
\footnote{Here we take the abbreviations such as
$\cos\alpha=c_\alpha$, $\sin\alpha=s_\alpha$, etc.}
\begin{eqnarray}
\label{eq:omix0}
\hspace{-0.5cm}
\vspace{ 1.0cm}
O &=&
\left( \begin{array}{ccc}
 -s_\alpha  &   c_\alpha   &  0   \\
  c_\alpha  &   s_\alpha   &  0   \\
  0         &      0       &  1   \\
  \end{array} \right)
\left( \begin{array}{ccc}
    c_\eta               &      0             &   s_\eta  \\
    0                    &      1             &   0   \\
   -s_\eta               &      0             &   c_\eta   \\
  \end{array} \right)
\left( \begin{array}{ccc}
    1               &      0       &   0   \\
    0               &   c_\omega   &   s_\omega   \\
    0               &  -s_\omega   &   c_\omega   \\
  \end{array} \right)  
\nonumber \\[5mm] &=&
\left( \begin{array}{ccc}
    -s_\alpha c_\eta  &  c_\alpha c_\omega + s_\alpha s_\eta s_\omega   &
    c_\alpha s_\omega - s_\alpha s_\eta c_\omega     \\
    c_\alpha c_\eta   &  s_\alpha c_\omega - c_\alpha s_\eta s_\omega   &
    s_\alpha s_\omega + c_\alpha s_\eta c_\omega     \\
    -s_\eta  &  -c_\eta s_\omega   &    c_\eta c_\omega  \\
  \end{array} \right)\,, 
\end{eqnarray}
introducing a CP-conserving (CPC) mixing angle $\alpha$ and two
CPV angles $\omega$ and $\eta$.
We recall that the mixing matrix $O$
relates the electroweak eigenstates $(\phi_1\,,\phi_2\,,a)$
to the mass eigenstates $(H_1\,,H_2\,,H_3)$ via
$$
(\phi_1\,,\phi_2\,,a)^T_\alpha = O_{\alpha i} (H_1\,,H_2\,,H_3)^T_i\,,
$$
with the ordering of $M_{H_1}\leq M_{H_2}\leq M_{H_3}$.
Assuming the lightest Higgs boson is purely CP even or
taking $s_\eta=0$ and $c_\eta=1$,
the mixing matrix takes the simpler form of
\begin{equation}
\label{eq:omix1}
\left.O\right|_{s_\eta=0,c_\eta=1} \ = \
\left( \begin{array}{ccc}
   -s_\alpha  &  c_\alpha c_\omega  & c_\alpha s_\omega   \\
  ~~c_\alpha  &  s_\alpha c_\omega  & s_\alpha s_\omega   \\
   ~0  &  -s_\omega   &    c_\omega   \\
  \end{array} \right)\,.
\end{equation}
Note that, in the CP-conserving case,
one of the heavy Higgs boson is purely CP odd and its coupling
to a pair of massive gauges bosons is identically vanishing.
We observe that $H_3$ is purely CP odd when $|c_\omega|=1$
while $H_2$ is CP odd when $|s_\omega|=1$.
Plugging the above expression of $O$ into Eq.~(\ref{eq:2hdmhvvetc}),
the couplings of three neutral Higgs bosons to a pair of massive
vector bosons are given by
\begin{eqnarray}
g_{_{H_1VV}}  =  s_{\beta-\alpha} \equiv \sqrt{1-\epsilon}\,, \ \ \
g_{_{H_2VV}}  =  c_{\beta-\alpha} c_\omega \equiv \delta_2\,, \ \ \
g_{_{H_3VV}}  =  c_{\beta-\alpha} s_\omega \equiv \delta_3\,,
\end{eqnarray}
with $\delta_2^2+\delta_3^2=\epsilon$. We note that
the two mixing angles are determined as follows
\begin{eqnarray}
\label{eq:2hdmaw}
s_\alpha &=& -\sqrt{1-\epsilon}\, c_\beta +\frac{\delta_2}{c_\omega}\, s_\beta\,, \ \ \
 c_\alpha = \sqrt{1-\epsilon}\, s_\beta +\frac{\delta_2}{c_\omega}\, c_\beta\,,
\nonumber \\[2mm]
c_\omega^2 &=& \frac{\delta_2^2}{\delta_2^2+\delta_3^2} =
\frac{\delta_2^2}{\epsilon}\,, \ \ \
s_\omega^2  =  \frac{\delta_3^2}{\delta_2^2+\delta_3^2} =\frac{\delta_3^2}{\epsilon}\,.
\end{eqnarray}
in terms of the couplings $\delta_2=g_{_{H_2VV}}$ and $\delta_3=g_{_{H_3VV}}$
together with $t_\beta$.
And then, the Yukawa couplings of
the three neutral Higgs bosons are determined by
\begin{equation}
g^S_{H_i\bar u u}=g^S_{H_i\bar d d}=g^S_{H_i\bar \ell \ell}=
O_{\phi_2i}/s_\beta\,; \ \ \
-g^P_{H_i\bar u u}=g^P_{H_i\bar d d}= g^P_{H_i \bar \ell \ell}
=O_{ai}/t_\beta\,,
\end{equation}
where $u$ and $d$ stand for the up- and down-type quarks, respectively, and
$\ell$ for three charged leptons.
To summarize, in the scenario under consideration, all
the Yukawa couplings of the two heavy Higgs bosons could be
fixed by giving their couplings to the massive vector bosons.
On the other hand, depending on ${\rm sign}[\delta_2/c_\omega]$,
all the Yukawa couplings of the lightest Higgs boson are determined by
$O_{\phi_2 1}/s_\beta=c_\alpha/s_\beta=\sqrt{1-\epsilon}\pm\sqrt\epsilon/t_\beta$
which, especially for large $t_\beta$,
approaches the SM value of $1$ as quickly as the
$g_{_{H_1VV}}=\sqrt{1-\epsilon}$ coupling when $\epsilon$ goes to zero.
This is the very reason we choose the type-I 2HDM
for our numerical study avoiding
conflicts with the current LHC Higgs
precision data~\cite{Choi:2020uum}.

\medskip

\begin{figure}[t!]
\begin{center}
\includegraphics[width=17.0cm]{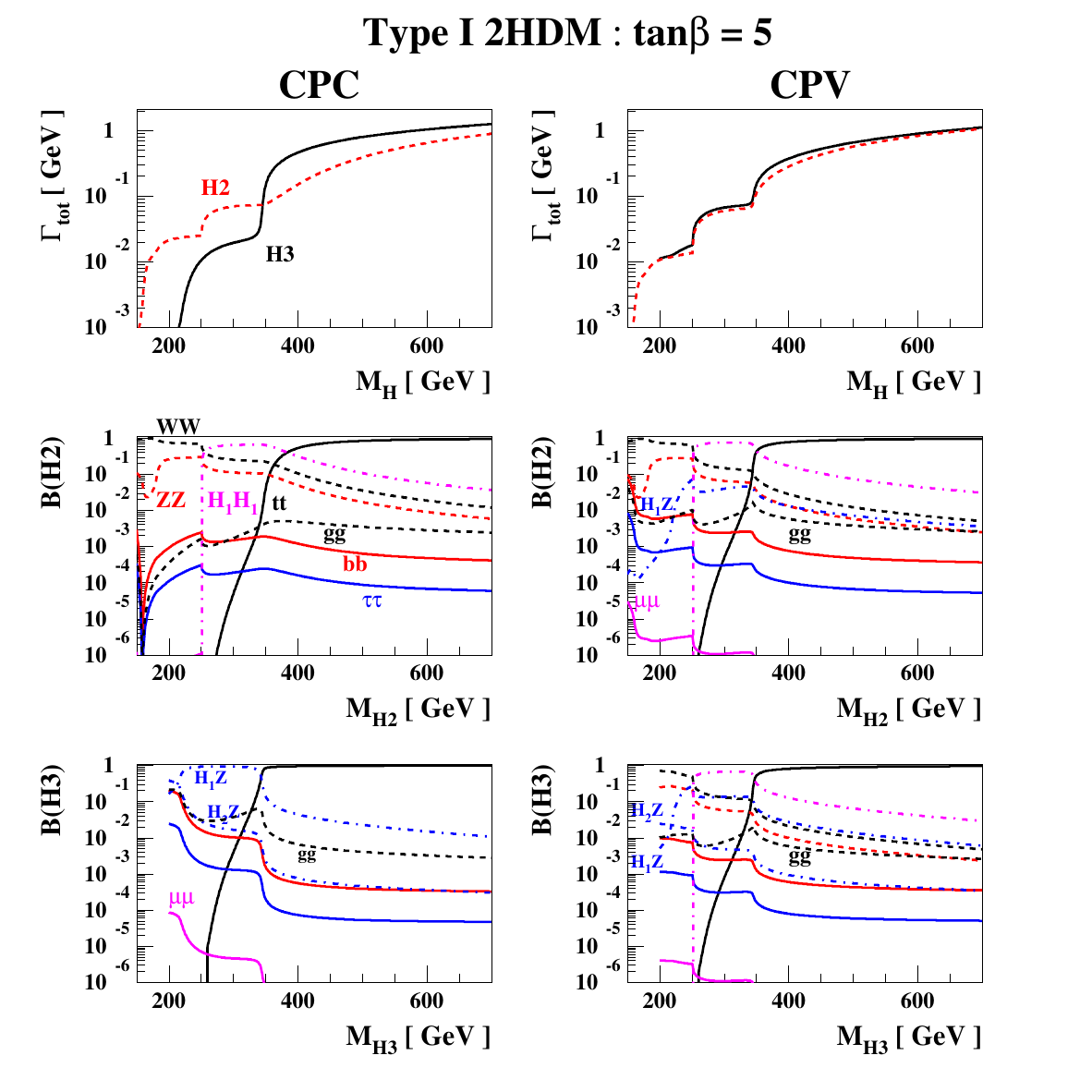}
\end{center}
\vspace{-1.0cm}
\caption{
The total decay widths and branching ratios of the two heavy Higgs bosons
in the type-I 2HDM taking $t_\beta=5$  and $|g_{_{H_2H_1H_1}}|=|g_{_{H_3H_1H_1}}|=0.1$.
In the left CPC panels, we take
$\delta_2^2=g_{_{H_2VV}}^2=2\epsilon_0\,(M_{H_1}^4/M_{H_2}^4)=\epsilon$ and
$\delta_3^2=g_{_{H_3VV}}^2=0$. While,
in the right CPV panels, we take
$\delta_2^2=g_{_{H_2VV}}^2=\epsilon_0\,(M_{H_1}^4/M_{H_2}^4)$ and
$\delta_3^2=g_{_{H_3VV}}^3=\epsilon_0\,(M_{H_1}^4/M_{H_3}^4)$.
In the both CPC and CPV cases, we take $\epsilon_0=0.05$.
}
\label{fig:2hdmh23tb05}
\end{figure}
\begin{figure}[t!]
\begin{center}
\includegraphics[width=17.0cm]{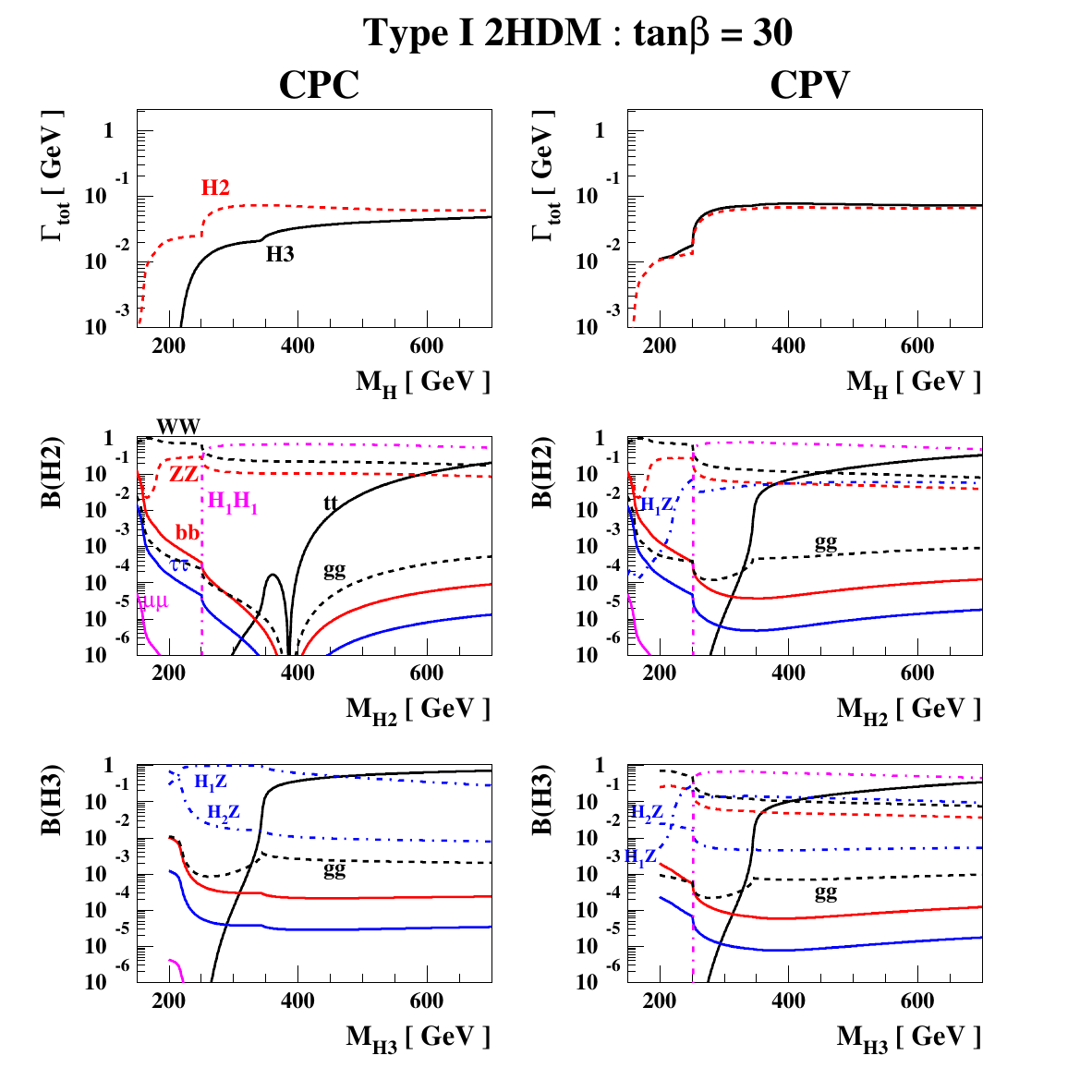}
\end{center}
\vspace{-1.0cm}
\caption{
The same as in Fig.~\ref{fig:2hdmh23tb05} but for
$t_\beta=30$.
}
\label{fig:2hdmh23tb30}
\end{figure}
For our numerical study, we vary $t_\beta$ but, for $\delta_{2,3}$,
we are taking
\begin{equation}
\label{eq:eps23}
g_{_{H_2VV}}=
\delta_{2} =  \sqrt\epsilon_{0}\,\left(\frac{M_{H_1}}{M_{H_2}}\right)^2\,, \ \ \
g_{_{H_3VV}}=
\delta_{3} =  \sqrt\epsilon_{0}\,\left(\frac{M_{H_1}}{M_{H_3}}\right)^2\,,
\end{equation}
reflecting the behavior of $\epsilon$ which is
suppressed by the quartic powers of the heavy
Higgs-boson masses at leading order~\cite{Choi:2020uum}.
With the above parameterizations of $\delta_{2,3}$
and taking $c_\omega>0$
we have $\delta_2/c_\omega=c_{\beta-\alpha}=\sqrt\epsilon > 0$
and $s_\omega^2=M_{H_{2}}^4/(M_{H_{2}}^4+M_{H_{3}}^4)$ leading to
a maximal CPV mixing between the two heavy Higgs bosons when they are degenerate.
For $\epsilon_{0}$ or the largest possible value of $\delta_{2,3}^2$,
we choose a value which is a little bit larger than
the lower $1\sigma$ error of $C_v$ in the {\bf CPC4} fit:
\footnote{See
Section \ref{sec:HiggcisionSub} and Table~\ref{tab:CPC} therein.
Note $C_v\equiv C_w=C_z$, see Eq.~(\ref{eq:Cnotations}).
}
\begin{equation}
\epsilon_0=0.05\,,
\end{equation}
having in mind the relation
$g_{_{H_1VV}}=\sqrt{1-\epsilon} \geq \sqrt{1-2\epsilon_0}\simeq 1-\epsilon_0$.
For the masses of Higgs bosons, we take
\begin{equation}
M_{H_1}=125.5 \ {\rm GeV}\,, \ \ \
M_{H_2}=M_{H_3} - 50 \ {\rm GeV}\,, \ \ \
M_{H_3} < M_{H^\pm} \sim 1~{\rm TeV}\,,
\end{equation}
with $M_{H_3}$ varied. This choice may result in
the simpler decay pattern
by forbidding or suppressing the decay channels of
$H_{2,3}\to H^\pm W^{\mp*}$,
$H_{2,3}\to H^\pm H^\mp$,
$H_{3}\to H_1 H_2$,
$H_{3}\to H_2 H_2$, etc.
By assuming very heavy charged Higgs boson,
also neglected are the contributions from the charged-Higgs-boson loops
to the decay processes of the heavy neutral Higgs bosons into
$\gamma\gamma$ and $Z\gamma$.
\footnote{For the details of the contributions from
the charged-Higgs-boson loops to the neutral Higgs boson decays into
two photons in the 2HDM and the MSSM, see Appendix \ref{app:cubicself}.}
Finally, for $H_{2,3}\to H_1H_1$,
we take $|g_{_{H_2H_1H_1}}|=|g_{_{H_3H_1H_1}}|=0.1$.
For the rigorous treatment of
the cubic $H_{2}H_1H_1$ and $H_{3}H_1H_1$ self-couplings
expressed in terms of
the masses of charged and neutral Higgs bosons and the elements
of the mixing matrix $O$, see Appendix \ref{app:cubicself}.

\medskip

In Fig.~\ref{fig:2hdmh23tb05}, we show the decay widths and
branching ratios of the two heavy Higgs bosons
in the type-I 2HDM taking $t_\beta=5$.
For CPC, $H_3$ is taken to be CP odd with $g_{_{H_3VV}}^2=0$
and, accordingly, the decays of $H_3$ into $WW$, $ZZ$, and $H_1H_1$
are forbidden. Incidentally, we note that $H_2\to H_1 Z$ decay is also
forbidden since $g_{_{H_1H_2Z}}^2=g_{_{H_3VV}}^2$.
On the other hand, for CPV, there are no forbidden decay modes
as long as they are kinematically allowed.
In the CPC case, the total decay widths of
CP-even $H_2$ and CP-odd $H_3$
are largely enhanced at the $H_1H_1$ and $t\bar t$ thresholds, respectively.
While, in the CPV case, both of the thresholds contribute to
the total decay widths as shown in the upper panels
of Fig.~\ref{fig:2hdmh23tb05}.

\medskip

For CPC, we further observe that the couplings of
the CP-even state of $H_2$ to fermions are
identically vanishing when $g^S_{H_2\bar f f}=0$.
It does happen at $s_\alpha=0$ or $\epsilon=1/(1+t_\beta^2)$,
see Eq.~(\ref{eq:2hdmaw}).
This explains why there are dips for the
fermionic decay modes of the CP-even $H_2$ state
at $M_{H_2}= [2\epsilon_0(1+t_\beta^2)]^{1/4}M_{H_1}
\simeq 70\,\sqrt{t_\beta}$ GeV
\footnote{In CPC, note that we take
$\delta_2^2=2\epsilon_0\,(M_{H_1}^4/M_{H_2}^4)=\epsilon$ with
$\delta_3^2=0$.}
as found in the middle-left panels of Fig.~\ref{fig:2hdmh23tb05} and
Fig.~\ref{fig:2hdmh23tb30}
around $M_{H_2}=160$ GeV and $385$ GeV, respectively.
We note that the branching ratios of fermionic decay modes
$H_{2,3}\to t\bar t, b\bar b, \tau\tau, \mu\mu$ are smaller for
the larger value of $t_\beta$ since the corresponding
decay widths are suppressed by
the factor of $\sim 1/t_\beta^2$.
For large values of $M_{H_{2,3}}$,
the numerical results are consistent with the observation that
the decay width of fermionic decay modes is proportional to $M_{H_{2,3}}$
while that of bosonic decay ones is {\it inversely}
proportional to $M_{H_{2,3}}$ especially with the parameterization
of Eq.~(\ref{eq:eps23})
for $H_{2,3}\to VV\,,H_{1} Z$. On the other hand,
the decay width $\Gamma(H_{3}\to H_{2} Z)$ is not suppressed
by the heavy mass $M_{H_3}$
because of  the relation $g_{_{H_2H_3Z}}^2=g_{_{H_1VV}}^2$.
In our numerical study,
it is suppressed since we have taken the small mass difference
between $H_3$ and $H_2$ of $M_{H_3}-M_{H_2}=50\, {\rm GeV} <M_Z$.
Otherwise, it may increase in proportion to $M_{H_3}^3$.

\section{Decays of a Charged Higgs Boson}
\label{sec:decays_of_a_charged_higgs_boson}

The effective couplings of the charged
Higgs boson $H^\pm$ to quarks and leptons are
described by the interaction Lagrangian:
\begin{equation} \label{Hud}
{\cal L}_{H^\pm f_\uparrow f_\downarrow}\ =\ \sqrt{2}\
H^+\, \overline{f_\uparrow}\,\left(\, \frac{m_{f_\uparrow}}{v}\,
g_{f_\uparrow}\, P_L\ +\frac{m_{f_\downarrow}}{v}\, g_{f_\downarrow}\, P_R\,
\right)\, {f_\downarrow}\ +\ {\rm h.c.}\,,
\end{equation}
where $P_{L,R}=(1\mp\gamma_5)/2$ and
$({f_\uparrow},{f_\downarrow})=(t,b)\,,(c,s)\,,(\nu_\tau,\tau)\,,(\nu_\mu,\mu)$, etc.
The masses of the up- and down-type fermions are denoted by
$m_{f_\uparrow}$ and ${m}_{f_\downarrow}$, respectively.
On the other hand, the interaction of the charged Higgs boson with  a massive
gauge boson $W$ and a neutral Higgs boson $H$
is given by
\begin{equation}
{\cal L}_{HH^\pm W^\mp} = -\frac{g}{2} \,  g_{_{HH^+ W^-}}\,
W^{-\mu} (H\, i\!\stackrel{\leftrightarrow}{\partial}_\mu
H^+)\, +\, {\rm h.c.}\,,
\end{equation}
with the convention $X\stackrel{\leftrightarrow}{\partial}_\mu Y
\equiv X (\partial_\mu Y) - (\partial_\mu X) Y$.
With these effective interactions,
in this section, we study the charged Higgs decays at LO
except for the QCD corrections considered in the decay modes into quarks,
concentrating on mainstream instead of being complete.
For comprehensive studies of decays of a charged Higgs boson in BSM models,
we refer to, for example, Refs.
\cite{Akeroyd:2016ymd,Arhrib:2010ju,Arbey:2017gmh,Arhrib:2018ewj}.

\medskip

When a charged Higgs boson decays into quarks, the decay width is given by
\footnote{For $H^+\to t^{(*)} \bar b$, note that the QCD corrections are not valid
in the threshold region due to the top-quark mass effects. For them,
we refer to~\cite{Spira:2016ztx} and references there in.}
\begin{eqnarray}
\label{eq:gchff}
\Gamma(H^+\to {f_\uparrow} {f_\downarrow}) &=&
N_C^{f_\uparrow f_\downarrow}\frac{M_{H^\pm}}{8\pi v^2}
\lambda^{1/2}\left(1,\kappa_{f_\uparrow},\kappa_{f_\downarrow}\right)\,
\Bigg\{\left(1-\kappa_{f_\uparrow}-\kappa_{f_\downarrow}\right)
\left(m_{f_\uparrow}^2 g_{f_\uparrow}^2 +
m_{f_\downarrow}^2 g_{f_\downarrow}^2\right)
\nonumber \\
&&
-4\sqrt{\kappa_{f_\uparrow}\kappa_{f_\downarrow}}\,
m_{f_\uparrow} m_{f_\downarrow} g_{f_\uparrow} g_{f_\downarrow}
\Bigg\} \, \left(1+\delta_{\rm QCD} \right)\,,
\end{eqnarray}
including the QCD correction factor $\delta_{\rm QCD}$
with $\kappa_{{f_\uparrow},{f_\downarrow}}
=M^2_{{f_\uparrow},{f_\downarrow}}/M_{H^\pm}^2$ and
$N_C^{f_\uparrow f_\downarrow}=3$.
Note that the $\overline{\rm MS}$ quark masses such as
$\overline{m}_{t,c}(M_{H^\pm})$ and $\overline{m}_{b,s}(M_{H^\pm})$
are taken for $m_{f_\uparrow}$ and $m_{f_\downarrow}$,
respectively.  While, for the decays into leptons,
the charged lepton pole masses are used for $m_{f_\downarrow}$
together with $m_{f_\uparrow}=0$,
$N_C^{f_\uparrow f_\downarrow}=1$, and $\delta_{\rm QCD}=0$.
A charged Higgs boson may decay into a lighter neutral
Higgs boson $H$ and a massive gauge boson $W$ and
the LO decay width is given by
\cite{Cahn:1988ru,Moretti:1994ds,Djouadi:1995gv}
\begin{equation}
\label{eq:gchhw}
\Gamma^{\rm LO}(H^+\rightarrow H W^{+*}) =
\frac{G_FM_{H^\pm}^3|g_{_{HH^+W^-}}|^2}{8\sqrt{2}\pi}
\int_0^{(\sqrt{\omega_\pm}-\sqrt{\omega})^2}{\rm d}x\,
\frac{\epsilon_W\lambda^{3/2}(\omega_\pm,\omega,x)}
{\omega_\pm^3\pi[(x-1)^2+\epsilon_W^2]}\,,
\end{equation}
with $\omega=M_{H}^2/M_W^2$ and $\omega_\pm=M_{H^\pm}^2/M_W^2$.
When $M_{H^\pm}$ is larger than the sum $M_{H}+ M_W$,
it reduces to
\begin{equation}
\Gamma^{\rm LO}(H^+\to H W^{+}) \ = \
\frac{G_F M_{H^\pm}^3}{8\sqrt{2}\pi} |g_{_{HH^+W^-}}|^2
\lambda^{3/2}(1,\kappa_H,\kappa_W)\,,
\end{equation}
where $\kappa_H=M_{H}^2/M_{H^\pm}^2$ and $\kappa_W=M_W^2/M_{H^\pm}^2$.
The electroweak corrections within the 2HDM framework
have been calculated, specifically
for the process of $H^\pm\to W^\pm h, W^\pm H, W^\pm A$
\cite{Akeroyd:1998uw,Akeroyd:2000xa,Krause:2016oke}.
They are of moderate size and numerically stable
if a process- and gauge-independent
renormalization scheme is chosen~\cite{Krause:2016oke}.
Note that the package {\tt 2HDECAY}~\cite{Krause:2018wmo}
is available for
the calculation of full ELW one-loop corrections to
both neutral and charged Higgs decays in the CP-conserving 2HDM.

\medskip

Finally, at LO, the decay widths of a charged Higgs boson
$H^+$ into a chargino $\wt{\chi}_j^+$ and a neutralino $\wt{\chi}_i^0$
are given by
\begin{eqnarray}
\Gamma^{\rm LO}(H^+\to \wt{\chi}_j^+ \wt{\chi}_i^0) &=&
\frac{g^2 M_{H^\pm} \lambda^{1/2}(1,\kappa_i,\kappa_j)}{16\pi} \\
&\times&
\left[(1\!-\!\kappa_i\!-\!\kappa_j)
(|g^S_{H^+\wt{\chi}_i^0\wt{\chi}_j^-}|^2+|g^P_{H^+\wt{\chi}_i^0\wt{\chi}_j^-}|^2)
-2\sqrt{\kappa_i\kappa_j}
(|g^S_{H^+\wt{\chi}_i^0\wt{\chi}_j^-}|^2-|g^P_{H^+\wt{\chi}_i^0\wt{\chi}_j^-}|^2)\right]\,,
\nonumber
\end{eqnarray}
with $\kappa_i=m_{\wt{\chi}_i^0}^2/M_{H^\pm}^2$ and
$\kappa_j=m_{\wt{\chi}_j^\pm}^2/M_{H^\pm}^2$
and the LO decay widths into a pair of sfermions by
\begin{equation}
\Gamma^{\rm LO}(H^+\to \wt{f}_i\wt{f^\prime}_j^*) =
N_C^{ff^\prime}\frac{v^2\,|g_{H^+\wt{f}_i^*\wt{f^\prime}_j}|^2}{16\pi M_{H^\pm}}
\lambda^{1/2}(1,\kappa_i,\kappa_j)\,,
\end{equation}
where $\kappa_{i}=M_{\wt{f}_{i}}^2/M_{H^\pm}^2$,
$\kappa_{j}=M_{\wt{f^\prime}_{j}}^2/M_{H^\pm}^2$, and
$N_C^{ff^\prime}=3$ and $1$ for squarks and sleptons, respectively.

\medskip

For a numerical example, we take the 2HDMs in which the relevant couplings
are given by
\footnote{See Eq.~(\ref{eq:chff.2hdm}) and Table~\ref{tab:2hdtype}.}
\begin{eqnarray}
H^+\to t\bar b\,, c\bar s \ &:& \ g_{f_\uparrow}=\frac{1}{t_\beta}\,, \ \
g_{f_\downarrow}=-\frac{1}{t_\beta}\,({\rm I\,,III})\,,  \ t_\beta\,({\rm II\,,IV})\,;
\nonumber \\
H^+\to \nu \tau^+\,, \nu \mu^+ \ &:& \ g_{f_\uparrow}=0\,, \ \ \ \,
g_{f_\downarrow}=-\frac{1}{t_\beta}\,({\rm I\,,IV})\,,  \ t_\beta\,({\rm II\,,III})\,;
\nonumber \\
H^+\to H W^{+} \ &:& \ |g_{_{HH^+ W^-}}|^2 = 1 - g_{_{HVV}}^2\,
({\rm I\,,II\,,III\,,IV})\,,
\end{eqnarray}
depending on the 2HDM type as denoted by I, II, III or IV.
For the decay $H^+\to H W^{+}$,  we set
$M_H=125.5$ GeV and $|g_{_{HH^+ W^-}}|^2 = 0.1$.
\footnote{For this,
we take $g_{_{HVV}} \simeq 0.95$ adopting
a little bit larger value than
the lower $1\sigma$ error of $C_v$ in the {\bf CPC4} fit, see
Section \ref{sec:HiggcisionSub} and Table~\ref{tab:CPC} there in.}
For the $\overline{\rm MS}$ mass of the strange quark, we take the
approximation $\overline m_s(\mu)=\overline m_c (\mu)/11.72$~\cite{PDG2020}.
On the other hand, we neglect the effects of
its pole mass which is relevant to the $H^+\to c\bar s$ decay only
through the kinematical factor $\kappa_s$, see Eq.~(\ref{eq:gchff}),
and too tiny to influence the numerical results.
%

\medskip

For the charged Higgs--boson decay $H^+\to t\bar b$, we take
the contribution of the off-shell top quark into account
\cite{Moretti:1994ds,Djouadi:1995gv}:
\begin{eqnarray}
\label{eq:gchtb}
\Gamma^{\rm LO}(H^\pm\to t(p_t)\,\bar b\to b \bar b W^+)
&=& N_C^{tb}\,\frac{1}{v^2}\frac{g^2 M_{H^\pm}}{2^{9}\pi^3}\,
\int_{M_W^2}^{(M_{H^\pm}-M_b)^2}\,\lambda_{H^\pm}^{1/2}\,
p_t^2\left(1-\frac{M_W^2}{p_t^2}\right)^2
\left(2+\frac{p_t^2}{M_W^2}\right)\nonumber \\[2mm]
&&\hspace{-5.0cm}\times\,
\frac{\left(1-\alpha_t-\alpha_{\bar b}\right)
\left[\overline m_t^2(M_{H^\pm}) g_t^2 + \overline m_b^2(M_{H^\pm}) g_b^2\right]
-4\sqrt{\alpha_t\alpha_{\bar b}}\,\overline m_t(M_{H^\pm}) \overline m_b(M_{H^\pm})
g_t g_b}{(p_t^2-M_t^2)^2+M_t^2\Gamma_t^2}\, dp_t^2\,,
\end{eqnarray}
where the kinematical $b$-quark mass is neglected in the $t\to b W^+$
decay process and
the triangle function $\lambda_{H^\pm}$ is given by
\begin{equation}
\lambda_{H^\pm}=1+\alpha_t^2+\alpha_{\bar b}^2-2\alpha_t-2\alpha_{\bar b}
-2\alpha_t\alpha_{\bar b}\,,
\end{equation}
with $\alpha_t=p_t^2/M_{H^\pm}^2$ and $\alpha_{\bar b}=M_b^2/M_{H^\pm}^2$.
When $M_{H^\pm} > M_t+M_b$,
using Eqs.~(\ref{eq:delta}) and (\ref{eq:gamtb}), we have
\begin{equation}
\Gamma^{\rm LO}(H^\pm\to t\,\bar b\to b \bar b W^+) =
\Gamma^{\rm LO}(H^\pm\to t\,\bar b) \,
\frac{\Gamma^{\rm LO}(t\to b W)}{\Gamma_t}\,,
\end{equation}
in a factorized form.

\medskip

\begin{figure}[t!]
\begin{center}
\includegraphics[width=8.4cm]{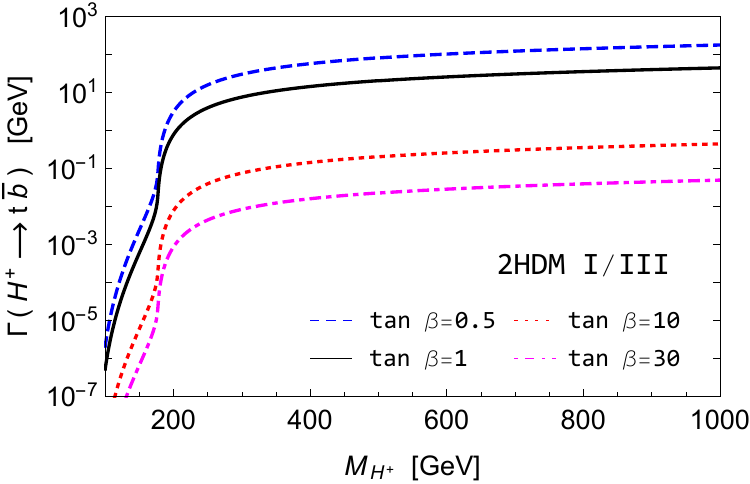}
\includegraphics[width=8.4cm]{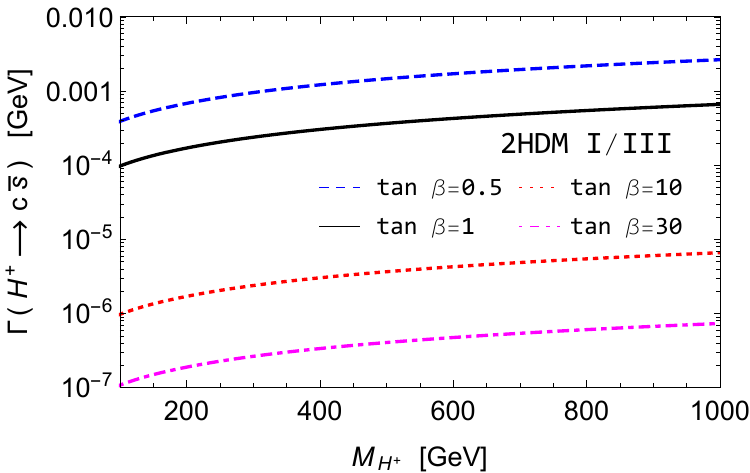}
\includegraphics[width=8.4cm]{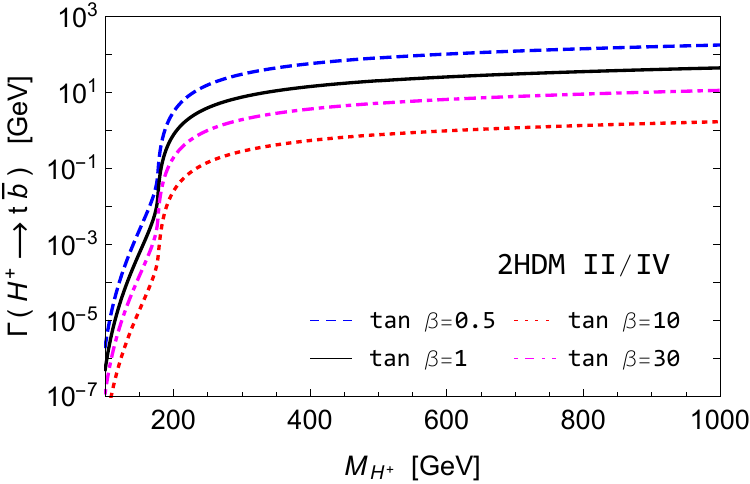}
\includegraphics[width=8.4cm]{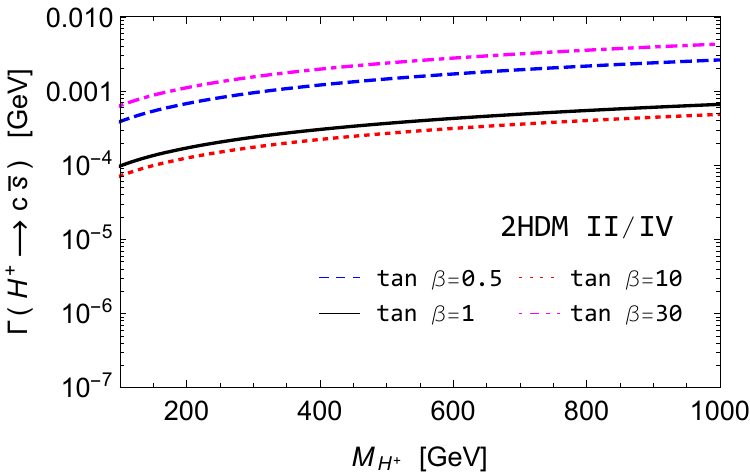}
\end{center}
\caption{
(Upper)
Decay widths of a charged Higgs boson with a mass $M_{H^\pm}$ into
$t\bar b$ (left) and $c\bar s$ (right) in the type-I and
type-III 2HDMs. QCD corrections are taken into account.
For $\tan\beta$, we take four values of
$0.5$ (blue dashed), $1$ (black solid), $10$ (red dotted),
and $30$ (magenta dash-dotted).
(Lower)
The same as in the upper panels but  in the type-II and
type-IV 2HDMs.
}
\label{fig:ch2qqp}
\end{figure}
\begin{figure}[t!]
\begin{center}
\includegraphics[width=8.4cm]{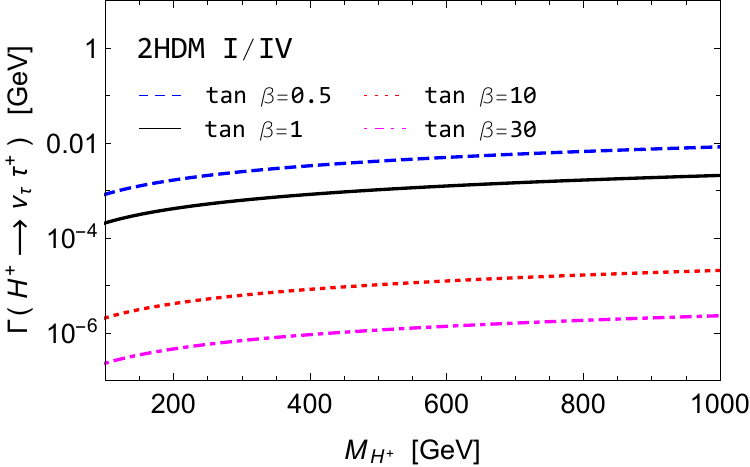}
\includegraphics[width=8.4cm]{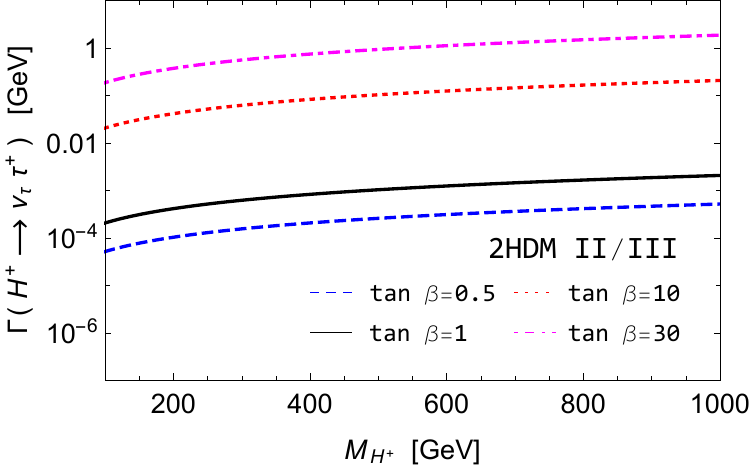}
\end{center}
\caption{
(Upper)
Decay widths of a charged Higgs boson with a mass $M_{H^\pm}$ into $\tau^+\nu$
in the type-I/IV 2HDMs (left) and in the type-II/III ones (right).
For $\tan\beta$, we take four values of
$0.5$ (blue dashed), $1$ (black solid), $10$ (red dotted),
and $30$ (magenta dash-dotted).
}
\label{fig:ch2nul}
\end{figure}
\begin{figure}[t!]
\begin{center}
\includegraphics[width=8.4cm]{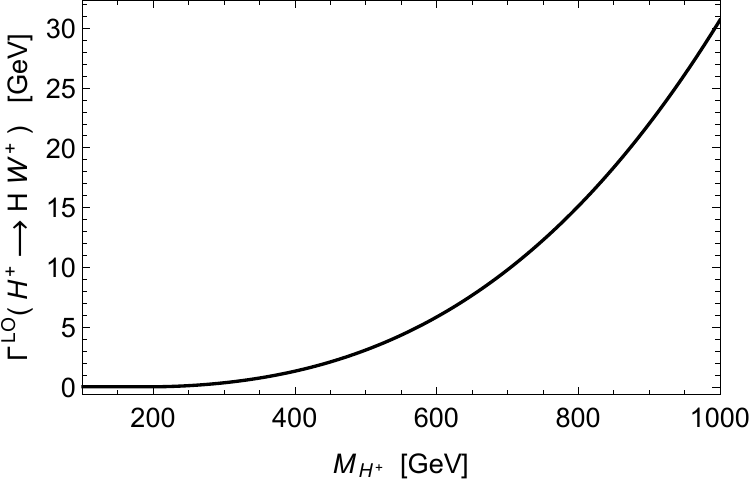}
\includegraphics[width=8.4cm]{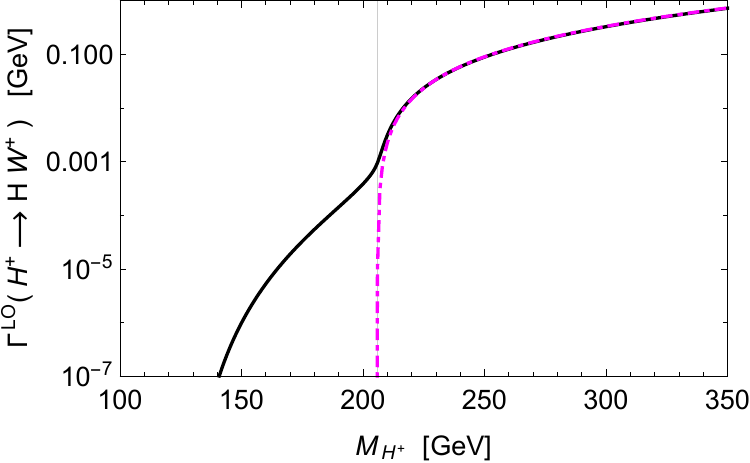}
\end{center}
\caption{
Decay widths of a charged Higgs boson with a mass $M_{H^\pm}$ into $HW^+$
taking $M_H=125.5$ GeV and $|g_{_{HH^+ W^-}}|^2 = 0.1$.
In the right panel, we magnify the low $M_H$ region
and compare with the two-body decay width (dash-dotted magenta line).
The vertical line locates
the position $M_{H^\pm}=M_H+M_W$.
}
\label{fig:ch2hw}
\end{figure}

In the upper panels of Fig.~\ref{fig:ch2qqp}, we show the decay widths
of a charged Higgs boson into two quarks, $t^{(*)}\bar b$ (left)
\footnote{At $M_{H^\pm}=500$ GeV, we switch from the three-body decay width
$\Gamma^(H^+\to b\bar b W^+)$ to the two-body decay width $\Gamma(H^+\to t\bar b)$
because of the same reasons as in $H\to t^*\bar t^*$ described
previously.} and $c\bar s$ (right)
in the type-I and type-III 2HDMs taking account of the QCD corrections.
In these models, $g_{t,c}=-g_{b,s}=1/t_\beta$ and the top-quark and
charm-quark contributions dominate and the decay widths scale
as $1/t_\beta^2$.
In the lower panels of Fig.~\ref{fig:ch2qqp}, we consider the type-II and 
type-IV 2HDMs
in which $g_{t,c}=1/t_\beta$ and $g_{b,s}=t_\beta$.
For low values of $t_\beta$, the top-quark and charm-quark
contributions dominate and the decay widths scale as $1/t_\beta^2$.
On the other hand, for high values of $t_\beta$,
the bottom-quark and strange-quark contributions are enhanced by the
factor of $t_\beta^2$ while
the top-quark and charm-quark contributions are suppressed
by the factor of $1/t_\beta^2$.
The bottom-quark and strange-quark contributions
start to dominate when $t_\beta$ is larger than
$\sqrt{\overline m_t/\overline m_b} \simeq 7.7$ and
$\sqrt{\overline m_c/\overline m_s}\simeq 3.4$, respectively.
When the bottom-quark and strange-quark contributions dominate,
the decay widths scale as $(t_\beta/60)^2$ and $(t_\beta/12)^2$, respectively,
compared to the case with $t_\beta=1$ and these factors
are responsible for the significant change of the decay widths for
high $t_\beta$ values, as can be checked by comparing the lines with
$t_\beta=1$ and those with $t_\beta=10$ and $30$.

\medskip

In Fig.~\ref{fig:ch2nul}, we show the LO decay widths
of a charged Higgs boson with mass $M_{H^\pm}$ into $\tau^+\nu$
in the type-I/IV 2HDMs (left) and in the type-II/III ones (right)
taking $t_\beta=0.5$ (blue dashed), $1$ (black solid), $10$ (red dotted),
and $30$ (magenta dash-dotted).
In the type-I/IV and type-II/III 2HDMs, the decay widths scale as
$g^2_\tau=1/t^2_\beta$ and $t^2_\beta$, respectively.
Compared to the case of $H^+\to\tau^+\nu$,
the decay widths $\Gamma(H^+\to\mu^+\nu)$ are simply suppressed by
the large factor of $m_\tau^2/m_\mu^2\sim 300$.

\medskip

In Fig.~\ref{fig:ch2hw}, we show the LO
decay widths of a charged Higgs boson with a mass $M_{H^\pm}$ into $HW^+$
taking $M_H=125.5$ GeV and $|g_{_{HH^+ W^-}}|^2 = 0.1$.
In the right panel, we magnify the low $M_H$ region covering
the case with a virtual $W^{+*}$ and we compare the decay width
with the prediction of the two-body decay width (dash-dotted magenta line).
We note that they are nearly identical for $M_{H^\pm}> M_H+M_W$,
as expected.

\medskip

\begin{figure}[t!]
\begin{center}
\includegraphics[width=7.2cm]{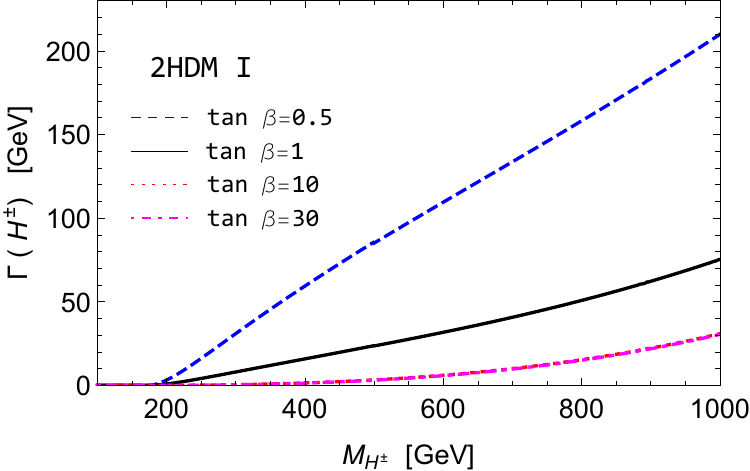}
\includegraphics[width=7.2cm]{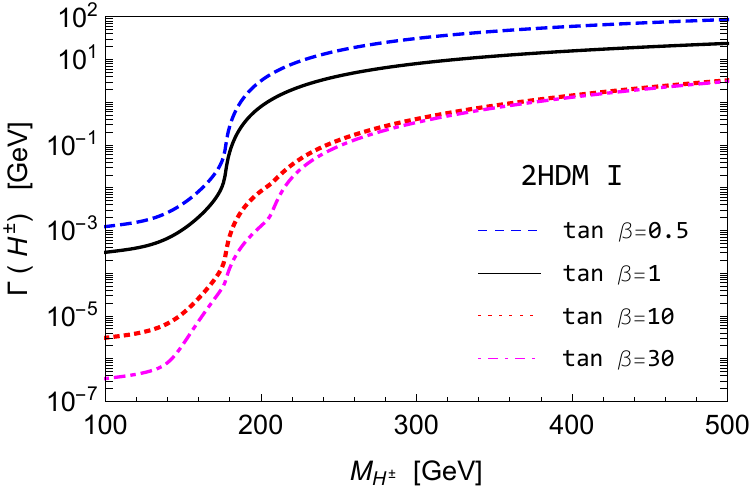}
\includegraphics[width=7.2cm]{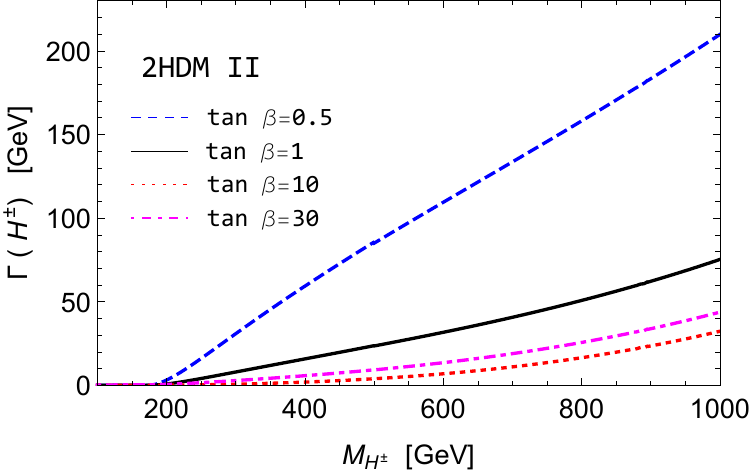}
\includegraphics[width=7.2cm]{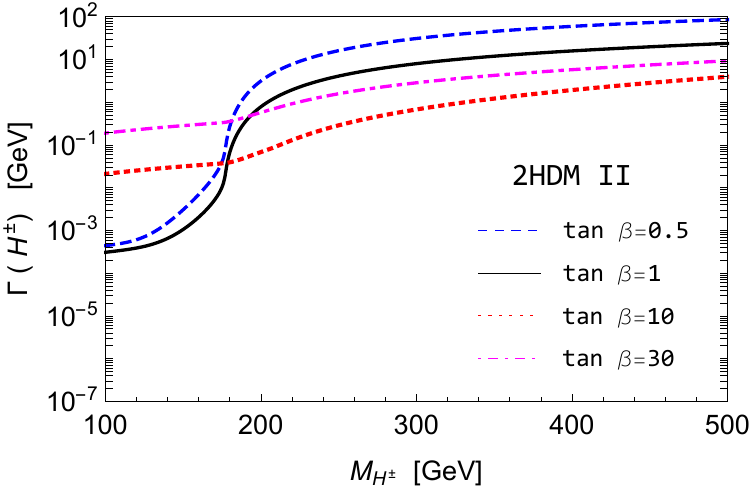}
\includegraphics[width=7.2cm]{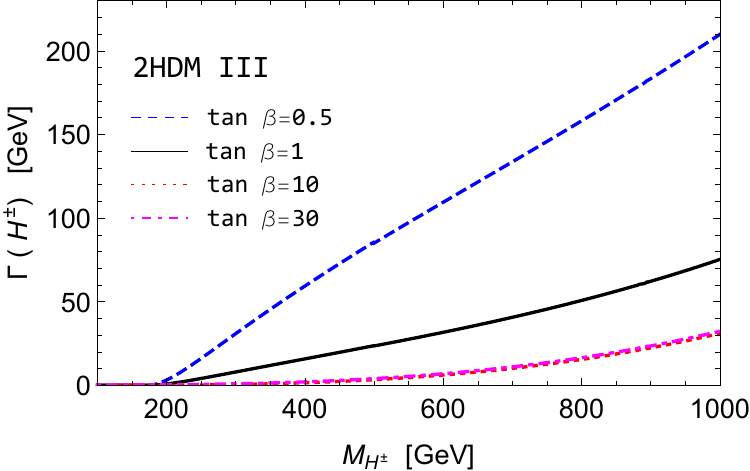}
\includegraphics[width=7.2cm]{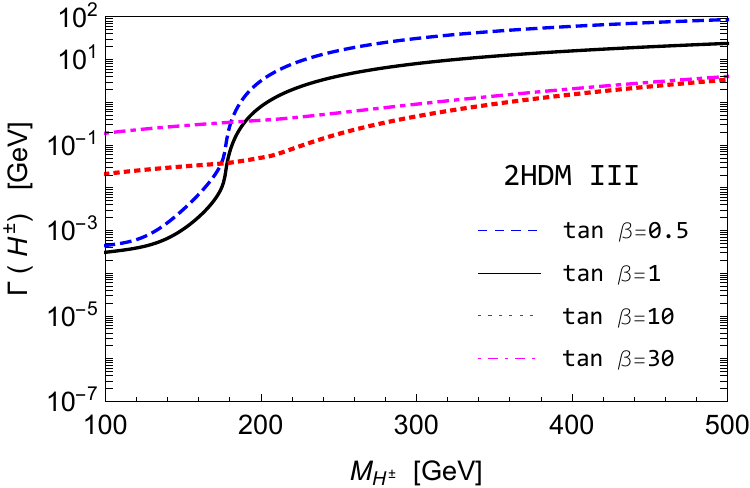}
\includegraphics[width=7.2cm]{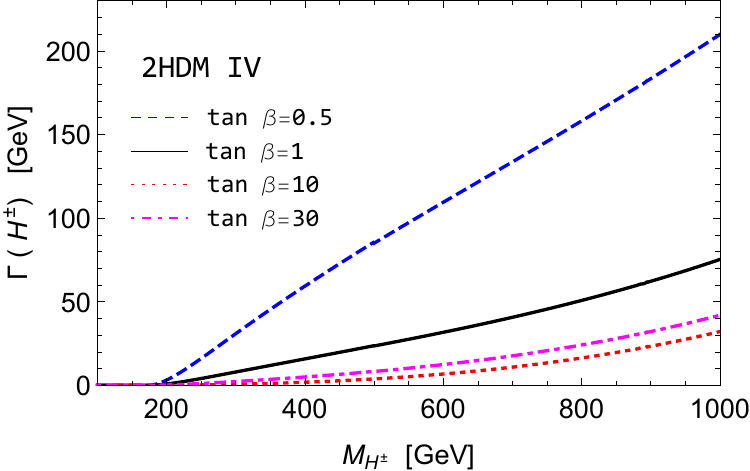}
\includegraphics[width=7.2cm]{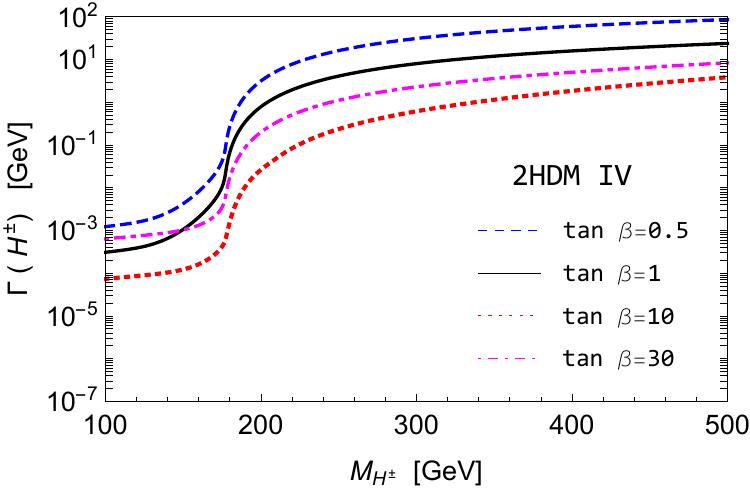}
\end{center}
\caption{
Total decay widths of a charged Higgs boson with a mass $M_{H^\pm}$
in the type-I, type-II, type-III, and type-IV 2HDMs from top to bottom.
We are taking $\tan\beta=0.5$ (blue dashed),
$1$ (black solid), $10$ (red dotted), and $30$ (magenta dash-dotted).
In the right panels, we magnify the low $M_H$ region.
}
\label{fig:chgamtot}
\end{figure}
\begin{figure}[t!]
\begin{center}
\includegraphics[width=7.2cm]{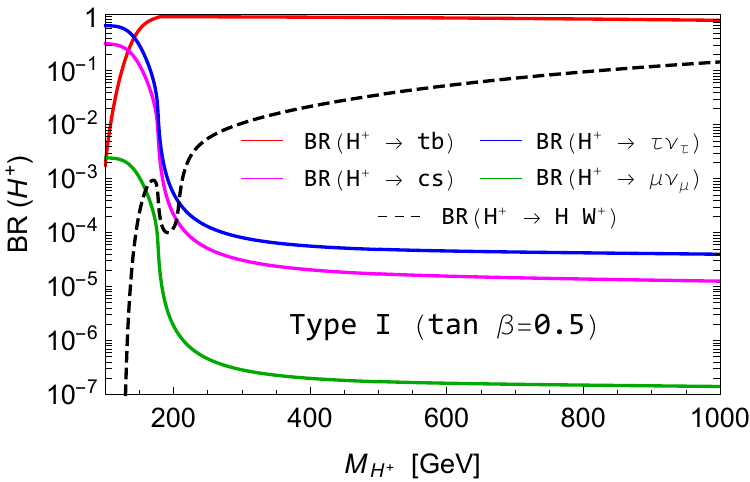}
\includegraphics[width=7.2cm]{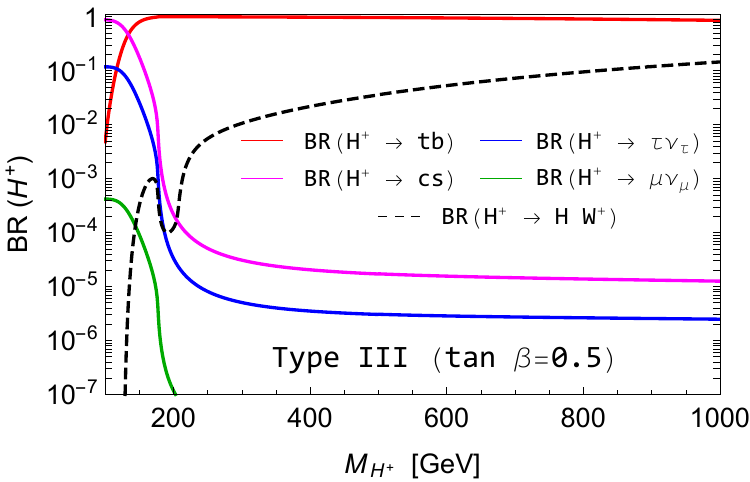}
\includegraphics[width=7.2cm]{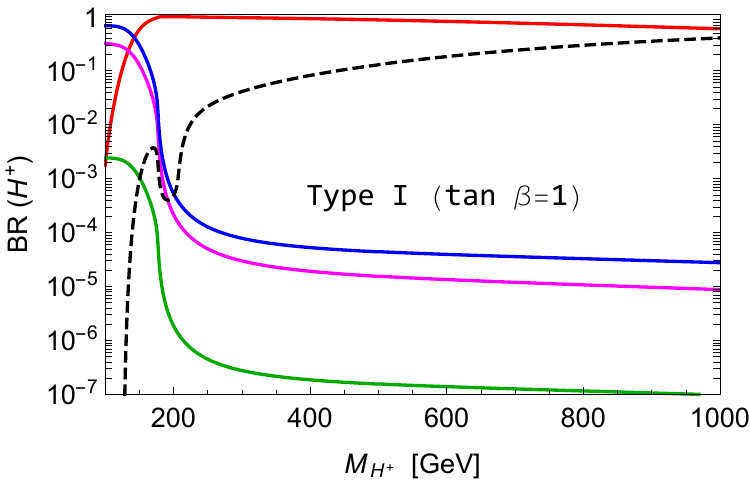}
\includegraphics[width=7.2cm]{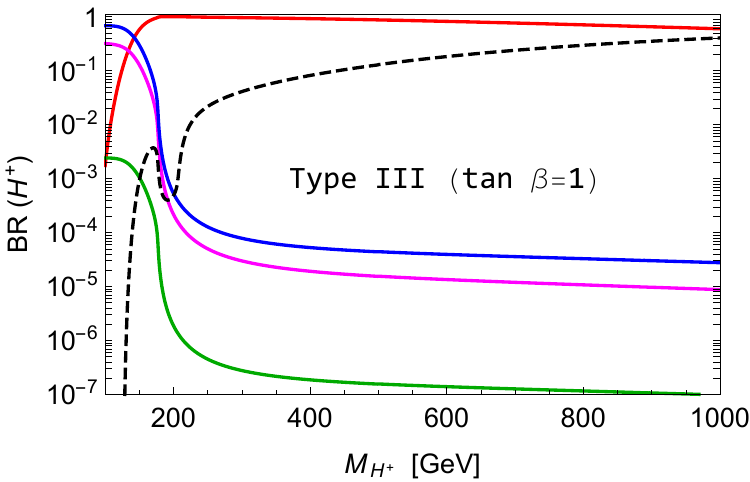}
\includegraphics[width=7.2cm]{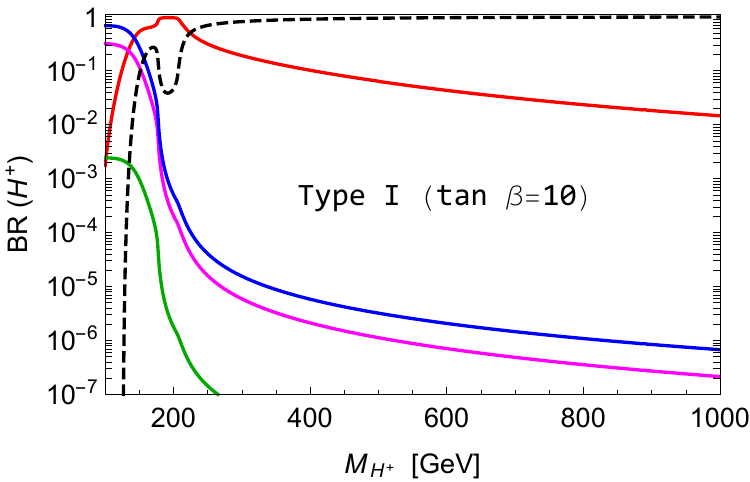}
\includegraphics[width=7.2cm]{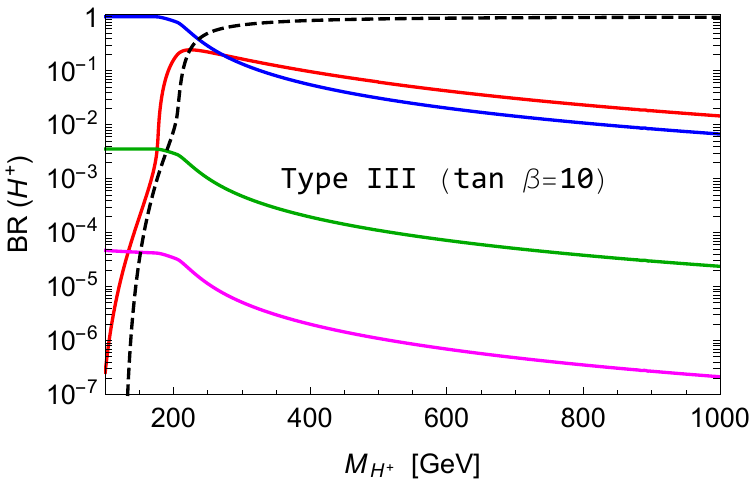}
\includegraphics[width=7.2cm]{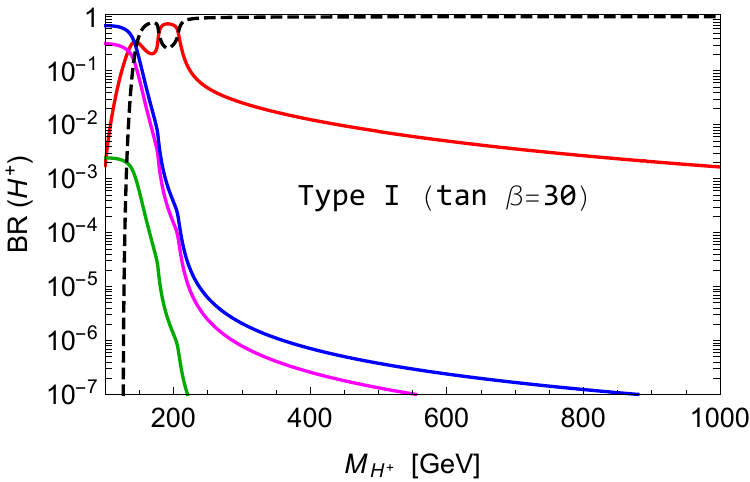}
\includegraphics[width=7.2cm]{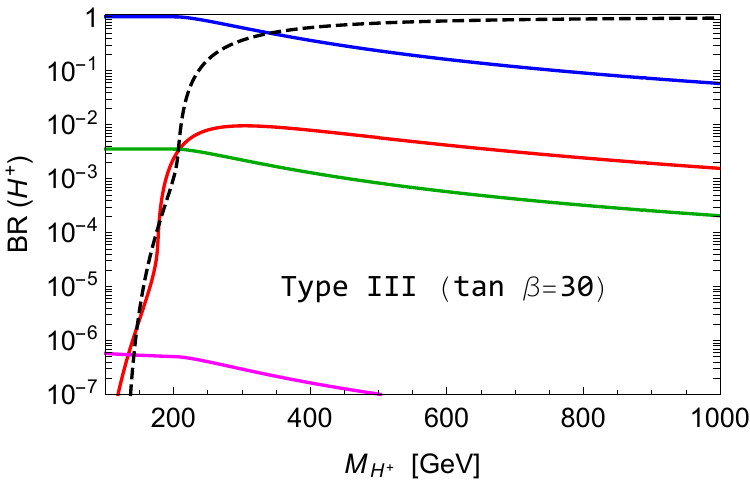}
\end{center}
\caption{
Branching ratios of a charged Higgs boson with a mass $M_{H^\pm}$
in the type-I (left) and type-III (right) 2HDMs taking
$\tan\beta=0.5$, $1$, $10$, and $30$ from top to bottom.
We consider 5 decay modes of
$H^+\to t\bar b$ (red solid), $H^+\to c\bar s$ (magenta solid),
$H^+\to \tau\nu$ (blue solid), $H^+\to \mu\nu$ (green solid),
and $H^+\to H W^+$ (black dashed) with $M_H=125.5$ GeV
and $|g_{_{HH^+ W^-}}|^2 = 0.1$.
}
\label{fig:chbr13}
\end{figure}
\begin{figure}[t!]
\begin{center}
\includegraphics[width=7.2cm]{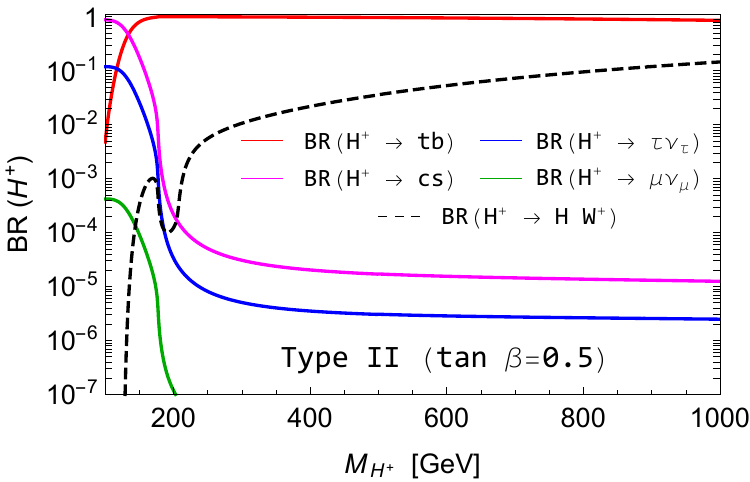}
\includegraphics[width=7.2cm]{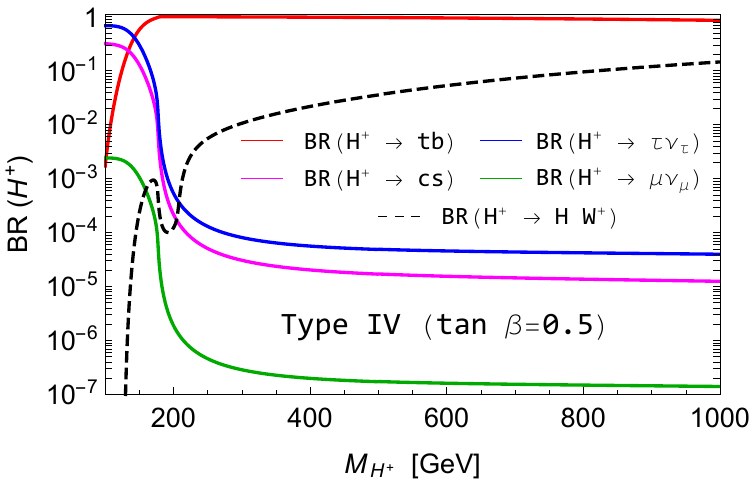}
\includegraphics[width=7.2cm]{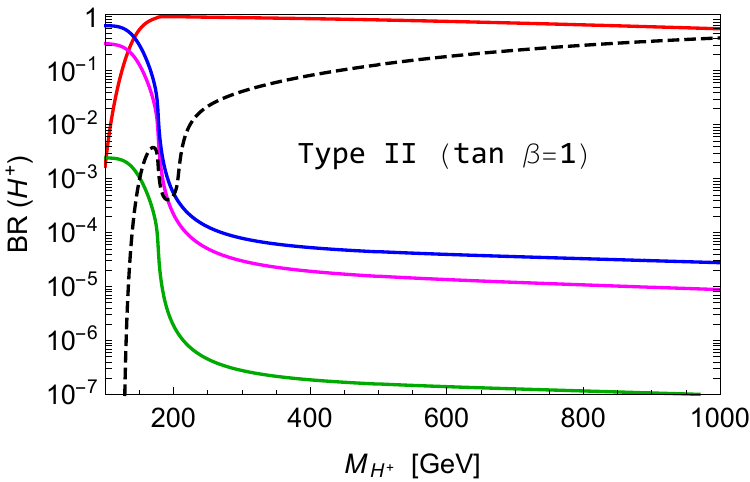}
\includegraphics[width=7.2cm]{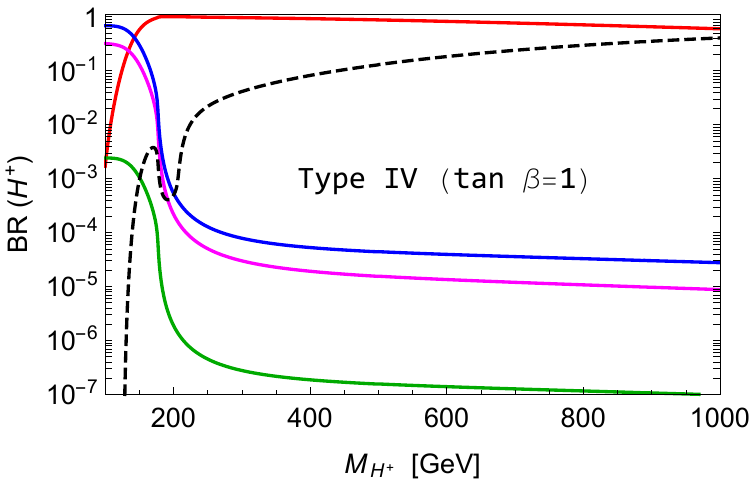}
\includegraphics[width=7.2cm]{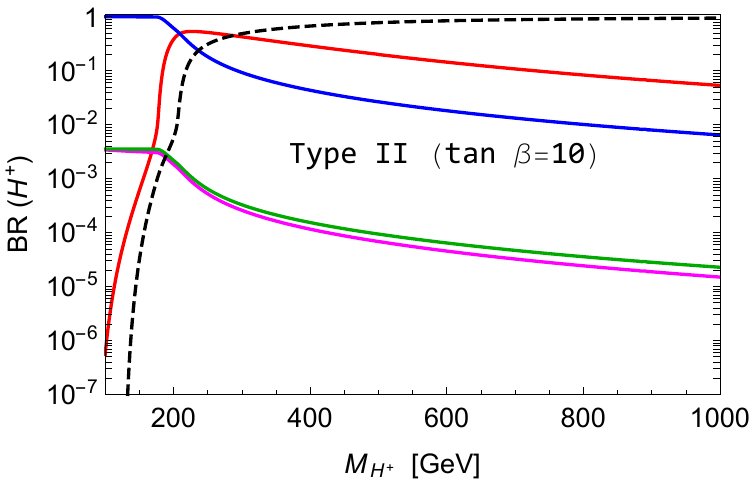}
\includegraphics[width=7.2cm]{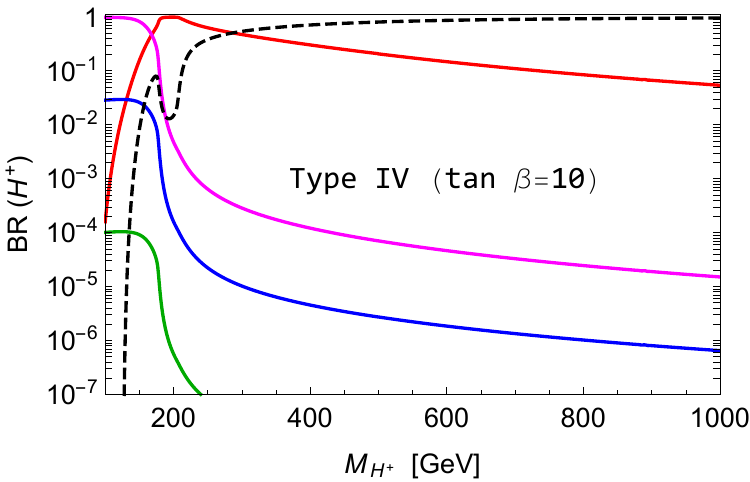}
\includegraphics[width=7.2cm]{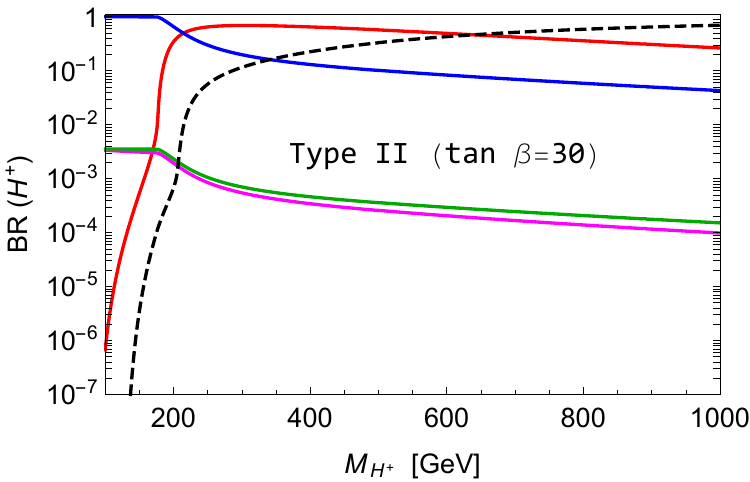}
\includegraphics[width=7.2cm]{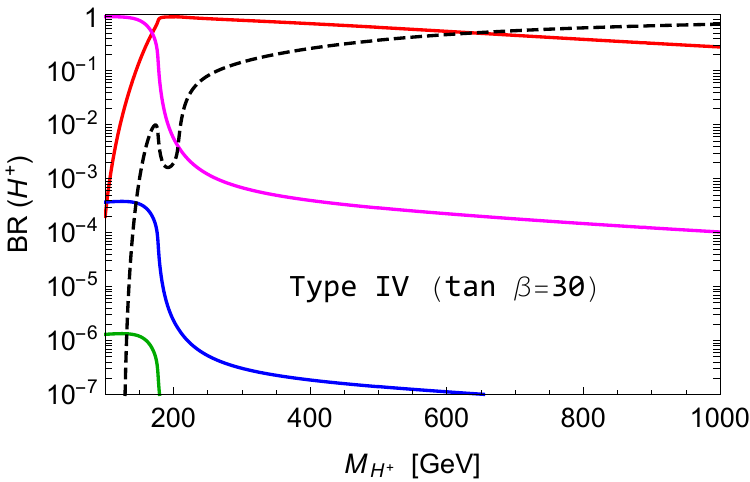}
\end{center}
\caption{The same as in Fig.~\ref{fig:chbr13} but for
the type-II (left) and type-IV (right) 2HDMs.
}
\label{fig:chbr24}
\end{figure}

In Fig.~\ref{fig:chgamtot}, we show the total decay widths of
a charged Higgs boson for four values of $t_\beta$:
$\tan\beta=0.5$ (blue dashed),
$1$ (black solid), $10$ (red dotted), and $30$ (magenta dash-dotted).
We sum over the 5 decay modes of
$H^+\to t\bar b$, $H^+\to c\bar s$,
$H^+\to \tau\nu$, $H^+\to \mu\nu$,
and $H^+\to H W^+$ with $M_H=125.5$ GeV and
$|g_{_{HH^+ W^-}}|^2 = 0.1$.

\medskip

Firstly, we consider the total charged-Higgs
decay width in the case of a heavy charged Higgs boson, see the left
panels of Fig.~\ref{fig:chgamtot}.
We note that the top-quark contributions to
$\Gamma(H^+\to t\bar b)$ scale as $1/t_\beta^2$, independently of
the 2HDM types. As the value of $t_\beta$ grows,
the total decay width monotonically decreases
in the type-I and type-III 2HDMs. In contrast, in the
type-II and type-IV 2HDMs, the decay width reaches the minimum around
$t_\beta = 8$, where the top-quark and bottom-quark contributions
become comparable, and then it starts to increases as $t_\beta$ grows
further.
Based on these observations, we can work out the behavior
of the total decay width of a heavy charged Higgs boson for large values of
$t_\beta$: the dominant contribution comes from
the $H^+\to H W^+$ decay mode with the subleading/competing contributions
from $H^+\to t\bar b$.
We identify that there exists an additional subleading
contribution from $H^+\to \tau\nu$ for $t_\beta=10$ and $30$
in the type-II and type-III 2HDMs.
On the other hand, for low values of $t_\beta$, the $H^+\to t\bar b$ decay
mode dominates the total decay width
with the subleading contribution from $H^+\to H W^+$
which amounts to about 30 GeV at $M_{H^\pm}=1$ TeV
taking $M_H=125.5$ GeV and $|g_{_{HH^+ W^-}}|^2 = 0.1$, see Fig.~\ref{fig:ch2hw}.

\medskip

Secondly, we consider the case of a charged Higgs boson lighter
than $t$ quark, see the right panels of Fig.~\ref{fig:chgamtot}.
For large values of $t_\beta$, the $H^+\to \tau\nu$ decay mode dominates
in the type-II and type-III 2HDMs.
In the type-I 2HDM, the $H^+\to \tau\nu$ decay mode still dominates
with the subleading contributions from $H^+\to c\bar s$
though both of them are suppressed by $1/t_\beta^2$.
In the type-IV 2HDM, the $H^+\to c\bar s$ decay mode dominates
where the strange-quark contribution is enhanced by $t_\beta^2$.
For small values of $t_\beta$, the $H^+\to \tau\nu$ decay mode
mostly dominates, leading to the larger decay widths for $t_\beta=0.5$
in the type-I and type-IV 2HDMs in which $|g_\tau|=1/t_\beta$.

\medskip

In Fig.~\ref{fig:chbr13}, we show the
branching ratios of a charged Higgs boson varying its mass
between 100 GeV and 1 TeV in the type-I and type-III 2HDMs
taking $\tan\beta=0.5$, $1$, $10$, and $30$ from top to bottom.
We consider 5 decay modes of
$H^+\to t\bar b$ (red solid), $H^+\to c\bar s$ (magenta solid),
$H^+\to \tau\nu$ (blue solid), $H^+\to \mu\nu$ (green solid),
and $H^+\to H W^+$ (black dashed) with $M_H=125.5$ GeV.
In the type-I and type-III models, the normalized
charged Higgs couplings to quarks are the same because
$g_{t,c}=|g_{b,s}|=1/t_\beta$ while those to leptons
are given by $g_{\mu,\tau}=-1/t_\beta$ (type I) and $g_{\mu,\tau}=t_\beta$ (type III).
We observe that, especially for large values of $t_\beta$,
the heavy charged Higgs boson dominantly decays into $H W^+$ (black dashed)
since $\Gamma(H^+ \to t\bar b)$ is suppressed as $t_\beta$ grows.
And, when the $H^+\to H W^+$ decay mode dominates,
the leptonic branching ratios are enhanced by $t_\beta^2$ in the type-III 2HDM.
In the type-III 2HDM, we note that
${B}(H^+ \to \tau\nu)$ becomes comparable to ${B}(H^+ \to t\bar b)$
around $t_\beta \sim 10$ and it becomes larger as $t_\beta$ grows.

\medskip

In Fig.~\ref{fig:chbr24}, we show the
branching ratios of a charged Higgs boson for its mass
between 100 GeV and 1 TeV in the type-II and type-IV 2HDMs
taking $\tan\beta=0.5$, $1$, $10$, and $30$ from top to bottom.
We consider the same 5 decay modes as in Fig.~\ref{fig:chbr13}.
In the type-II and type-IV models, the charged Higgs couplings
to the (right-handed) up-type quarks are again given by $g_{t,c}=1/t_\beta$
but those to the down-type fermions are
by $g_{b,s}=g_{\mu,\tau}=t_\beta$ (type II) and
$g_{b,s}=1/|g_{\mu,\tau}|=t_\beta$ (type IV).
We observe that, especially for large values of $t_\beta$,
the heavy charged Higgs boson dominantly decays into $H W^+$ (black dashed)
eventually. But, compared to the type-I and type-III 2HDMs
where the top-quark contributions to the $H^+ \to t\bar b$ decay
always dominate and the decay width decreases as $t_\beta$ grows,
the dominance of the $H^+\to HW^+$ decay mode develops
rather slowly because the bottom-quark contributions take over
the dominance around $t_\beta=8$ and the partial width
$\Gamma(H^+ \to t\bar b)$ increases as $t_\beta$ grows in the type-II and
type-IV 2HDMs.
The leptonic decay widths are also enhanced by $t_\beta^2$ in the type-II 2HDM
and we find that, specifically for $t_\beta=30$, the three decay
modes of $H^+ \to t\bar b$, $H^+\to H W^+$, and $H^+ \to \tau\nu$
are competing in the region of $350\lsim M_{H^\pm}/{\rm GeV}\lsim 600$.

\medskip

For a light charged Higgs boson, it mostly decays into $\tau\nu$
and/or $c\bar s$ before the $H^+ \to t^*\bar b$ decay channel opens
and starts to dominate.
Exceptions occur in the type-I and type-III 2HDMs when $t_\beta$ is large,
see the lowest panels of Fig.~\ref{fig:chbr13}.
Specifically, for $t_\beta=30$ in the type-I model,
the charged Higgs boson dominantly decays into
$H W^{+*}$ in the narrow region of $M_{H^\pm}$ between 150 GeV and 180 GeV.

\medskip

We confirm that the behavior of the branching ratio of
each decay mode
versus $M_{H^\pm}$ does not depend on the 2HDM type
when $t_\beta=1$ as it should be: see the $\tan\beta=1$ panels in
Figs.~\ref{fig:chbr13} and \ref{fig:chbr24}.

\medskip
\begin{figure}[t!]
\centering
\includegraphics[height=3.1in,angle=0]{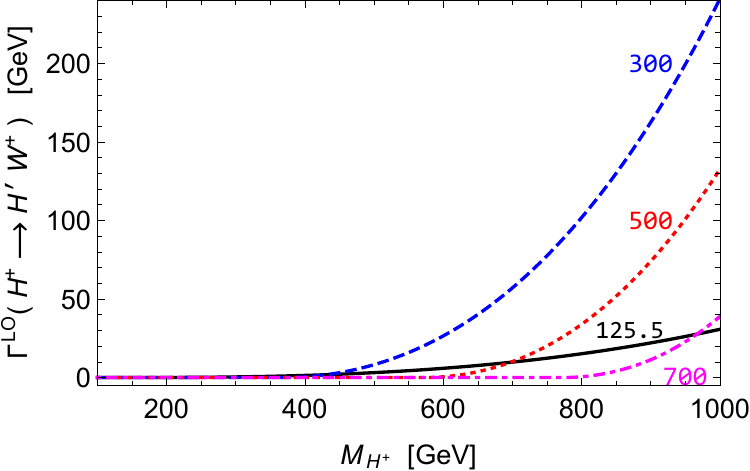}
\caption{\label{fig:CH2HeavyHW}
Decay widths of a charged Higgs boson with a mass $M_{H^\pm}$ into
$H^\prime W^+$ for three values of
$M_{H^\prime}=300$ GeV (blue dashed),
$500$ GeV (red dotted), and $700$ GeV (magenta dash-dotted)
taking $|g_{_{H^\prime H^+ W^-}}|^2 = 1$.
For comparisons, $\Gamma^{\rm LO}(H^+\to HW^+)$
with $M_H=125.5$ GeV and $|g_{_{HH^+ W^-}}|^2 = 0.1$
is also shown in black solid line,
see Fig.~\ref{fig:ch2hw}.
}
\end{figure}
Before closing this section, we address the case in which
there exists another neutral Higgs boson $H^\prime$
with its mass smaller than $M_{H^\pm}$ and the decay
$H^+\to H^\prime W^{+}$ is kinematically allowed.
Here $H^\prime$ stands for a mixture of
CP-even and CP-odd states in general which, in the CP-conserving case,
could be either of them.
In the 2HDM framework, identifying the lightest neutral Higgs boson
as the SM-like $H$ with $M_H=125.5$ GeV and
$|g_{_{H H^+ W^-}}|^2=1-g_{_{HVV}}^2 = 0.1$ as already considered,
the charged Higgs coupling to the heavier neutral Higgs boson $H^\prime$ and
a charged vector boson $W$ takes almost the maximum value of $1$
due to the sum rules
given in Eq.~(\ref{eq:2hdmsumrule}).
Furthermore, the decay width grows by the cubic powers of the
charged Higgs-boson mass.
These combined properties result in a large width for the decay mode
$H^\pm\to H^\prime W^\pm$ when ${H^\pm}$ is heavy enough.
When $M_{H^\pm}\gsim 950$ GeV, we observe that the decay width
$\Gamma^{\rm LO}(H^+\to H^\prime W^+)$
is larger than 200 GeV for a $H^\prime$ boson with $M_{H^\prime}=300$ GeV,
see the blue dashed line in Fig.~\ref{fig:CH2HeavyHW}.
Even for $H^\prime$ as heavy as 700 GeV, we note that
the decay width could be comparable to
$\Gamma^{\rm LO}(H^+\to HW^+)$  when $M_{H^\pm}\sim 1 $ TeV
as denoted by the magenta dash-dotted line in the same figure.
%

\section{Higgcision of the Higgs Boson Discovered at the LHC}
\label{sec:Higgcision}

Ever since the discovery of a SM-like Higgs boson in the year 2012
\cite{Aad:2012tfa,Chatrchyan:2012ufa},
the era of Higgs-boson precision studies,
which was termed as Higgcision
in Refs.$\,$\cite{Cheung:2013kla,Cheung:2014noa,Cheung:2018ave},
has begun.
In this section we present highlights
from global fits of the Higgs
boson couplings to all the 7 TeV, 8 TeV, and 13 TeV data available
up to the Summer 2018,  based on the
works~\cite{Cheung:2013kla,Cheung:2014noa,Cheung:2018ave}.
For several different types of approaches with their own
merits on the global fits of the Higgs couplings, see, for example,
Refs.~\cite{deFlorian:2016spz,Duhrssen:2003tba,Duhrssen:2004cv,Lafaye:2009vr,
Espinosa:2012ir,Azatov:2012bz,Carmi:2012yp,LHCHiggsCrossSectionWorkingGroup:2012nn,
Gonzalez-Alonso:2014eva,Greljo:2015sla,Giudice:2007fh,Willenbrock:2014bja,
Brivio:2017vri,Dawson:2018dcd,Grzadkowski:2010es}.
For some other early works based on model-independent approach,
we refer to
Refs.~\cite{Klute:2012pu,Carmi:2012zd,Low:2012rj,Ellis:2012hz,Espinosa:2012im,
Carmi:2012in,Banerjee:2012xc,Bonnet:2012nm,Plehn:2012iz,Djouadi:2012rh,
Dobrescu:2012td,Cacciapaglia:2012wb,Belanger:2012gc,Moreau:2012da,
Corbett:2012dm,Corbett:2012ja,Masso:2012eq}.

\medskip

Assuming generation independence for the normalized Yukawa couplings of
$g^{S,P}_{H\bar f f}$,
we use the following $C$ notations for the couplings in the global
fits:
\begin{eqnarray}
\label{eq:Cnotations}
&&
C_u^S=g^S_{H\bar uu}\,, \ \
C_d^S=g^S_{H\bar dd}\,, \ \
C_\ell^S=g^S_{H\bar ll}\,; \ \
C_w=g_{_{HWW}}\,, \ \
C_z=g_{_{HZZ}}\,; \nonumber \\
&&
C_u^P=g^P_{H\bar uu}\,, \ \
C_d^P=g^P_{H\bar dd}\,, \ \
C_\ell^P=g^P_{H\bar ll}\,.
\end{eqnarray}
We further keep the custodial symmetry between the $W$ and $Z$ bosons
which leads to the relation $C_v \equiv C_w = C_z$.
%
%
We emphasize that one can carry out
the precision study of the LHC Higgs data using the
model-independent approach taken in Section
\ref{sec:decays_of_a_generic_neutral_higgs_boson} for
decays of a generic neutral Higgs boson
by matching the relevant couplings and fitting parameters
as in Eq.~(\ref{eq:Cnotations}).

\subsection{Higgs signal strengths}
\label{subsec:higgs_signal_strengths_higgsicion}

Each theoretical signal strength can be written in a product form as
\begin{equation}
\widehat\mu({\cal P},{\cal D}) \simeq
\widehat\mu({\cal P})\ \widehat\mu({\cal D})\,,
\end{equation}
where ${\cal P}={\rm ggF}, {\rm VBF}, {\rm VH}, {\rm ttH}$ denote the
production mechanisms and ${\cal D}=\gamma\gamma , ZZ^{(*)}, WW^{(*)}, b\bar{b},
\tau^+\tau^-$ the decay channels, which are experimentally clean and/or dominant
for $M_H\simeq 125\,{\rm GeV}$.
The factorization assumption is valid only when
the production and decay processes are well separated like as in the
resonant $s$-channel Higgs production in the NWA.
By factorizing them, non-resonant and interference effects are
inevitably neglected.
More explicitly, at LO, the production signal strengths are
given in terms of the relevant form factors and couplings by
\begin{eqnarray}
\label{eq:production_signastrength}
\widehat\mu({\rm ggF}) &=&
\frac{\left|S^g(M_H)\right|^2+\left|P^g(M_H)\right|^2}
{\left|S^g_{\rm SM}(M_H)\right|^2}\,, \nonumber \\[2mm]
\widehat\mu({\rm VBF}) &=& \widehat\mu({\rm VH}) = g_{_{HWW,HZZ}}^2\,, \nonumber
\\[2mm]
\widehat\mu({\rm ttH}) &=& \left(g^S_{H\bar{t}t}\right)^2
+\left(g^P_{H\bar{t}t}\right)^2\,,
\end{eqnarray}
and the decay signal strengths by
\begin{equation}
\widehat\mu({\cal D}) = \frac{B(H\to {\cal D})}{B(H_{\rm SM}\to {\cal D})}\,,
\end{equation}
with the branching fraction of each decay mode defined by
\begin{equation}
\label{eq:dgam}
B(H\to {\cal D})=\frac{\Gamma(H\to{\cal D})}
{\Gamma_{\rm tot}(H)+\Delta\Gamma_{\rm tot}}\,.
\end{equation}
Note that an arbitrary non-SM contribution $\Delta\Gamma_{\rm tot}$
to the total decay width is introduced.
We observe
$\Gamma_{\rm tot}(H)$ becomes the SM total decay width
when $g^S_{H\bar{f}f}=1$,
$g^P_{H\bar{f}f}=0$,
$g_{_{HWW,HZZ}}=1$, and
$\Delta S^{\gamma,g,Z\gamma}= \Delta P^{\gamma,g,Z\gamma}=0$.
Note that the LO relations in Eq.~(\ref{eq:production_signastrength})
are most reliable when higher order corrections
to a BSM production cross section and those to the corresponding
SM one are the same and so they are canceled out in the BSM-to-SM ratios.
Otherwise they are valid at LO strictly.
%

\medskip

\begin{table}[t!]
\caption{\label{tab:tev}
{\bf (Tevatron: 1.96 TeV)}
The signal strengths data from Tevatron (10.0 fb$^{-1}$ at 1.96 TeV).}
\begin{center}
\begin{footnotesize}
\begin{tabular}{cccccccr}
\hline\hline
Channel & Signal strength $\mu$ & $M_H$(GeV) & \multicolumn{4}{c}{Production mode}  &
$\chi^2_{\rm SM}$(each)\\
        & c.v $\pm$ error       &            & ggF & VBF & VH & ttH & \\
\hline
\multicolumn{8}{c}{Tevatron (Nov. 2012)} \\
\hline
Combined $H\to \gamma\gamma$\cite{tevatron_aa_ww} & $6.14^{+3.25}_{-3.19}$   & 125 & 78\%
& 5\% & 17\% & - & 2.60 \\
Combined $H\to WW^{(\ast)}$\cite{tevatron_aa_ww}  & $0.85^{+0.88}_{-0.81}$        & 125 &
78\% & 5\% & 17\% & - & 0.03 \\
VH tag $H\to bb$\cite{tevatron_bb}        & $1.59^{+0.69}_{-0.72}$        & 125 & -    &
-   & 100\% & - & 0.67 \\
\hline
&&&&&&& $\chi^2_{\rm SM}$(subtot): 3.30 \\
\hline\hline
\end{tabular}
\end{footnotesize}
\end{center}
\end{table}
\begin{table}[t!]
\caption{\label{tab:78all}
{\bf (LHC: 7$\oplus$8 TeV)}
Combined ATLAS and CMS data on signal strengths from Table 8 of
Ref.~\cite{Khachatryan:2016vau}.}
\begin{center}
\begin{tabular}{c|ccccc}
\hline\hline
 & \multicolumn{5}{c}{Decay mode} \\
\hline
Production mode & $H\to \gamma\gamma$ & $H\to Z Z^{(\ast)} $ & $H\to W W^{(\ast)} $
& $H\to bb$ & $H\to \tau^+\tau^-$  \\
\hline
ggF & $1.10^{+0.23}_{-0.22}$  & $1.13^{+0.34}_{-0.31}$ & $0.84^{+0.17}_{-0.17}$
    & -                       & $1.0^{+0.6}_{-0.6}$       \\
VBF & $1.3^{+0.5}_{-0.5}$     & $0.1^{+1.1}_{-0.6}$    & $1.2^{+0.4}_{-0.4}$
    & -                       & $1.3^{+0.4}_{-0.4}$       \\
WH  & $0.5^{+1.3}_{-1.2}$     & -                      & $1.6^{+1.2}_{-1.0}$
    & $1.0^{+0.5}_{-0.5}$     & $-1.4^{+1.4}_{-1.4}$      \\
ZH  & $0.5^{+3.0}_{-2.5}$     & -                      & $5.9^{+2.6}_{-2.2}$
    & $0.4^{+0.4}_{-0.4}$     & $2.2^{+2.2}_{-1.8}$       \\
ttH  & $2.2^{+1.6}_{-1.3}$    & -                      & $5.0^{+1.8}_{-1.7}$
     & $1.1^{+1.0}_{-1.0}$    & $-1.9^{+3.7}_{-3.3}$      \\
\hline
&&&&& $\chi^2_{\rm SM}$(subtot): 19.93\\
\hline\hline
\end{tabular}
\end{center}
\end{table}
\begin{table}[h!]
\caption{\label{tab:all13}
{\bf (LHC: 13 TeV)}
Combined ATLAS and CMS (13 TeV) data on signal strengths.
The $\mu^{\rm dec}_{\rm combined}$ ($\mu^{\rm prod}_{\rm combined}$)
represents the combined signal strength
for a specific decay (production) channel by summing all the
production (decay) modes,
and $\chi^2_{\rm min}$ are the corresponding minimal
$\chi^2$ values.  In the VH/WH row,
the production mode for $H\to\gamma\gamma$ and $H\to ZZ^{(*)}$
is VH while it is
WH for $H\to WW^{(*)}$ and $H\to \tau^+\tau^-$;
for the remaining
decay mode $H\to b \bar b$, we combine the two signal strengths from
WH and VH, see Table XII in Ref.~\cite{Cheung:2018ave}.}
\begin{center}
\begin{footnotesize}
\begin{tabular}{c|ccccc|cc}
\hline\hline
 & \multicolumn{5}{c}{Decay mode} \\
\hline
Production mode & $H\to \gamma\gamma$ & $H\to Z Z^{(\ast)} $ & $H\to W W^{(\ast)} $
& $H\to bb$ & $H\to \tau^+\tau^-$
& $\mu^{\rm prod}_{\rm combined}$ & $\chi^2_{\rm SM}(\chi^2_{\rm min})$  \\
\hline
ggF & $1.02^{+0.12}_{-0.11}$  & $1.09^{+0.11}_{-0.11}$ &  $1.29^{+0.16}_{-0.16}$
    & $2.51^{+2.43}_{-2.01}$  & $1.06^{+0.40}_{-0.37}$ &  $1.11^{+0.07}_{-0.07}$
    & 5.42(3.15)    \\
VBF & $1.23^{+0.32}_{-0.31}$  & $1.51^{+0.59}_{-0.59}$ &  $0.54^{+0.32}_{-0.31}$
    & -                       & $1.15^{+0.36}_{-0.34}$ &  $1.02^{+0.18}_{-0.18}$
    & 7.53(7.51)    \\
VH/WH & $1.42^{+0.51}_{-0.51}$  & $0.71^{+0.65}_{-0.65}$ & $3.27^{+1.88}_{-1.70}$
        & $1.07^{+0.23}_{-0.22}$  & $3.39^{+1.68}_{-1.54}$ & $1.15^{+0.20}_{-0.19}$
        & 7.05(6.44) \\
ZH  & -  & - &  $1.00^{+1.57}_{-1.00}$ & $1.20^{+0.33}_{-0.31}$
    & $1.23^{+1.62}_{-1.35}$   & $1.19^{+0.32}_{-0.30}$
    & 0.45(0.02)    \\
ttH & $1.36^{+0.38}_{-0.37}$  & $0.00^{+0.53}_{-0.00}$ &  - & $0.91^{+0.45}_{-0.43}$
    & -                       & $0.93^{+0.24}_{-0.24}$
    & 5.96(5.86)    \\
ttH (excl.) & $1.39^{+0.48}_{-0.42}$ &  - &$1.59^{+0.44}_{-0.43}$ & $0.77^{+0.36}_{-0.35}$
            & $0.87^{+0.73}_{-0.73}$ &  $1.16^{+0.22}_{-0.22}$
            & 4.17(3.62)   \\
\hline
$\mu^{\rm dec}_{\rm combined}$ & $1.10^{+0.10}_{-0.10}$  & $1.05^{+0.11}_{-0.11}$
                     & $1.20^{+0.14}_{-0.13}$  & $1.05^{+0.19}_{-0.19}$
                     & $1.15^{+0.24}_{-0.23}$  & $1.10^{+0.06}_{-0.06}$
                      \\
$\chi^2_{\rm SM}$($\chi^2_{\rm min}$) & 6.83(5.72)  & 9.13(8.88)
             & 9.48(7.32) & 1.56(1.51) & 3.58(3.20)  & & 30.58(27.56) \\
\hline\hline
\end{tabular}
\end{footnotesize}
\end{center}
\end{table}
On the experimental side,
we use the direct Higgs signal strength data collected at the Tevatron and the LHC.
Specifically,
we use 3 signal strengths measured at the Tevatron, see Table~\ref{tab:tev}.
At the LHC with the center-of-mass energies of 7 and 8 (7$\oplus$8) TeV,
the signal strengths obtained from a combined ATLAS and CMS
analysis~\cite{Khachatryan:2016vau} are used, see Table~\ref{tab:78all}.
%
On the other hand, the 13 TeV data are
still given separately by ATLAS and CMS and
in different production and decay channels.
\footnote{For the details of the 13 TeV data sets
used, see Appendix B of Ref.~\cite{Cheung:2018ave} and references therein.}
Under this situation,
to derive the combined signal strengths of various channels,
we use a simple $\chi^2$ method assuming that
each distribution is Gaussian. The results
are shown in Table~\ref{tab:all13}.
%
%

\subsection{Constraints on the couplings of Higgs boson weighing $125.5$ GeV}
\label{sec:HiggcisionSub}

In this subsection, we present a few representative results
obtained from performing LO analysis of the direct Higgs data
collected at the Tevatron and the LHC by considering
CP-conserving (CPC) scenarios only.
Note that, in the most general {\bf CPC} scenario, one may vary
all the 7 parameters of $C_u^S$, $C_d^S$, $C_\ell^S$,$C_{v}$, $\Delta S^g$,
$\Delta S^\gamma$, and $\Delta \Gamma_{\rm tot}$ while taking
vanishing pseudoscalar couplings and form factors as
$C_u^P=C_d^P=C_\ell^P=\Delta P^{\gamma}=\Delta P^g=0$.

\medskip

\begin{table}[t!]
\caption{\label{tab:CPC}
{\bf (CPC)}
The best-fitted values in various CP conserving fits and the
corresponding $\chi^2$
per degree of freedom ({\it dof}) and goodness
of fit. The $p$-value for each fit hypothesis against the SM null
hypothesis is also shown.
For the SM, we obtain $\chi^2=53.81$, $\chi^2/dof=53.81/64$, and
so the goodness of fit $=0.814$. From Ref.~\cite{Cheung:2018ave}. }
\vspace{-0.5cm}
\begin{flushleft}
\begin{footnotesize}
\begin{tabular}{c|ccc|cccc}
\hline\hline
Cases & {\bf CPC1} & {\bf CPC2}   & {\bf CPC4} & \multicolumn{4}{c}{\bf CPCN4} \\[1mm]
\hline  \\[-3mm]
Varying      & $\Delta\Gamma_{\rm tot}$ & $\Delta S^\gamma$
      & $C^S_u,~C^S_d,$
      & \multicolumn{4}{c}{$C^S_u,C_v$}   \\[1mm]
Parameters &   & $\Delta S^g$
           & $C^S_\ell,~C_v$
           & \multicolumn{4}{c}{$\Delta S^\gamma$, $\Delta S^g$}   \\[1mm]
\hline \\[-3mm]
$C^S_u$           & 1 & 1 &  $1.001^{+0.056}_{-0.055}$
                  & $1.042^{+0.077}_{-0.081}$ & $1.042^{+0.078}_{-0.081}$
                  & $-1.042^{+0.081}_{-0.078}$ & $-1.042^{+0.081}_{-0.078}$ \\[1mm]
$C^S_d$           & 1 & 1 &  $0.962^{+0.101}_{-0.101}$ & $1$ & $1$ & $1$ & $1$ \\[1mm]
$C^S_\ell$        & 1 & 1 &  $1.024^{+0.093}_{-0.093}$ & $1$ & $1$ & $1$ & $1$
\\[1mm]
$C_v$             & 1 & 1 &  $1.019^{+0.044}_{-0.045}$
                  & $1.027^{+0.034}_{-0.036}$ & $1.027^{+0.034}_{-0.036}$
                  & $1.028^{+0.034}_{-0.036}$ & $1.028^{+0.034}_{-0.036}$ \\[1mm]
$\Delta S^\gamma$ & 0                      & $-0.226^{+0.32}_{-0.32}$ & 0
                  & $-0.129^{+0.37}_{-0.37}$   & $-0.129^{+0.37}_{-0.37}$
                  & $3.524^{+0.41}_{-0.42}$  & $3.523^{+0.41}_{-0.42}$  \\[1mm]
$\Delta S^g$      & 0                      & $0.016^{+0.025}_{-0.025}$ & 0
                  & $-0.021^{+0.057}_{-0.055}$  & $-1.34^{+0.066}_{-0.065}$
                  & $0.095^{+0.055}_{-0.057}$  & $1.414^{+0.066}_{-0.066}$ \\[1mm]
$\Delta \Gamma_{\rm tot}$ (MeV) & $-0.285^{+0.18}_{-0.17}$  & 0 & 0 & 0 & 0 & 0 &
0\\[1mm]
\hline \\[-3mm]
$\chi^2/dof$ & 51.44/63 & 51.87/62 & 50.79/60 & \multicolumn{4}{c}{50.96/60}   \\[1mm]
goodness of fit    & 0.851    & 0.817     & 0.796    & \multicolumn{4}{c}{0.791}
\\[1mm]
$p$-value    & 0.124 & 0.379    & 0.554    & \multicolumn{4}{c}{0.583}   \\[1mm]
\hline\hline
\end{tabular}
\end{footnotesize}
\end{flushleft}
\end{table}

Our goal is to provide constraints on
the couplings of the neutral Higgs boson, which was discovered at the LHC,
without much loss of generality
when it is interpreted in various frameworks beyond the SM.
Accordingly, we consider the four CPC fits listed in Table \ref{tab:CPC}
in which the second row explicitly shows the varying parameters of each fit.

\medskip

\begin{figure}[t!]
\centering
\includegraphics[height=2.6in,angle=0]{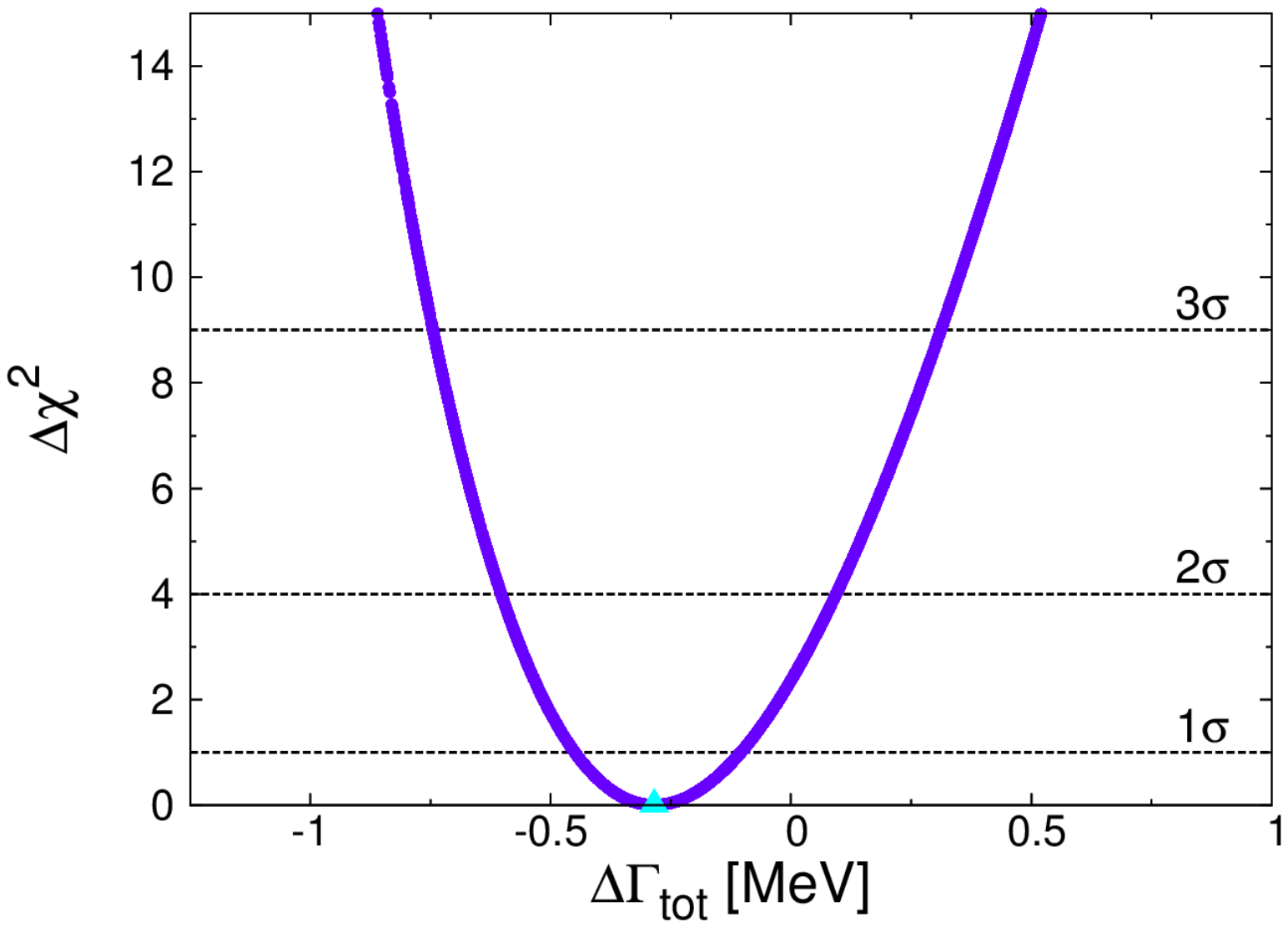}
\caption{\label{fig:CPC1}
{\bf CPC1}:
$\Delta \chi^2$ from the minimum versus $\Delta \Gamma_{\rm tot}$
with only $\Delta \Gamma_{\rm tot}$ varying in the fit.
From Ref.~\cite{Cheung:2018ave}.
}
\end{figure}
\begin{figure}[t!]
\centering
\includegraphics[height=1.6in,angle=0]{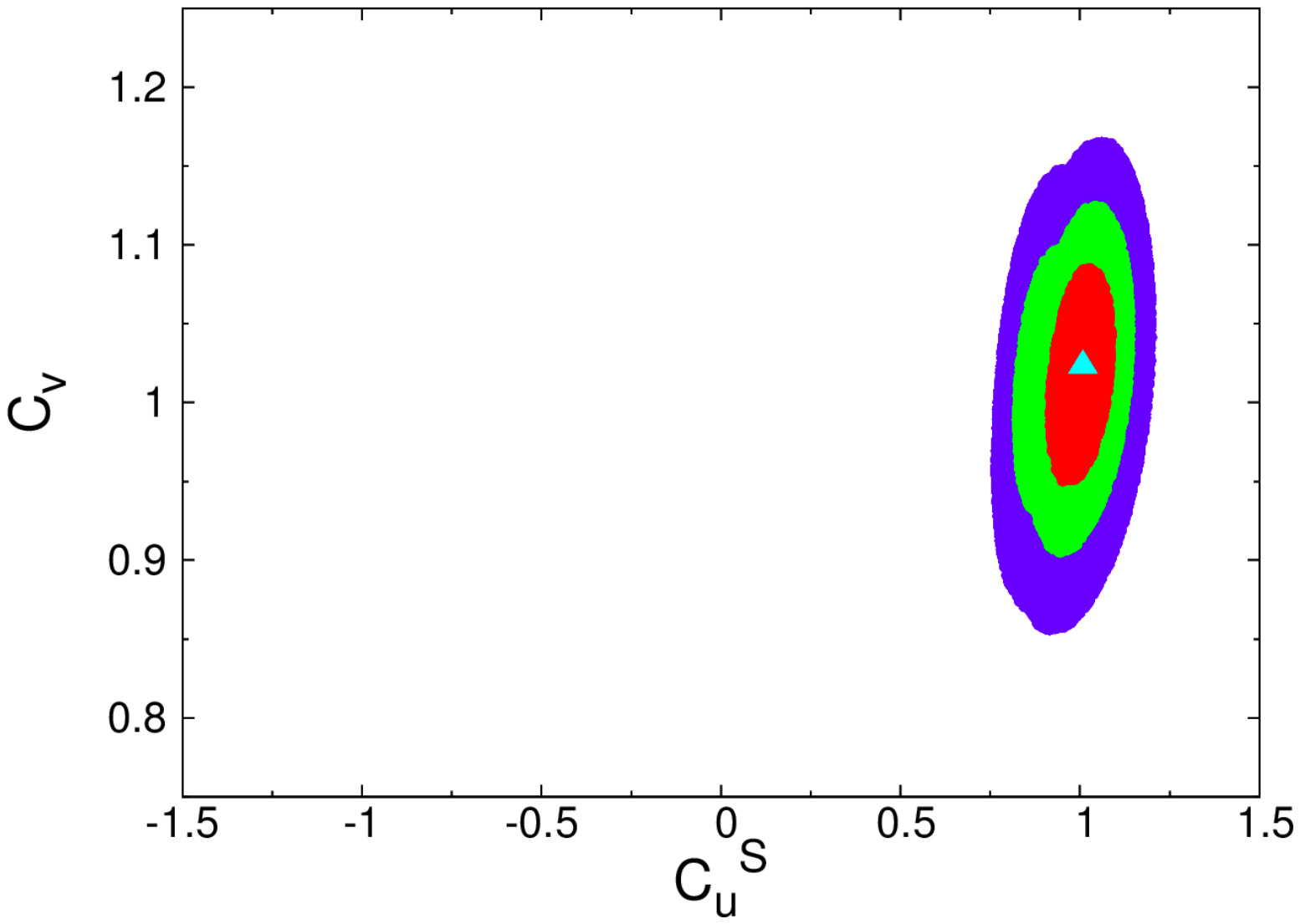}
\includegraphics[height=1.6in,angle=0]{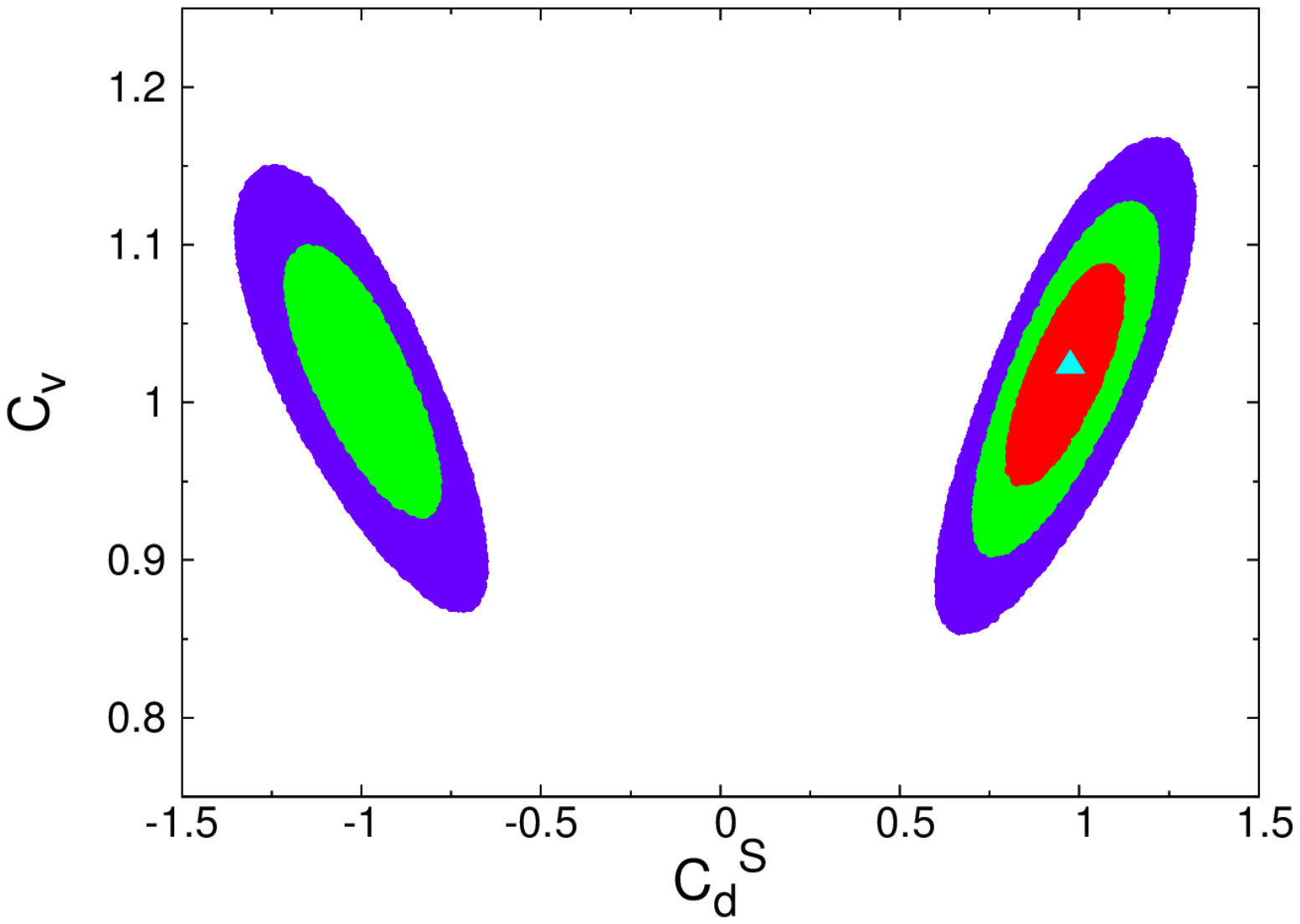}
\includegraphics[height=1.6in,angle=0]{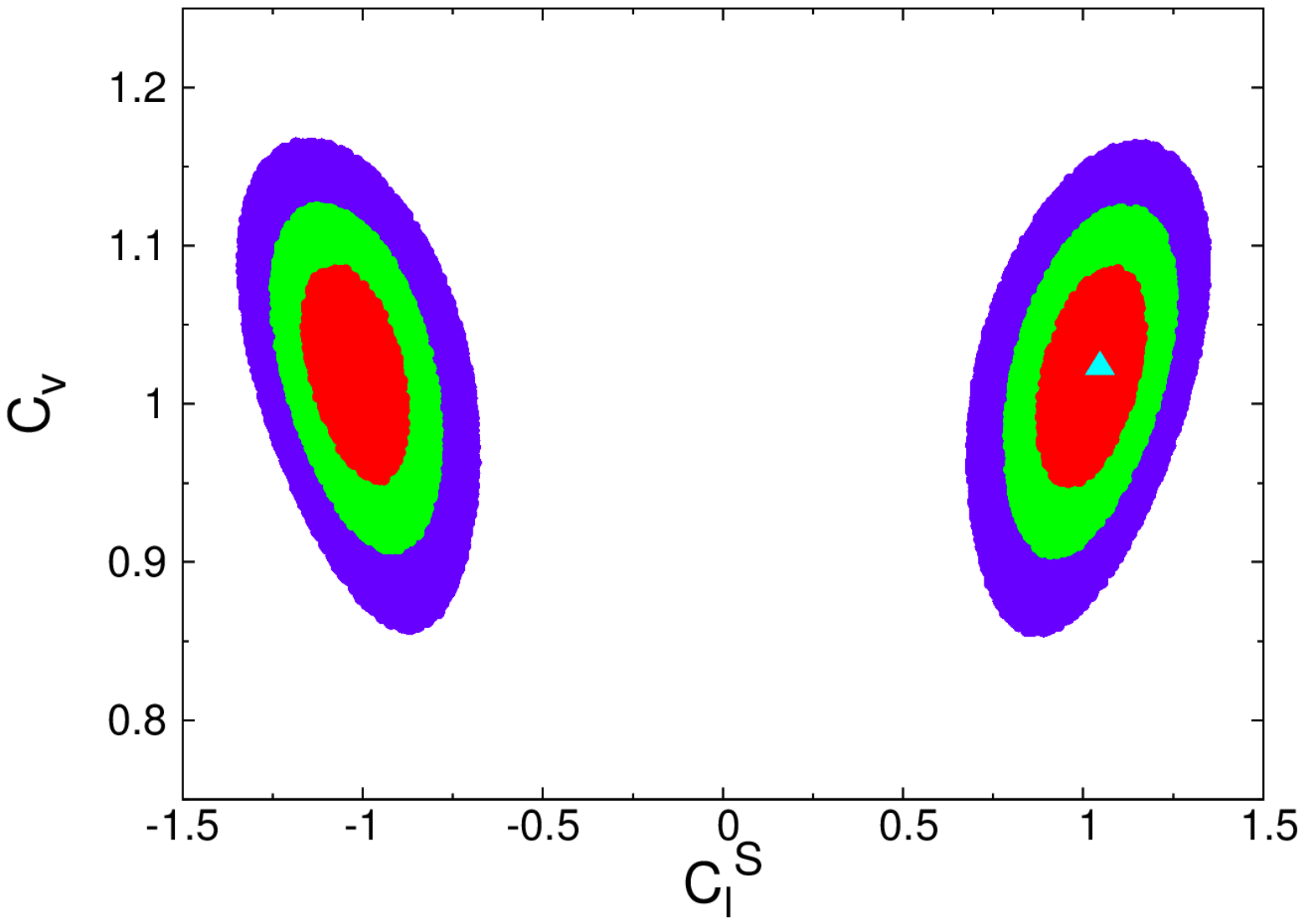}
\includegraphics[height=1.6in,angle=0]{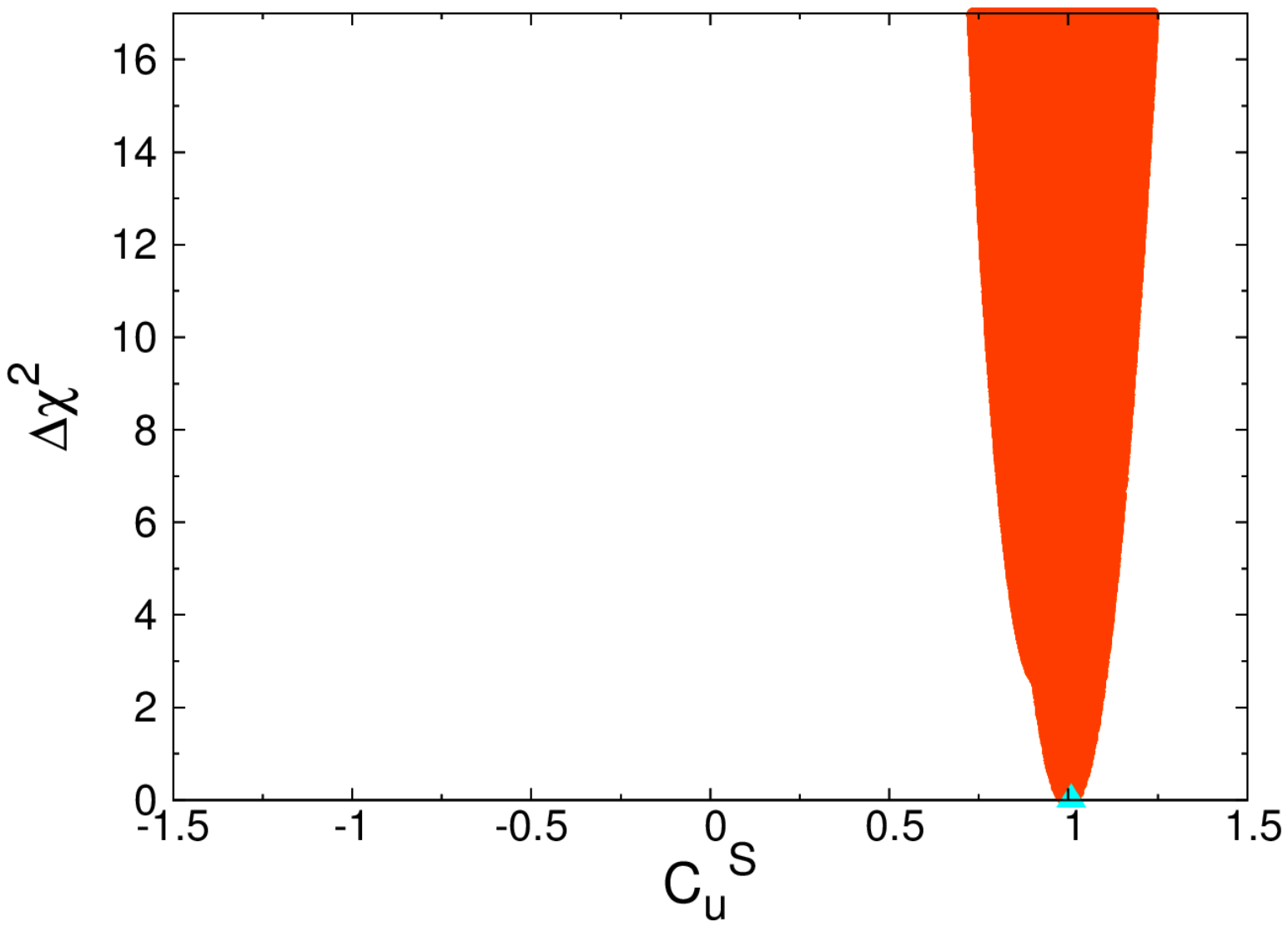}
\includegraphics[height=1.6in,angle=0]{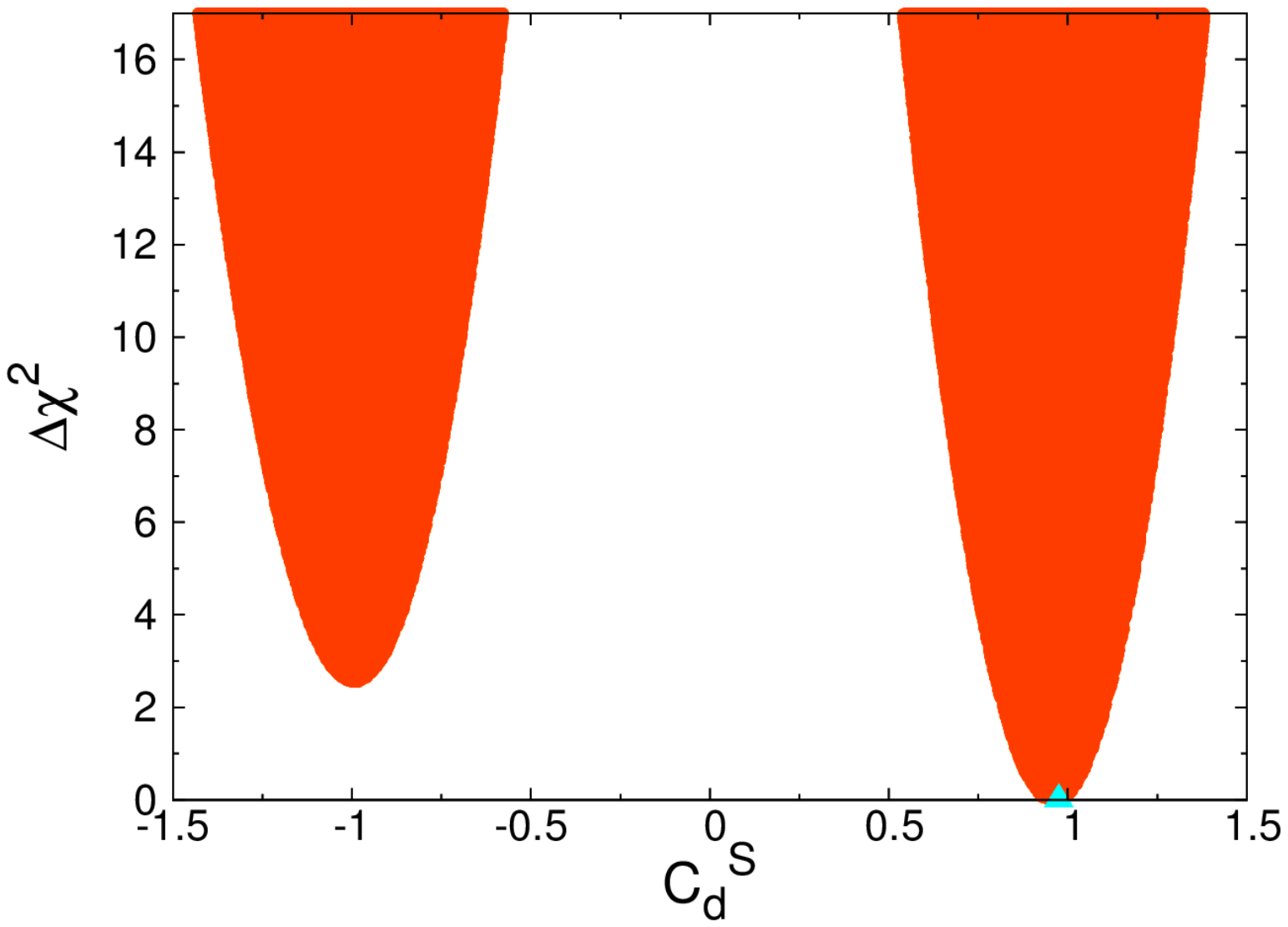}
\includegraphics[height=1.6in,angle=0]{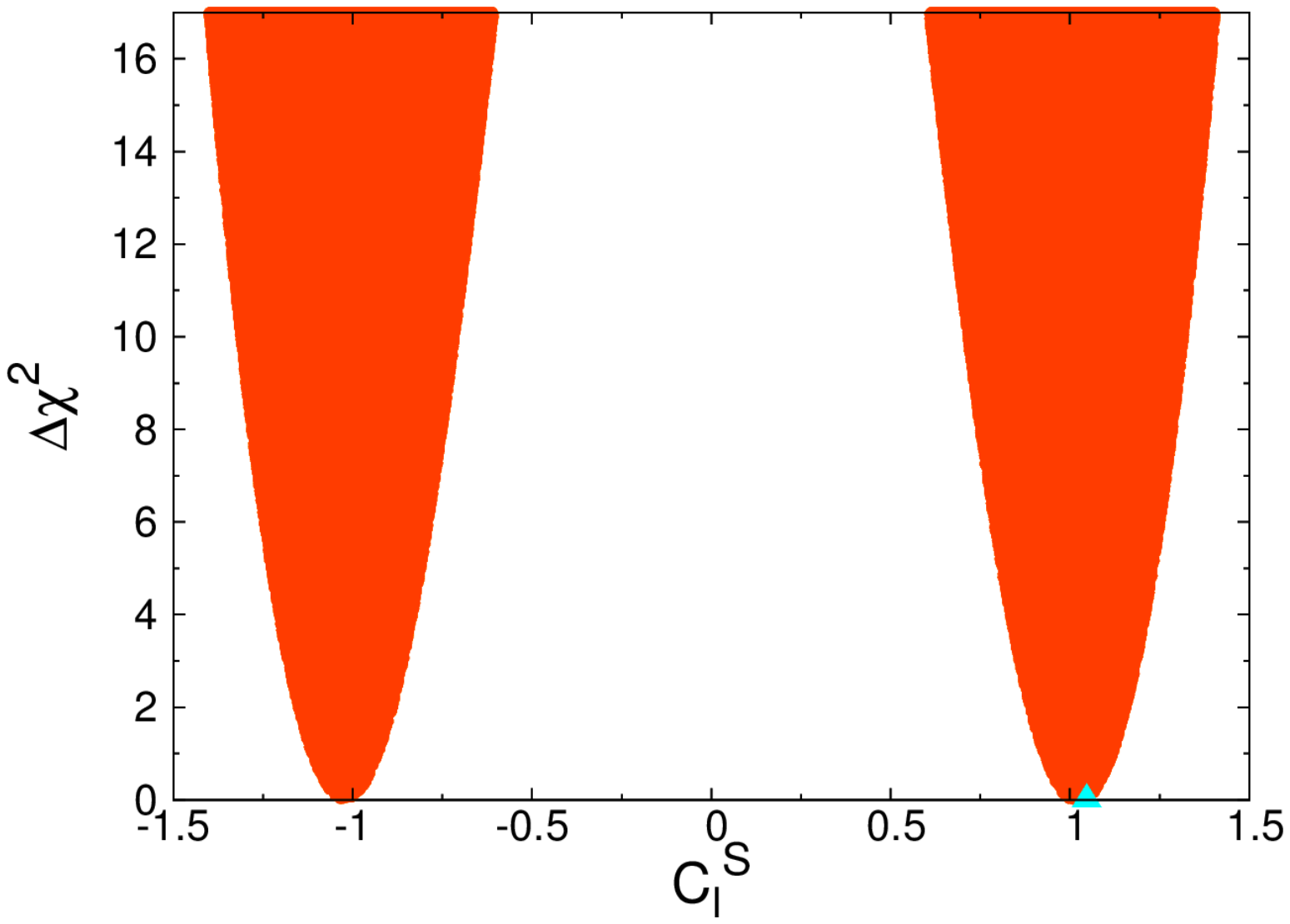}
\caption{\label{fig:CPC4}
{\bf CPC4}: (Upper)
The confidence-level (CL) regions of the fit by varying $C_v$, $C_u^S$,
$C_d^S$, and $C_\ell^S$. The contour regions shown are for
$\Delta \chi^2 \le 2.3$ (red), $5.99$
(red+green), and $11.83$ (red+green+blue) above the minimum, which correspond to
confidence levels of 68.3\%, 95\%, and 99.7\%, respectively.
The best-fit points are denoted by triangles.
(Lower)
$\Delta \chi^2$ from the minimum versus Yukawa couplings.
From Ref.~\cite{Cheung:2018ave}.
}
\end{figure}
Referring to Ref.~\cite{Cheung:2018ave} for more detailed explanations,
we offer highlights of each fit as follows:
\begin{enumerate}
\item[$\bullet$]{\bf CPC1}: The best-fit value for
the residual total decay width $\Delta \Gamma_{\rm tot}$ is
$\Delta \Gamma_{\rm tot} = -0.285\,^{+0.18}_{-0.17} \; {\rm MeV}$
which is $1.6\,\sigma$ below zero.
At 95\% confidence level (CL), on the other hand,
\begin{equation}
\Delta \Gamma_{\rm tot}=-0.285^{+0.38}_{-0.32}\ \ {\rm MeV}\,,
\end{equation}
as shown in Fig.~\ref{fig:CPC1}.
Using the upper error as the upper limit,
we obtain the constraint $\Delta \Gamma_{\rm tot} \leq 0.38$ MeV
simply taking the central value equal to zero which
can be translated to the following 95\% CL constraint on the
branching ratio of the Higgs boson decays into non-SM particles:
$B(H \to {\rm nonstandard}) \leq 8.4\% $.
\item[$\bullet$]{\bf CPC2}:
The best-fit point $(\Delta S^\gamma, \Delta S^g) = (-0.226,0.016)$
of the form factors indicates an increase of 3.4\% and 2.4\% in
the absolute values  of the scalar $H\gamma\gamma$ and
$Hgg$ form factors of $|S^\gamma|$ and $|S^g|$, respectively.
We note that the error of $\Delta S^g$ is $\pm 0.025$,
which is numerically smaller than the SM bottom-quark contribution of
$-0.043$ to the real part of $S^g$, see Eq.~(\ref{eq:hgg}),
alerting that we have already reached the sensitivity to probe
the sign of the bottom-quark Yukawa coupling in gluon fusion.
\item[$\bullet$]{\bf CPC4}:
We observe that the possibility for the top-quark Yukawa coupling to be
negative has been entirely ruled out as shown
clearly in the left upper and lower panels of Fig.~\ref{fig:CPC4}.
And, as already anticipated in the {\bf CPC2} fit, the bottom-quark Yukawa coupling
$C_d^S$ prefers the positive sign to the negative one, see the middle panels
of Fig.~\ref{fig:CPC4}. It is more clear from the
middle lower panel that the point $C_d^S = -1$ has $\Delta \chi^2 > 2$ above
the minimum at $C_d^S = +1$.
The current data precision is yet insufficient for
showing any preference for the sign of tau-Yukawa coupling, as shown
in the right panels of Fig.~\ref{fig:CPC4}.
\footnote{For this reason, we have considered the minimum
around $C^S_\ell=+1$ only in Table \ref{tab:CPC}.}
\item[$\bullet$]{\bf CPCN4}:
In this fit, there are 4 degenerate minima with $\Delta S^g \sim 0\,,\mp 1.4$
for $C_u^S \sim \pm 1$ which could be understood from the relation
$|S^g| \sim |0.7\,  C^u_S + \Delta S^g| \sim 0.7$, see Table \ref{tab:CPC}.
Note that $\Delta S^\gamma$ can compensate the sign change in
$C_u^S$ allowing it to be about $-1$ with $\Delta S^\gamma \sim +3.5$.
This could be understood by noting  the relation
$|S^\gamma| \sim |-8.3\, C_v + 1.8 \ C^u_S + \Delta S^\gamma| \sim 6.5$.
We further observe that the negative top-quark Yukawa coupling is allowed
only when there exist additional particles running in the
$H$-$\gamma$-$\gamma$ loop with the size of contributions
equal to two times the SM top-quark contribution
within about 10 \%.
This tuning on the couplings and form factors will become more
and more severe as more data are accumulated at the LHC.
\end{enumerate}

\subsection{Implications beyond the Standard Model}
\label{subsec:implications_beyond_the_standard_model}
\begin{figure}[t!]
\begin{center}
\includegraphics[width=8.3cm]{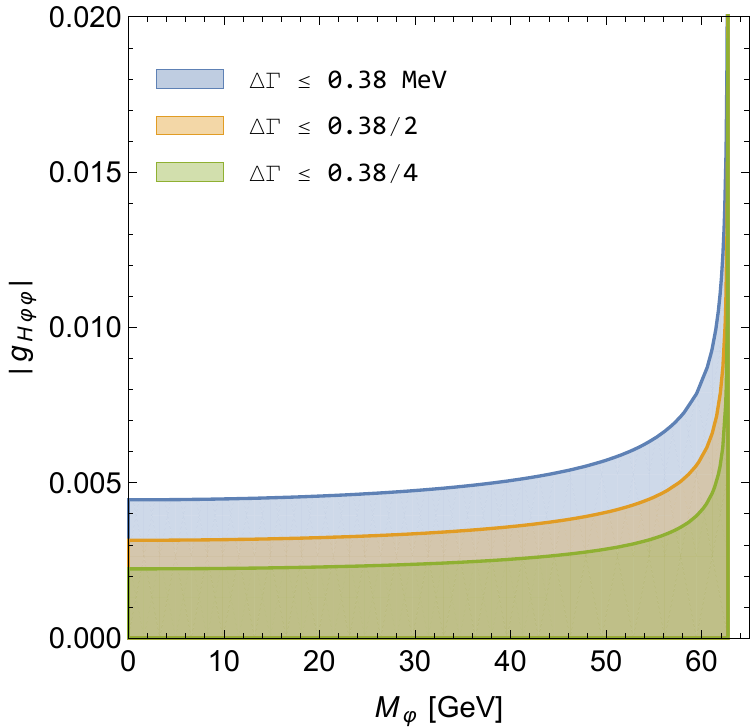}
\includegraphics[width=8.0cm]{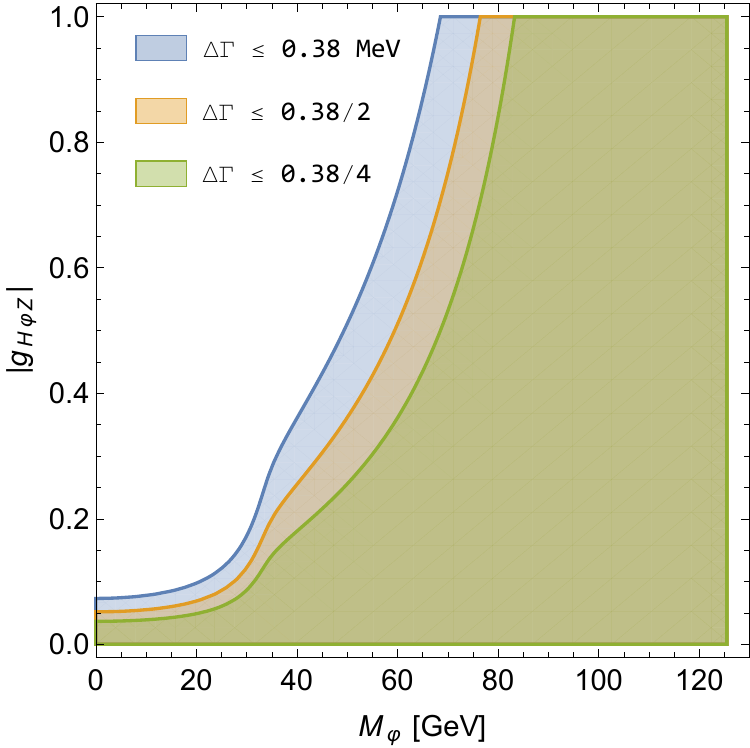}
\end{center}
\vspace{-0.5cm}
\caption{
(Left) The allowed parameter space on the
($M_\varphi$, $|g_{H\varphi\varphi}|$)
plane from $\Delta\Gamma_{\rm tot}\leq 0.38$ MeV at  95\% CL.
Also shown are future prospects assuming two and four times stronger constraints
on $\Delta\Gamma_{\rm tot}$.  (Right) The same as in the left panel but for
($M_\varphi$, $|g_{H\varphi Z}|$).}
\label{fig:cxsm}
\end{figure}
In the cxSM where the SM is extended by adding a complex SU(2)$_L$ singlet,
there could be a light Higgs boson $\varphi$ mainly from the singlet sector and
the 125.5 GeV Higgs boson $H$ may couple to a pair of them through
the singlet-doublet mixing term in the potential.
When kinematically allowed or $2M_\varphi<125.5$ GeV, $H$ decays into a pair of
light scalars and, in this case, $\Delta\Gamma_{\rm tot}\leq 0.38$
MeV may provide constraints on the mass $M_\varphi$ of the scalar
particle and the absolute value of the coupling $g_{H\varphi\varphi}$ at
95\% CL. We find $|g_{H\varphi\varphi}| \lsim 0.005$ for $M_\varphi\lsim 40$ GeV,
see the left panel of Fig.~\ref{fig:cxsm}.
In 2HDMs, assuming that the scalar $\varphi$ is CP odd and
the Higgs boson $H$ with $M_H = 125.5\, {\rm GeV}$ is CP even,
we consider another decay $H\to\varphi Z^*$ to constrain the coupling $g_{H\varphi Z}$
depending on $M_\varphi$. We find $|g_{H\varphi Z}| \lsim 0.5$ for
$M_\varphi\lsim 50$ GeV, see the right panel of Fig.~\ref{fig:cxsm}.

\medskip

\begin{figure}[t!]
\begin{center}
\includegraphics[width=8.3cm]{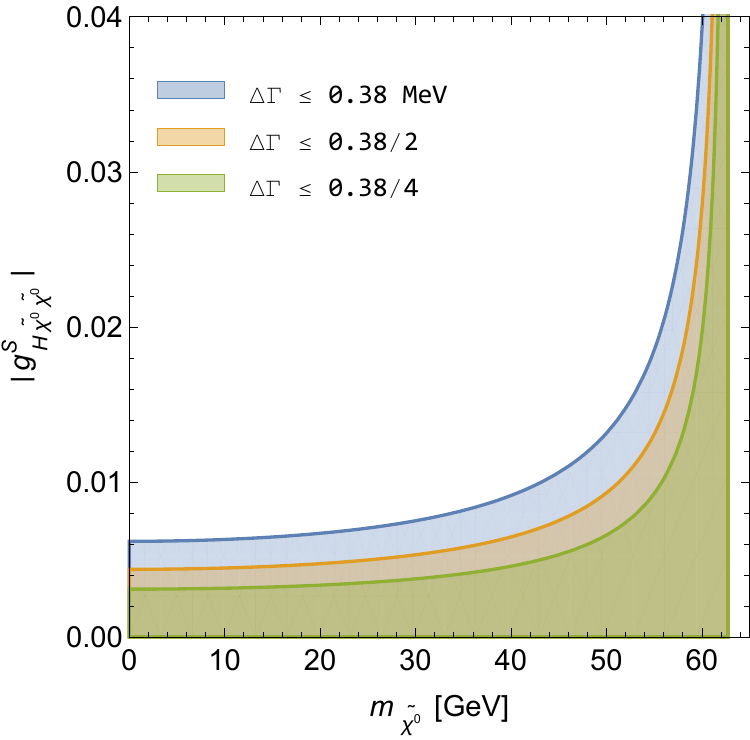}
\end{center}
\vspace{-0.5cm}
\caption{
The allowed parameter space on the
($m_{\widetilde\chi_1^0}$, $|g^S_{H\widetilde\chi_1^0\widetilde\chi_1^0}|$)
plane from $\Delta\Gamma_{\rm tot}\leq 0.38$ MeV at  95\% CL.
Also shown are future prospects assuming
two and four times stronger constraints on $\Delta\Gamma_{\rm tot}$.}
\label{fig:x0x0}
\end{figure}

In the MSSM, there still exists a room for the lightest neutralino
${\widetilde\chi_1^0}$ to be light with its mass two times smaller
than $M_H=125.5\, {\rm GeV}$~\cite{Barman:2020zpz}.
In this case, again, $\Delta\Gamma_{\rm tot}\leq 0.38$ MeV may provide 95\% CL constraints on
the mass $m_{\widetilde\chi_1^0}$ and the
absolute value of the coupling
$g^S_{H\widetilde\chi_1^0\widetilde\chi_1^0}$ assuming $H$ is purely CP even.
We find $|g^S_{H\widetilde\chi_1^0\widetilde\chi_1^0}| \lsim 0.01$ for
$m_{\widetilde\chi_1^0}\lsim 45$ GeV, see Fig.~\ref{fig:x0x0}.

\medskip

\begin{figure}[t!]
\begin{center}
\includegraphics[width=8.3cm]{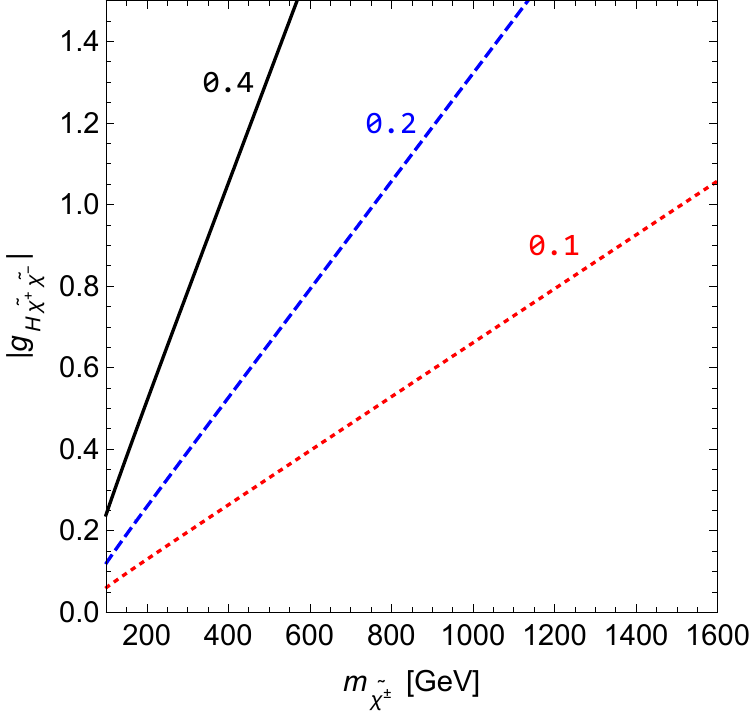}
\includegraphics[width=8.3cm]{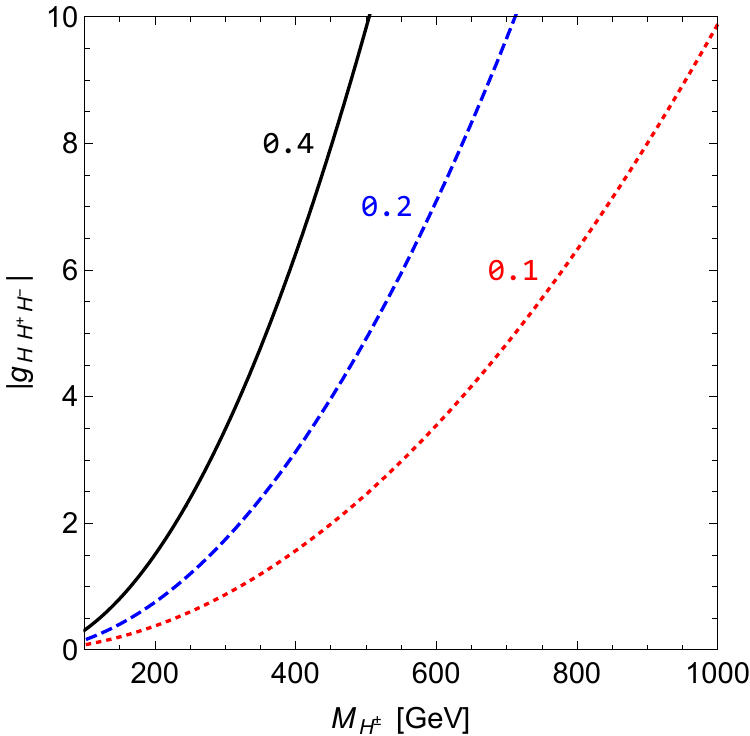}
\end{center}
\vspace{-0.5cm}
\caption{
(Left) Contour lines for $|\Delta S^\gamma| = 0.4\,,0.2\,,0.1$ from left to
right on the ($m_{\widetilde\chi_1^\pm}$,
$|g_{H\widetilde\chi_1^+\widetilde\chi_1^-}|$) plane assuming
the lighter-chargino-loop contributions dominate $\Delta S^\gamma$.
(Right) The same as in the left panel but on the ($M_{H^\pm}$, $|g_{HH^+H^-}|$)
plane now assuming $\Delta S^\gamma$ is
dominated by the contributions from charged-Higgs loops.
For the reference value of $|\Delta S^\gamma| = 0.4$, we are taking
the $1\sigma$ error of the {\bf CPCN4} fit.  }
\label{fig:dsa}
\end{figure}
\begin{figure}[t!]
\begin{center}
\includegraphics[width=8.3cm]{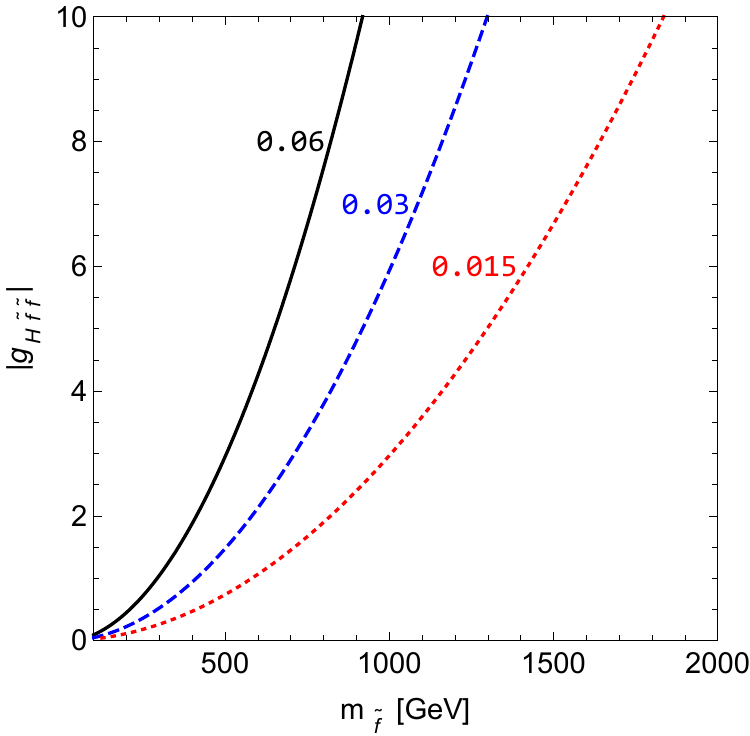}
\end{center}
\vspace{-0.5cm}
\caption{
Contour lines for $|\Delta S^g| =0.06\,,0.03\,,0.015$ from left to right
on the ($m_{\widetilde t_1}$,
$|g_{H\widetilde t_1^* \widetilde t_1}|$) plane
assuming $\Delta S^g$ is
dominated by the contributions from the lighter-stop loops.
For the reference value of $|\Delta S^g| = 0.06$, we are taking
the $1\sigma$ error of the {\bf CPCN4} fit.  }
\label{fig:dsg}
\end{figure}

From Table \ref{tab:CPC}, we find
$|\Delta S^\gamma|\lsim 0.4$ and $|\Delta S^g|\lsim 0.06$
by taking the errors in the {\bf CPCN4} fit.
Exploiting these constraints, we derive
constraints on the $H$ couplings to a pair of charged Higgs bosons, lighter
charginos, and the lightest stops/sbottoms. For the SUSY contributions to
$\Delta S^\gamma$ and $\Delta S^g$, we refer to Appendix \ref{app:dsdp}.

\medskip

In the left panel of Fig.~\ref{fig:dsa}, we show the contour lines
for $|\Delta S^\gamma| = 0.4\,,0.2\,,0.1$  assuming it is dominated by
the contributions from the lighter-chargino loops and considering
only them. We find that the mass of the lighter chargino
is constrained
to be $m_{\widetilde\chi_1^\pm}\gsim 390$ GeV by the current LHC Higgs
data when the relevant coupling is assumed to be 1.
The lower-bound constraint on $m_{\widetilde\chi_1^\pm}$
linearly increases as the bound on $|\Delta S^\gamma|$ becomes stronger.
In the right panel of Fig.~\ref{fig:dsa}, we show the same
contour lines now assuming that
$\Delta S^\gamma$ is dominated by the charged-Higgs loops.
In the 2HDM and the MSSM, keeping the most significant three contributions
when $M_{H_{2,3}}\gsim v$,
we find that the $g_{_{H_1H^+H^-}}$ coupling is given by
\footnote{See the first relation in
Eq.~(\ref{eq:appNHCHCH}) in Appendix~\ref{app:cubicself}.}
\begin{eqnarray}
g_{_{H_1H^+H^-}} & \sim  &
30\, \left(\frac{M_{H^\pm}}{\rm 1\,TeV}\right)^2 +
5 \left[2\left(\frac{t_\beta}{30}\right) -
\left(\frac{400\,{\rm GeV}}{M_{23}}\right)^2
\left(\frac{t_\beta^2-1}{t_\beta^2}\right)
\left(\frac{t_\beta}{30}\right)^2 \right]
\left(\frac{1+t_\beta^2}{t_\beta^2}\right)
\left[\frac{\real(m_{12}^2{\rm e}^{i\xi})}{(100\,{\rm GeV})^2}\right]\,
\nonumber \\[2mm] && \hspace{2.62cm}
- 14\, \left(\frac{400\,{\rm GeV}}{M_{23}}\right)^2\,
\left(\frac{1+t_\beta^2}{t_\beta^2}\right)
\left(\frac{t_\beta}{30}\right)^2\,\real{(\lambda_7 {\rm e}^{i \xi})}\,,
\end{eqnarray}
under the assumption that $H_1$ is purely CP even state.
Note that the second and the third terms contain contributions
enhanced by the factors of $t_\beta$ and $t_\beta^2$
and, for the $t_\beta^2$-enhanced contributions,
we are taking $M_{H_2}\sim M_{H_3}\equiv M_{23}$
and $c_{\beta-\alpha}=\sqrt\epsilon=\sqrt{\delta_2^2+\delta_3^2}=
\sqrt{2\epsilon_0}\,M_{H_1}^2/M_{23}^2$
with $\epsilon_0=0.05$
as in the subsubsection~\ref{sec:heavyHdecaysInCPV2HDMs},
see Eq.~(\ref{eq:eps23}).
It is worthwhile to note that
$\Delta S^\gamma_1 (H^\pm) \to 1/3$ when
$M_{H^\pm}^2\,,M_{H_{2,3}}^2 \gg M_{H_1}^2\,,v^2\,,\real({m_{12}^2{\rm e}^{i\xi}})$,
see Eq.~(\ref{eq:appNHCHCH}) and the two equations following it.
The constraint on $M_{H^\pm}$ becomes
two times stronger when the bound on $|\Delta S^\gamma|$ becomes
four times stronger.

\medskip

In Fig.~\ref{fig:dsg}, we show the contour lines
for $|\Delta S^g| = 0.06\,,0.03\,,0.015$  assuming that it is
dominated by the contributions from the lighter-stop loops.
The current LHC Higgs data constrain the mass of the lighter stop as
$m_{\widetilde t_1}\gsim 1$ TeV only when the relevant coupling is as large
as 10 and, otherwise, it is weaker or much weaker than several direct bounds
on the lighter-stop mass~\cite{PDG2020}.
\footnote{See also https://pdglive.lbl.gov/DataBlock.action?node=S046STP.}

\section{Conclusions}
\label{sec:conclusions}

In this review, we have calculated and discussed in detail all the
decay widths and branching fractions of Higgs bosons in the frameworks of the SM
and its BSM extensions such as cxSM, 2HDMs with natural flavor conservation,
and MSSM.  We have allowed for CP-violating
complex phases as generally as possible
which induce CP-violating mixing among the neutral Higgs bosons.
We have included the relevant
higher order corrections, which are dominated by QCD corrections,
as much as possible.

\medskip

Firstly, the comprehensive analytical results presented in this
review can be applied for
not only presenting but also understanding all the
details of calculating the decay widths of the
neutral Higgs boson discovered at the LHC by
allowing for nontrivial pseudoscalar as well as scalar couplings of the
Higgs boson to fermions pairs.
In the SM limit, it turned out that our numerical results
solely based on the analytic expressions and
supplemental materials provided in this review
show good consistency with those
presented by, for example, the LHC Higgs Cross Section Working
Group implementing the state-of-the-art theoretical calculations
\cite{deFlorian:2016spz}.
%
%
We expect that the analytic expressions for LO decay widths
and QCD corrections together with the SM ELW corrections
presented in this review could be used for analyzing
the decays of neutral Higgs bosons appearing in BSM models
with the precision comparable to that achieved through the
full-fledged theoretical calculations.

\medskip

The second application of our analytical results
was to the neutral 2HDM
Higgs sector in the presence of nontrivial CP-violating phases of the
complex quartic couplings. Specifically,
we have taken the scenario where the lightest
Higgs boson is purely CP even while the two heavier Higgs bosons
are CP-mixed states exhibiting
maximal CP violation when they are degenerate.
To illustrate the typical decay patterns of the
CP-violating heavy neutral Higgs bosons, we
have taken the type-I 2HDM
and we have contrasted them with those
of purely CP-even or CP-odd neutral heavy Higgs bosons.
Incidentally, we also have presented
the decay pattern of a charged Higgs boson appearing in the framework
of 2HDMs.

\medskip

Thirdly, we have presented the constraints on
the couplings of the Higgs boson weighing about 125 GeV
obtained from implementing the global fits
to all the Tevatron and LHC data available up to the summer in 2018.
The global fits were based on several
scenarios of the couplings and form factors for the main Higgs production
and decay modes.
Generally, the constraints  turned out to be already tight
on the tree-level couplings with the possibility that the loop-induced
couplings could deviate sizably from the SM predictions.

\medskip

Finally, we emphasize that, even with this comprehensive review, there
remain lots of improvements for the QCD and electroweak corrections
to be made so as to match the expected ever-increasing precision
measurements from upgraded LHC and future high energy collider experiments.
Furthermore, to draw a more unified picture for the genuine BSM physics,
many BSM scenarios considered seriously at present and
expected to be developed fully in a concrete form
in the near future have to be included.

\section*{Acknowledgment}
We thank Giampiero Passarino, Christian Sturm, and Sandro Uccirati
for helping us to implement the electroweak corrections
to the $H\to gg$ and $H\to\gamma\gamma$ decays and
Julien Baglio, Thi Nhung Dao and Margarete Muhlleitner
for comments on {\tt NMSSMCALCEW}.
We thank
Abdesslam Arhrib,
Eri Asakawa,
Gabriela Barenboim,
Francesca Borzumati,
Cristian Bosch,
Marcela Carena,
Jung Chang,
Kingman Cheung,
Kiwoon Choi,
Debajyoti Choudhury,
Byung-chul Chung,
Brian Cox,
Manuel Drees,
Birgit Eberle,
John Ellis,
Christoph Englert,
Jeffrey Forshaw,
Ayres Freitas,
Benedikt Gaissmaier,
Kaoru Hagiwara,
Tie-Jiun Hou,
Ran Huo,
Jan Kalinowski,
Pyungwon Ko,
Yi Liao,
M. Luisa L\'opez-Iba\~nez,
Chih-Ting Lu,
David Miller,
James Monk,
Margarete Muhlleitner,
Junya Nakamura,
Chan Beom Park,
Yvonne Peters,
Apostolos Pilaftsis,
Christian Schwanenberger,
Stefano Scopel,
Eibun Senaha,
Jeonghyeon Song,
Wan Young Song,
Michael Spira,
Yue-Lin Sming Tsai,
Po-Yan Tseng,
Oscar Vives,
Carlos Wagner, and
Peter Zerwas
for fruitful collaborations.
This work was supported by the National Research Foundation (NRF) of Korea
Grant No. NRF-2016R1E1A1A01943297 (J.S.L. and J.P.) and
No. NRF-2018R1D1A1B07051126 (J.P.).
The work of S.Y.C was supported
in part by Basic Science Research Program through the NRF of Korea
Grant No. NRF-2016R1D1A3B01010529 and in part by the CERN-Korea theory
collaboration.
%

\section*{Appendices}
\label{sec:appendices}

\def\theequation{\Alph{section}.\arabic{equation}}

This section consists of six appendices.
Appendix~\ref{app:smpara} is for
a summary of the SM parameters used for the numerical estimates of the Higgs
decay widths and a description of the running of the strong coupling
constant and quark masses.
Appendix~\ref{app:dsdp} is for the supersymmetric contributions
to the loop-induced couplings of the Higgs boson to two gluons, two photons
and $Z\gamma$. Appendix~\ref{app:csfcpf} is for the QCD corrections to the
partial width of the Higgs-boson decay to two photons.
We work out the relations among
the parameters of the most general 2HDM in Appendix~\ref{app:2hdminput}
and we apply them for deriving cubic Higgs-boson self-couplings
in Appendix~\ref{app:cubicself}.
Finally, in Appendix~\ref{app:bsmtools} we briefly
introduce a few well-developed numerical packages for calculating precise
SM and full BSM-dependent ELW corrections.

\begin{appendix}

\setcounter{equation}{0}
\section{Standard Model Parameters}
\label{app:smpara}

In this appendix, we summarize the SM parameters used for the estimation of
decay widths of Higgs bosons. And we also show the running of the strong coupling
constant and quark masses.

\subsection{Input parameters}
\label{subsec:input_parameters}

The SM parameters used for the estimation of decay widths of Higgs bosons are
\cite{PDG2020,deFlorian:2016spz}:
\footnote{
For the precision measurement of the Fermi constant $G_F$,
see Refs.~\cite{Webber:2010zf,Tishchenko:2012ie}.}
\begin{itemize}
\item\underline{Gauge coupling strengths}
\begin{eqnarray}
\alpha_s(M_Z)&=&0.118 \pm 0.0015\,, \nonumber \\
\alpha(0)&=&1/137.035999\,, \nonumber \\
\alpha(M_W)&=&1/128\,.
\end{eqnarray}
\item\underline{Electroweak parameters}
\begin{eqnarray}
M_W&=&(80.385 \pm 0.015)~\mathrm{GeV}\,,  \ \ \ \ \ \
\Gamma_W=(2.085 \pm 0.042)~\mathrm{GeV}\,;  \nonumber \\
M_Z&=&(91.1876 \pm 0.0021)~\mathrm{GeV}\,,  \ \ \
\Gamma_Z=(2.4952 \pm 0.0023)~\mathrm{GeV}\,;  \nonumber \\
G_F&=&1.1663787(6)\times 10^{-5}~\mathrm{GeV}^{-2}\,.
\end{eqnarray}
The vev $v$ {of the SM Higgs field} is given by $v=\left(\sqrt{2}G_F\right)^{-1/2}
\simeq 246.22$ GeV
with $G_F=\sqrt{2}g^2/8M_W^2$ and $v=2M_W/g$.
For the square of the sine of the weak mixing angle,
we adopt the so-called on-shell scheme
in which the tree-level relation $s_W^2=1-M^2_W/M^2_Z$
is promoted to define the renormalized $s_W^2$ to all orders in perturbation
theory~\cite{Sirlin:1980nh}:
$s^2_W=0.22290$, $c_W^2=1-s^2_W=0.77710$,
$g(M_W)=e/s_W\simeq 0.66366$ and
$g^\prime(M_W)=e/c_W\simeq 0.35544$
with $e=e(M_W)=2\sqrt{\pi\alpha(M_W)}=0.31333$.
\item\underline{Lepton masses}
\begin{equation}
M_{\mu}=(105.6583715 \pm 0.00000035) ~\mathrm{MeV}\,, \ \ \
M_{\tau}=(1776.82 \pm 0.16) ~\mathrm{MeV}\,.
\end{equation}
\item\underline{Quark masses}
\begin{eqnarray}
\label{eq:appQmasses}
M_t&=&(172.5 \pm 1)~\mathrm{GeV}\,,  \nonumber \\
\overline{m}_b(\overline{m}_b)&=&(4.18 \pm 0.03)~\mathrm{GeV}\,, \nonumber \\
\overline{m}_c(3\,\mathrm{GeV})&=&(0.986 \pm 0.026)~\mathrm{GeV}\,.
\end{eqnarray}
Note that the pole mass is used for $t$ quark while, for $b$ and $c$ quarks,
$\overline{\rm MS}$ masses are used.
\footnote{The electron
and $u,d$-quark masses are not included as their masses are too tiny to
influence the numerical analyses made in this work.}
\end{itemize}

\subsection{Running of the strong coupling constant and quark masses}
\label{subsec:running_of_the_strong_coupling_and_quark_masses}

\begin{figure}[t!]
\begin{center}
\includegraphics[width=10.0cm]{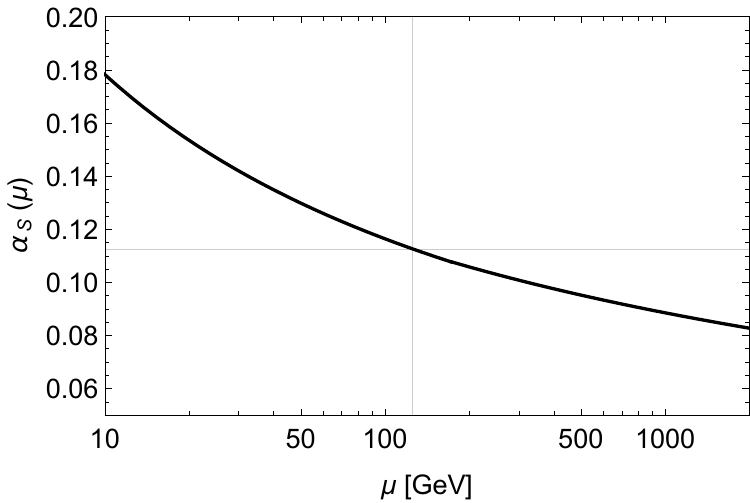}
\includegraphics[width=10.0cm]{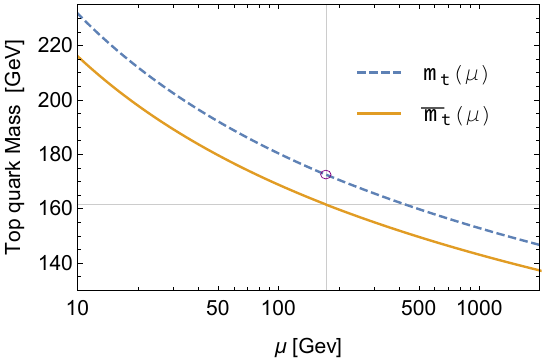}
\\ \hspace{0.5cm}
\includegraphics[width=9.5cm]{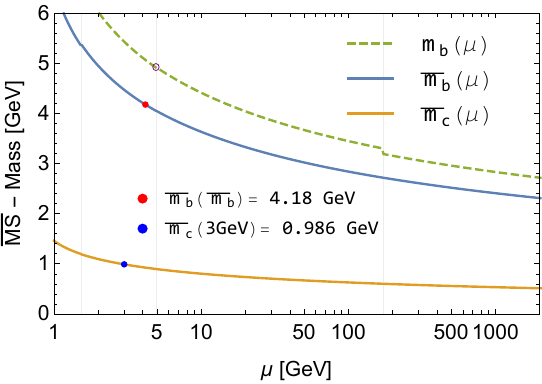}
\end{center}
\vspace{-0.5cm}
\caption{
Running of the strong coupling constant $\alpha_s(\mu)$ (upper)
and the $\overline{\rm MS}$ quark masses
$\overline{m}_t(\mu)$ (middle) and
$\overline{m}_{b,c}(\mu)$ (lower).
In the upper frame, the vertical line
locates the position of $M_H=125.5$ GeV.
In the middle and lower frames, we also show $m_{t,b}(\mu)$
used in the calculations of $\Gamma(H\to\gamma\gamma)$ with
the expressions given in Eq.~(\ref{eq:mqforhaa}).
The vertical lines
locate the positions of the pole masses of $M_t$ and $M_{b,c}$.
And, in the lower frame, the input
values for $b$ and $c$ quark masses are denoted by bullets,
see Eq.~(\ref{eq:appQmasses}).
The open circles in the middle and lower frames denote the positions
$m_t(\mu=M_t)=M_t$  and $m_b(\mu=M_b)=M_b$, respectively,
with $M_t=172.5$ GeV and $M_b=4.93$ GeV.
}
\label{fig:runasmq}
\end{figure}

We neglect the running of the SU(2)$_L$ and U(1)$_Y$ electroweak
couplings and the leptons masses, measured experimentally
with great precision.
For the running of the strong coupling strength $\alpha_s(\mu)$
and the $\overline{\rm MS}$ quark masses $\overline{m}_q(\mu)$,
we use the most recent version of {\tt RunDec}~\cite{Chetyrkin:2000yt,Herren:2017osy}
in which five-loop corrections of the QCD beta function and four-loop decoupling
effects are included. The results are shown in Fig.~\ref{fig:runasmq}
and Table~\ref{tab:runasmq}.
We note that
$\alpha_s(125.5\,{\rm GeV})=0.1126$ and $\overline{m}_t(M_t)=161.5$ GeV.
For the pole mass of $b$ quark, the three-loop conversion relation is taken
to give $M_b=4.93$ GeV~\cite{PDG2020}. On the other hand, for
the pole mass of $c$ quark, we take the relation between the on-shell
charm-quark and bottom-quark masses, giving
$M_c=M_b-3.41~{\rm GeV}=1.52$ GeV
\cite{PDG2020,deFlorian:2016spz,Bauer:2004ve}.
\begin{table}[t!]
\caption{\label{tab:runasmq}
Running of
$\alpha_s(\mu)$, $\overline{m}_c(\mu)$, $\overline{m}_b(\mu)$,  and
$\overline{m}_t(\mu)$.
$M_q^+$ and $M_{q-}$ are introduced
for decoupling effects from matching
the (effective) theory with $N_F-1$ with the (full) theory with $N_F$
at the scale $M_q$.
}
\begin{center}
\begin{tabular}{c|cccc}
\hline
$\mu$ [GeV] & $\alpha_s(\mu)$ & $\overline{m}_c(\mu)$ [GeV] & $\overline{m}_b(\mu)$ [GeV] &
$\overline{m}_t(\mu)$ [GeV] \\
\hline
$1.52^+\,(M_c^+)$ &
$3.48\times 10^{-1}$ & $1.19\times 10^{0}\,~$ & $5.40\times 10^{0}$ & $3.21\times 10^{2}$ \\
$3$ &
$2.54\times 10^{-1}$ & $9.86\times 10^{-1}$ & $4.47\times 10^{0}$ & $2.66\times 10^{2}$ \\
$4.18\,(\overline{m}_b)$ &
$2.25\times 10^{-1}$ & $9.21\times 10^{-1}$ & $4.18\times 10^{0}$ & $2.48\times 10^{2}$ \\
$4.93_-\,(M_{b-})$ &
$2.13\times 10^{-1}$ & $8.94\times 10^{-1}$ & $4.06\times 10^{0}$ & $2.41\times 10^{2}$ \\
$4.93^+\,(M_b^+)$ &
$2.14\times 10^{-1}$ & $8.92\times 10^{-1}$ & $4.06\times 10^{0}$ & $2.41\times 10^{2}$ \\
$10$ &
$1.78\times 10^{-1}$ & $8.00\times 10^{-1}$ & $3.64\times 10^{0}$ & $2.16\times 10^{2}$ \\
$20$ &
$1.53\times 10^{-1}$ & $7.32\times 10^{-1}$ & $3.33\times 10^{0}$ & $1.98\times 10^{2}$ \\
$30$ &
$1.42\times 10^{-1}$ & $7.00\times 10^{-1}$ & $3.18\times 10^{0}$ & $1.89\times 10^{2}$ \\
$40$ &
$1.35\times 10^{-1}$ & $6.79\times 10^{-1}$ & $3.09\times 10^{0}$ & $1.84\times 10^{2}$ \\
$50$ &
$1.30\times 10^{-1}$ & $6.65\times 10^{-1}$ & $3.02\times 10^{0}$ & $1.80\times 10^{2}$ \\
$60$ &
$1.26\times 10^{-1}$ & $6.54\times 10^{-1}$ & $2.97\times 10^{0}$ & $1.77\times 10^{2}$ \\
$70$ &
$1.23\times 10^{-1}$ & $6.44\times 10^{-1}$ & $2.93\times 10^{0}$ & $1.74\times 10^{2}$ \\
$80$ &
$1.20\times 10^{-1}$ & $6.37\times 10^{-1}$ & $2.90\times 10^{0}$ & $1.72\times 10^{2}$ \\
$90$ &
$1.18\times 10^{-1}$ & $6.30\times 10^{-1}$ & $2.87\times 10^{0}$ & $1.70\times 10^{2}$ \\
$100$ &
$1.16\times 10^{-1}$ & $6.25\times 10^{-1}$ & $2.84\times 10^{0}$ & $1.69\times 10^{2}$ \\
$110$ &
$1.15\times 10^{-1}$ & $6.20\times 10^{-1}$ & $2.82\times 10^{0}$ & $1.67\times 10^{2}$ \\
$120$ &
$1.13\times 10^{-1}$ & $6.15\times 10^{-1}$ & $2.80\times 10^{0}$ & $1.66\times 10^{2}$ \\
$125.5$ &
$1.13\times 10^{-1}$ & $6.13\times 10^{-1}$ & $2.79\times 10^{0}$ & $1.66\times 10^{2}$ \\
$130$ &
$1.12\times 10^{-1}$ & $6.11\times 10^{-1}$ & $2.78\times 10^{0}$ & $1.65\times 10^{2}$ \\
$140$ &
$1.11\times 10^{-1}$ & $6.08\times 10^{-1}$ & $2.76\times 10^{0}$ & $1.64\times 10^{2}$ \\
$150$ &
$1.10\times 10^{-1}$ & $6.04\times 10^{-1}$ & $2.75\times 10^{0}$ & $1.63\times 10^{2}$ \\
$160$ &
$1.09\times 10^{-1}$ & $6.01\times 10^{-1}$ & $2.73\times 10^{0}$ & $1.62\times 10^{2}$ \\
$170$ &
$1.08\times 10^{-1}$ & $5.98\times 10^{-1}$ & $2.72\times 10^{0}$ & $1.62\times 10^{2}$ \\
$172.5_-\,(M_{t-})$ &
$1.08\times 10^{-1}$ & $5.98\times 10^{-1}$ & $2.72\times 10^{0}$ & $1.62\times 10^{2}$ \\
$172.5^+\,(M_t^+)$ &
$1.08\times 10^{-1}$ & $5.98\times 10^{-1}$ & $2.72\times 10^{0}$ & $1.61\times 10^{2}$ \\
$180$ &
$1.07\times 10^{-1}$ & $5.96\times 10^{-1}$ & $2.71\times 10^{0}$ & $1.61\times 10^{2}$ \\
$190$ &
$1.06\times 10^{-1}$ & $5.93\times 10^{-1}$ & $2.70\times 10^{0}$ & $1.60\times 10^{2}$ \\
$200$ &
$1.06\times 10^{-1}$ & $5.91\times 10^{-1}$ & $2.69\times 10^{0}$ & $1.60\times 10^{2}$ \\
$300$ &
$1.01\times 10^{-1}$ & $5.74\times 10^{-1}$ & $2.61\times 10^{0}$ & $1.55\times 10^{2}$ \\
$400$ &
$9.76\times 10^{-2}$ & $5.62\times 10^{-1}$ & $2.56\times 10^{0}$ & $1.52\times 10^{2}$ \\
$500$ &
$9.52\times 10^{-2}$ & $5.54\times 10^{-1}$ & $2.52\times 10^{0}$ & $1.50\times 10^{2}$ \\
$1000$ &
$8.85\times 10^{-2}$ & $5.29\times 10^{-1}$ & $2.41\times 10^{0}$ & $1.43\times 10^{2}$ \\
\hline
\end{tabular}
\end{center}
\end{table}

\medskip

Specifically for the loop-induced decay $H\to\gamma\gamma$,
we use the running masses $m_{t,b}(\mu)$ of which
the expressions are given at the four-loop level
by~\cite{Gorishnii:1990zu,Vermaseren:1997fq}
\begin{equation}
\label{eq:mqforhaa}
m_q(\mu) = M_q \, \Bigg[\frac{\alpha_s(\mu)}{\alpha_s(M_q)}\Bigg]
^{12/ (33 - 2\, N_F)}\,
\frac
{1+c_1^q\left[\frac{\alpha_s(\mu)}{\pi}\right]+c_2^q\left[\frac{\alpha_s(\mu)}{\pi}\right]^2
+c_3^q\left[\frac{\alpha_s(\mu)}{\pi}\right]^3}
{1+c_1^q\left[\frac{\alpha_s(M_q)}{\pi}\right]+c_2^q\left[\frac{\alpha_s(M_q)}{\pi}\right]^2
+c_3^q\left[\frac{\alpha_s(M_q)}{\pi}\right]^3}\,,
\end{equation}
where the numerical values of the six dimensionless coefficients are
\begin{eqnarray}
c_1^b&=&1.17549\,,  \ \ \
c_2^b = 1.50071\,,  \ \ \
c_3^b = 0.172478\,;  \nonumber \\[0mm]
c_1^t&=&1.39796\,,  \ \ \
c_2^t = 1.79348\,,  \ \ \
c_3^t =-0.683433\,.
\end{eqnarray}
Note that $m_q(M_q)=M_q$ as denoted by open circles in
the middle and lower frames in Fig.~\ref{fig:runasmq}.

\setcounter{equation}{0}
\section{Supersymmetric Contributions to
the $Hgg$, $H\gamma\gamma$, and $HZ\gamma$ Form Factors}
\label{app:dsdp}

In this appendix, we present the contributions to
the loop-induced $Hgg$, $H\gamma\gamma$, and $HZ\gamma$
form factors from the triangle diagrams in which charginos, charged
and/or colored sfermions, and/or charged Higgs bosons are
running.

\medskip

In the minimal supersymmetric extension of the SM (MSSM),
the form factors of $\Delta S^{g,\gamma}$ and $\Delta P^{g,\gamma}$
denoting new MSSM contributions to the $Hgg, H\gamma\gamma$ vertices
are given by:
\begin{eqnarray}
\label{eq:appDSDPgg}
\Delta S^g_i&=&
-\sum_{\widetilde{f}_j=\widetilde{t}_1,\widetilde{t}_2,\widetilde{b}_1,\widetilde{b}_2}
g_{H_i\widetilde{f}^*_j\widetilde{f}_j}
\frac{v^2}{4m_{\widetilde{f}_j}^2} F_0(\tau_{i\widetilde{f}_j}) \,, \nonumber \\
\Delta P^g_i&=& 0 \,; \\[3mm]
\label{eq:appDSDPaa}
\Delta S^\gamma_i&=&\sqrt{2}\,g\sum_{f=\widetilde{\chi}^\pm_1,\widetilde{\chi}^\pm_2}
g^{S}_{H_i\bar{f}f}\,\frac{v}{m_f} F_{sf}(\tau_{if})
- \sum_{\widetilde{f}_j=\widetilde{t}_1,\widetilde{t}_2,\widetilde{b}_1,\widetilde{b}_2,
           \widetilde{\tau}_1,\widetilde{\tau}_2}
N_C^f\, Q_f^2g_{H_i\widetilde{f}^*_j\widetilde{f}_j}
\frac{v^2}{2m_{\widetilde{f}_j}^2} F_0(\tau_{i\widetilde{f}_j})
\nonumber \\
&& + g_{_{H_iH^+H^-}}\frac{v^2}{2 M_{H^\pm}^2} F_0(\tau_{iH^\pm})
\,, \nonumber \\
\Delta P^\gamma_i&=&\sqrt{2}\,g\sum_{f=\widetilde{\chi}^\pm_1,\widetilde{\chi}^\pm_2}
g^{P}_{H_i\bar{f}f} \,\frac{v}{m_f} F_{pf}(\tau_{if}) \,,
\end{eqnarray}
where $\tau_{ix}=M_{H_i}^2/4m_x^2$ with $i=1,2,3$
denoting three neutral Higgs bosons and $N_C^f=3$ for (s)quarks and $N_C^f=1$ for
status, respectively.
The form factor $F_0(\tau)$ is given by
\begin{equation}
\label{eq:f0}
F_0(\tau)=\tau^{-1}\,[-1+\tau^{-1}f(\tau)]\,,
\end{equation}
which takes the value of $1/3$ when $\tau=0$, see Fig.~\ref{fig:fplot}.

\medskip

On the other hand, the form factors of $\Delta S^{Z\gamma}$ and $\Delta P^{Z\gamma}$
denoting new MSSM contributions to the $HZ\gamma$ vertices may take
forms of~\cite{Djouadi:1996yq}
\begin{eqnarray}
\Delta S_i^{Z\gamma}\,
&=&
-\sqrt{2}\,\frac{g}{c_Ws_W} \sum_{j,k} v\,m_{\widetilde\chi_j^\pm}
f\left(m_{\widetilde\chi_j^\pm},m_{\widetilde\chi_k^\pm},m_{\widetilde\chi_k^\pm}
\right) v_{Z\widetilde\chi_j^+\widetilde\chi_k^-}
g^S_{H_i\widetilde\chi_k^+\widetilde\chi_j^-} \nonumber \\
&&-4
\sum_{f=t,b,\tau} \frac{N_C^f Q_f}{c_Ws_W}
\left[\sum_{j,k}
g_{H_i\widetilde{f}_j^*\widetilde{f}_k}
g_{Z\widetilde{f}_k^*\widetilde{f}_j}\,v^2\,
C_2(m_{\widetilde{f}_j},m_{\widetilde{f}_k},m_{\widetilde{f}_k})\right] \nonumber \\
&&+2\,
\frac{g_{H_{i}H^+H^-}}{c_Ws_W}\, \frac{v^2}{M_{H^\pm}^2}\,
I_1\left(\tau_{iH^\pm},\lambda_{H^\pm}\right) \,, \nonumber \\[3mm]
\Delta P_i^{Z\gamma}\,
&=&
-\sqrt{2}\,\frac{i\,g}{c_Ws_W} \sum_{j,k} v\,m_{\widetilde\chi_j^\pm}
g\left(m_{\widetilde\chi_j^\pm},m_{\widetilde\chi_k^\pm},m_{\widetilde\chi_k^\pm}
\right) v_{Z\widetilde\chi_j^+\widetilde\chi_k^-}
g^P_{H_i\widetilde\chi_k^+\widetilde\chi_j^-}\,,
\end{eqnarray}
where $\tau_{iH^\pm}=M_{H_i}^2/4M_{H^\pm}^2$ with $i=1,2,3$,
$\lambda_{H^\pm}=M_{Z}^2/4M_{H^\pm}^2$,
and $N_C^f=3$ for squarks and $N_C^f=1$ for staus, respectively.
For the explicit form of the three loop functions of
$f(m_1,m_2,m_2)$, $g(m_1,m_2,m_2)$, and
$C_2(m_1,m_2,m_2)$, we refer to~\cite{Djouadi:1996yq}. Note that
they implicitly depend on $M_{H_i}^2$ and $M_Z^2$.
For the Higgs couplings to
SUSY particles,
see subsubsection~\ref{sec:SUSYinteractions}
and the relevant $Z$-boson interactions are given by the
following Lagrangian terms:
\begin{itemize}
\item\underline{$Z$-sfermion-sfermion}
\begin{eqnarray}
{\cal L}_{Z\widetilde{f}\widetilde{f}} \ = \
-ig_Z\, g_{Z\widetilde{f}_j^*\widetilde{f}_i} \
\left(\widetilde{f}_j^* \stackrel{\leftrightarrow}{\partial_\mu}
\widetilde{f}_i\right) \ Z^\mu \,,
\end{eqnarray}
where $g_Z=e/(s_Wc_W)$ and
\begin{eqnarray}
g_{Z\widetilde{f}_j^*\widetilde{f}_i} \ = \
I_3^f U_{Lj}^{\widetilde{f} *} U_{Li}^{\widetilde{f}}
-Q_f s_W^2 \delta_{ij}\,,
\end{eqnarray}
with
$I_3^{\,u, \nu}=+1/2$ and $I_3^{\,d, e}=-1/2$.
\item\underline{$Z$-chargino-chargino}
\begin{eqnarray}
{\cal L}_{Z\widetilde\chi^+\widetilde\chi^-} = -\,g_Z\,
\overline{\widetilde\chi^-_i}\,\gamma^\mu\,
\left(v_{Z\chi^+_i\widetilde\chi^-_j} - a_{Z\chi^+_i\widetilde\chi^-_j}
\gamma_5\right)\,
\widetilde\chi^-_j\,Z_\mu \, ,
\end{eqnarray}
where  the vector and axial--vector couplings are given by
\begin{eqnarray}
v_{Z\chi^+_i\widetilde\chi^-_j}& = &
\frac{1}{4}\left[
\left(C_L\right)_{i2} \left(C_L\right)_{j2}^* +
\left(C_R\right)_{i2} \left(C_R\right)_{j2}^*
\right] \ - \ c_W^2 \,\delta_{ij}\,,
\nonumber \\
a_{Z\chi^+_i\widetilde\chi^-_j} & = &
\frac{1}{4}\left[
\left(C_L\right)_{i2} \left(C_L\right)_{j2}^* -
\left(C_R\right)_{i2} \left(C_R\right)_{j2}^*
\right]\,.
\end{eqnarray}
For completeness, we recall that the $Z$-boson couplings to the quarks and leptons are given by
\begin{eqnarray}
\label{eq:zff}
{\cal L}_{Z\bar{f}f} = -\,g_Z\,
\bar{f}\,\gamma^\mu\,\left(v_{Z\bar{f}f} - a_{Z\bar{f}f}
\gamma_5\right)\,f\,Z_\mu \, ,
\end{eqnarray}
with $v_{Z\bar{f}f}=I_3^f/2-Q_f s_W^2$ and $a_{Z\bar{f}f}=I_3^f/2$
in terms of the third component of  weak isospin $I^f_3$ and the electric
charge $Q_f$ of each fermion $f$.
\end{itemize}

\setcounter{equation}{0}
\section{QCD Corrections to $\Gamma(H\to\gamma\gamma)$:
$C_{sf}(\tau)$ and $C_{pf}(\tau)$}
\label{app:csfcpf}

The scaling factors $C_{sf}(\tau)$ and $C_{pf}(\tau)$ for the QCD corrections
to the decay width of a Higgs boson $H$ into two photons
might be given by~\cite{Spira:1995rr}
\begin{eqnarray}
C_{sf} (\tau;\rho) = C_1^H (\tau) + C_2^H (\tau)
\left[  \log \tau + \log\frac{4}{\rho^2}\right]\,; \ \ \
C_{pf} (\tau;\rho) = C_1^A (\tau) + C_2^A (\tau)
\left[  \log \tau + \log\frac{4}{\rho^2}\right]\,,
\end{eqnarray}
where $\tau = M_H^2/4 m_q^2 (\mu_q)$ with the renormalization scale $\mu_q= M_H/\rho$.
As demonstrated in Section \ref{sec:h2aa}, we take $\rho=2$.
We note again the running mass $m_q$ is normalized as $m_q(M_q)=M_q$.

\medskip

The $C_1^H (\tau)$ function is given via the following relation
\cite{Fleischer:2004vb,Harlander:2005rq}
\footnote{
Eqs.\,(10) and (12) in Ref.\,\cite{Fleischer:2004vb} contain typos,
see footnote 3 of Ref.~\cite{Harlander:2005rq} for locating them.
}
\begin{eqnarray}
\label{eq::c1h}
F_{sf}(\tau)\,C_1^H (\tau)  &=&
-\frac{2\, \theta \left(1+\theta+\theta^2+\theta^3\right) } {3\,{\left(1-\theta\right)}^5}
    \biggl[ 108\Li_4(\theta) + 144\Li_4(-\theta) - 64\Li_3(\theta)\ln\theta
\nonumber \\ &&\hspace{1.0cm}
      - 64\Li_3(-\theta)\ln\theta + 14\Li_2(\theta)\ln^2\theta + 8\Li_2(-\theta)\ln^2\theta
      + \frac{1}{12}\ln^4\theta
\nonumber \\ &&\hspace{1.0cm}
      + 4\,\zeta_2\ln^2\theta + 16\,\zeta_3\ln\theta + 18\,\zeta_4 \biggr]
\nonumber \\ &&
    +\frac{2\,\theta {\left(1+\theta\right)}^2}{3\,{\left(1-\theta\right)}^4}
    \biggl[ -32\Li_3(-\theta) + 16\Li_2(-\theta)\ln\theta -4\,\zeta_2\ln\theta
      \biggr]
\nonumber \\ &&
    - \frac{8\, \theta \left(7-2\,\theta+7\,\theta^2\right) } {3\,{\left(1-\theta\right)}^4}
    \Li_3(\theta) +\frac{16\, \theta \left(3-2\,\theta+3\,\theta^2\right) }
    {3\,{\left(1-\theta\right)}^4} \Li_2(\theta)\ln\theta
\nonumber \\ &&
    +\frac{4\, \theta \left(5-6\,\theta+5\,\theta^2\right) }
    {3\,{\left(1-\theta\right)}^4} \ln(1-\theta)\ln^2\theta
    +\frac{2\theta \left(3+25\,\theta-7\,\theta^2+3\,\theta^3\right) }
    {9\, {\left(1-\theta\right)}^5} \ln^3\theta
\nonumber \\ &&
    +\frac{8\, \theta \left(1-14\,\theta+\theta^2\right) } {3\,{\left(1-\theta\right)}^4}
    \zeta_3
    +\frac{8\, \theta^2 } {{\left(1-\theta\right)}^4} \ln^2\theta
    -\frac{8\, \theta \left(1+\theta\right) } {{\left(1-\theta\right)}^3}
    \ln\theta -\frac{40\, \theta} {3\,{\left(1-\theta\right)}^2}
    \,,
\end{eqnarray}
where $\theta$ is a $\tau$-dependent function defined by
\begin{equation}
  \theta\equiv\theta(\tau) =
  \frac{\sqrt{1-\tau^{-1}}-1}{\sqrt{1-\tau^{-1}}+1}\,.
  \label{eq::thetadef}
\end{equation}
The three values, $\zeta_2, \zeta_3$ and $\zeta_4$, of
the Riemann's zeta function are given by
\footnote{It is not yet
clear whether $\zeta_3$ is given in a compact form or not, unlike $\zeta_{2,4}$.}
\begin{equation}
\zeta_2 = \frac{\pi^2}{6}\,,\quad
\zeta_3 = 1.20205690\,,\quad
\zeta_4 = \frac{\pi^4}{90}\,,
\end{equation}
and the polylogarithm function is defined by a power series in
a complex variable $z$ as follows
\footnote{$\Li_n(z)$ has a branch cut discontinuity in the
complex $z$ plane running from 1 to $\infty$.}
\begin{equation}
\Li_n(z) = \sum_{k=1}^{\infty} \frac{z^k}{k^n}~,~~n=1, 2, 3, \cdots\,.
\end{equation}
For analytic continuation to the complex $\tau$ plane, the replacement
$\tau \to \tau+0\,i$ is understood.
The $C_2^H (\tau)$ function is given via~\cite{Harlander:2005rq}
\begin{equation}
F_{sf}(\tau)\,C_2^H(\tau) = 2\, \tau^{-2} \left[
\tau + \left(\tau - 2\right)f(\tau) - (\tau-1)\tau\,
\frac{{\rm d}f(\tau)}{{\rm d}\tau}\right]\,,
\label{eq::c2h}
\end{equation}
with the function $f(\tau)$ defined in Eq.$\,$(\ref{eq:f_function}).

\medskip

On the other hand,
the $C_1^A (\tau)$ and $C_2^A (\tau)$ functions are given via the
following relations~\cite{Harlander:2005rq}
\begin{eqnarray}
F_{pf}(\tau)\, C_1^A (\tau) & = &
    -\frac{ \theta \left(1+\theta^2\right) }
    {{\left(1-\theta\right)}^3\left(1+\theta\right)}
    \Bigg\{ 72\Li_4(\theta) + 96\Li_4(-\theta)
      -\frac{128}{3}\Big[\Li_3(\theta) + \Li_3(-\theta)\Big]\ln \theta
\nonumber \\ && \hspace{3.3cm}
      +\frac{28}{3}\Li_2(\theta) \ln^2 \theta +\frac{16}{3}\Li_2(-\theta) \ln^2 \theta
      +\frac{1}{18} \ln^4 \theta
\nonumber \\ && \hspace{3.3cm}
      +\frac{8}{3}\zeta_2 \ln^2 \theta +\frac{32}{3}\zeta_3\ln \theta +12\zeta_4
\Bigg\}
\nonumber \\ && \hspace{0.0cm}
    +\frac{ \theta }{ {\left(1-\theta\right)}^2}
    \biggl[ -\frac{56}{3}\Li_3(\theta) -\frac{64}{3}\Li_3(-\theta)
      +16\Li_2(\theta)\ln \theta
\nonumber \\ && \hspace{2.0cm}
      +\frac{32}{3}\Li_2(-\theta)\ln \theta +\frac{20}{3}\ln(1-\theta)\ln^2 \theta
      -\frac{8}{3}\zeta_2\ln \theta +\frac{8}{3}\zeta_3 \biggr]
\nonumber \\ && \hspace{0.0cm}
    +\frac{ 2\,\theta \left(1+\theta\right) }{{3\left(1-\theta\right)}^3}\, \ln^3 \theta
    \,; \\[0.2cm]
F_{pf}(\tau)\, C_2^A (\tau) & = &
2\, \tau^{-1} \left[ f(\tau) - \tau\, \frac{{\rm d}f(\tau)}{{\rm d}\tau}\right]\,,
\end{eqnarray}
respectively.

\setcounter{equation}{0}
\section{Input parameters for the most general 2HDM potential}
\label{app:2hdminput}

In this appendix, we work out the relations among
the parameters needed to fully specify the most general 2HDM potential,
the masses of neutral and charged Higgs bosons,
and the mixing matrix $O$.
%

\medskip

In subsection~\ref{sec:two_higgs_doublet_models},
we demonstrate that one needs all the elements of the following set of parameters
\begin{equation}
{\cal I}=\left\{
v \,, t_\beta\,, |m_{12}| \,;
\lambda_1\,, \lambda_2\,,\lambda_3\,,\lambda_4\,,
|\lambda_5|\,,|\lambda_6|\,,|\lambda_7|\,,
\phi_5+2\xi\,,\phi_6+\xi\,,\phi_7+\xi\,;{\rm sign}[\cos(\phi_{12}+\xi)]
\right\}\,,
\end{equation}
to fully specify the most general 2HDM scalar potential, see Eq.~(\ref{eq:2hdmpara}).
The set ${\cal I}$ contains 13 parameters plus 1 sign
with $\sin(\phi_{12}+\xi)$ being determined by the
third CP-odd tadpole condition in Eq.~(\ref{eq:2hdmtadpole}).
Alternatively to the set ${\cal I}$, one may use the following equivalent set
\begin{eqnarray}
{\cal I}^\prime&=&\{
v \,, t_\beta\,, \real(m_{12}^2{\rm e}^{i\xi}) \,;  \\[2mm]
&&\hspace{2mm}
\lambda_1\,, \lambda_2\,,\lambda_3\,,\lambda_4\,,
\real(\lambda_5{\rm e}^{2i\xi})\,,
\real(\lambda_6{\rm e}^{i\xi})\,,
\real(\lambda_7{\rm e}^{i\xi})\,,
\imag(\lambda_5{\rm e}^{2i\xi})\,,
\imag(\lambda_6{\rm e}^{i\xi})\,,
\imag(\lambda_7{\rm e}^{i\xi})\}\,. \nonumber
\end{eqnarray}
The above set ${\cal I}^\prime$
contains 10 parameters for the real and complex quartic couplings $\lambda_{1-7}$
and any 7 of them, in principle, can be traded with
the 4 masses of charged and neutral Higgs bosons and the 3 independent
angles of the $3\times 3$ orthogonal mixing matrix $O$ by
judiciously exploiting
Eq.~(\ref{eq:2hdm_mch2}) and the matrix relation
$O^T {\cal M}_0^2 O={\rm diag}(M_{H_1}^2,M_{H_2}^2,M_{H_3}^2)$.
\footnote{For ${\cal M}_0^2$, see Eq.~(\ref{eq:2hdmm0sq}) and
the two subsequent relations following it.}
By choosing
$\imag(\lambda_5{\rm e}^{2i\xi})$,
$\real(\lambda_6{\rm e}^{i\xi})$, and
$\real(\lambda_7{\rm e}^{i\xi})$ as independent input parameters,
one may use the following set of more physical parameters:
\begin{eqnarray}
{\cal P}=\{
v \,, t_\beta\,, \real(m_{12}^2{\rm e}^{i\xi}) \,;
M_{H_1}\,,M_{H_2}\,,M_{H_3}\,,M_{H^\pm}\,,\{O\}\,;
\imag(\lambda_5{\rm e}^{2i\xi})\,,
\real(\lambda_6{\rm e}^{i\xi})\,,
\real(\lambda_7{\rm e}^{i\xi})
\}\,.
\end{eqnarray}
Explicitly, we find that the 7 quartic couplings of
$\lambda_1$, $\lambda_2$, $\lambda_3$, $\lambda_4$,
$\real(\lambda_5{\rm e}^{2i\xi})$,
$\imag(\lambda_6{\rm e}^{i\xi})$, and
$\imag(\lambda_7{\rm e}^{i\xi})$ in the set ${\cal I}^\prime$
can be expressed in terms of
$M_{H_{1,2,3}}$, $M_{H^\pm}$, and the elements of the mixing matrix $O$
in the set ${\cal P}$ as follows:
\begin{eqnarray}
\label{eq:2hdmInputConversion}
&&\lambda_1 =
\frac{ s_\beta }{ 2 v^2 c_\beta^3 }\, \real(m_{12}^2{\rm e}^{i\xi})
+ \frac{ O_{\phi_1 1}^2 } { 2 v^2 c_\beta^2}\, M_{H_1}^2
+ \frac{ O_{\phi_1 2}^2 } { 2 v^2 c_\beta^2}\, M_{H_2}^2
+ \frac{ O_{\phi_1 3}^2 } { 2 v^2 c_\beta^2}\, M_{H_3}^2
- \frac{3 t_\beta} {4} \real{(\lambda_6 {\rm e}^{i \xi})}
+ \frac{t_\beta^3} {4} \real{(\lambda_7 {\rm e}^{i \xi})}\,,
\nonumber \\[3mm]
&&\lambda_2 =
\frac{ c_\beta}{ 2 v^2 s_\beta^3}\, \real(m_{12}^2{\rm e}^{i\xi})
+ \frac{ O_{\phi_2 1}^2 } { 2 v^2 s_\beta^2}\, M_{H_1}^2
+ \frac{ O_{\phi_2 2}^2 } { 2 v^2  s_\beta^2}\, M_{H_2}^2
+ \frac{ O_{\phi_2 3}^2 } { 2 v^2  s_\beta^2}\, M_{H_3}^2
+ \frac{1} {4 t_\beta^3} \real{(\lambda_6 {\rm e}^{i \xi})}
- \frac{3} {4 t_\beta} \real{(\lambda_7 {\rm e}^{i \xi})}\,,
\nonumber \\[3mm]
&&\lambda_3 =
\frac{1}{v^2 s_\beta c_\beta} \real(m_{12}^2{\rm e}^{i\xi})
+\frac{2}{v^2} M_{H^\pm}^2
+\frac{O_{\phi_1 1} O_{\phi_2 1} }{v^2  s_\beta c_\beta} \, M_{H_1}^2
+\frac{O_{\phi_1 2} O_{\phi_2 2} }{v^2  s_\beta c_\beta} \, M_{H_2}^2
+\frac{O_{\phi_1 3} O_{\phi_2 3} }{v^2  s_\beta c_\beta} \, M_{H_3}^2
\nonumber \\ &&\hspace{1cm}
-\frac{1}{2 t_\beta} \real{(\lambda_6 {\rm e}^{i \xi})}
-\frac{t_\beta}{2} \real{(\lambda_7 {\rm e}^{i \xi})}\,,
\nonumber \\[3mm]
&&\lambda_4 =
-\frac{ 1 }{v^2 s_\beta c_\beta}\, \real(m_{12}^2{\rm e}^{i\xi})
-\frac{2}{v^2} M_{H^\pm}^2
+\frac{O_{a1}^2}{v^2}\, M_{H_1}^2
+\frac{O_{a2}^2}{v^2}\, M_{H_2}^2
+\frac{O_{a3}^2}{v^2}\, M_{H_3}^2
-\frac{1}{2t_\beta} \real{(\lambda_6 {\rm e}^{i \xi})}
-\frac{t_\beta}{2} \real{(\lambda_7 {\rm e}^{i \xi})} \,,
\nonumber \\[3mm]
&&\real{(\lambda_5 {\rm e}^{2i \xi})}  =
-\frac{1}{v^2 s_{2\beta}}\, \real(m_{12}^2{\rm e}^{i\xi})
-\frac{O_{a1}^2}{2 v^2}\, M_{H_1}^2
-\frac{O_{a2}^2}{2 v^2}\, M_{H_2}^2
-\frac{O_{a3}^2}{2 v^2}\, M_{H_3}^2
-\frac{1}{4t_\beta} \real{(\lambda_6 {\rm e}^{i \xi})}
-\frac{t_\beta}{4} \real{(\lambda_7 {\rm e}^{i \xi})} \,,
\nonumber \\[3mm]
&&\imag{(\lambda_6 {\rm e}^{i \xi})}  =
-\frac{O_{\phi_1 1} O_{a1} }{v^2 c_\beta}\, M_{H_1}^2
-\frac{O_{\phi_1 2} O_{a2} }{v^2 c_\beta}\, M_{H_2}^2
-\frac{O_{\phi_1 3} O_{a3} }{v^2 c_\beta}\, M_{H_3}^2
-t_\beta \imag{(\lambda_5 {\rm e}^{2i \xi})}\,,
\nonumber \\[3mm]
&&\imag{(\lambda_7 {\rm e}^{i \xi})}  =
-\frac{O_{\phi_2 1} O_{a1} }{v^2 s_\beta}\, M_{H_1}^2
-\frac{O_{\phi_2 2} O_{a2} }{v^2 s_\beta}\, M_{H_2}^2
-\frac{O_{\phi_2 3} O_{a3} }{v^2 s_\beta}\, M_{H_3}^2
-\frac{1}{t_\beta} \imag{(\lambda_5 {\rm e}^{2i \xi})}\,.
\end{eqnarray}
We find that our results are consistent with those
presented in Refs.~\cite{Grzadkowski:2014ada,Grzadkowski:2018ohf}.
We note that $\lambda_2$ can be obtained from $\lambda_1$ or vice versa
by exchanging $c_\beta \leftrightarrow s_\beta$,
$\phi_1 \leftrightarrow \phi_2$, and $\lambda_6 \leftrightarrow \lambda_7$.
The same observation could be applied for
$\imag{(\lambda_6 {\rm e}^{i \xi})}$ and
$\imag{(\lambda_7 {\rm e}^{i \xi})}$ which are vanishing
when each Higgs boson is purely CP-even or CP-odd state
and $\imag{(\lambda_5 {\rm e}^{2i \xi})}=0$.
About $\lambda_3$ and $\lambda_4$, we note that
$\lambda_3+\lambda_4$ is independent of
$\real(m_{12}^2{\rm e}^{i\xi})$ and $M_{H^\pm}^2$,
the neutral Higgs mass contributions to
$\lambda_3$ ($\lambda_4$)  are involved with
only the CP-even (CP-odd) components of each Higgs boson, and
the contributions from
$\real{(\lambda_6 {\rm e}^{i \xi})}$ and
$\real{(\lambda_7 {\rm e}^{i \xi})}$ are in common.
In passing, we check that Eq.~(\ref{eq:2hdmL4L5}) for the difference
between $\lambda_4/2$ and $\real{(\lambda_5 {\rm e}^{2i \xi})}$ is satisfied
by noting the relation
$O_{a1}^2 M_{H_1}^2 + O_{a2}^2 M_{H_2}^2 + O_{a3}^2 M_{H_3}^2=M_A^2$.

\setcounter{equation}{0}
\section{Cubic Higgs-boson self-couplings in 2HDMs}
\label{app:cubicself}

In this appendix, we apply the relations among the 2HDM
input parameters obtained in Appendix~\ref{app:2hdminput}
to derive cubic Higgs-boson self-couplings
when the lightest Higgs boson is purely CP even as assumed in
subsubsection~\ref{sec:heavyHdecaysInCPV2HDMs} by taking
the following $O$ matrix:
\begin{equation}
\label{eq:2hdmOmixAppendix}
O \ = \
\left( \begin{array}{ccc}
   -s_\alpha  &  c_\alpha c_\omega  & c_\alpha s_\omega   \\
  ~~c_\alpha  &  s_\alpha c_\omega  & s_\alpha s_\omega   \\
   ~0  &  -s_\omega   &    c_\omega   \\
  \end{array} \right)\,.
\end{equation}
In this case,
using Eq.~(\ref{eq:2hdmSelf}),
Eq.~(\ref{eq:2hdmTriSelf}) for
the cubic self--couplings of the Higgs weak eigenstates,
and Eq.~(\ref{eq:2hdmInputConversion}) for the conversion relations,
we find the couplings of the heavy neural Higgs bosons $H_{2,3}$
to a pair of the lightest Higgs bosons are given by
\begin{eqnarray}
g_{_{H_2\, H_1\, H_1\,}} &=&
-\frac{ c_\omega}{ s_{2\beta}^2} \bigg[
 4 s_{2\beta}c_{\beta-\alpha}
+6 c_{2\beta} s_{\beta-\alpha} c_{\beta-\alpha}^2
-6 s_{2\beta}c_{\beta-\alpha}^3\bigg]
\left( \frac{\real(m_{12}^2{\rm e}^{i\xi})}{v^2} \right)
\nonumber \\
&&
-\frac{c_\omega} {  s_{2\beta}}
\bigg[ s_{2 \beta} c_{\beta-\alpha}
+2 c_{2\beta} s_{\beta-\alpha}  c_{\beta-\alpha}^2
-2s_{2 \beta} c_{\beta-\alpha}^3\bigg]
\left[ \left( \frac{M_{H_1}^2} {v^2} \right)
+\frac{1}{2} \left( \frac{M_{H_2}^2} {v^2} \right) \right]
\nonumber \\
&&
-\frac{ s_\omega c_{\beta-\alpha}^2}{s_{2\beta}} \,
\imag{(\lambda_5 {\rm e}^{2i \xi})}
-\frac{3 c_\omega s_{\beta-\alpha} c_{\beta-\alpha}^2 }{ 4 s_{\beta}^2} \,
\real{(\lambda_6 {\rm e}^{i \xi})}
+\frac{3 c_\omega s_{\beta-\alpha} c_{\beta-\alpha}^2 }{ 4 c_{\beta}^2} \,
\real{(\lambda_7 {\rm e}^{i \xi})}\,,
\nonumber \\[2mm]
g_{_{H_3\, H_1\, H_1\,}} &=&
-\frac{ s_\omega}{ s_{2\beta}^2} \bigg[
 4 s_{2\beta} c_{\beta-\alpha}
+6 c_{2\beta} s_{\beta-\alpha} c_{\beta-\alpha}^2
-6 s_{2\beta}c_{\beta-\alpha}^3\bigg]
\left( \frac{\real(m_{12}^2{\rm e}^{i\xi})}{v^2} \right)
\nonumber \\
&&
-\frac{s_\omega} {  s_{2\beta}}
\bigg[ s_{2 \beta} c_{\beta-\alpha}
+2c_{2\beta} s_{\beta-\alpha} c_{\beta-\alpha}^2
-2s_{2 \beta} c_{\beta-\alpha}^3\bigg]
\left[ \left( \frac{M_{H_1}^2} {v^2} \right)
+\frac{1}{2} \left( \frac{M_{H_3}^2} {v^2} \right) \right]
\nonumber \\
&&
+\frac{ c_\omega c_{\beta-\alpha}^2}{s_{2\beta}} \,
\imag{(\lambda_5 {\rm e}^{2i \xi})}
-\frac{3 s_\omega s_{\beta-\alpha} c_{\beta-\alpha}^2 }{ 4 s_{\beta}^2} \,
\real{(\lambda_6 {\rm e}^{i \xi})}
+\frac{3 s_\omega s_{\beta-\alpha} c_{\beta-\alpha}^2 }{ 4 c_{\beta}^2} \,
\real{(\lambda_7 {\rm e}^{i \xi})}\,.
\end{eqnarray}
We note that
$g_{_{H_2\, H_1\, H_1\,}}/c_\omega = g_{_{H_3\, H_1\, H_1\,}}/s_\omega$ when
$\imag{(\lambda_5 {\rm e}^{2i \xi})}=0$ and $M_{H_2}=M_{H_3}$ are taken
and the contributions proportional to the input quartic couplings
are suppressed by the factor of $c_{\beta-\alpha}^2$.
We further note that the couplings $g_{_{H_2\, H_1\, H_1\,}}$
and $g_{_{H_3\, H_1\, H_1\,}}$ are vanishing in the limit
where $g_{_{H_2VV}}=g_{_{H_3VV}}=0$ or $c_{\beta-\alpha}=0$.
Otherwise, they are non-vanishing.
To be more specific,
as in subsubsection~\ref{sec:heavyHdecaysInCPV2HDMs}, we take
\begin{equation}
\delta_{2} =  \sqrt\epsilon_{0}\,\left(\frac{M_{H_1}}{M_{H_2}}\right)^2\,, \ \ \
\delta_{3} =  \sqrt\epsilon_{0}\,\left(\frac{M_{H_1}}{M_{H_3}}\right)^2\,,
\end{equation}
together with Eq.~(\ref{eq:2hdmOmixAppendix}).
These specific choices of $\delta_{2,3}$
fully fix all the elements of the $3\times 3$ orthogonal
mixing matrix $O$ in terms of the masses of the neutral Higgs bosons and
the parameter $\epsilon_0$ through
\footnote{Note we are taking $c_\omega>0$ and $s_\omega>0$.}
\begin{eqnarray}
s_\alpha &=& -\sqrt{1-\epsilon}\, c_\beta +\sqrt\epsilon\, s_\beta\,, \ \ \
c_\alpha =    \sqrt{1-\epsilon}\, s_\beta +\sqrt\epsilon\, c_\beta\,,
\nonumber \\[2mm]
c_\omega &=& \frac{M_{H_{3}}^2}{\sqrt{M_{H_{2}}^4+M_{H_{3}}^4}}\,, \ \ \
s_\omega  =  \frac{M_{H_{2}}^2}{\sqrt{M_{H_{2}}^4+M_{H_{3}}^4}}\,,
\end{eqnarray}
where
\begin{equation}
\epsilon=\epsilon_0\,\left(\frac{M_{H_1}^4}{M_{H_2}^4}+
\frac{M_{H_1}^4}{M_{H_3}^4} \right)\,.
\end{equation}
Incidentally, we note $s_{\beta-\alpha}=\sqrt{1-\epsilon}$ and
$c_{\beta-\alpha}=\sqrt{\epsilon}$.
Taking $\epsilon_0=0.05$, $M_{H_1}=125.5$ GeV, and $M_{H_2}\sim M_{H_3}$,
we find the couplings of the heavy neural Higgs bosons
$H_{2,3}$ to a pair of the lightest Higgs bosons are given by
\begin{eqnarray}
g_{_{H_3\, H_1\, H_1\,}} \simeq g_{_{H_2\, H_1\, H_1\,}}
\simeq -0.03-\frac{0.23}{s_{2\beta}}
\left(\frac{\real(m_{12}^2{\rm e}^{i\xi})}{M_{H_3}^2} \right)\,,
\end{eqnarray}
keeping the two most significant contributions:
the first term comes from
$-s_\omega c_{\beta-\alpha}\,\frac{{M_{H_3}^2}}{2{v^2}}$
and the second one from
$-4\frac{s_\omega}{s_{2\beta}} c_{\beta-\alpha}\,
\frac{\real(m_{12}^2{\rm e}^{i\xi})}{v^2}$.
We observe that it is easy to achieve
$|g_{_{H_3\, H_1\, H_1\,}}| = |g_{_{H_2\, H_1\, H_1\,}}|=0.1$
as taken in subsubsection~\ref{sec:heavyHdecaysInCPV2HDMs}
especially for
$\left|\real(m_{12}^2{\rm e}^{i\xi})\right|/s_{2\beta} \sim M_{H_3}^2$.
Incidentally, we find
\begin{eqnarray}
g_{_{H_1\, H_1\, H_1\,}} &=&
\frac{2}{s_\beta^2}(s_{2\beta} s_{\beta-\alpha} c_{\beta-\alpha}^2
+ c_{2\beta} c_{\beta-\alpha}^3 )
\left( \frac{\real(m_{12}^2{\rm e}^{i\xi})}{v^2} \right)
\nonumber \\
&&+
\frac{1}{2\, s_{2\beta}} \bigg[
  s_{2\beta} s_{\beta-\alpha} +
2 s_{2\beta} s_{\beta-\alpha} c_{\beta-\alpha}^2 +
2 c_{2\beta} c_{\beta - \alpha}^3 \bigg]
\left( \frac{M_{H_1}^2} {v^2} \right)
\nonumber \\
&&+
 \frac{ \, c_{\beta-\alpha}^3 } { 4 s_\beta^2 } \,
\real{(\lambda_6 {\rm e}^{i \xi})}
- \frac{ \, c_{\beta-\alpha}^3 } { 4 c_\beta^2 } \,
\real{(\lambda_7 {\rm e}^{i \xi})}\,,
\end{eqnarray}
which becomes the SM coupling of $M_{H_1}^2/2v^2$ when $c_{\beta-\alpha}=0$,
see Eq.~(\ref{eq:SMPotential}).

\medskip

Finally, we address the contributions from the charged-Higgs-boson loops
to the decay processes of neutral Higgs bosons into two photons
in the 2HDM and the MSSM which are mentioned in
subsubsection~\ref{sec:heavyHdecaysInCPV2HDMs} and
subsection~\ref{subsec:implications_beyond_the_standard_model}.
Using Eq.~(\ref{eq:2hdmSelf}), Eq.~(\ref{eq:2hdmTriSelfCH}),
and Eq.~(\ref{eq:2hdmInputConversion}),
we obtain the following couplings of neutral Higgs bosons to
a pair of charged Higgs bosons:
\begin{eqnarray}
\label{eq:appNHCHCH}
g_{_{H_1\, H^+\, H^-\,}} &=&
2 s_{\beta-\alpha} \left( \frac{M_{H^\pm}^2}{v^2} \right)
+\frac{4}{s_{2\beta}^2}
(s_{2\beta}s_{\beta-\alpha} + c_{2\beta}c_{\beta-\alpha})
\left( \frac{\real(m_{12}^2{\rm e}^{i\xi})}{v^2} \right)
\nonumber \\ &&
+\frac{1}{\, s_{2\beta}} \left(
s_{2\beta} s_{\beta-\alpha} + 2 c_{2\beta} c_{\beta - \alpha} \right)
\left( \frac{M_{H_1}^2} {v^2} \right)
+\frac{ \, c_{\beta-\alpha} } { 2 s_\beta^2 } \,
\real{(\lambda_6 {\rm e}^{i \xi})}
- \frac{ \, c_{\beta-\alpha} } { 2 c_\beta^2 } \,
\real{(\lambda_7 {\rm e}^{i \xi})}\,,
\nonumber \\ [2mm]
g_{_{H_2\, H^+\, H^-\,}} &=&
2 c_\omega c_{\beta-\alpha} \left( \frac{M_{H^\pm}^2}{v^2} \right)
-\frac{4 c_\omega}{s_{2\beta}^2}(c_{2\beta}s_{\beta-\alpha} - s_{2\beta}c_{\beta-\alpha})
\left( \frac{\real(m_{12}^2{\rm e}^{i\xi})}{v^2} \right)
\nonumber \\ &&
-\frac{c_\omega}{\, s_{2\beta}} \left(
2 c_{2\beta} s_{\beta-\alpha} -s_{2\beta} c_{\beta - \alpha} \right)
\left( \frac{M_{H_2}^2} {v^2} \right)
-\frac{ 2 s_\omega } { s_{2\beta} } \,
\imag{(\lambda_5 {\rm e}^{2 i \xi})}
\nonumber \\ &&
-\frac{ c_\omega s_{\beta-\alpha} } { 2 s_\beta^2 } \,
\real{(\lambda_6 {\rm e}^{i \xi})}
+\frac{ c_\omega s_{\beta-\alpha} } { 2 c_\beta^2 } \,
\real{(\lambda_7 {\rm e}^{i \xi})}  \,,
\nonumber \\ [2mm]
g_{_{H_3\, H^+\, H^-\,}} &=&
2 s_\omega c_{\beta-\alpha} \left( \frac{M_{H^\pm}^2}{v^2} \right)
-\frac{4 s_\omega}{s_{2\beta}^2}
(c_{2\beta}s_{\beta-\alpha} - s_{2\beta}c_{\beta-\alpha})
\left( \frac{\real(m_{12}^2{\rm e}^{i\xi})}{v^2} \right)
\nonumber \\ &&
-\frac{s_\omega}{\, s_{2\beta}} \left(
 2 c_{2\beta} s_{\beta-\alpha} - s_{2\beta} c_{\beta - \alpha} \right)
\left( \frac{M_{H_3}^2} {v^2} \right)
+\frac{ 2 c_\omega } { s_{2\beta} } \,
\imag{(\lambda_5 {\rm e}^{2 i \xi})}
\nonumber \\ &&
-\frac{ s_\omega s_{\beta-\alpha} } { 2 s_\beta^2 } \,
\real{(\lambda_6 {\rm e}^{i \xi})}
+\frac{ s_\omega s_{\beta-\alpha} } { 2 c_\beta^2 } \,
\real{(\lambda_7 {\rm e}^{i \xi})} \,.
\end{eqnarray}
In the 2HDM as well as in the MSSM,
the contributions from the charged-Higgs-boson loops to the neutral Higgs
boson decays into two photons enter through
the form factor~\footnote{See, Eq.~(\ref{eq:appDSDPaa}).}
\begin{equation}
\Delta S^\gamma_i (H^\pm)=  g_{_{H_iH^+H^-}}\frac{v^2}{2 M_{H^\pm}^2}
F_0\left(\frac{M_{H_i}^2}{4M_{H^\pm}^2}\right)\,.
\end{equation}
In the infinite charged-Higgs-boson mass limit, using $F_0(0)=1/3$, we find
\begin{equation}
\Delta S^\gamma_1 (H^\pm) = s_{\beta-\alpha}/3\,, \ \ \
\Delta S^\gamma_2 (H^\pm) = c_\omega c_{\beta-\alpha}/3\,, \ \ \
\Delta S^\gamma_3 (H^\pm) = s_\omega c_{\beta-\alpha}/3\,.
\end{equation}
We note the non-decoupling feature of $\Delta S^\gamma_1 (H^\pm)$
\cite{Djouadi:1996yq,Arhrib:2004ak,Bhattacharyya:2013rya}
which does not vanish even when all the heavier Higgs bosons including the
charged one are decoupled.
On the other hand, $\Delta S^\gamma_{2,3} (H^\pm)$
are vanishing when the heavy neutral Higgs bosons are decoupled
or when $s_{\beta-\alpha}\to 1$ and $c_{\beta-\alpha}\to 0$.

\setcounter{equation}{0}
\section{Packages for electroweak corrections}
\label{app:bsmtools}
In this appendix, we introduce various  numerical packages for calculating
precise SM and full BSM-dependent ELW corrections.

\medskip

\begin{table}
\caption{
Two packages of {\texttt{HDECAY}} and {\texttt{PROPHECY4F}}
for the SM ELW corrections.
In each package line, decay modes are checked when
the SM ELW corrections to them could be calculated by use of it.}
\label{tab:SM-Tools}
\begin{center}
\begin{tabular}{|c|c|c|c|c|c|}
\hline
Package$\,\backslash$Decay Mode \/\  & $H\to f\bar{f}$  & $H\to WW/ZZ \to 4f$  & $H \to
gg$
& $H\to \gamma\gamma$ & $H\to Z \gamma$ \\
\hline
\hline
\multirow{2}{*}{\texttt{HDECAY}}
 & \yes &  \yes  & \yes   &  \yes &     \\ 
 & NLO & Approx. NLO  &  NLO
& NLO  &   \\ \hline  
\multirow{2}{*}{\texttt{PROPHECY4F}}
 & & \yes & & &  \\
 & & NLO & & &  \\ \hline
\end{tabular}
\end{center}
\end{table}
To begin with, in Table \ref{tab:SM-Tools},
we list the two packages of
\texttt{HDECAY}~\cite{Djouadi:1997yw,Djouadi:2018xqq} and
\texttt{PROPHECY4F}\,\cite{Bredenstein:2006rh,Bredenstein:2006nk,Bredenstein:2006ha,
Altenkamp:2017ldc,Altenkamp:2017kxk,Altenkamp:2018bcs,Denner:2018opp}
for precise SM ELW corrections.
We note that
\texttt{HDECAY} is commonly used except for the decay mode
$H\to V V \to 4f$ for which \texttt{PROPHECY4F}
provides the complete ${\cal O}(\alpha)$ electroweak
corrections to the Higgs decays into four fermions through intermediate
$W$ and $Z$ bosons.
And,  as far as we have configured, there exist no
reliable theoretical calculations and/or
numerical packages
implementing the ELW corrections to the radiative $H\to Z\gamma$ decay.
Incidentally, for the other radiative decays of $H\to gg$
and $H\to\gamma\gamma$, we refer to
Refs.~\cite{Actis:2008ts,Actis:2008ug,Actis:2008uh,Passarino:2007fp}
for theoretical calculations.


\begin{table}
\caption{
Packages for BSM-dependent ELW corrections.
In each package line, BSM models are checked when
ELW corrections in those models are implemented in the package.
The last column is to note that \texttt{PROPHECY4F} is exclusively
for the $H\to VV \to 4f$ decays and, for other packages, to
indicate the level of precision at which the ELW corrections
are calculated.
}
\label{tab:BSM-Tools}
\begin{center}
\setlength{\tabcolsep}{10pt} 
\renewcommand{\arraystretch}{1.8} 
\begin{tabular}{|c|c|c|c|c|c|}
\hline
Package$\,\backslash$BSM Model  & cxSM  & 2HDM & MSSM & NMSSM & Remark \\
\hline
\hline
\texttt{PROPHECY4F}
 &  \yes  &  \yes  &   &     &  $H \to WW/ZZ \to 4f$     \\ \hline \hline
\texttt{2HDECAY}
  &   &  \yes  &   &    & Full one-loop   \\ \hline
\texttt{H-COUP}
 &  &  \yes  &   &     & NLO  \\ \hline \hline
\texttt{HFOLD}
   &    &   &  \yes &   & Full one-loop    \\ \hline
\texttt{FeynHiggs}
  &    &   &  \yes &   & Full one-loop \\ \hline
\texttt{NMSSMCALCEW}
    &     &    &   &  \yes   & Full one-loop    \\ \hline
\end{tabular}
\end{center}
\end{table}
In Table \ref{tab:BSM-Tools}, we list various packages for
ELW corrections in the BSM models considered in this review.
Additionally,
as an example of BSM models containing Higgs sectors beyond the
cxSM and 2HDM/MSSM,  the
Next-to-MSSM (NMSSM)~\cite{Maniatis:2009re,Ellwanger:2009dp,Cheung:2010ba}
is included.
First of all, exclusively for the $H\to VV \to 4f$ processes,
\texttt{PROPHECY4F} can be used for the ELW corrections in the cxSM and 2HDM.
In passing, we note that the full NLO corrections in
the real singlet extension of the SM  (rxSM)
\cite{McDonald:1993ex,Burgess:2000yq,OConnell:2006rsp,
BahatTreidel:2006kx,Barger:2007im,He:2007tt,Davoudiasl:2004be}
and the SM4G (SM with the fourth generation) are also implemented
in \texttt{PROPHECY4F}.

\medskip

For the calculations of the ELW corrections in 2HDM,
the packages \texttt{2HDECAY}~\cite{Krause:2018wmo}  and
\texttt{H-COUP}~\cite{Kanemura:2017gbi,Kanemura:2019slf} can be used.
We comment that the package \texttt{H-COUP} can be used
for the NLO ELW corrections in the rxSM
and inert doublet models as well.
For the MSSM ELW corrections at the full one-loop level precision,
one can use the packages \texttt{HFOLD}~\cite{Frisch:2010gw} and
\texttt{FeynHiggs}~\cite{Bahl:2018qog}.
On the other hand, in the NMSSM,
the package \texttt{NMSSMCALCEW}~\cite{Baglio:2019nlc} provides
the full one-loop ELW corrections.

\medskip

Before closing this appendix, we introduce two useful webpages
containing significantly extensive lists of numerical
packages or tools for various purposes from  model building to event
generation, etc:\\[2mm]
1) Supersymmetry Les Houches Accord : http://skands.physics.monash.edu/slha/~, \\[2mm]
2) HEPForge Projects : https://www.hepforge.org/projects~. \\

\end{appendix}


\end{document}